\documentclass[twocolumn,times,tighten]{aastex631}

\hypersetup{linkcolor=blue,citecolor=blue,filecolor=cyan,urlcolor=blue}

\def\smallplot#1{\centering \leavevmode
\includegraphics[width=89.5mm]{#1} }

\def\largeplot#1{\centering \leavevmode
\includegraphics[width=150mm]{#1} }

\def\smallplottwo#1#2{\centering \leavevmode
\includegraphics[width=89.5mm]{#1}
 \hfil \includegraphics[width=89.5mm]{#2} }

\def\smallplotfour#1#2#3#4{\centering \leavevmode
\includegraphics[width=89.5mm]{#1} \includegraphics[width=89.5mm]{#2}
\hfil \includegraphics[width=89.5mm]{#3} \includegraphics[width=89.5mm]{#4}}

\def\smallplotsix#1#2#3#4#5#6{\centering \leavevmode
\includegraphics[width=89.5mm]{#1} \includegraphics[width=89.5mm]{#2}
\hfil \includegraphics[width=89.5mm]{#3} \includegraphics[width=89.5mm]{#4}
\hfil \includegraphics[width=89.5mm]{#5} \includegraphics[width=89.5mm]{#6}}

\def\xsmallplotfour#1#2#3#4{\centering \leavevmode
\includegraphics[width=80mm]{#1} \includegraphics[width=80mm]{#2}
\hfil \includegraphics[width=80mm]{#3} \includegraphics[width=80mm]{#4}}

\def\largeplottwo#1#2{\centering \leavevmode
\includegraphics[width=150mm]{#1}
 \hfil \includegraphics[width=150mm]{#2} }

\graphicspath{{./}{figures/}}

\received{June 1, 2021}
\revised{June 30, 2021}
\accepted{July 5, 2021}
\submitjournal{ApJS}

\shorttitle{A new catalog of AGB stars in our Galaxy}
\shortauthors{Suh} 

\begin{document}

\title{A New Catalog of Asymptotic Giant Branch Stars in Our Galaxy}

\correspondingauthor{Kyung-Won Suh}
\email{kwsuh@chungbuk.ac.kr}

\author[0000-0001-9104-9763]{Kyung-Won Suh}
\affiliation{Department of Astronomy and Space Science, Chungbuk National University, Cheongju-City, 28644, Republic of Korea}

\begin{abstract}
We present a new catalog of 11,209 O-rich AGB stars and 7172 C-rich AGB stars
in our Galaxy identifying more AGB stars in the bulge component and
considering more visual carbon stars. For each object, we cross-identify the
IRAS, AKARI, MSX, WISE, 2MASS, and AAVSO counterparts. We present the new
catalog in two parts: one is based on the IRAS PSC for brighter or more
isolated objects; the other one is based on the ALLWISE source catalog for
less bright objects or objects in crowded regions. We present various
infrared two-color diagrams (2CDs) for the sample stars. We find that the
theoretical dust shell models can roughly explain the observations of AGB
stars on the various IR 2CDs. We investigate IR properties of SiO and OH
maser emission sources in the catalog. For Mira variables in the sample
stars, we find that the IR colors get redder for longer pulsation periods. We
also study infrared variability of the sample stars using the WISE
photometric data in the last 12 yr: the ALLWISE multiepoch data and the
Near-Earth Object WISE Reactivation (NEOWISE-R) 2021 data release. We
generate light curves using the WISE data at W1 and W2 bands and compute the
Lomb-Scargle periodograms for all of the sample stars. From the WISE light
curves, we have found useful variation parameters for 3710 objects in the
catalog, for which periods were either known or unknown in previous works.
\end{abstract}



\section{Introduction} \label{sec:intro}

Asymptotic giant branch (AGB) stars are believed to be low-to-intermediate-mass
stars (0.5 - 10 $M_{\odot}$) in the last evolutionary phase evolving into
post-AGB stars and planetary nebulae (\citealt{siess2006};
\citealt{hofner2018}). Almost all AGB stars are long-period variables (LPVs)
with large amplitude pulsations. As the early-phase AGB stars evolve into the
thermally pulsing AGB (TP-AGB) phase, they produce dust grains more effectively
and show higher mass-loss rates (e.g., \citealt{jimenez2015}).

Based on chemistry of the photosphere and/or the circumstellar dust envelope,
AGB stars are classified as O-rich AGB (OAGB) or C-rich AGB (CAGB). The
spectral energy distributions (SEDs) of OAGB stars show 10 and 18 $\mu$m
features due to amorphous silicate dust. Low mass-loss rate OAGB (LMOA) stars
with thin dust envelopes show the emission features and high mass-loss rate
OAGB (HMOA) stars with thick dust envelopes show the absorption features at the
same wavelengths (e.g., \citealt{suh1999}). To reproduce the detailed SEDs of
LMOA stars, amorphous alumina (Al$_2$O$_3$; \citealt{suh2016}) and Fe-Mg oxides
\citep{thposch2002} dust grains are also necessary. For CAGB stars, SiC and MgS
grains as well as featureless amorphous carbon (AMC) dust can reproduce the
SEDs (e.g., \citealt{suh2000}; \citealt{hony2002}).

CAGB stars are generally believed to be the evolutionary successors of OAGB
stars. When M-type OAGB stars of intermediate mass range (1.55 $M_{\odot}$
$\leq$ M $<$ 4 $M_{\odot}$: for solar metallicity; \citealt{groenewegen1995})
go through carbon dredge-up processes and so the C/O ratio is larger than one,
O-rich dust formation ceases and the stars become visual carbon stars in the
AGB phase. The visual carbon stars evolve into infrared carbon stars with thick
C-rich dust envelopes and high mass-loss rates (e.g., \citealt{suh2000}). It is
generally thought that S stars (more specifically, intrinsic S stars) are in
the intermediate phase between OAGB and CAGB stars (e.g., \citealt{sk2011}).
Though this M-S-C evolutionary sequence could not be right for all AGB stars
(e.g., \citealt{ck1990}), there is much evidence that supports this idea (e.g.,
\citealt{suh2020}).

The Infrared Astronomical Satellite (IRAS), Infrared Space Observatory (ISO),
Midcourse Space Experiment (MSX), AKARI, Two-Micron All-Sky Survey (2MASS), and
Wide-field Infrared Survey Explorer (WISE) have provided various IR
observational data, which have been useful to identify and study new AGB stars.
Using various IR two-color diagrams (2CDs), we can study properties of the
central stars and dust envelopes for a large sample of AGB stars (e.g.,
\citealt{sevenster2002}; \citealt{sk2011}). Thanks to the optical gravitational
lensing experiment (OGLE) projects (\citealt{sus13a}), a larger number of LPVs
in the Galactic bulge are identified and studied.

In 2009, WISE \citep{wright2010} started mapping the sky. The ALLWISE
multiepoch photometry table obtained in 2009-2010 provided the photometric data
at four bands (3.4, 4.6, 12, and 22 $\mu$m; W1, W2, W3, and W4). And the
Near-Earth Object WISE Reactivation (NEOWISE-R) mission (\citealt{mainzer2014})
has been providing photometric data at W1 and W2 bands for last seven yr (14
epochs; 2021 data release), two in every year between 2014 and 2020. The
ALLWISE multiepoch photometry table and the NEOWISE-R data may allow
characterization of the periodic variations of IR emission from AGB stars at
the W1 and W2 bands in last 12 yr.

In this paper, we present a new catalog of AGB stars in our Galaxy, identifying
more AGB stars in the bulge component and considering more visual carbon stars.
Section~\ref{sec:sample} presents the new catalog in two parts: one is based on
the IRAS PSC and the other is based on the ALLWISE source catalog. For each
object, we cross-identify the IRAS, AKARI, MSX, WISE, and 2MASS counterparts.
Section~\ref{sec:models} describes the theoretical radiative transfer models
for dust shells around AGB stars. Section~\ref{sec:2cds} presents various
infrared 2CDs using the IR photometric data compared with theoretical models.
In Section~\ref{sec:cnumber}, we compare the number distribution of observed IR
colors for different classes (or subgroupes) of AGB stars to find differences
in the IR properties. Section~\ref{sec:spacial} describes spacial distributions
of the AGB stars in our Galaxy. Using the new sample stars, we present infrared
properties of known pulsating variables in Section~\ref{sec:pul}. In
Section~\ref{sec:neo}, we study infrared variability of the sample stars using
the WISE photometric data at the W1 and W2 bands in the last 12 yr: the ALLWISE
multiepoch data that were acquired between 2009 and 2010 and the NEOWISE-R 2021
data release that was acquired from 2013 until the end of 2020.
Section~\ref{sec:catalog} presents the catalog data. Finally,
Section~\ref{sec:sum} summarizes results of the paper.

\begin{table*}
\centering
\caption{Sample of AGB stars based on the IRAS PSC (AGB-IRAS) \label{tab:tab1}}
\begin{tabular}{llllllll}
\hline \hline
Class     &Subgroup & Reference & Number & Selected & Duplicate & Added-Excluded & Remaining$^1$ \\
\hline
OAGB-IRAS & OI-SH &\citet{sh2017} & 3828  & 3828 & 0  & 3$^2$-5$^3$ & 3826 (1560)  \\
OAGB-IRAS & OI-UR &\citet{urago2020} & 42$^4$  & 37$^5$ & 0  & 0 & 37 (9)\\
OAGB-IRAS & OI-JB &\citet{jimenez2015} & 37  & 37 & 32$^6$  & 0 & 5 (0)\\
OAGB-IRAS & OI-JB &\citet{blommaert2018} & 8  & 8 & 7$^6$  & 0 & 1 (0) \\
OAGB-IRAS & OI-ME &\citet{messineo2018} &  571  & 166$^7$  & 17$^6$  & -22$^8$ & 127 (6)\\
OAGB-IRAS & OI-ST &\citet{stroh2019} &  1427  & 735$^7$  & 59$^9$+3$^{10}$  & 0 & 673 (32)\\
OAGB-IRAS & OI-OG &\citet{sus13a} &  6039$^{11}$ & 1116$^{12}$ & 59$^6$ & 0 & 1057 (1057) \\
OAGB-IRAS & OI-WU &\citet{wu2018} & 44   &  44  & 25$^6$  & 0 & 19 (19)\\
OAGB-IRAS & OI-AM$^{13}$ &This work (IRAS PSC)  & 894   & 894  & 731$^{14}$  & 0 & 163 (163) \\
OAGB-IRAS & OI-all&- & - & - & -  & -  & 5908 (2846) \\
\hline
CAGB-IRAS & CI-SH &\citet{sh2017} &  1168  & 1168 & 0  & 1$^{15}$-3$^2$-1$^{16}$ & 1165 (340) \\
CAGB-IRAS & CI-UR &\citet{urago2020} & 42$^4$  & 5$^5$ & 0  & 0 & 5 (0)\\
CAGB-IRAS & CI-GC &\citet{alksnis2001} & 6891  & 3379$^{12}$ & 925$^{17}$+15$^{18}$+20$^{19}$  & -2$^{20}$ & 2417 (99)\\
CAGB-IRAS & CI-OG &\citet{sus13a} & 168$^{11}$ & 9$^{12}$ & 0  & 0 & 9 (9) \\
CAGB-IRAS & CI-all&- & -  & - & -  & - & 3596 (447) \\
\hline
S stars & SI   &\citet{sk2011} &  329 & 329  & 0  & 1$^{21}$  & 330 \\
silicate carbon stars  & SCI  &\citet{ks2014} &  29  & 29  & 0  & 0  & 29 \\
\hline
\end{tabular}
\begin{flushleft}
$^1$number in parentheses denotes number of Miras in AAVSO (version 2021-04-19; \citealt{watson2021}).
$^2$three objects in CI-SH are SiO maser sources (OAGB stars) without clear CAGB evidences.
$^3$YSOs or RSGs (see Tables~\ref{tab:tab3}).
$^4$the number of the original IRAS PSC sources (108) minus duplicate OI-SH sources (66)
$^5$color-selected O-AGB or C-AGB stars (see Section~\ref{sec:iras}).
$^6$in OI-SH.
$^7$SiO maser sources with positive IRAS PSC counterparts.
$^8$RSGs.
$^9$58 objects in OI-SH and one object in OI-JB.
$^{10}$IRAS 17105-3746 is in CI-SH (IRAS LRS type C; SiO maser source), IRAS 17001-3651 is an S star (SI),
IRAS 15575-5238 is a silicate carbon star (SCI).
$^{11}$color-selected OAGB or CAGB stars from the OGLE3 sample of Miras in the Galactic bulge (see Section~\ref{sec:galb}).
One exception is IRAS 18100-2808, which is an OAGB star (IRAS LRS type E) with the CAGB color.
$^{12}$sources with positive IRAS PSC counterparts.
$^{13}$AAVSO Miras with IRAS LRS type E (see Section~\ref{sec:iras}).
$^{14}$694 in OI-SH, 7 in OI-OG, 6 in OI-ST, 1 in OI-WU, 19 in SI, and 4 in SCI.
$^{15}$IRAS 18027-1316 is a typical visual carbon star.
$^{16}$IRAS 06176-1036 is a planetary nebula (Red Rectangle).
$^{17}$in CI-SH
$^{18}$OAGB stars (in OI-SH).
$^{19}$two S stars (SI) and 18 silicate carbon stars (SCI).
$^{20}$duplicate objects.
$^{21}$IRAS 19354+5005 is an intrinsic S star.
\end{flushleft}
\end{table*}

\begin{table*}
\centering
\caption{Sample of AGB stars based on the ALLWISE source catalog (AGB-WISE)\label{tab:tab2}}
\begin{tabular}{llllllll}
\hline \hline
Class     &Subgroup & Reference & Number & Selected & Duplicate & Excluded & Remaining$^1$ \\
\hline
OAGB-WISE & OW-ME & \citet{messineo2018} & 571  & 196$^2$ & 0  & 21$^3$+18$^4$ & 157 (6) \\
OAGB-WISE & OW-ST & \citet{stroh2019} & 1427  & 285$^2$ & 0  & 54$^4$ & 231 (5) \\
OAGB-WISE & OW-OG & \citet{sus13a} & 6039$^5$ & 4923$^6$ & 3$^7$ & 7$^8$ & 4913 (4910$^9$) \\
OAGB-WISE & OW-all & - & - & -  & - &  -   & 5301 (4921)\\
\hline
CAGB-WISE & CW-GC & \citet{alksnis2001} & 6891  & 3512$^6$ & 0  & 95$^4$ & 3417 (43)\\
CAGB-WISE & CW-OG & \citet{sus13a} & 168$^5$ & 159$^6$ & 0 &  0 & 159 (157$^{10}$)\\
CAGB-WISE & CW-all & - & - & -  & - & - & 3576 (193)\\
\hline
\end{tabular}
\begin{flushleft}
$^1$number in parentheses denotes number of Miras in AAVSO.
$^2$SiO maser sources without positive IRAS PSC counterparts.
$^3$RSGs.
$^4$no WISE counterparts.
$^5$color-selected OAGB or CAGB stars from the OGLE3 sample of Miras in the Galactic bulge (see text).
$^6$sources without positive IRAS PSC identification.
$^7$in OW-ST.
$^8$duplicate objects in cross-matching ALLWISE sources.
$^9$For OGLE3 Mira objects, two objects are classified into SRA and one into SR in AAVSO.
$^{10}$For OGLE3 Mira objects, one object is classified into SRA and one into RCB: in AAVSO.
\end{flushleft}
\end{table*}

\begin{table}
\caption{Five Objects excluded from the list of OAGB stars in \citet{sh2017} (OI-SH)\label{tab:tab3}}
\centering
\begin{tabular}{llll}
\hline \hline
IRAS PSC  & Other name &Remark &Reference 	\\
\hline
05380-0728 & HBC 494 & YSO & \citet{cieza2018}\\
07209-2540 & VY CMa & RHG & \citet{humphreys2021}\\
16547-4247 & - & YSO & \citet{zapata2019} \\
18272+0114 & OO Ser & YSO & \citet{hodapp2012}   \\
19312+1950 & - & YSO or RSG   & \citet{cordiner2016}       \\
\hline
\end{tabular}
\end{table}

\begin{table}
\scriptsize
\caption{IR bands and zero magnitude flux values\label{tab:tab4}}
\centering
\begin{tabular}{lllll}
\hline \hline
Band &$\lambda_{ref}$ ($\mu$m)	&ZMF (Jy) &Telescope &Reference 	\\
\hline
K[2.2]  &2.159	&	666.7	&	2MASS	& \citet{cohen2003}\\
W1[3.4]	&3.35	&	306.682	&	WISE    & \citet{jarrett2011}	\\
W2[4.6]	&4.60	&	170.663	&	WISE    & \citet{jarrett2011}	\\
MA[8.3]  &8.28	&	58.49	&	MSX    & \citet{egan2003}	\\
AK[9]   &9	    &	56.262	&	AKARI    & \citet{murakami2007}	\\
IR[12]	&12     &   28.3 	&	IRAS     & \citet{beichman1988}	\\
W3[12]$^1$	&12 (11.56)	& 28.3 (29.045)	&	WISE & \citet{jarrett2011}\\
MC[12.1]  &12.13	&	26.51	&	MSX    &  \citet{egan2003}       	\\
MD[14.7]  &14.65	&	18.29	&	MSX    & \citet{egan2003}	\\
AK[18]   &18	&	12.001	&	AKARI    & \citet{murakami2007}	\\
ME[21.3]  &21.34	&	8.8  	&	MSX    & \citet{egan2003}	\\
W4[22]	&22.08	&	8.284	&	WISE    & \citet{jarrett2011}	\\
IR[25]	&25 & 6.73 	&	IRAS      & \citet{beichman1988}	\\
IR[60]	&60 & 1.19 	&	IRAS      & \citet{beichman1988}	\\
\hline
\end{tabular}
\begin{flushleft}
$^1$For W3[12], we use a new reference wavelength and zero magnitude flux for theoretical models
(original values are given in parentheses; see section 4.2 in ~\citealt{suh2020}).
\end{flushleft}
\end{table}

\section{Sample Stars\label{sec:sample}}

A catalog of AGB stars for 3003 OAGB, 1168 CAGB, 329 S-type, and 35 silicate
carbon stars in our Galaxy was presented by \citet{sk2011} based on the IRAS
PSC. \citet{ks2012} presented a revised list of 3373 OAGB stars considering new
SiO maser sources. \citet{ks2014} presented a revised sample of 29 silicate
carbon stars. \citet{sh2017} presented the revised catalog of 3828 OAGB and
1168 CAGB stars (version 2017; \url{http://web.chungbuk.ac.kr/~kwsuh/agb.htm}).

In this work, we present a new catalog of 11,209 OAGB stars and 7172 CAGB stars
in our Galaxy, identifying more AGB stars in the bulge component and
considering more visual carbon stars (see Tables~\ref{tab:tab1} and
\ref{tab:tab2}).

Because IRAS has a large beam size and limited sensitivity, it is not possible
to find appropriate counterparts for a major portion of newly identified AGB
stars, most of which are less bright or smaller objects in crowded regions.
Therefore, we present the new revised catalog in two parts: one is based on the
IRAS PSC for brighter or more isolated objects; the other is based on the
ALLWISE source catalog for less bright objects or objects in crowded regions,
for which the IRAS could not observe properly.

Table~\ref{tab:tab1} lists the class name, subgroup name, original reference,
and numbers of selected objects for AGB-IRAS objects. Table~\ref{tab:tab2}
lists the class name, subgroup name, original reference, and numbers of
selected objects for AGB-WISE objects. While IRAS PSC data are available only
for the 5908 OAGB-IRAS and 3596 CAGB-IRAS objects, the ALLWISE data are
available for most of the AGB-IRAS objects as well as 5301 OAGB-WISE and 3576
CAGB-WISE objects.

Table~\ref{tab:tab1} lists the objects whose original references are based on
the IRAS PSC (subgroup names: SH, UR, JB, WU, and AM), all of which are
compiled into the AGB-IRAS catalog (see Sections \ref{sec:iras} and
\ref{sec:newoagb}). Table~\ref{tab:tab3} lists five objects excluded from the
list of OAGB Stars (OI-SH) in \citet{sh2017}.

Tables \ref{tab:tab1} also lists a major portion of the objects whose original
references are not based on the IRAS PSC (subgroup names: ME, ST, OG, and GC),
which can be either AGB-IRAS or AGB-WISE objects depending on the existence of
the positive IRAS PSC counterpart (see Section~\ref{sec:agb-c}). The objects
that can be positively identified with the IRAS PSC sources are compiled into
the AGB-IRAS catalog (Table \ref{tab:tab1}). All the other objects with ALLWISE
counterparts, which cannot be positively identified with the IRAS PSC sources,
are compiled into the AGB-WISE catalog (Table \ref{tab:tab2}).

The sample of 18,381 Galactic AGB stars is composed of 8407 Mira variables (see
Tables~\ref{tab:tab1} and \ref{tab:tab2}) according to the American Association
of Variable Star Observers (AAVSO) international variable star index (VSX;
version 2021 April 19; \citealt{watson2021}). We have also considered the
General Catalogue of Variable Stars (GCVS version 5.1; \citealt{samus2017}). In
this work, we use the AAVSO which includes all of the LPVs in the GCVS list as
well as new lists from other observations.

\begin{figure*}
\centering
\smallplottwo{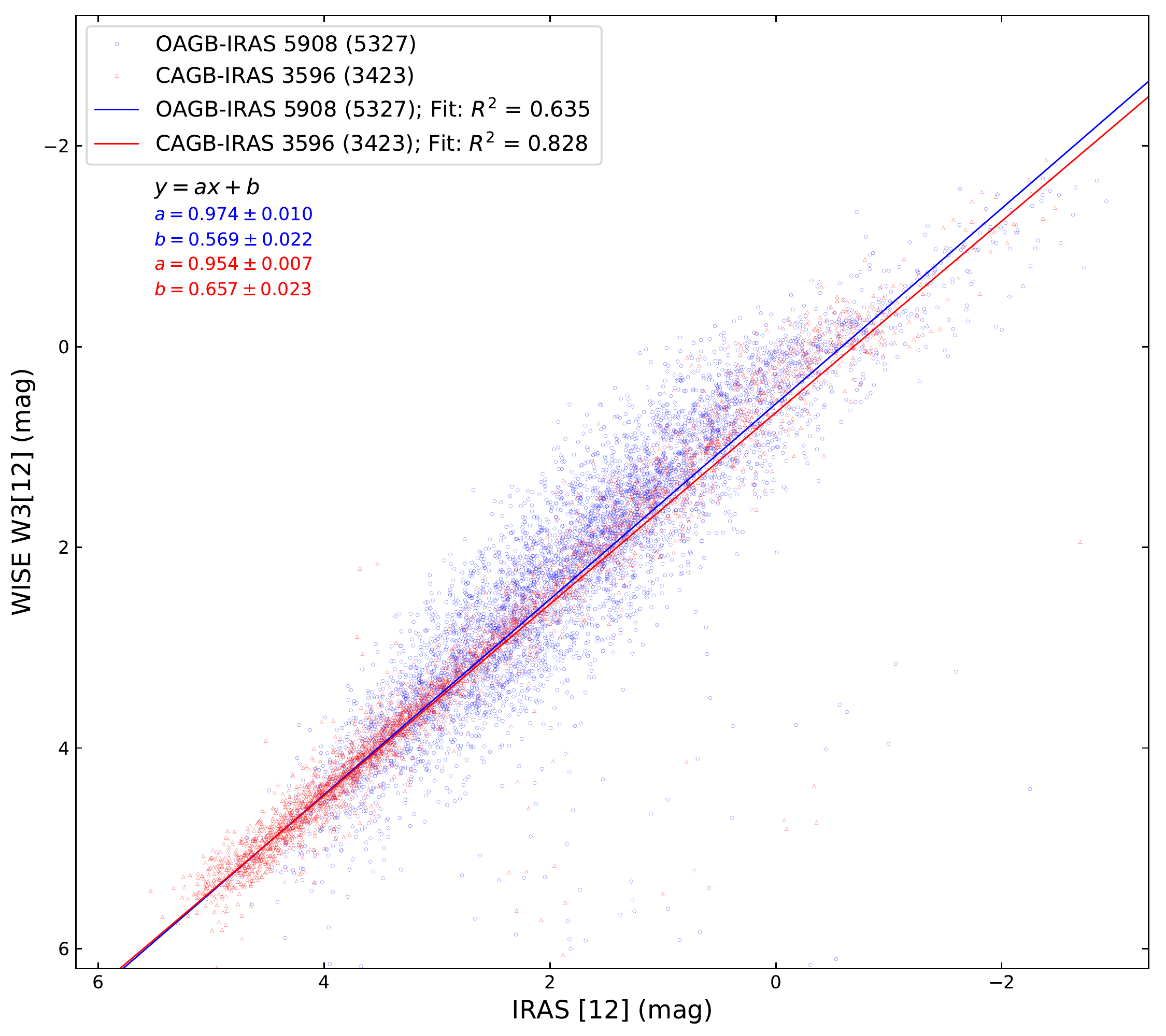}{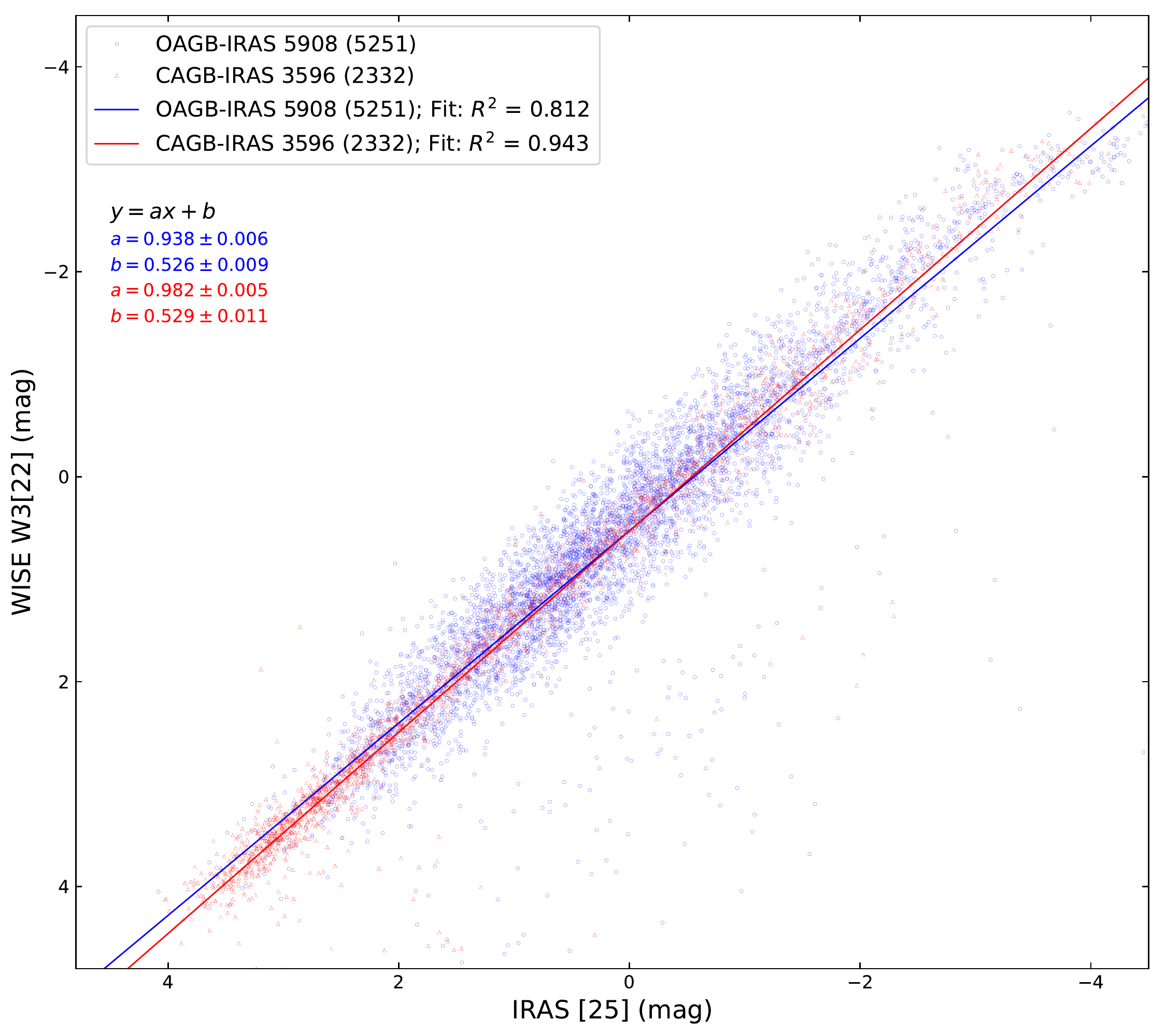}\caption{Comparison of IRAS and WISE fluxes for AGB-IRAS objects (see Table~\ref{tab:tab1}).
The number in parentheses denotes the number of the plotted objects with good-quality observed data.
See Section~\ref{sec:agb-c}.}
\label{f1}
\end{figure*}

\begin{figure*}
\centering
\largeplot{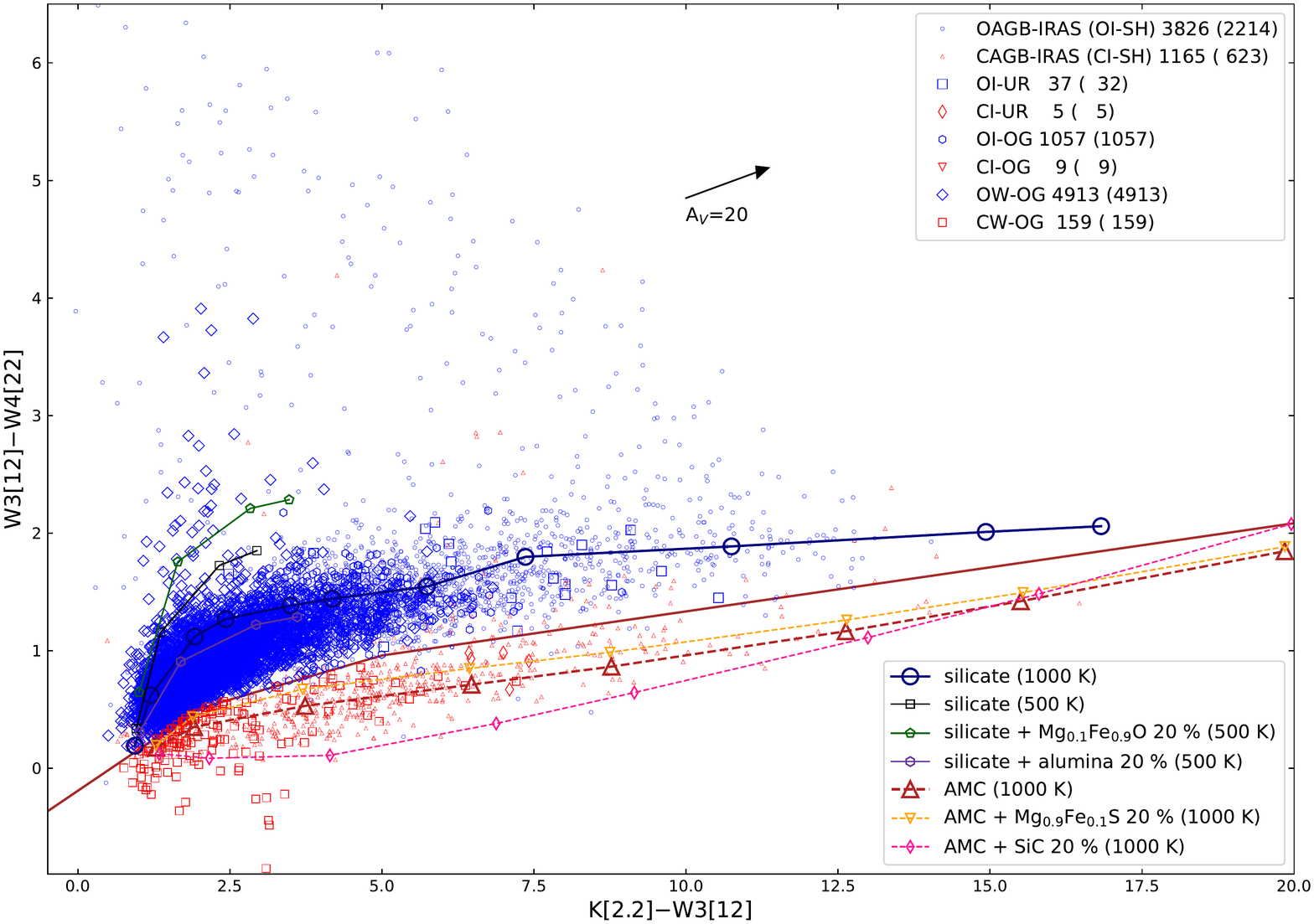}\caption{A WISE-2MASS 2CD for color-selected AGB-IRAS (OI-UR, CI-UR, OI-OG, and CI-OG) and AGB-WISE (OW-OG and CW-OG) stars
(see Tables~\ref{tab:tab1} and \ref{tab:tab2}) compared with theoretical models (see Section~\ref{sec:models}).
The brown line roughly distinguishes between OAGB and CAGB stars.
For OAGB models (silicate $T_c$ = 1000 K): $\tau_{10}$ = 0.001, 0.01, 0.05, 0.1, 0.5, 1, 3, 7, 15, 30, and 40 from left to right.
For CAGB models (AMC $T_c$ = 1000 K): $\tau_{10}$ = 0.001, 0.01, 0.1, 0.5, 1, 2, 3, and 5 from left to right.
For each subgroup, the number of objects is shown.
The number in parentheses denotes the number of the plotted objects on the 2CD with good-quality observed colors.}
\label{f2}
\end{figure*}

\begin{figure*}
\centering
\largeplot{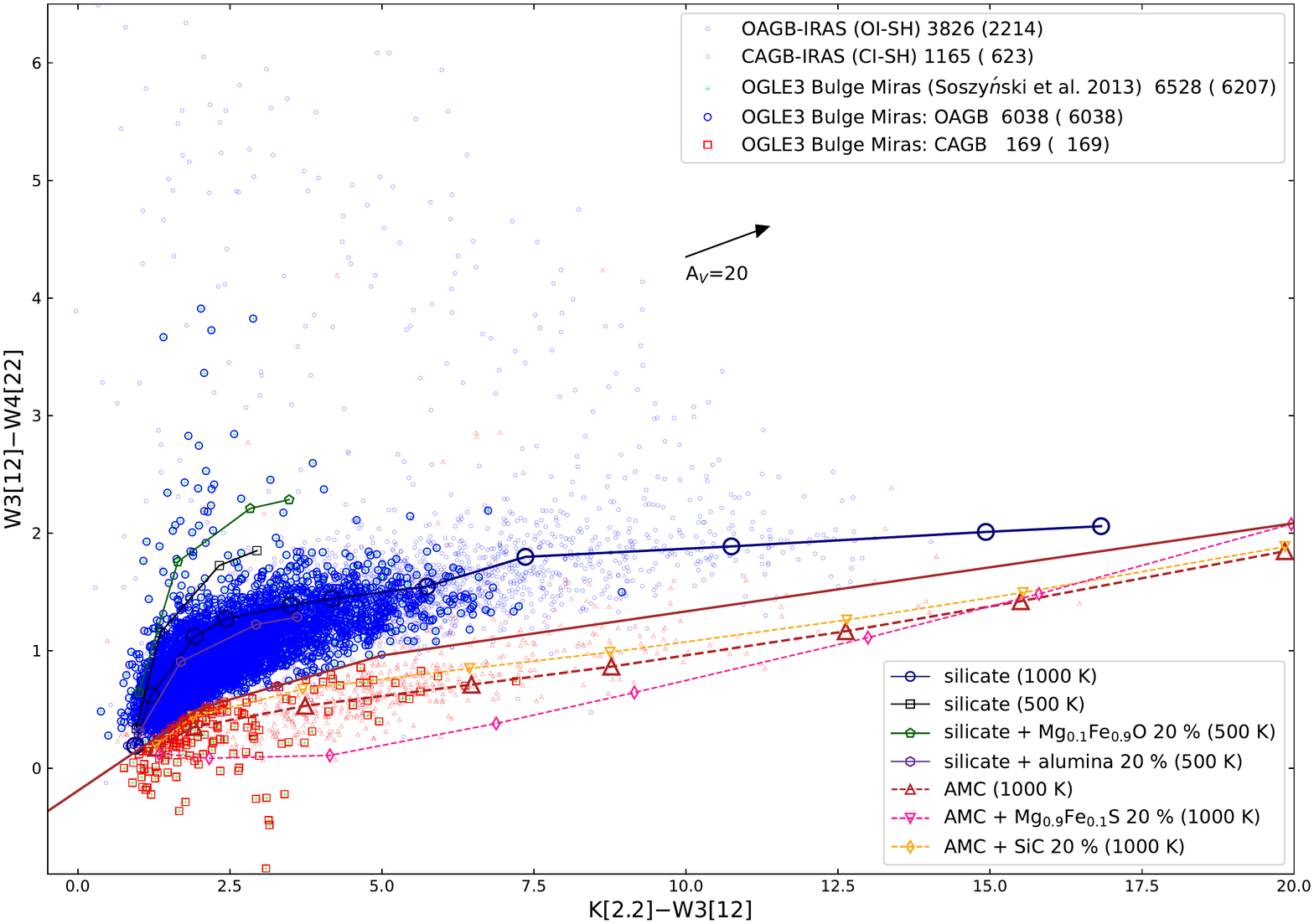}\caption{A WISE-2MASS 2CD for color-selected OGLE3 Miras (see Section~\ref{sec:galb}) compared with theoretical models (see Section~\ref{sec:models}).
The brown line roughly distinguishes between OAGB and CAGB stars.
For OAGB models (silicate $T_c$ = 1000 K): $\tau_{10}$ = 0.001, 0.01, 0.05, 0.1, 0.5, 1, 3, 7, 15, 30, and 40 from left to right.
For CAGB models (AMC $T_c$ = 1000 K): $\tau_{10}$ = 0.001, 0.01, 0.1, 0.5, 1, 2, 3, and 5 from left to right.
For each subgroup, the number of objects is shown.
The number in parentheses denotes the number of the plotted objects on the 2CD with good-quality observed colors.
Note that the 321 objects without good-quality observed colors are not plotted and they are not compiled into the new catalog.}
\label{f3}
\end{figure*}

\subsection{Infrared photometric data for the sample stars\label{sec:photdata}}

IRAS \citep{beichman1988} conducted an all-sky survey and the point source
catalog (PSC) provided  photometric data at four bands (12, 25, 60, and 100
$\mu$m). MSX (\citealt{egan2003}) surveyed the Galactic plane with higher
sensitivity and spatial resolution at four MIR bands (8.28, 12.13, 14.65, and
21.34 $\mu$m) for 441,879 sources. AKARI \citep{murakami2007} provided PSC data
at two bands (9 and 18 $\mu$m) and bright-source catalog (BSC) data at four
bands (65, 90, 140, and 160 $\mu$m). 2MASS \citep{cutri2003} provided fluxes at
J (1.24 $\mu$m), H (1.66 $\mu$m), and K (2.16 $\mu$m) bands. The field of view
(FOV) pixel sizes of the IRAS, MSX, AKARI PSC, AKARI BSC, and 2MASS images are
0$\farcm$75x(4$\farcm$5-4$\farcm$6), 18$\farcs$3, 10$\arcsec$, 30$\arcsec$, and
2$\arcsec$, respectively.

WISE \citep{wright2010} surveyed the entire sky. The ALLWISE source catalog
provided the photometric data at 3.4 $\mu$m (W1), 4.6 $\mu$m (W2), 12 $\mu$m
(W3), and 22 $\mu$m (W4) bands. The FOV pixel sizes are 2$\farcs$75,
2$\farcs$75, 2$\farcs$75, and 5$\farcs$5, and the 5$\sigma$ photometric
sensitivities are 0.068, 0.098, 0.86, and 5.4 mJy for the four WISE bands.

Though IRAS and AKARI data have been very useful for studying AGB stars in our
Galaxy (e.g., \citealt{sk2011}), the number of the cross-identified objects for
the new sample stars was very limited because of the relatively large beam
sizes and weak sensitivities. On the other hand, the WISE data would be more
useful for studying dim objects or objects in crowded regions.

In this paper, we use only good-quality observational data at all wavelength
bands for the IRAS, 2MASS, WISE, AKARI, and MSX photometric data (q = 3 for
IRAS and AKARI; q = A for 2MASS; q = A or B for WISE; q = 3 or 4 for MSX).

Table~\ref{tab:tab4} lists the IR bands used in this work. For each band, the
reference wavelength ($\lambda_{ref}$) and zero magnitude flux (ZMF) value are
also shown. The color index is defined by
\begin{equation}
M_{\lambda 1} - M_{\lambda 2} = - 2.5 \log_{10} {{F_{\lambda 1} / ZMF_{\lambda 1}} \over {F_{\lambda 2} / ZMF_{\lambda 2}}}
\end{equation}
where $ZMF_{\lambda i}$ is the ZMF at given wavelength ($\lambda i$) (see
Table~\ref{tab:tab4}). We may obtain the color indices from given fluxes from
observations (e.g., IRAS, AKARI, and MSX photometric data are in Jy unit) or
from theoretical model SEDs using the reference (or effective or isophotal)
wavelength and ZMF specified in the the telescope system manual.

\begin{table}
\centering
\caption{Cross-identified IR sources and AAVSO objects for AGB-IRAS sample stars\label{tab:tab5}}
\begin{tabular}{llllll}
\hline \hline
Subgroup$^1$ & IRAS & AKARI & ALLWISE & MSX & AAVSO \\
\hline
OI-SH & 3826 & 3673 & 3820 & 2394 & 2538 \\
OI-UR & 37   & 35   & 37   & 28   & 18\\
OI-JB & 6    & 6    & 6    & 6    & 5\\
OI-ME & 127  & 120  & 127  & 127  & 26\\
OI-ST & 673  & 651  & 673  & 673  & 255\\
OI-OG & 1057 & 1011 & 1057 & 949  & 1057\\
OI-AM & 163  & 160  & 127  & 102  & 163\\
OI-all& 5908 & 5675 & 5722 & 4279 & 4081\\
\hline
CI-SH & 1165 &1149 & 1164  & 753  &890\\
CI-UR & 5    &5    & 5     & 4    &1\\
CI-GC & 2417 &2393 & 2413  & 1848 &2117\\
CI-OG & 9    &8    & 9     & 7    &9\\
CI-all& 3596 &3555 & 3591  & 2612 &3017\\
\hline
\end{tabular}
\begin{flushleft}
$^1$See Table~\ref{tab:tab1}.
The numbers of the cross-identified 2MASS sources are the same as the numbers of IRAS PSC sources.
\end{flushleft}
\end{table}

\begin{table}
\centering
\caption{Cross-identified IR sources and AAVSO objects for AGB-WISE sample stars\label{tab:tab6}}
\begin{tabular}{llllll}
\hline \hline
Subgroup$^1$ & ALLWISE & 2MASS & AKARI & MSX & AAVSO\\
\hline
OW-ME & 157  & 157  & 120  & 157  & 15\\
OW-ST & 231  & 231  & 190  & 231  & 29\\
OW-OG & 4913 & 4913 & 3159 & 4171 & 4913  \\
OW-all& 5301 & 5301 & 3469 & 4559 & 4957\\
\hline
CW-GC & 3417 & 3187 & 1987  & 1627 & 1388\\
CW-OG & 159  & 159  & 70    & 121  & 159\\
CW-all& 3576 & 3346 & 2057  & 1748 & 1547\\
\hline
\end{tabular}
\begin{flushleft}
$^1$See Table~\ref{tab:tab2}.
\end{flushleft}
\end{table}

\subsection{Cross-matches\label{sec:agb-c}}

For all IRAS PSC sources, we have found the AKARI, 2MASS, ALLWISE, MSX, and
AAVSO counterparts by using following method. We cross-identify the AKARI PSC
or BSC counterpart by finding the nearest source within 60$\arcsec$ using the
position given in the IRAS PSC (version 2.1). Then, we cross-identify the
2MASS, WISE, and MSX counterparts using the position of the available AKARI PSC
or BSC counterpart. Only when there is no AKARI counterpart, we have used the
position of the IRAS PSC.

For the objects whose original references are not based on the IRAS PSC
(subgroup names: ME, ST, OG, and GC), we have found the positive IRAS PSC
counterparts using the following method. We find the nearest IRAS PSC, AKARI
PSC, ALLWISE, 2MASS, MSX, and AAVSO counterparts within 5$\arcsec$-60$\arcsec$
(depending on the beam size of the telescope) using the position information in
the original reference. Because there could be multiple ALLWISE sources for one
IRAS PSC source (with a larger beam size), we need to compare the ALLWISE
counterpart obtained from the IRAS PSC position (see the previous paragraph)
with the counterpart obtained from the position in the original reference. If
the ALLWISE (or AKARI or MSX) counterparts obtained from the two positions
(from the IRAS PSC and the original reference) are the same, the IRAS PSC
source would be a more reliable counterpart. These objects with the positive
IRAS PSC counterparts are compiled into the AGB-IRAS catalog (OI-ME, OI-ST,
OI-OG, CI-GC, and CI-OG; see Table~\ref{tab:tab1}). All other objects with
ALLWISE counterparts, which cannot be positively identified with the IRAS PSC
sources, are compiled into the AGB-WISE catalog (OW-ME, OW-ST, OW-OG, CW-GC,
and CW-OG; see Table~\ref{tab:tab2}).

For WISE data, multiple sample objects may have the same cross-matched ALLWISE
source. So we have checked all of the duplicate cross-matched objects and
selected only one nearest object for the one ALLWISE source (OW-OG and CW-GC in
Table~\ref{tab:tab2}).

Figure~\ref{f1} compares the IRAS PSC and WISE fluxes for all of the AGB-IRAS
sample stars (see Table 1). The overall comparison is fairly consistent for
most objects. Some objects show large deviations (the IRAS flux is much larger
than the WISE flux) possibly due to the contamination by the larger beam size
of IRAS.

Tables~\ref{tab:tab5} and \ref{tab:tab6} list the cross-matched counterparts
for all AGB-IRAS and AGB-WISE sample stars.

\subsection{Catalogs based on IRAS PSC\label{sec:iras}}

Based on the catalog of AGB stars presented by \citet{sh2017} (3828 OAGB and
1168 CAGB stars), we have revised it using new available literature. We have
excluded five objects from the OAGB list (OI-SH): four young stellar objects
(YSOs) or red supergiants (RSGs) and a red hypergiant (RHG) (see
Tables~\ref{tab:tab3}). We have moved from CI-SH to OI-SH three objects which
are SiO maser sources without clear CAGB evidences (see Tables~\ref{tab:tab1}).

The IRAS Low Resolution Spectrograph (LRS; $\lambda$ = 8$-$22 $\mu$m) data have
been very useful to identify important dust features of AGB stars.
\citet{kwok1997} used the IRAS LRS data for 11,224 IRAS PSC source to identify
new OAGB and CAGB stars. In the IRAS LRS, OAGB stars with silicate dust
envelopes are classified into type E (10 $\mu$m in emission) or A (10 $\mu$m in
absorption). CAGB stars with SiC grains are classified into type C (11.3 $\mu$m
in emission). There are 715 objects that are classified into group C: 713
objects are in the list of CI-SH (see Table~\ref{tab:tab1}), one object (IRAS
13136-4426) is an S star (in SI), and the other one (IRAS 22306+5918) is a
composite object.

Figure~\ref{f2} shows a WISE-2MASS 2CD using W3[12]$-$W4[22] versus
K[2.2]$-$W3[12] for the sample of AGB stars from \citet{sh2017} (3828 OAGB
stars in OI-SH and 1168 CAGB stars in CI-SH), which can be used to find a rough
guide line that distinguishes between OAGB and CAGB stars. The brown line
roughly distinguishes between OAGB and CAGB stars.

\citet{urago2020} presented a list 108 IRAS PSC objects that are believed to be
Mira variables because of the regular variability of the light curves at K
band. Sixty-six objects are duplicate with \citet{sh2017} (64: OI-SH;
\textbf{2}: CI-SH) and 42 objects remain. The 42 objects are classified into
OAGB or CAGB using the guide line on the IR 2CD in Figure~\ref{f2}, which
resulted in 32 OAGB and 5 CAGB objects. The remaining 5 objects do not have the
available colors to plot on the 2CD, but all of them are likely to be OAGB
stars because they show O-rich characters in the IRAS LRS or other IR colors.
Therefore, there are 37 OAGB (OI-UR) and 5 CAGB (CI-UR) stars in the sample
(see Figure~\ref{f2} and Table~\ref{tab:tab1}).

Figure~\ref{f2} also shows the color-selected AGB-IRAS (OI-OG and CI-OG) and
AGB-WISE (OW-OG and CW-OG) stars from \citet{sus13a} (see
Section~\ref{sec:galb}).

From all of the Mira variables with known periods in the AAVSO, we identified
10,753 IRAS PSC counterparts. Among them, there are 9 IRAS LRS type A sources
(all objects are in OI-SH) and 894 IRAS LRS type-E sources (731 sources are
already in OAGB-IRAS). The remaining 163 Miras, which are IRAS LRS type E
sources, are likely to be new OAGB stars in our Galaxy because no object shows
any evidences for CAGB or S stars. We compile the 163 AAVSO Miras, which are
IRAS LRS type-E sources, into the new catalog (OAGB-IRAS: OI-AM).

\subsection{New OAGB stars\label{sec:newoagb}}

\citet{jimenez2015} and \citet{blommaert2018} studied extremely reddened AGB
stars in the Galactic bulge and presented lists of OAGB stars. These OAGB stars
are compiled into the catalog based on IRAS PSC (OI-JB in OAGB-IRAS; see
Table~\ref{tab:tab1}).

SiO maser sources are generally believed to be associated with OAGB stars
(mostly Mira variables) with a thin silicate dust envelope (e.g.,
\citealt{deguchi2012}; \citealt{stroh2019}). But, two CAGB stars (IRAS LRS type
C objects in CI-SH) are SiO maser sources: IRAS 01022+6542
(\citealt{deguchi2012}) and IRAS 17105-3746 (\citealt{stroh2019}).

\citet{messineo2018} performed the SiO maser survey of late-type stars in the
inner Galaxy. \citet{stroh2019} performed the Bulge Asymmetries and Dynamical
Evolution (BAaDE) SiO Maser Survey at 86 GHz with the Atacama Large
Millimeter/submillimeter Array (ALMA) and presented a list of 1427 sources with
useful information. The OAGB stars (SiO maser sources) from
\citet{messineo2018} and \citet{stroh2019} are compiled into two separate
catalogs based on the IRAS PSC (OI-ME and OI-ST in OAGB-IRAS) and the ALLWISE
source catalog (OW-ME and OW-ST in OAGB-WISE) (see Section~\ref{sec:agb-c}).

\citet{wu2018} performed SiO maser survey toward off-plane OAGB stars around
the orbital plane of the Sagittarius stellar stream and detected SiO maser
emission from 44 targets, which are IRAS PSC sources.

We have compiled these OAGB objects into the new OAGB catalog (OI-JB, OI-ME,
OW-ME, OI-ST, OW-ST, and OI-WU; see Tables~\ref{tab:tab1} and \ref{tab:tab2}).

\subsection{LPVs in the Galactic bulge\label{sec:galb}}

The OGLE projects detected many LPVs in the bulge of our Galaxy. The fifteenth
part of the OGLE-III Catalog of Variable Stars (OIII-CVS) contains 232,406 LPVs
detected in the OGLE-II and OGLE-III fields toward the Galactic bulge. The
sample consists of 6528 Miras, 33,235 semiregular variables (SRVs) and 192,643
OGLE small amplitude red giants (OSARGs) (\citealt{sus13a}).

A giant star is believed to evolve from a OSARG to an SRV and finally to a
Mira, limiting the number of its excited modes and increasing its pulsating
periods and amplitudes (\citealt{swu13b}). Though most of SRVs and a majority
of OSARGs could also be in the AGB phase, most Mira variables are surely in the
AGB phase and they are usually oscillating in the fundamental mode and occupy a
single sequence in the period-luminosity diagram (\citealt{swu13b}). In this
work, we consider the 6528 OGLE3 Miras in the Galactic bulge for the new
catalog of AGB stars.

Using the guide line on the WISE-2MASS 2CD (see Figure~\ref{f2}), we may
roughly distinguish between OAGB and CAGB stars for the sample of the 6528
OGLE3 Miras in the Galactic bulge (\citealt{sus13a}). Figure~\ref{f3} shows the
2CD for the sample of 6528 OGLE3 Miras in the Galactic bulge
(\citealt{sus13a}), but 321 objects without good-quality observed colors
(W3[12]$-$W4[22] and K[2.2]$-$W3[12]) are not plotted and they are not compiled
into the new catalog. The brown line roughly distinguishes between OAGB and
CAGB stars. We find that 6038 and 169 objects can be classified into OAGB and
CAGB, respectively. One exception is IRAS 18100-2808, which is an OAGB star
(IRAS LRS type E) with the CAGB color. Therefore, 6039 and 168 objects can be
compiled into the OAGB and CAGB catalogs, respectively.

These OAGB and CAGB stars with the positive IRAS PSC counterparts (see
Section~\ref{sec:agb-c}) are compiled into the AGB-IRAS catalog (OI-OG and
CI-OG; see Table~\ref{tab:tab1}) and others into the AGB-WISE catalog (OW-OG
and CW-OG; see Table~\ref{tab:tab2}).

Figure~\ref{f2} shows the WISE-2MASS 2CD for the color-selected sample stars
from \citet{sus13a} in AGB-IRAS (OI-OG, and CI-OG) and AGB-WISE (OW-OG and
CW-OG) catalogs. The 2CD also shows the color selected AGB-IRAS objects (OI-UR
and CI-UR) from \citet{urago2020} (see Section~\ref{sec:iras}).

\subsection{Visual carbon stars\label{sec:vcs}}

\citet{alksnis2001} presented general catalog of galactic carbon stars (3rd
edition) which consist of 6891 entries, most of which are believed to be visual
carbon stars in the AGB phase. Most CAGB stars in previous catalog (CI-SH;
\citealt{sh2017}) are infrared carbon stars with relatively thick C-rich dust
envelopes.

Objects in \citet{alksnis2001} are compiled into two separate catalogs based on
the IRAS PSC and the ALLWISE source catalog (CI-GC or CW-GC in
Tables~\ref{tab:tab1} and \ref{tab:tab2}; see Section~\ref{sec:agb-c}).

Compared with other subgroups, the basic properties (e.g., detailed IR SEDs or
pulsations) of the objects from \citet{alksnis2001} (in CI-GC or CW-GC) are
less known, so it is possible that a greater portion of the objects from CI-GC
or CW-GC could not be in the AGB phase. Some objects could be in the post-AGB
phase or other stages of stellar evolution.

\begin{figure}
\centering
\smallplottwo{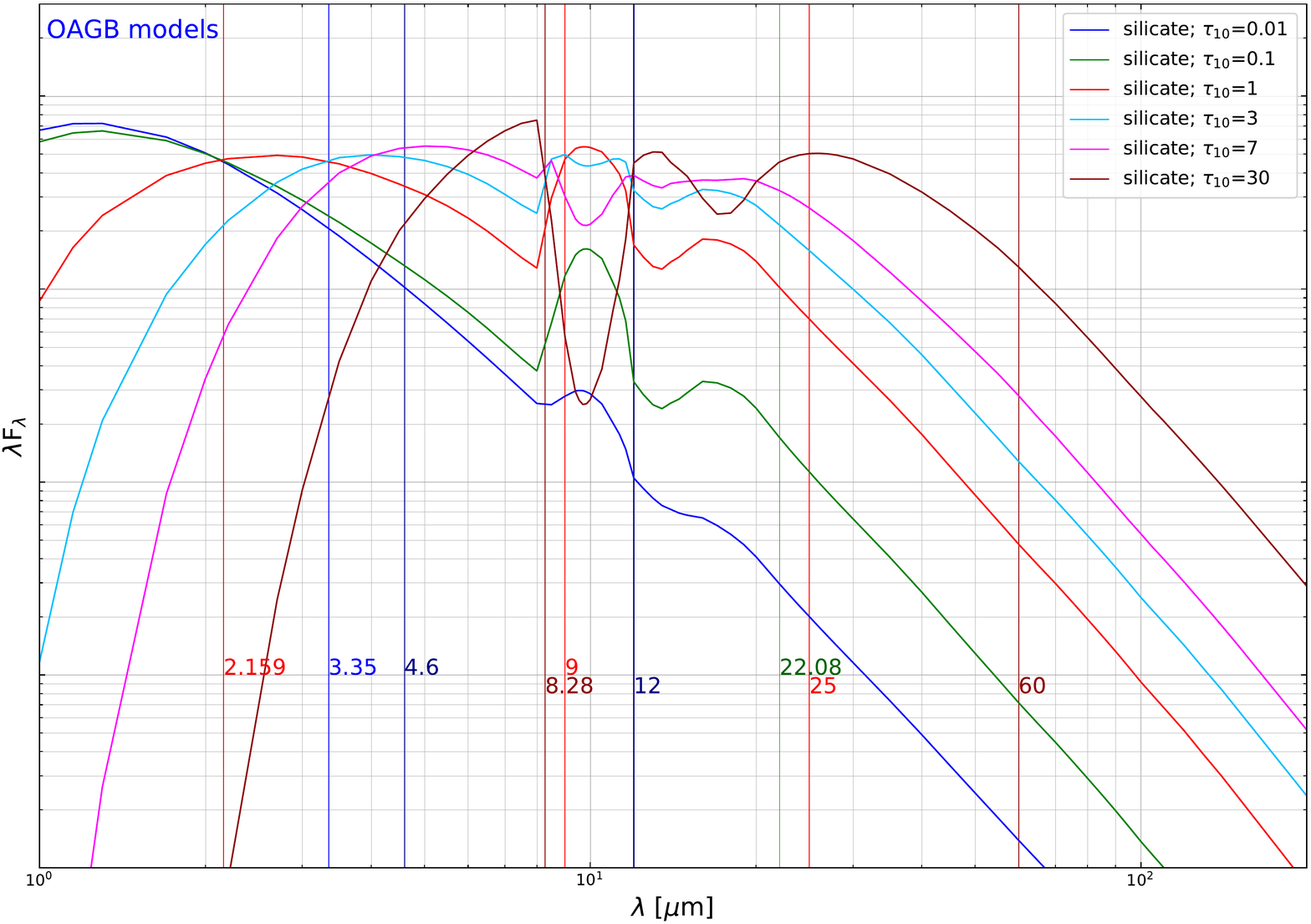}{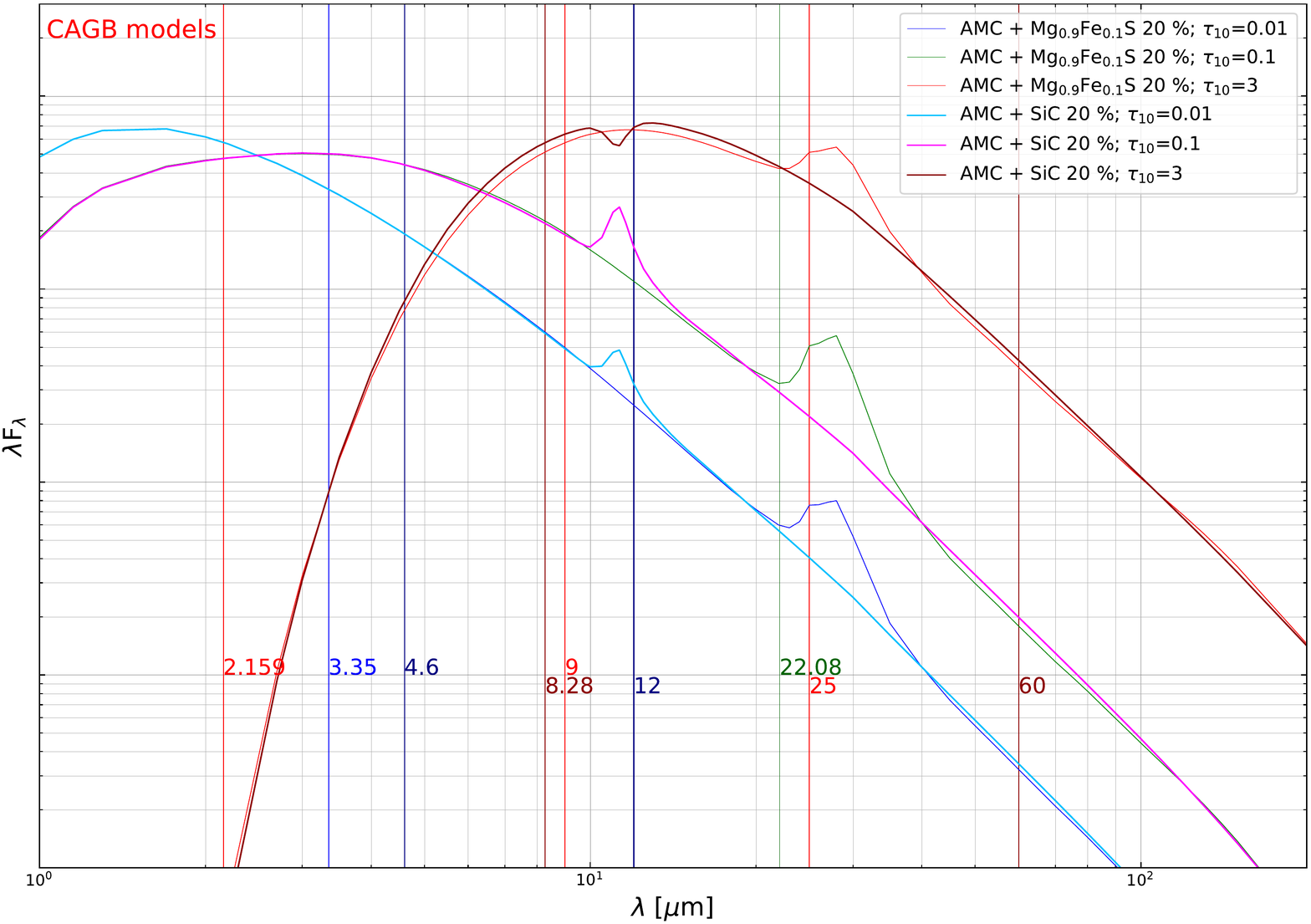}\caption{Theoretical model SEDs for OAGB stars (silicate; $T_c$=1000 K)
and CAGB stars (AMC; $T_c$=1000 K) for a number of dust optical depths (see~\ref{sec:models}).
The reference wavelengths for major IR bands are also indicated (see Table~\ref{tab:tab4}).}
\label{f4}
\end{figure}

\section{Theoretical Dust Shell Models\label{sec:models}}

We use the radiative transfer code DUSTY (\citealt{ivezic1997}) for a
spherically symmetric dust shell around a central star, which is a blackbody.
We use the models of \citet{suh2020} adapted for the new IR bands. In this
paper, we briefly describe the theoretical models. See Section 4 in
\citet{suh2020} for details about the models and their limitations.

For all models, we use a continuous power law ($\rho \propto r^{-2}$) dust
density distribution and assume that the dust formation temperature ($T_c$) is
1000 K. For LMOA stars, we also use $T_c$=500 K. The inner radius of the dust
shell is set by the $T_c$ and the outer radius of the dust shell is taken to be
$10^4$ times the inner radius. We use 10 $\mu$m as the fiducial wavelength of
the dust optical depth ($\tau_{10}$). The radii of spherical dust grains are
assumed to be 0.1 $\mu$m uniformly.

For OAGB stars, we use optical constants of warm (SILW) and cold (SILC)
silicate dust from \citet{suh1999}. We also use amorphous alumina
(\citealt{suh2016}) and Fe$_{0.9}$Mg$_{0.1}$O (\citealt{henning1995}) dust for
OAGB stars. For CAGB stars, we use the optical constants of AMC and SiC dust
grains from \citet{suh2000} and \citet{pegouri1988}, respectively. We also use
Mg$_{0.9}$Fe$_{0.1}$S (\citealt{begemann1994}) for CAGB stars.
Table~\ref{tab:tab7} summarizes the model parameters.

\begin{table}
\caption{Dust shell model parameters\label{tab:tab7}}
\centering
\begin{tabular}{lllll}
\hline \hline
Model &dust &$\tau_{10}$ & $T_c$ (K) & $T_*$$^1$ (K)  \\
\hline
LMOA  &O-rich$^2$  &0.001, 0.01, 0.05, 0.1 & 500 & 3000  \\
LMOA  &SILW  &0.001, 0.01, 0.05, 0.1 & 1000 & 3000  \\
LMOA  &SILW  &0.5, 1, 3  & 1000 & 2500   \\
CAGB  &C-rich$^3$   &0.001, 0.01, 0.1, 0.5 & 1000 & 2500  \\
CAGB  &C-rich$^3$  &1, 2, 3, 5   & 1000 & 2000  \\
HMOA  &SILC  &7, 15, 30, 40 & 1000   & 2000  \\
\hline
\end{tabular}
\begin{flushleft}
\scriptsize
$^1$the black body temperature of the central star
$^2$SILW, SILW + alumina (from \citealt{suh2016}), and SILW + Fe$_{0.9}$Mg$_{0.1}$O (from \citealt{henning1995}).
$^3$AMC, AMC + SiC, and AMC + Mg$_{0.9}$Fe$_{0.1}$S (from \citealt{begemann1994}).
\end{flushleft}
\end{table}

Figure~\ref{f4} shows model SEDs for AGB stars ($T_c$=1000 K) for major dust
optical depths. For OAGB models, silicate dust features at 10 and 18 $\mu$m are
shown for various dust optical depths ($\tau_{10}$). For CAGB models, SiC dust
features at 11.3 $\mu$m and Mg$_{0.9}$Fe$_{0.1}$S dust features at 28 $\mu$m
are shown for different dust optical depths. The reference wavelengths for
major IR bands are also indicated. We may obtain the theoretical model color
indices from the model SEDs using the ZMF values at given reference wavelength
(see Table~\ref{tab:tab4} and Section~\ref{sec:photdata}).

\begin{figure*}
\centering
\largeplottwo{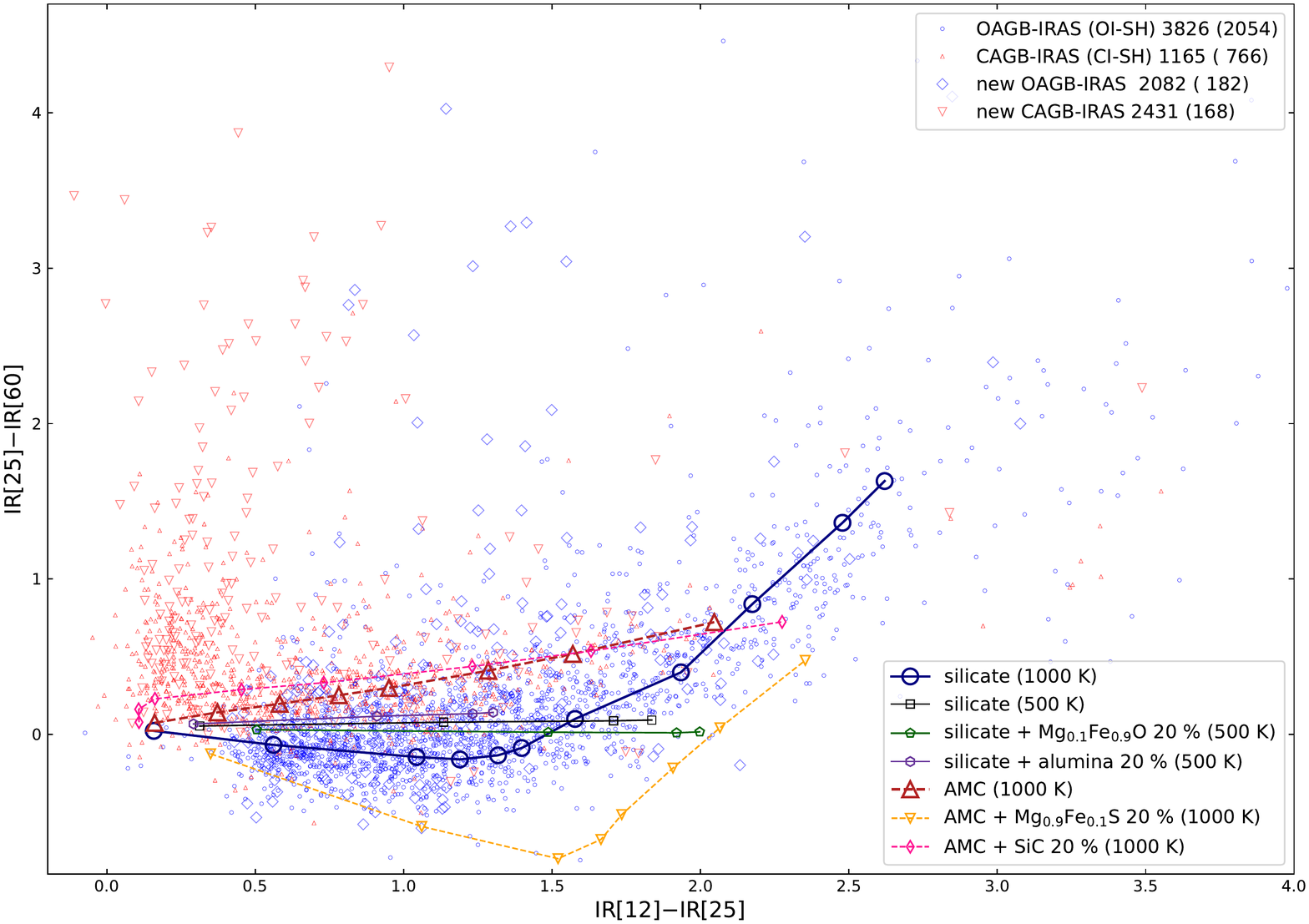}{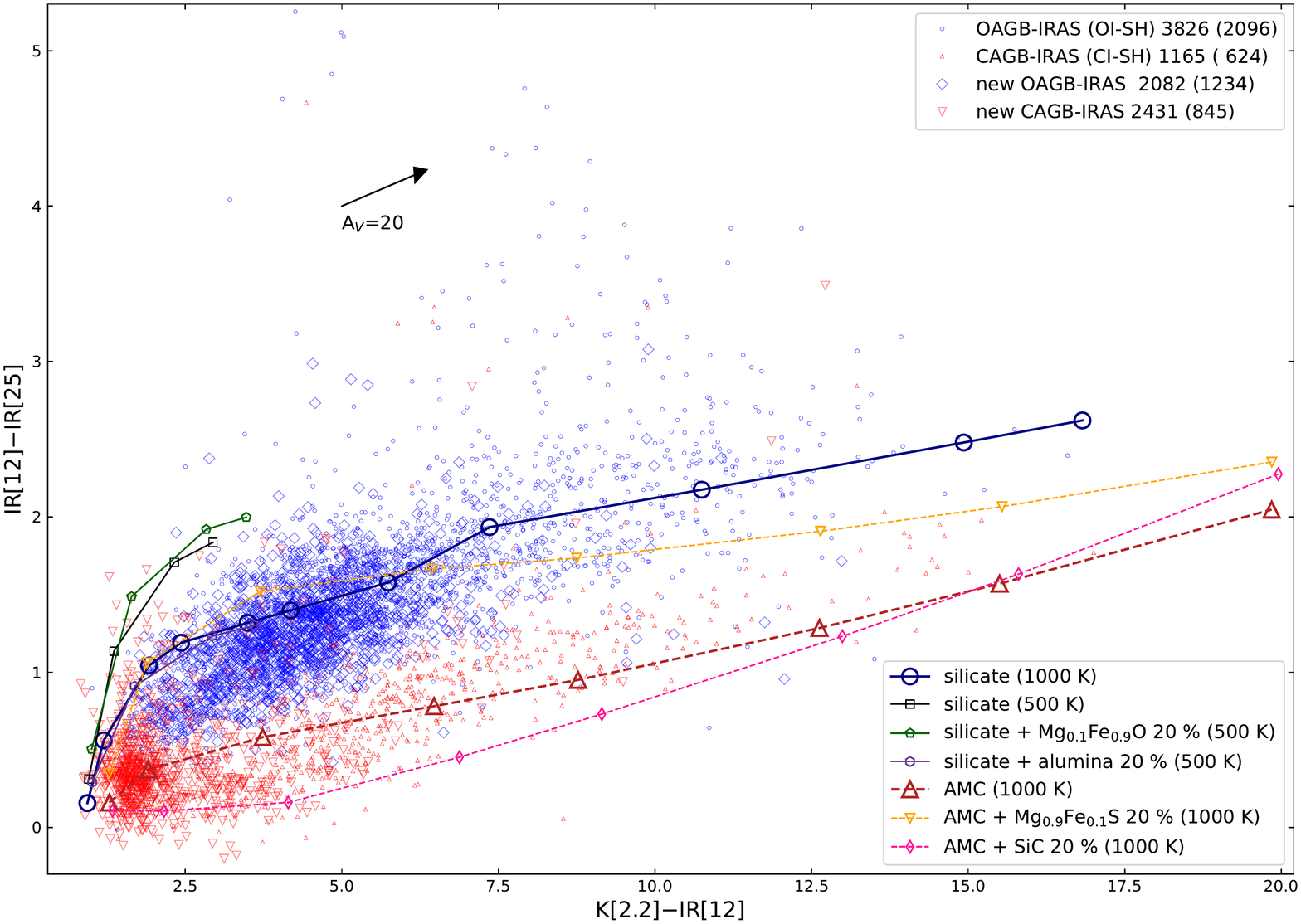}\caption{IRAS-2MASS 2CDs for all IRAS-AGB stars (see Table~\ref{tab:tab1}) in our Galaxy compared with theoretical models (see Section~\ref{sec:models}).
For OAGB models (silicate $T_c$ = 1000 K): $\tau_{10}$ = 0.001, 0.01, 0.05, 0.1, 0.5, 1, 3, 7, 15, 30, and 40 from left to right.
For CAGB models (AMC $T_c$ = 1000 K): $\tau_{10}$ = 0.001, 0.01, 0.1, 0.5, 1, 2, 3, and 5 from left to right.
For each class, the number of objects is shown.
The number in parentheses denotes the number of the plotted objects on the 2CD with good-quality observed colors.}
\label{f5}
\end{figure*}

\begin{figure*}
\centering
\largeplottwo{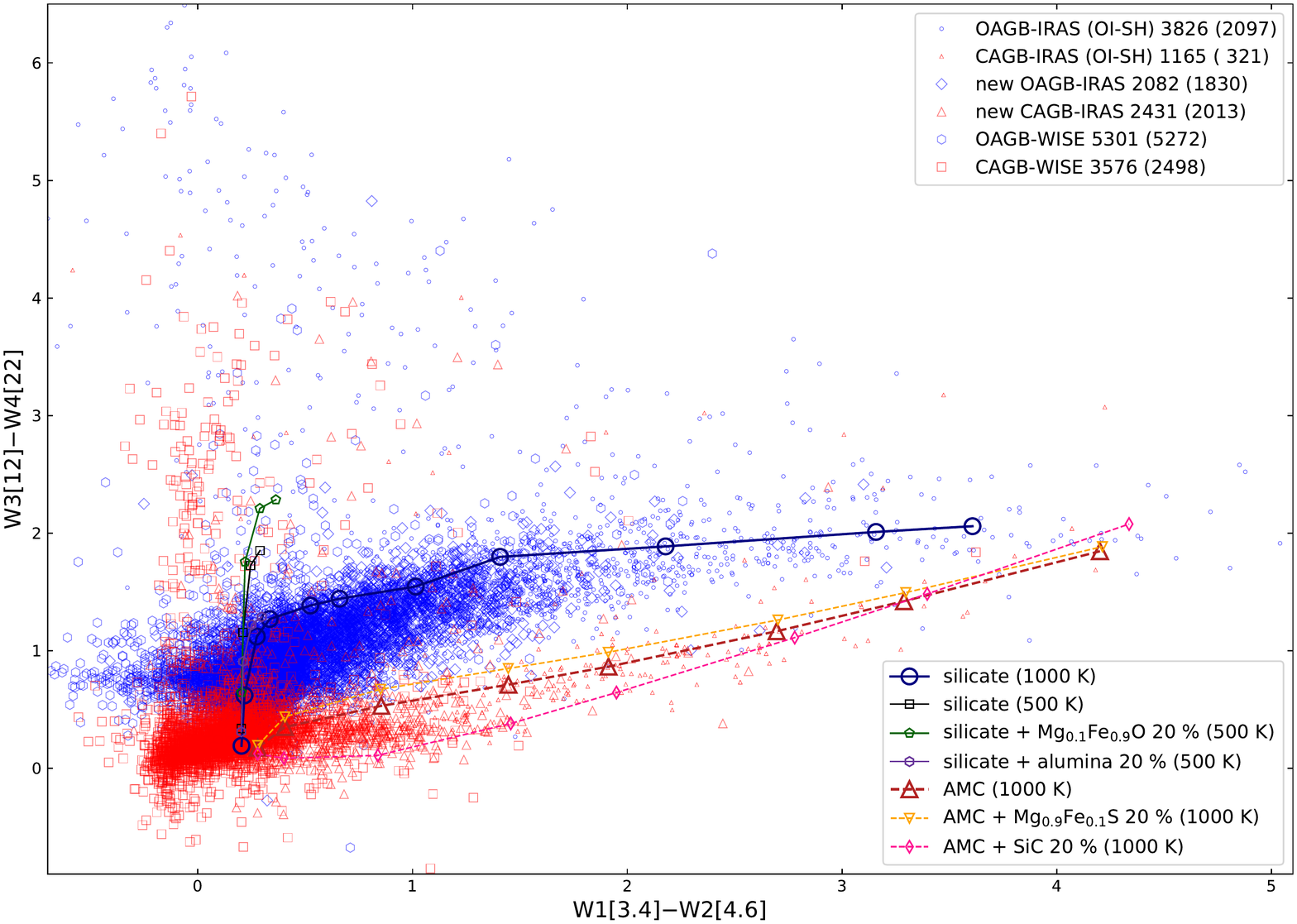}{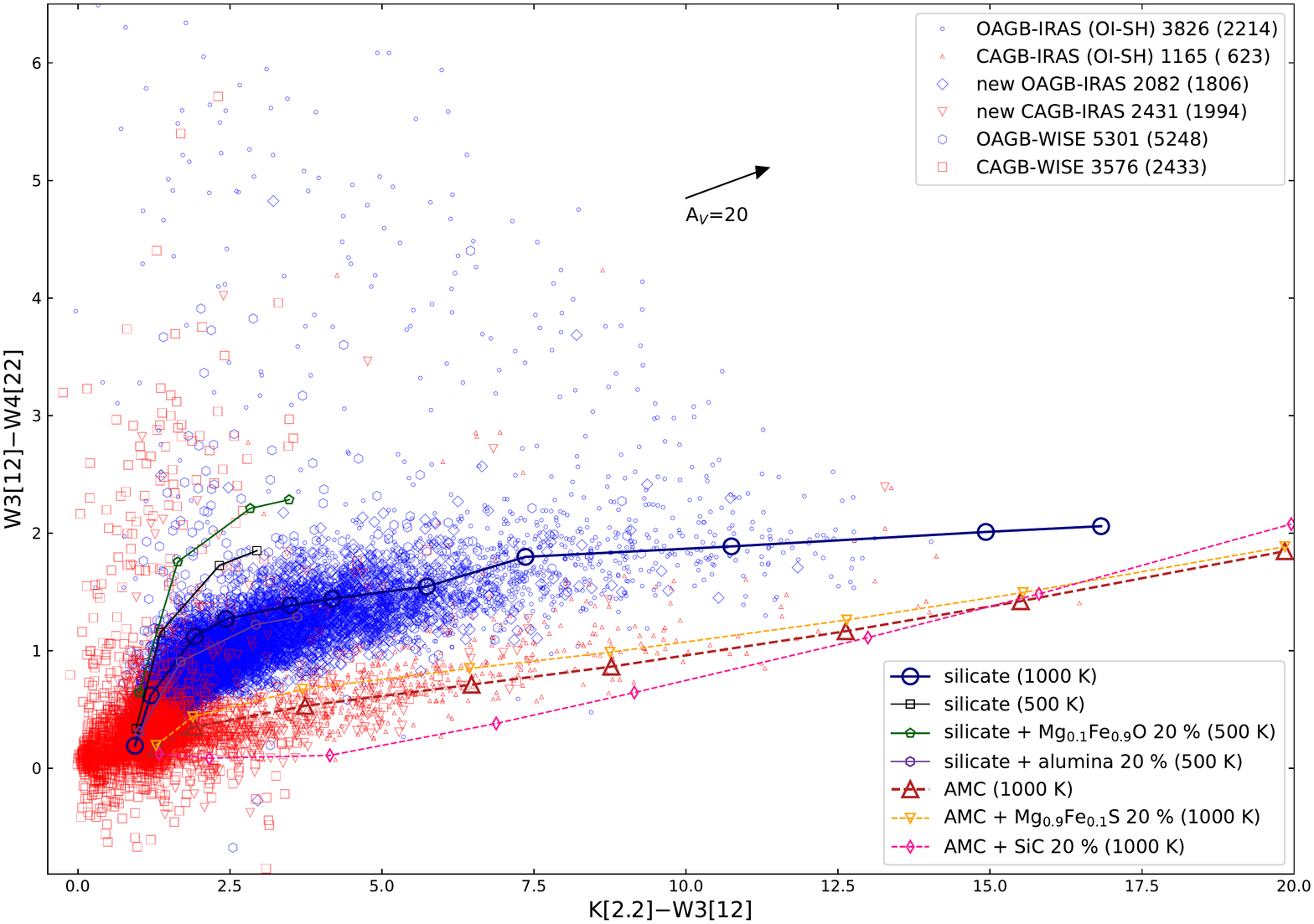}\caption{WISE-2MASS 2CDs for all IRAS-AGB and WISE-AGB stars (see Tables~\ref{tab:tab1} and \ref{tab:tab2}) in our Galaxy compared with theoretical models (see Section~\ref{sec:models}).
For OAGB models (silicate $T_c$ = 1000 K): $\tau_{10}$ = 0.001, 0.01, 0.05, 0.1, 0.5, 1, 3, 7, 15, 30, and 40 from left to right.
For CAGB models (AMC $T_c$ = 1000 K): $\tau_{10}$ = 0.001, 0.01, 0.1, 0.5, 1, 2, 3, and 5 from left to right.
For each class, the number of objects is shown.
The number in parentheses denotes the number of the plotted objects on the 2CD with good-quality observed colors.}
\label{f6}
\end{figure*}

\begin{figure*}
\centering
\largeplot{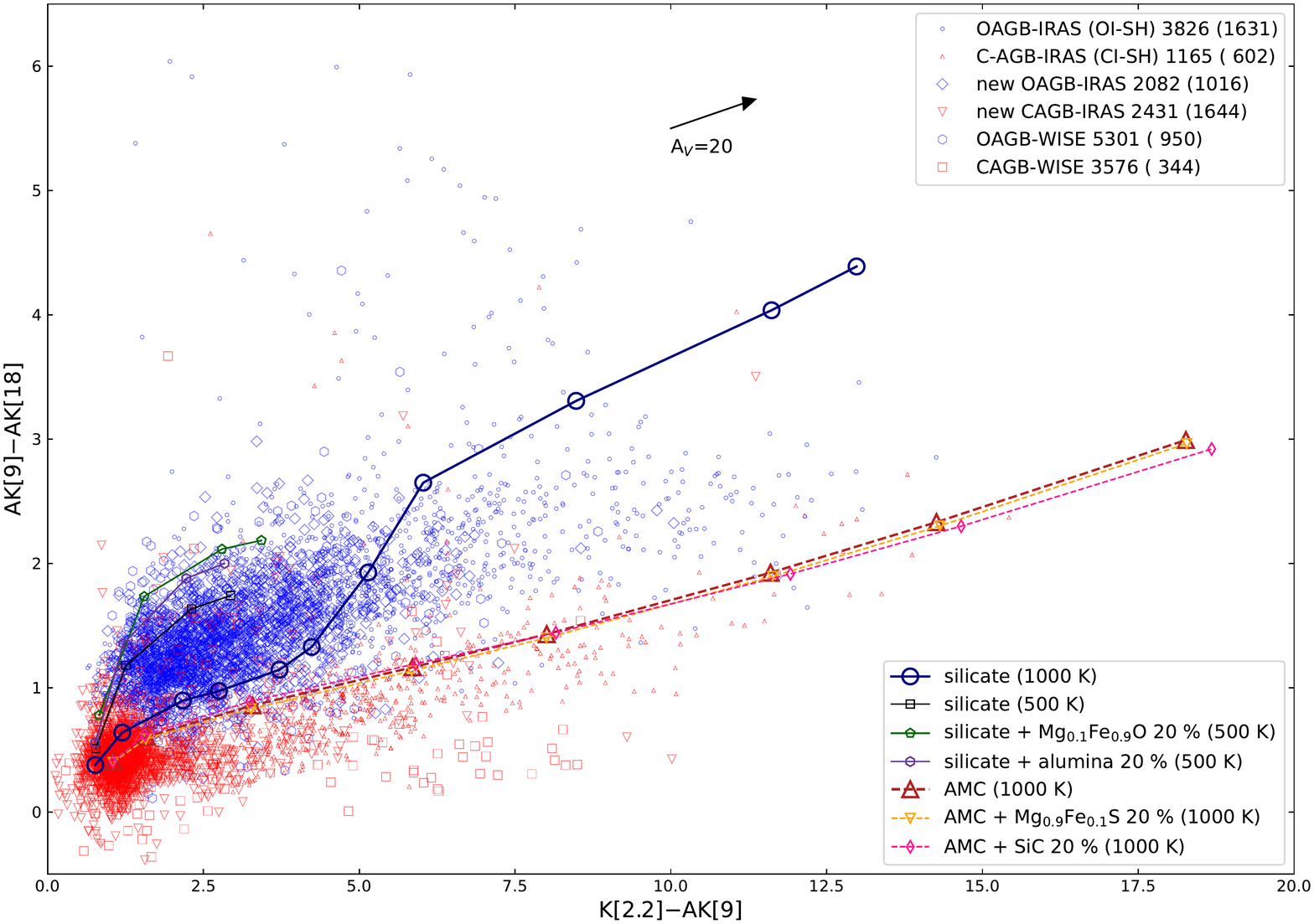}\caption{An AKARI-2MASS 2CD for all IRAS-AGB and WISE-AGB stars (see Tables~\ref{tab:tab1} and \ref{tab:tab2}) in our Galaxy compared with theoretical models (see Section~\ref{sec:models}).
For OAGB models (silicate $T_c$ = 1000 K): $\tau_{10}$ = 0.001, 0.01, 0.05, 0.1, 0.5, 1, 3, 7, 15, 30, and 40 from left to right.
For CAGB models (AMC $T_c$ = 1000 K): $\tau_{10}$ = 0.001, 0.01, 0.1, 0.5, 1, 2, 3, and 5 from left to right.
For each class, the number of objects is shown.
The number in parentheses denotes the number of the plotted objects on the 2CD with good-quality observed colors.}
\label{f7}
\end{figure*}

\begin{figure*}
\centering
\largeplottwo{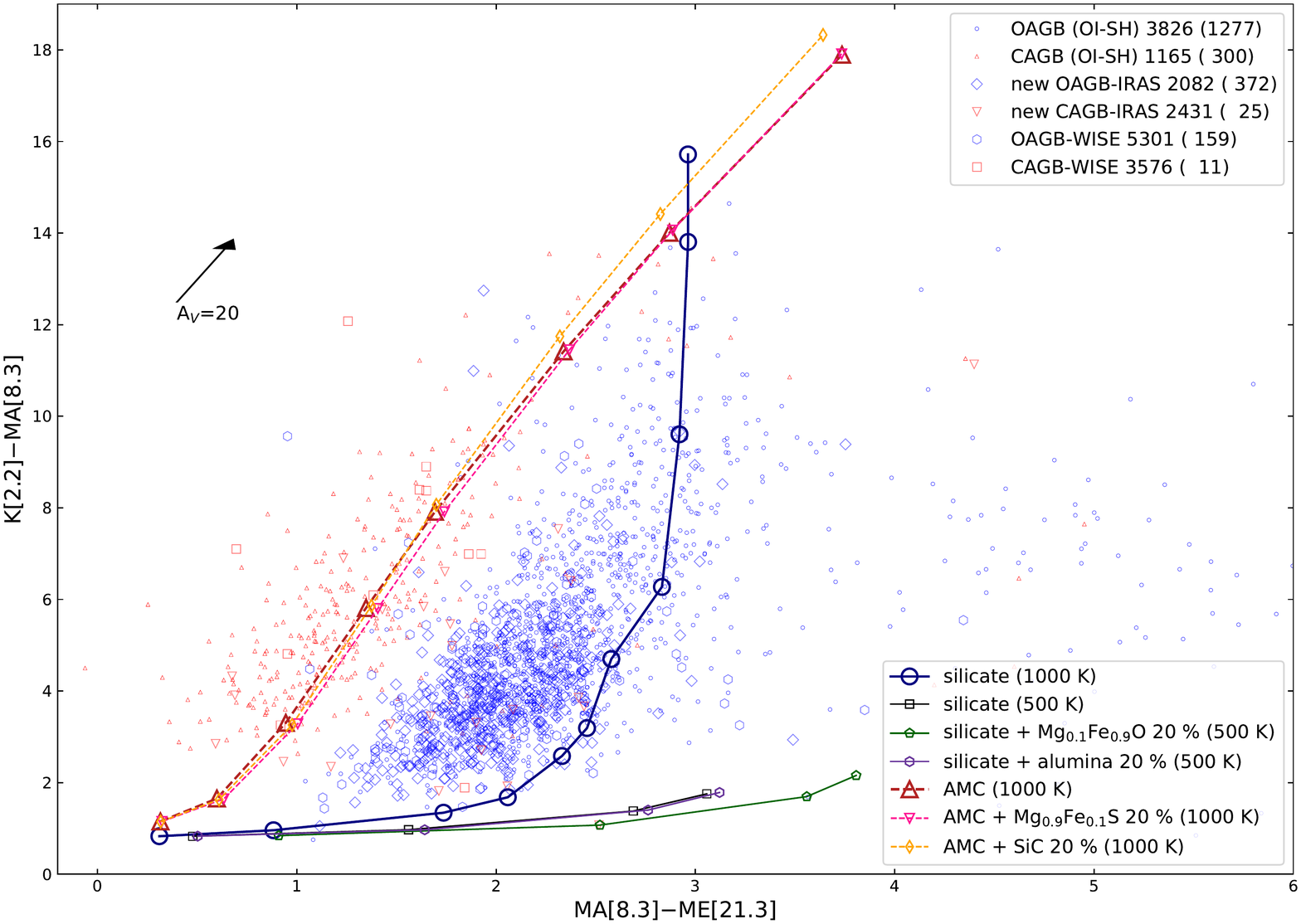}{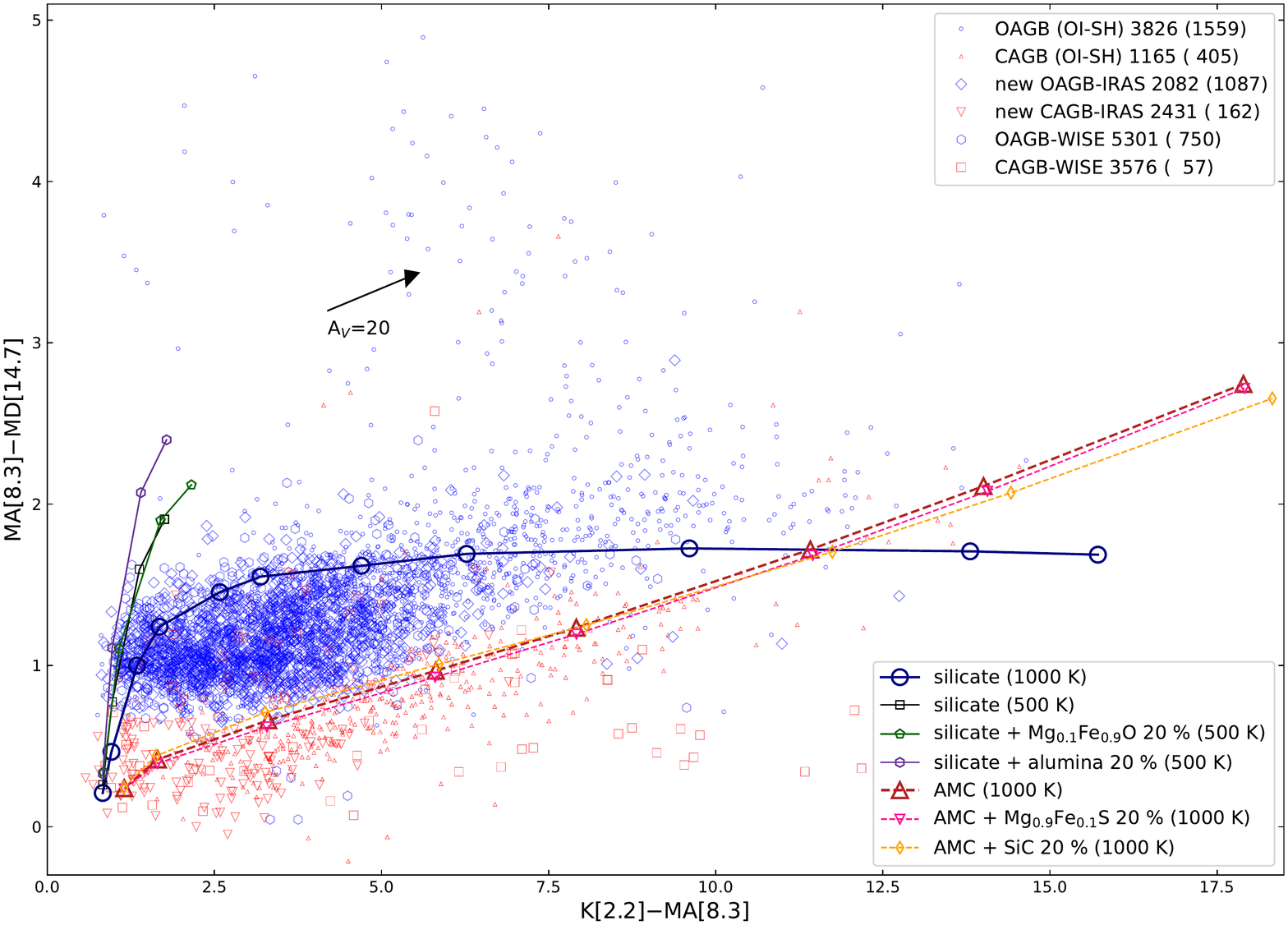}\caption{MSX-2MASS 2CDs for all IRAS-AGB and WISE-AGB stars (see Tables~\ref{tab:tab1} and \ref{tab:tab2}) in our Galaxy compared with theoretical models (see Section~\ref{sec:models}).
For OAGB models (silicate $T_c$ = 1000 K): $\tau_{10}$ = 0.001, 0.01, 0.05, 0.1, 0.5, 1, 3, 7, 15, 30, and 40 from left to right.
For CAGB models (AMC $T_c$ = 1000 K): $\tau_{10}$ = 0.001, 0.01, 0.1, 0.5, 1, 2, 3, and 5 from left to right.
For each class, the number of objects is shown.
The number in parentheses denotes the number of the plotted objects on the 2CD with good-quality observed colors.}
\label{f8}
\end{figure*}

\section{Infrared Two-Color Diagrams - Comparison between Theory and Observations\label{sec:2cds}}

Although the photometric fluxes are less useful than a full SED, the large
number of observations at various wavelength bands can be used to form a 2CD,
which can be compared with theoretical models. IR 2CDs are useful to
statistically distinguish various properties of AGB stars and post-AGB stars
(e.g., \citealt{sk2011};\citealt{suh2015}). Table~\ref{tab:tab4} lists the IR
bands used for the IR 2CDs presented in this work. In this work, we use only
good-quality observational data at all wavelength bands (see
Section~\ref{sec:photdata}) for plotting IR 2CDs.

Figures~\ref{f5} -~\ref{f8} show various IR 2CDs using different combinations
of observed IR colors. We compare the observations with the theoretical dust
shell models (see Section~\ref{sec:models}) for AGB stars. We find that the
theoretical dust shell model can roughly reproduce the observations of AGB
stars on the IR 2CDs using the dust opacity functions of amorphous silicate and
amorphous carbon with a mixture of other dust species.

To consider the Galactic extinction processes suggested by \citet{gordon2009}
(for the wavelength range from visual to NIR bands) and \citet{ct2006} (from
NIR to MIR bands), we plot reddening vectors for IR 2CD using NIR data (see
Figures~\ref{f5}-~\ref{f8}).

Generally, the stars that have thick dust shells with large dust optical depths
are located in the upper-right regions on the IR 2CDs. On all of the IR 2CDs,
we also plot the sequences of theoretical dust shell models at increasing dust
optical depth for AGB stars (see Section~\ref{sec:models}).

Because the 10 $\mu$m silicate feature changes from emission to absorption when
the dust optical depth becomes larger, there is a change in the slope of the
theoretical model line for OAGB stars.

We will discuss the meanings of these 2CDs in the following subsections by
comparing the observations with the theoretical models.

\subsection{IRAS and 2MASS 2CDs\label{sec:iras-2cd}}

The upper panel of Figure~\ref{f5} plots AGB stars in an IRAS 2CD using
IR[25]$-$IR[60] versus IR[12]$-$IR[25]. We find that the basic theoretical
model tracks can roughly explain the observed points. This 2CD has been widely
used since \citet{van der Veen1988} (note that the authors did not make
zero-magnitude calibrations for their 2CD) divided this 2CD into eight regions
of different classes of heavenly bodies. \citet{sevenster2002} used this 2CD to
explain the properties of observed points of AGB stars and post-AGB stars.
Using theoretical dust shell models for AGB stars and post-AGB stars,
\citet{suh2015} presented possible evolutionary tracks from AGB stars to
post-AGB stars and to planetary nebulae on this 2CD.

On the IRAS 2CD using IR[25]$-$IR[60] versus IR[12]$-$IR[25], CAGB stars are
distributed along a curve in the shape of a 'C'. A group of stars in the
upper-left region consists of visual carbon stars (most objects in OI-GC; see
Table~\ref{tab:tab1}) that show excessive flux at 60 $\mu$m, due to the remnant
of an earlier phase when the stars were OAGB stars (e.g. \citealt{ck1990}). A
group of stars in the lower region, which extends to the right side, consists
of infrared carbon stars. The infrared carbon stars on the right side have
thick dust envelopes with large dust optical depths.

The lower panel plots an 2MASS-IRAS 2CD using IR[12]$-$IR[60] versus
K[2.2]$-$IR[12]. The separation between OAGB and CAGB stars is clearer on this
2CD. If we consider Galactic extinction (see the reddening vector on the 2CD),
there would be more observed points of OAGB and CAGB stars that would fit the
theoretical models well.

\subsection{WISE and 2MASS 2CDs\label{sec:wise}}

Figure~\ref{f6} shows 2CDs using WISE and 2MASS colors. The upper panel of
Figure~\ref{f6} shows an WISE 2CDs using W3[12]$-$W4[22] versus
W1[3.4]$-$W2[4.6]. The lower panel of Figure~\ref{f6} shows a WISE-2MASS 2CD
using W3[12]$-$W4[22] versus K[2.2]$-$W3[12].

Generally, the theoretical dust shell models for OAGB and CAGB stars can
reproduce the observed points fairly well on these IR 2CDs, but for the
W1[3.4]$-$W2[4.6] color, the theoretical models with small dust optical depths
do not reproduce the observations well. On these 2CDs, the theoretical dust
shell models for CAGB stars with AMC, SiC, and Mg$_{0.9}$Fe$_{0.1}$S dust
grains can reproduce a wider range of observed W3[12]$-$W4[22] colors.

There is a group of CAGB-WISE objects (visual carbon stars in CI-GC and CW-GC)
that show different aspects from infrared carbon stars. The objects are in the
upper-left regions of the 2CDs that show bluer W1[3.4]$-$W2[4.6] or
K[2.2]$-$W3[12] colors but redder W3[12]$-$W4[22] colors, which would be due to
detached O-rich dust shells (that are remnants of an earlier phase when the
stars were OAGB stars).

\subsection{AKARI, MSX, and 2MASS 2CDs\label{sec:msx}}

Figure~\ref{f7} shows an AKARI-2MASS 2CD using AK[9]$-$AK[18] versus
K[2.2]$-$AK[9]. On this 2CD, the theoretical dust shell models for OAGB stars
with thin detached dust shells ($T_c$=500 K) with warm silicate, amorphous
alumina, and Fe$_{0.9}$Mg$_{0.1}$O dust are more useful to explain the observed
points. We find that the theoretical dust shell models for CAGB stars can
reproduce only a narrow range of AK[9]$-$AK[18] colors.

Figure~\ref{f8} shows MSX-2MASS 2CDs for all IRAS-AGB and WISE-AGB objects.
Though good-quality MSX data are available only for a portion of the sample
stars, these 2CDs clearly divide between OAGB and CAGB. Compared with other
colors, the opacity used for the theoretical models do not reproduce the MSX
colors using MA[8.3], MD[14.7], and ME[21.3] bands well. Again, the theoretical
dust shell models for CAGB stars can reproduce only a narrow range of
MA[8.3]$-$ME[21.3] and MA[8.3]$-$MD[14.7] colors.

The upper panel of Figure~\ref{f8} shows a 2CD using K[2.2]$-$MA[8.3] versus
MA[8.3]$-$ME[21.3]. \citet{lewis2020} used this 2CD to discuss the line that
separates between CAGB and OAGB stars. We also find that the separation between
OAGB and CAGB is relatively clear on this 2CD. If we consider Galactic
extinction, there would be more observed points of OAGB stars that would fit
the OAGB model with thin detached dust shells. The lower panel of
Figure~\ref{f8} shows a 2CD using MA[8.3]$-$MD[14.7] versus K[2.2]$-$MA[8.3].

\begin{figure*}
\centering
\smallplotfour{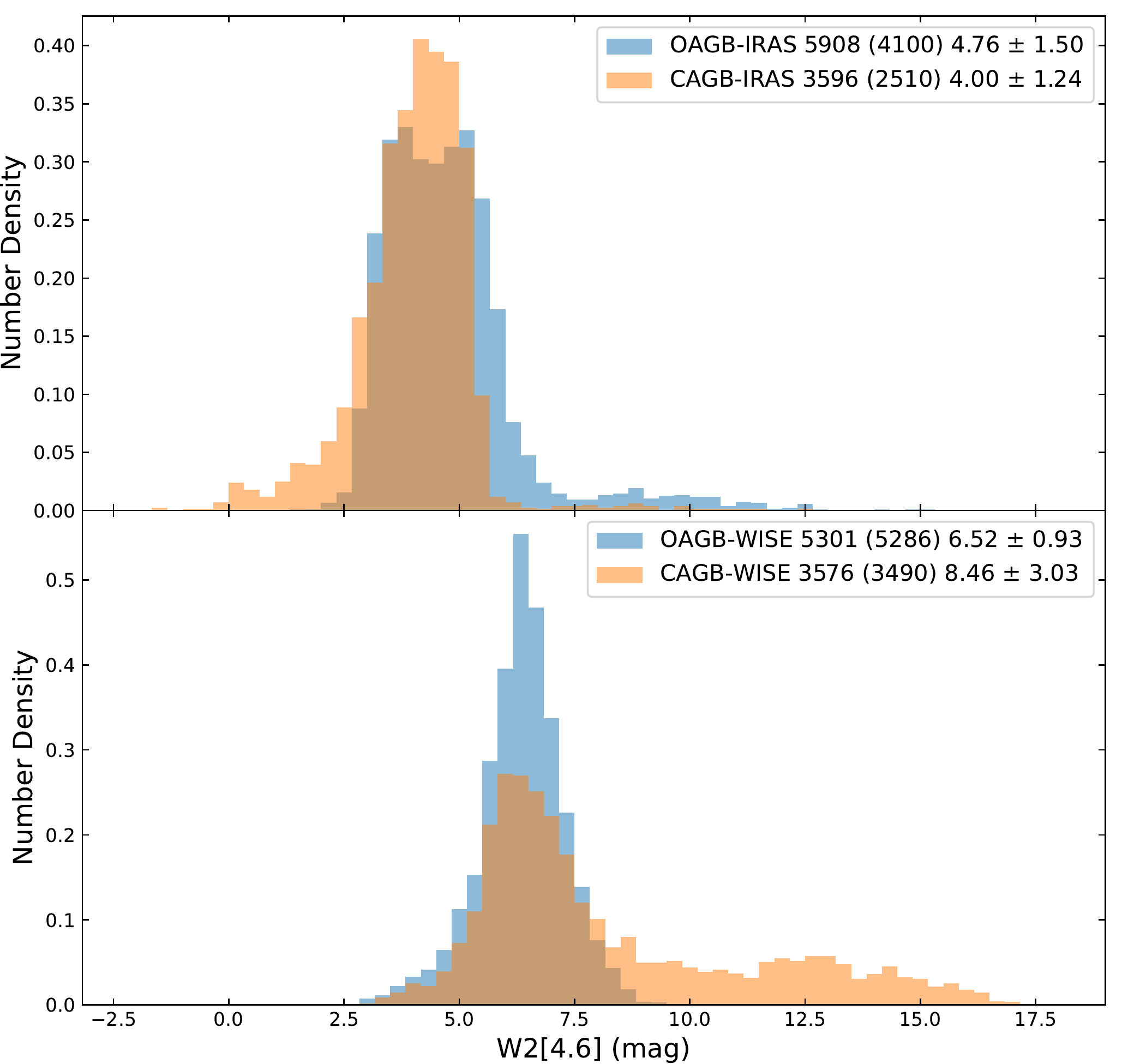}{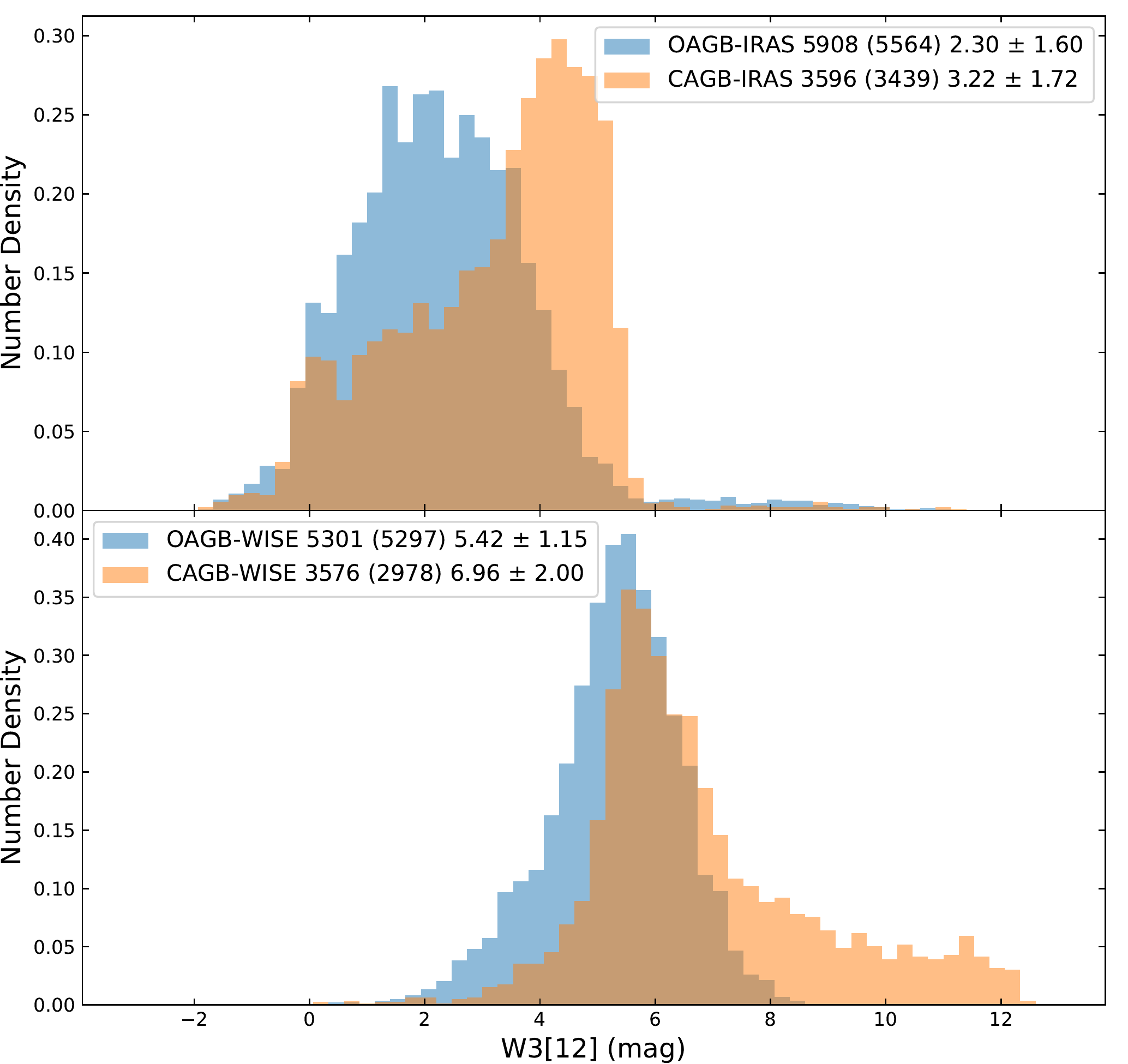}{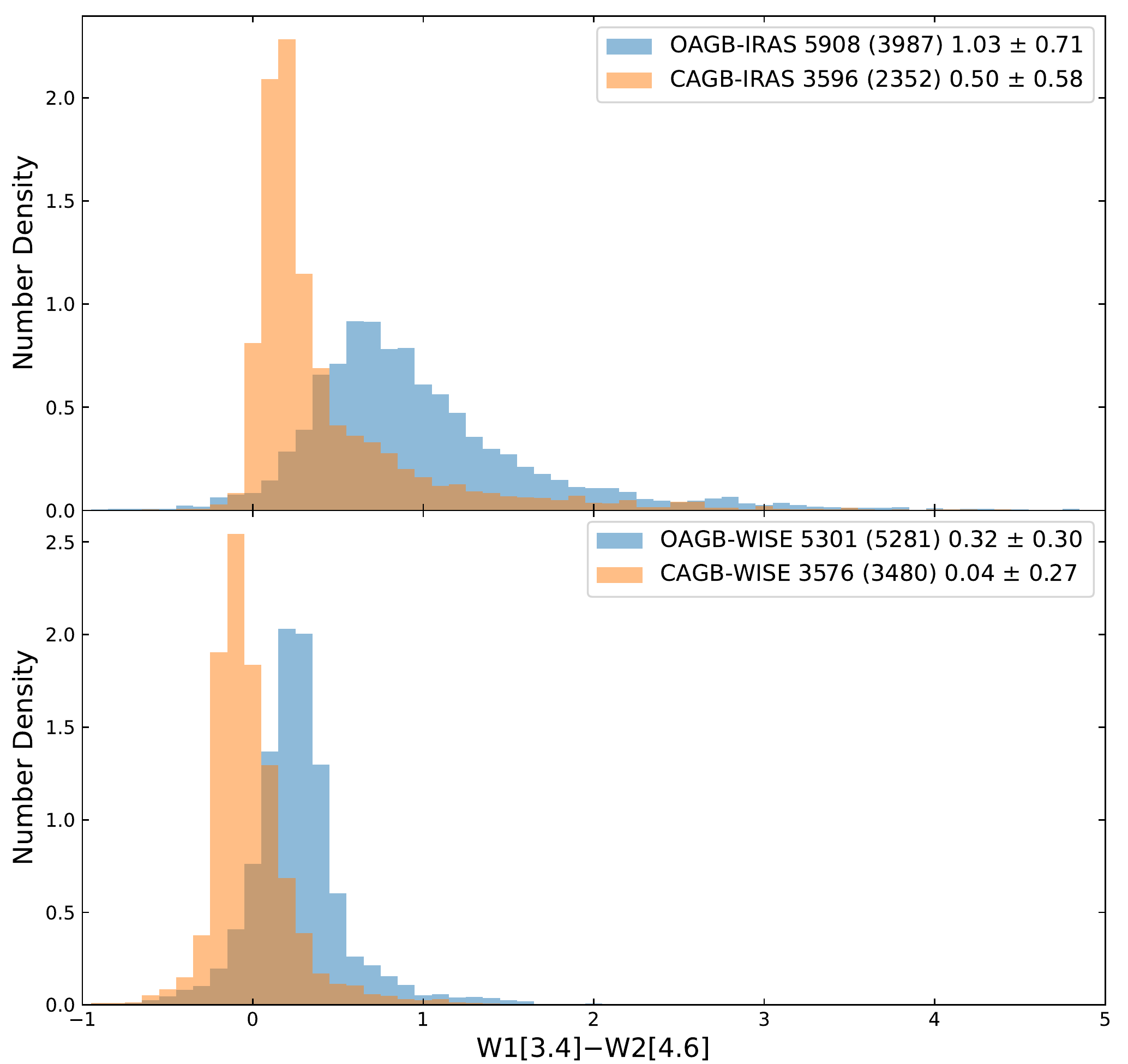}{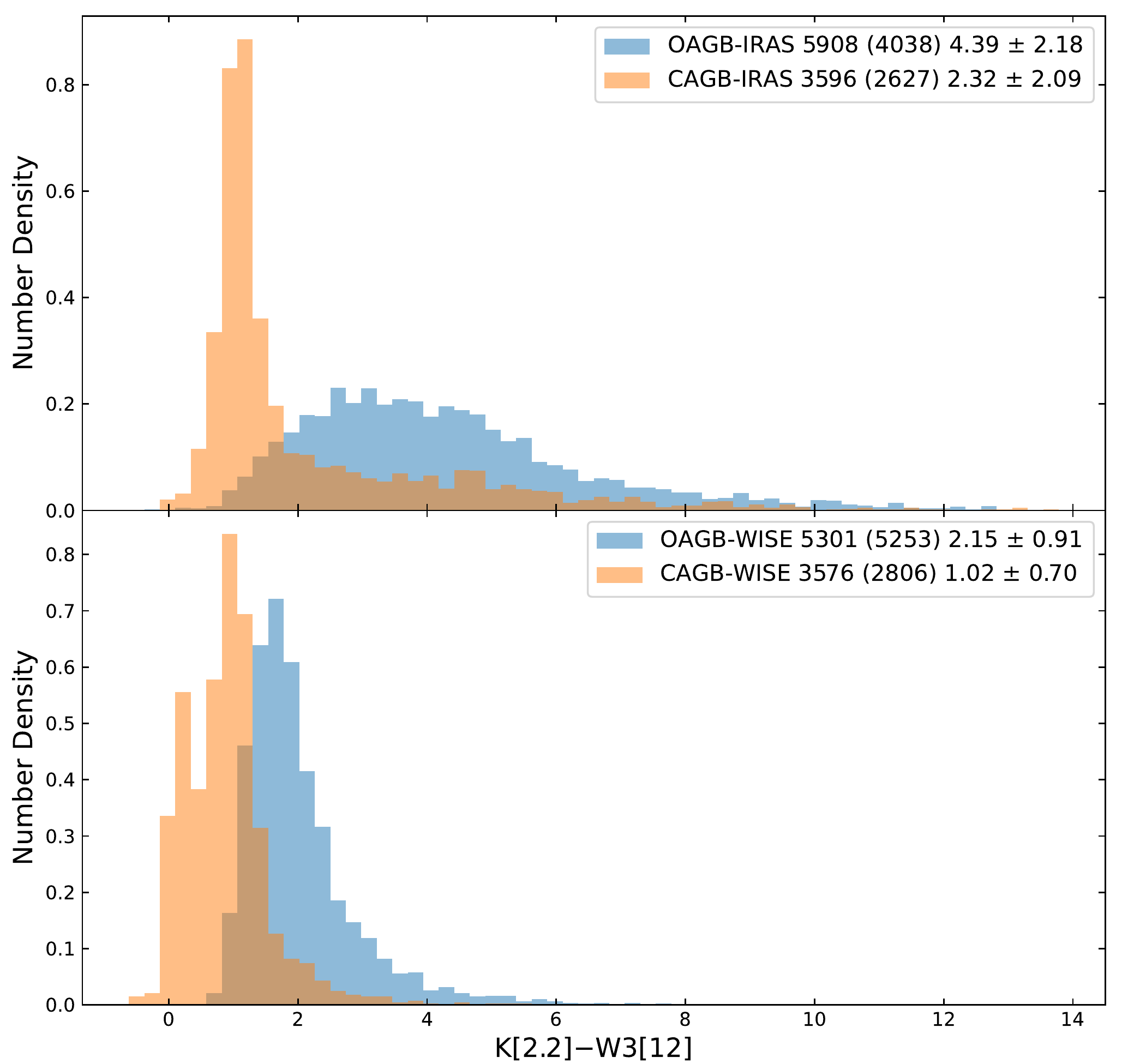}\caption{Number density distributions of observed IR magnitudes and colors for OAGB and CAGB stars in AGB-IRAS and AGB-WISE catalogs.
For each class, the number of objects is shown. The number in parentheses denotes the number of plotted objects with good-quality observed data.
The averaged values and standard deviations of the observed data are also shown.}
\label{f9}
\end{figure*}

\begin{figure*}
\centering
\smallplottwo{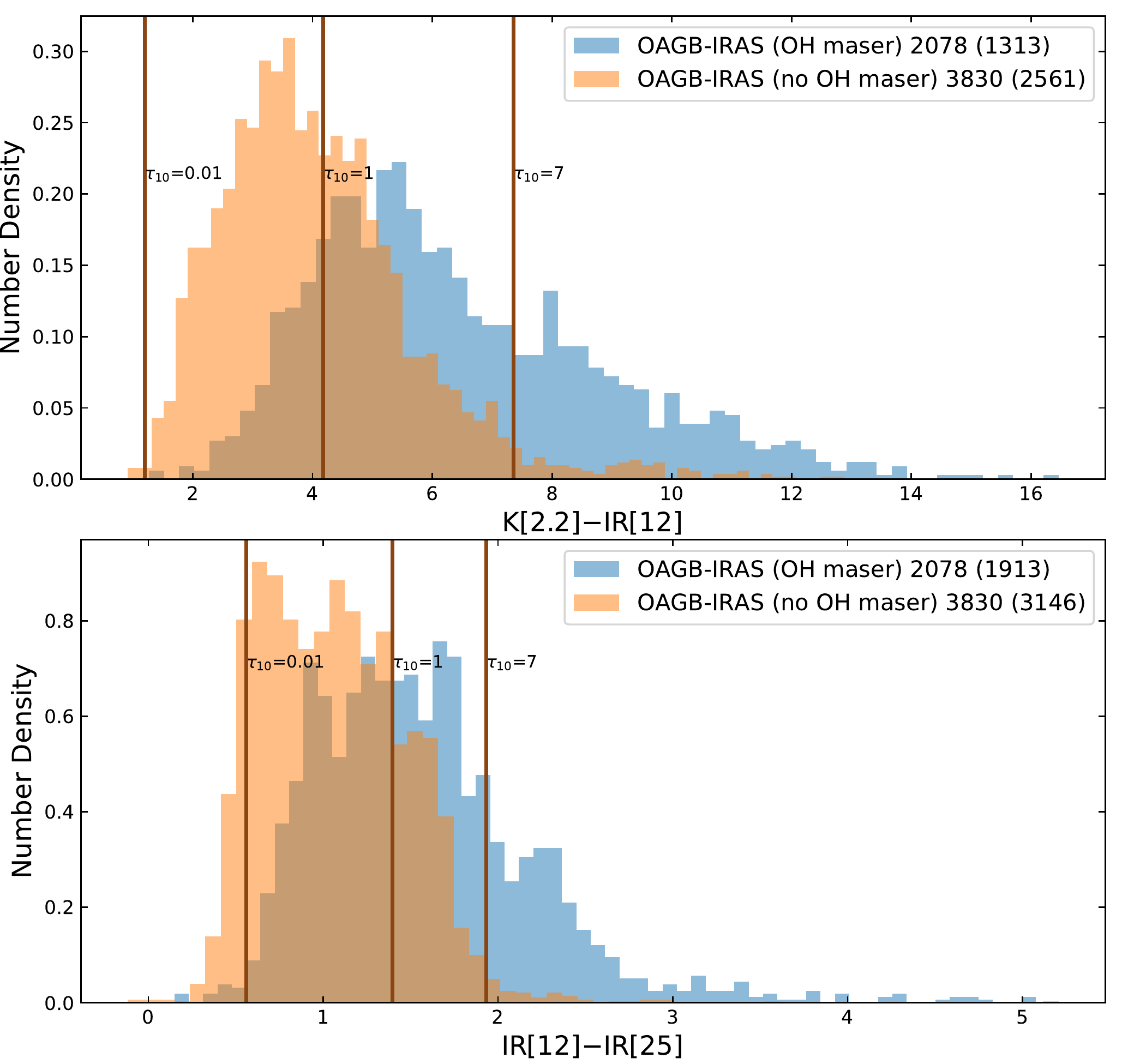}{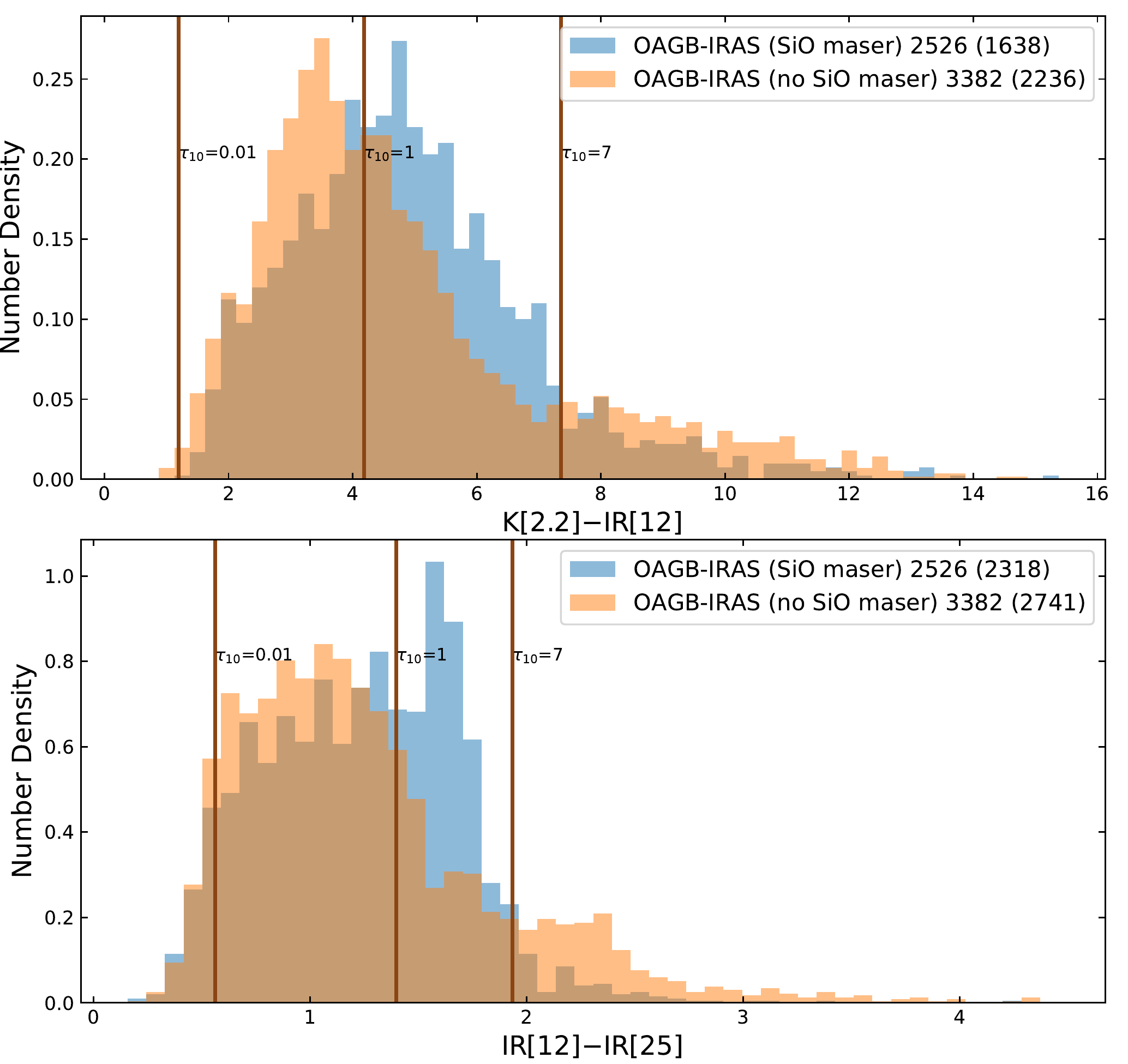}\caption{Number density distributions of observed IR colors for OAGB-IRAS sample stars and OH and SiO maser sources.
The group of objects, for which OH (or SiO) maser emission was detected, is denoted by 'OH (or SiO) maser' in parentheses. And the
group of objects, for which the maser emission was not detected or not observed, is denoted by 'no OH (or SiO) maser' in parentheses.
The vertical brown lines indicate theoretical OAGB model colors for three dust shell optical depths $\tau_{10}$.
For each class, the number of objects is shown. The number in parentheses denotes the number of plotted objects with good-quality observed colors.}
\label{f10}
\end{figure*}

\begin{figure*}
\centering
\smallplottwo{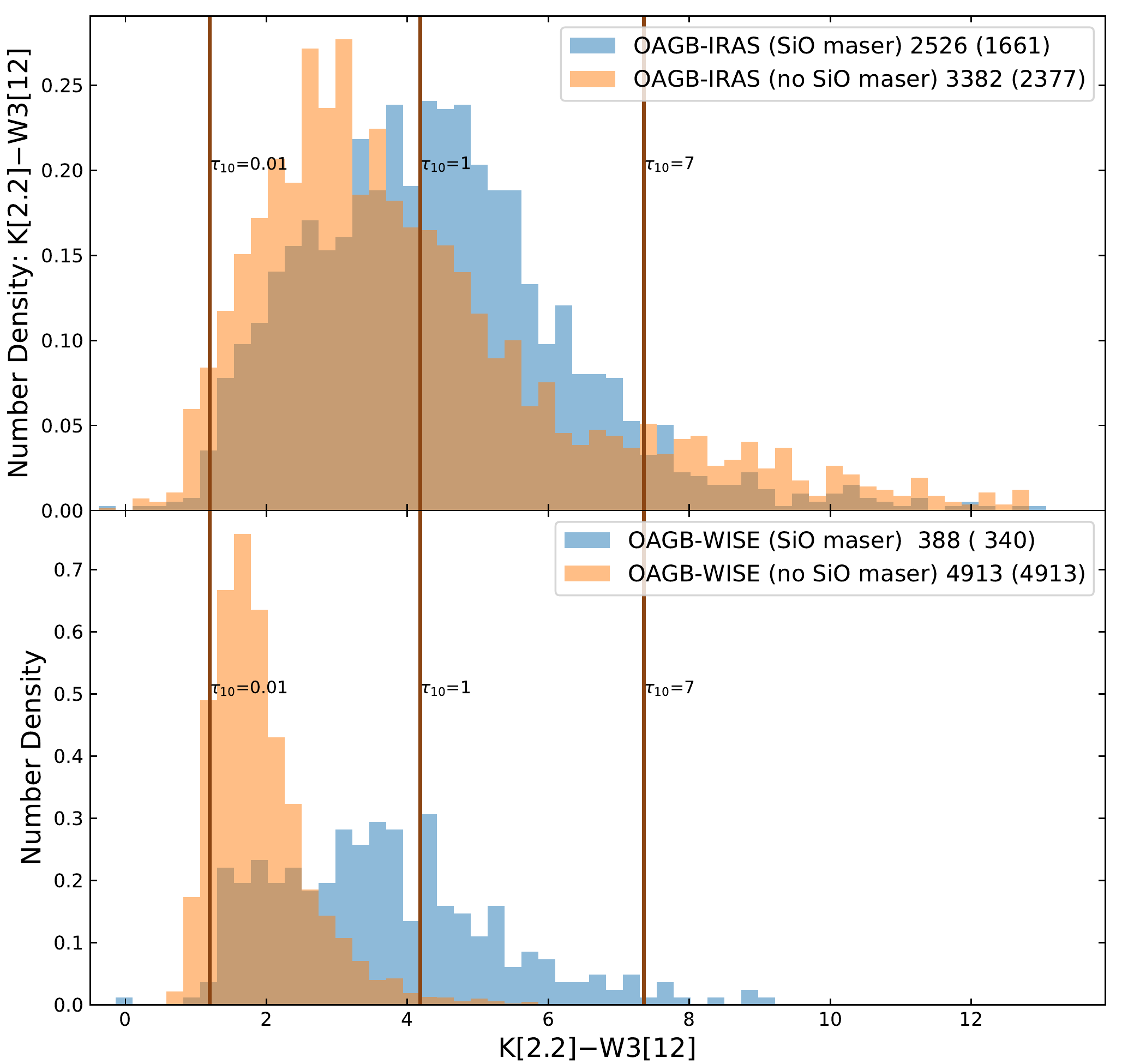}{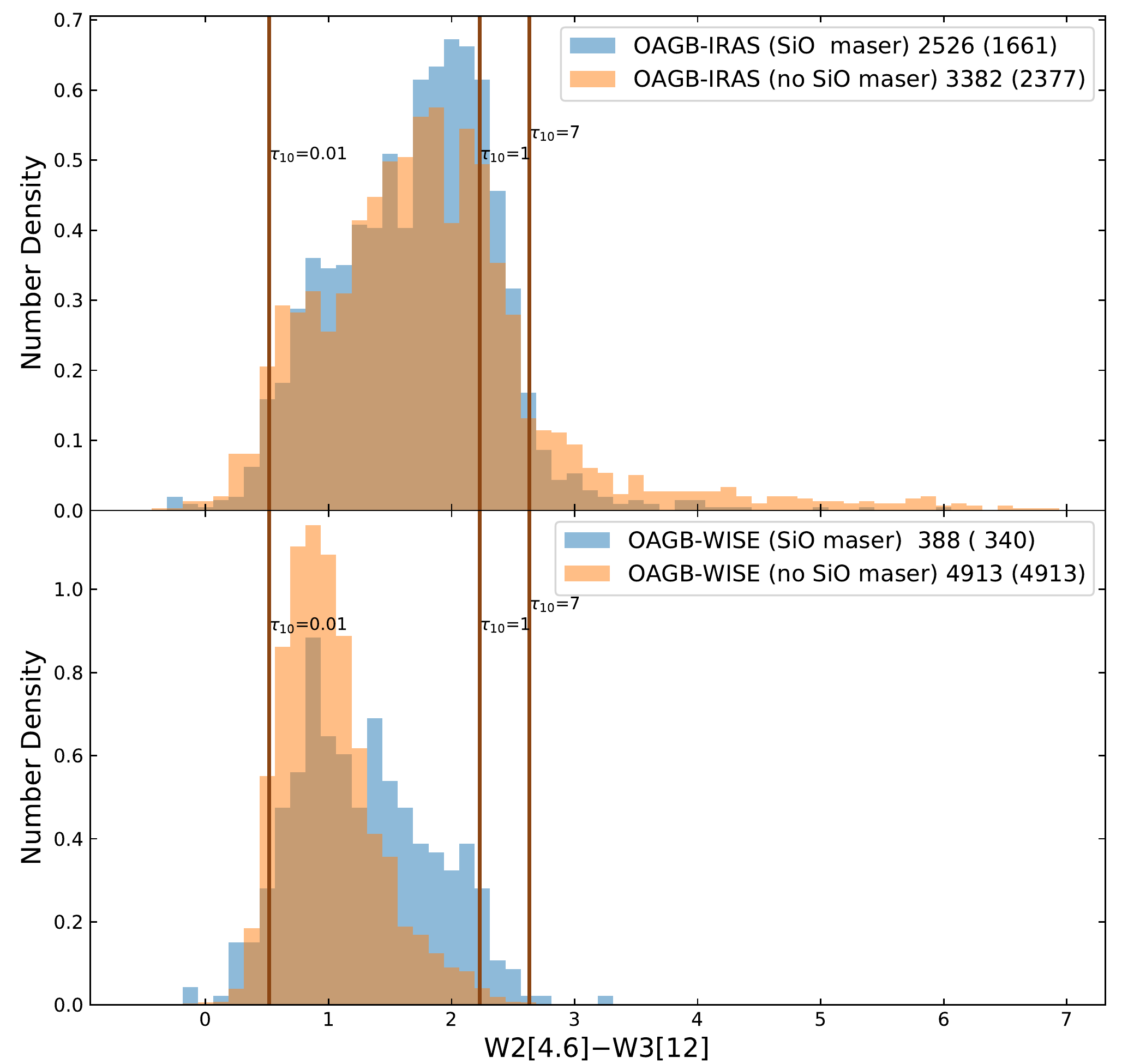}\caption{Number density distributions of observed IR colors for OAGB stars (OAGB-IRAS and OAGB-WISE) and SiO maser sources.
The group of objects, for which SiO maser emission was detected, is denoted by 'SiO maser' in parentheses. And the
group of objects, for which the maser emission was not detected or not observed, is denoted by 'no SiO maser' in parentheses.
The vertical brown lines indicate theoretical OAGB model colors for three dust shell optical depths $\tau_{10}$.
For each class, the number of objects is shown. The number in parentheses denotes the number of plotted objects with good-quality observed colors.}
\label{f11}
\end{figure*}

\begin{figure*}
\centering
\smallplotfour{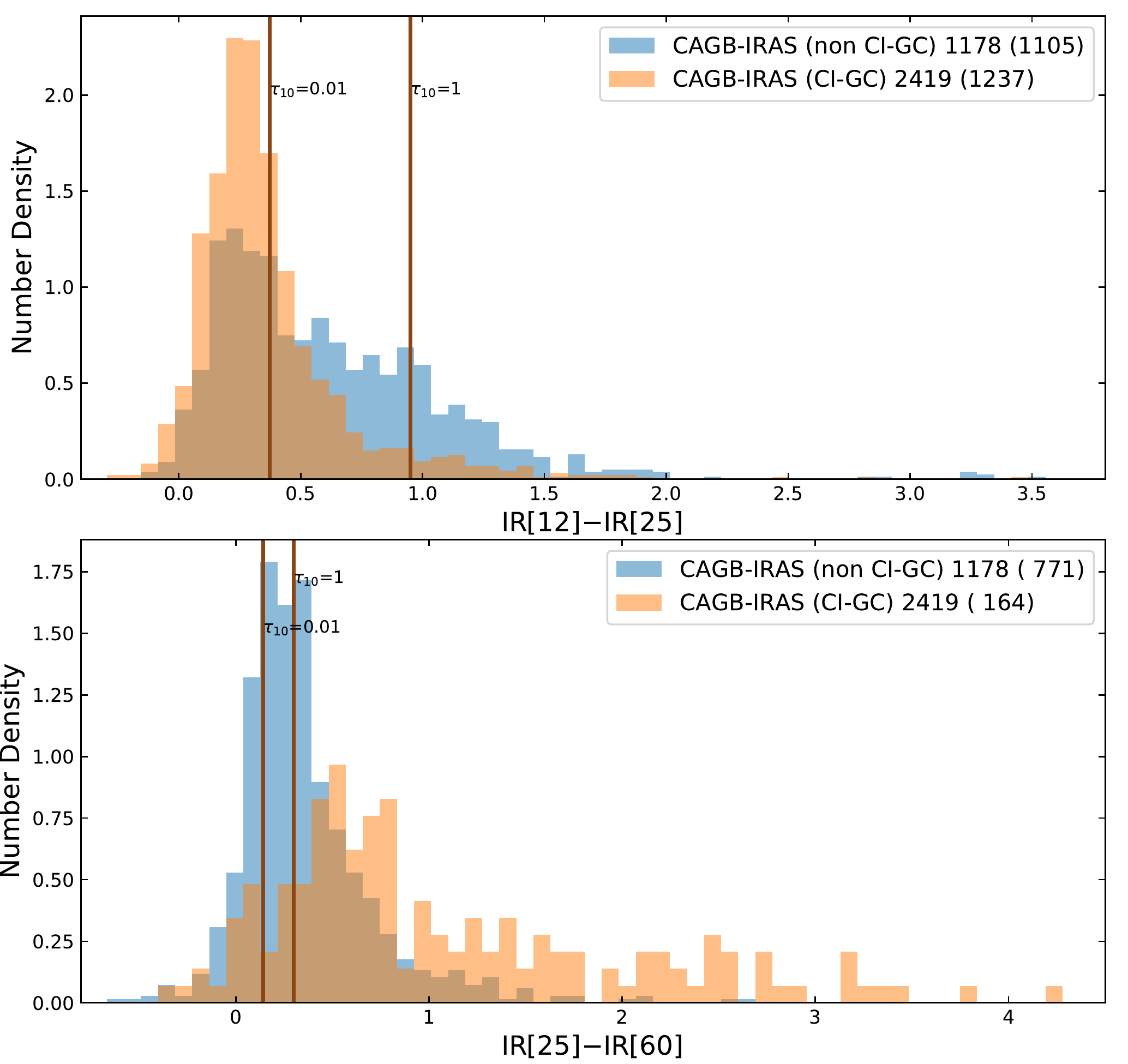}{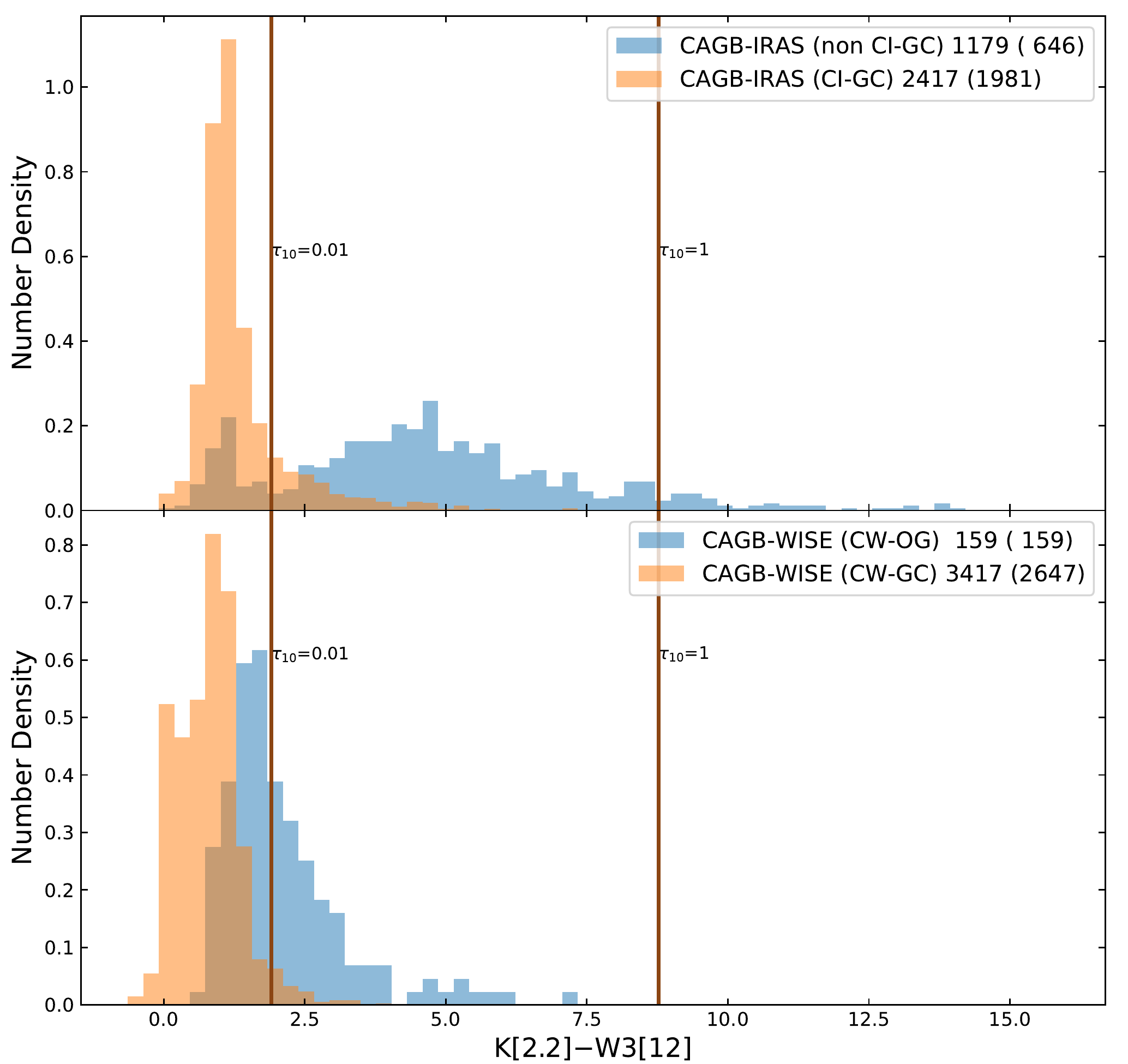}{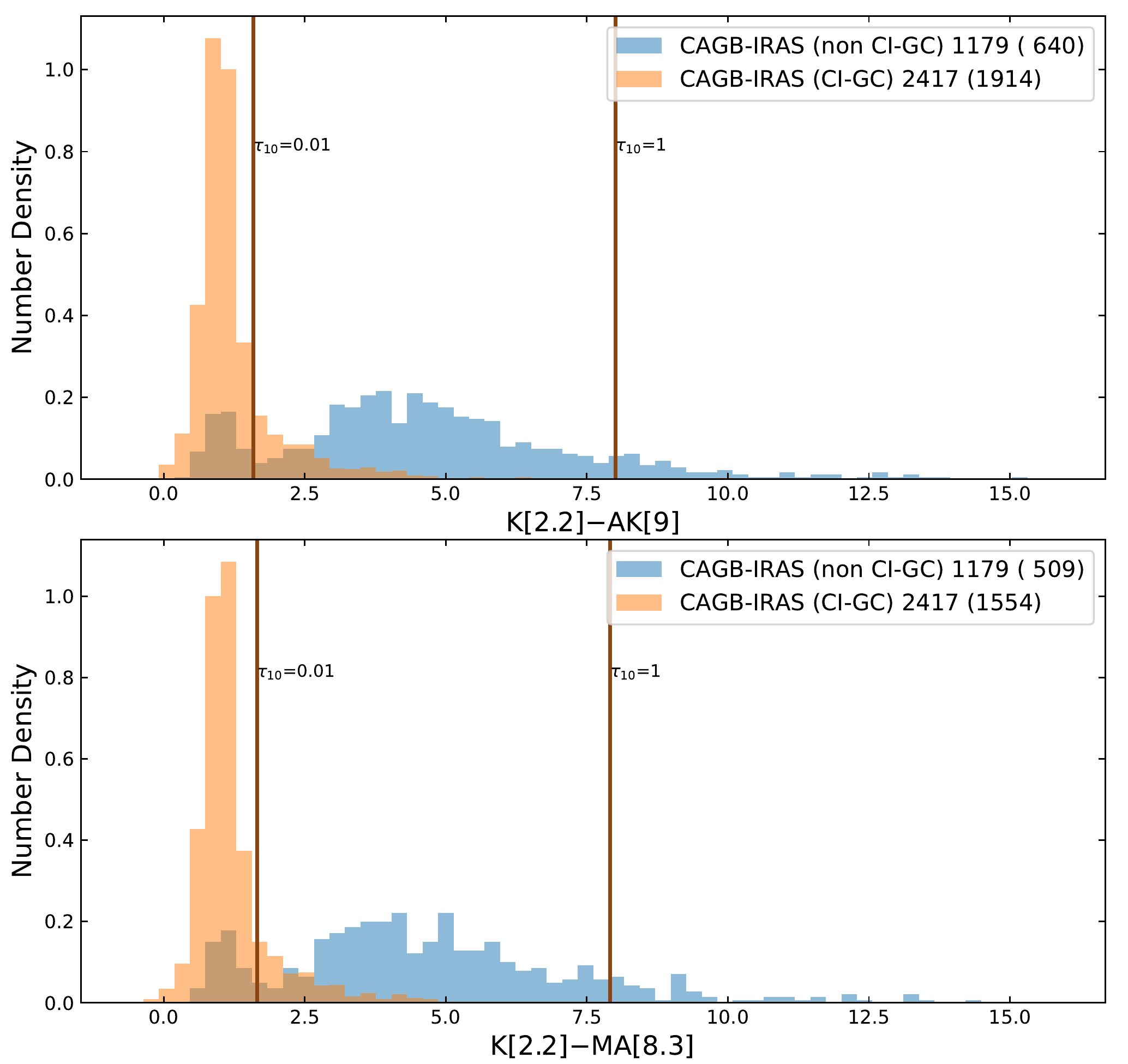}{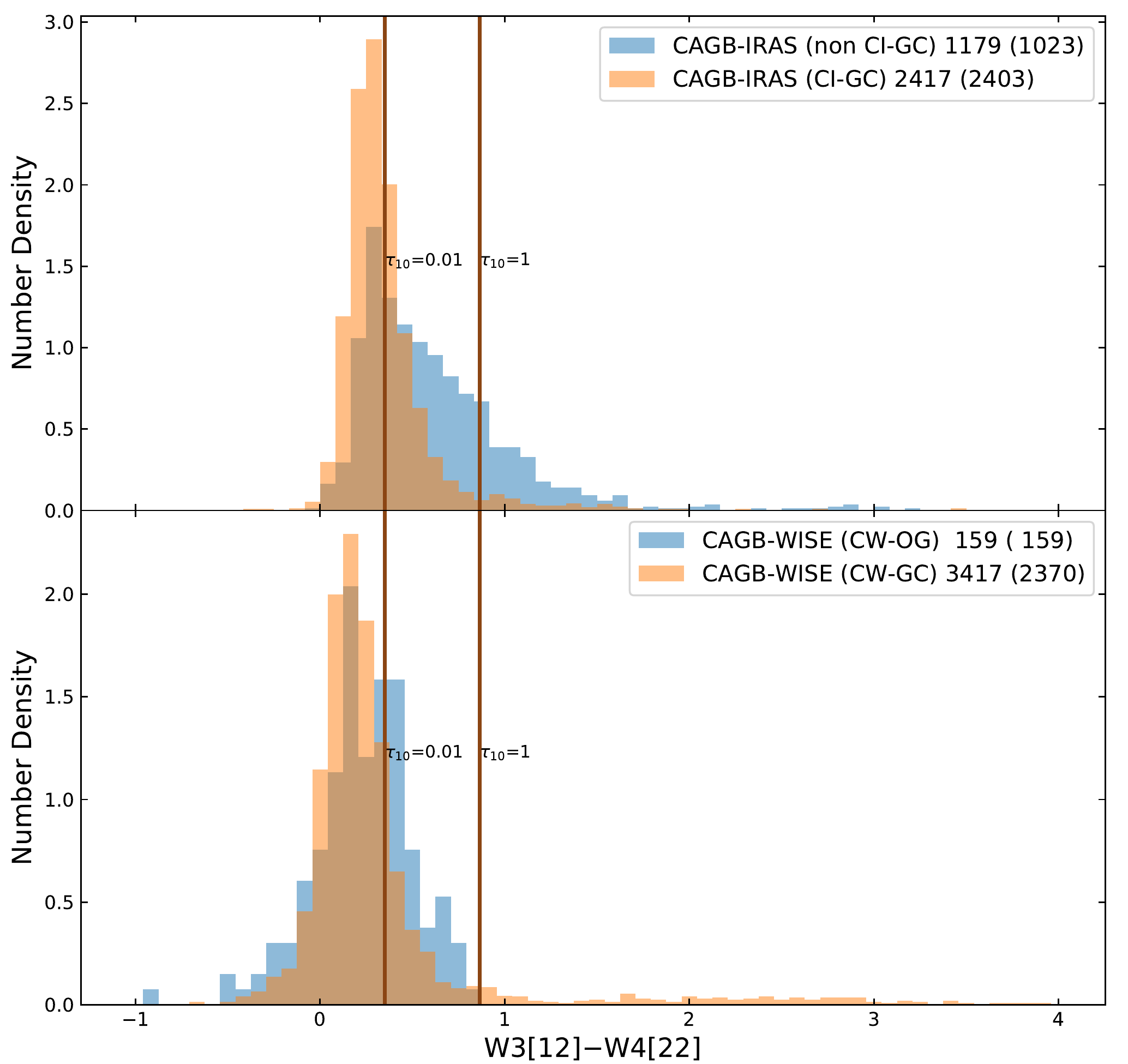}\caption{Number density distributions of observed IR colors for CAGB stars (CAGB-IRAS and CAGB-WISE) and visual carbon stars (CI-GC and CW-GC).
The vertical brown lines indicate theoretical CAGB model colors for two dust shell optical depths $\tau_{10}$.
For each class, the number of objects is shown. The number in parentheses denotes the number of plotted objects with good-quality observed colors.}
\label{f12}
\end{figure*}

\begin{figure*}
\centering
\smallplottwo{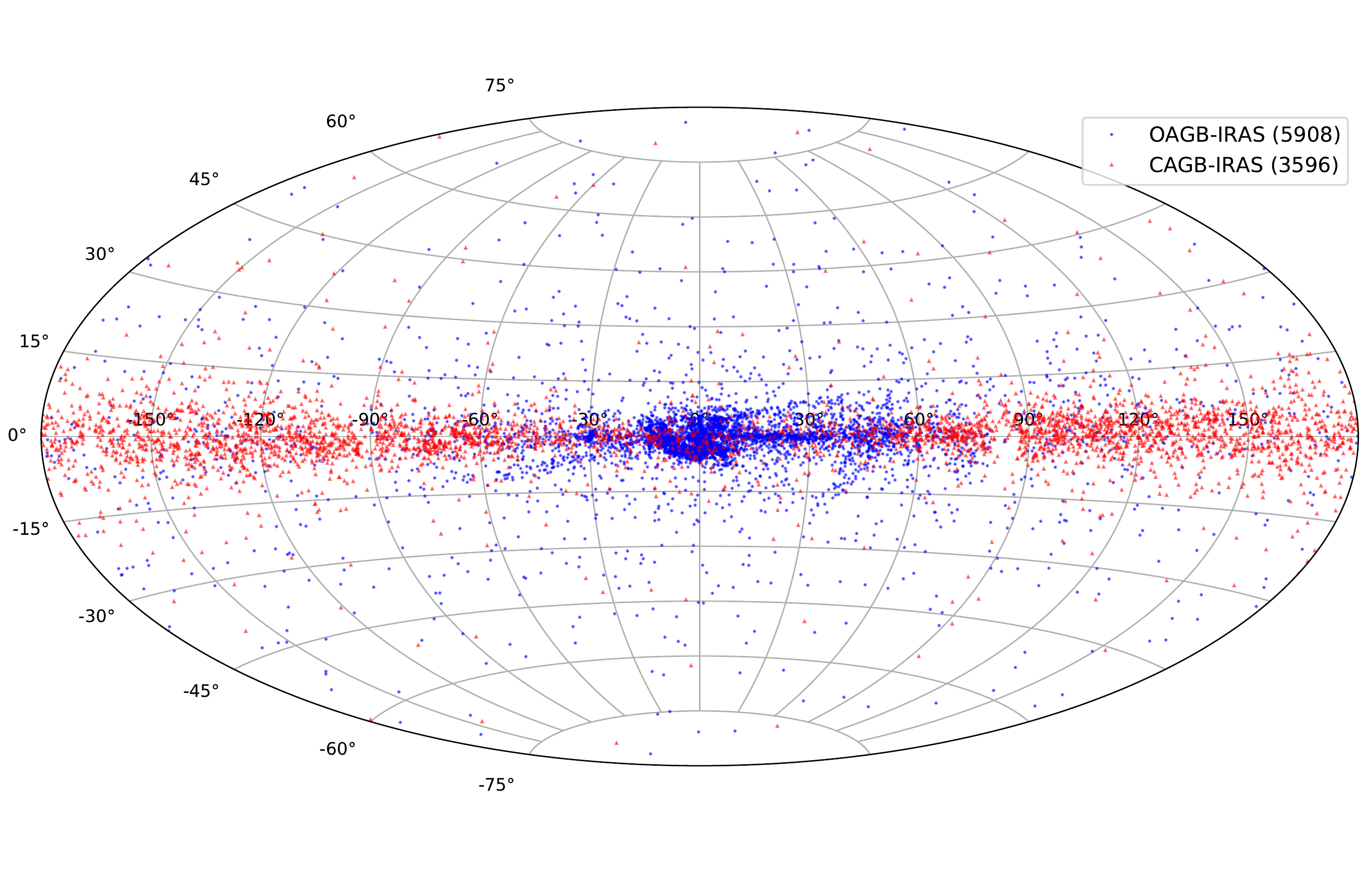}{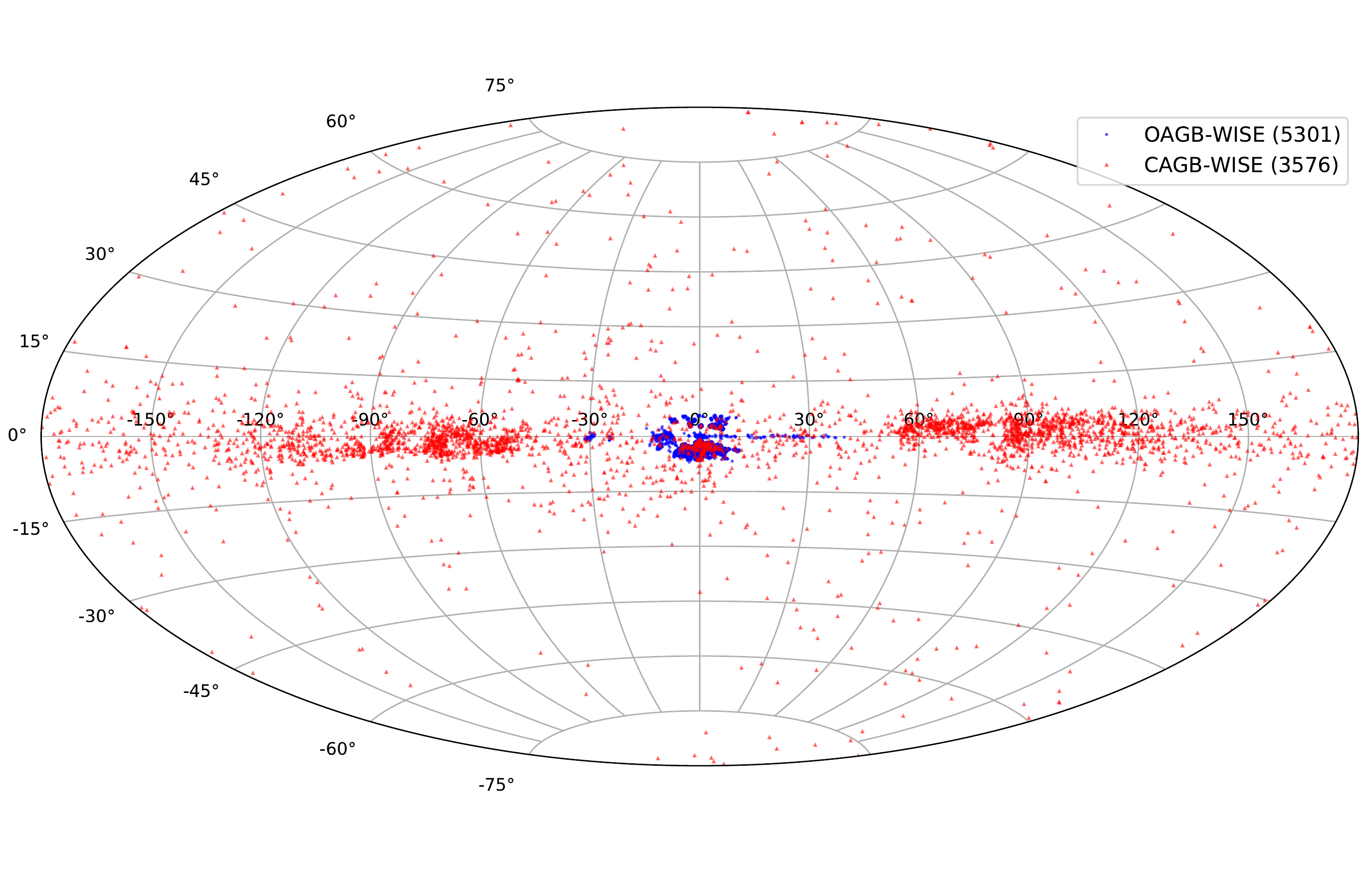}\caption{Spacial distributions AGB stars (AGB-IRAS and AGB-WISE) in Galactic coordinate.}
\label{f13}
\end{figure*}

\begin{figure*}
\centering
\smallplottwo{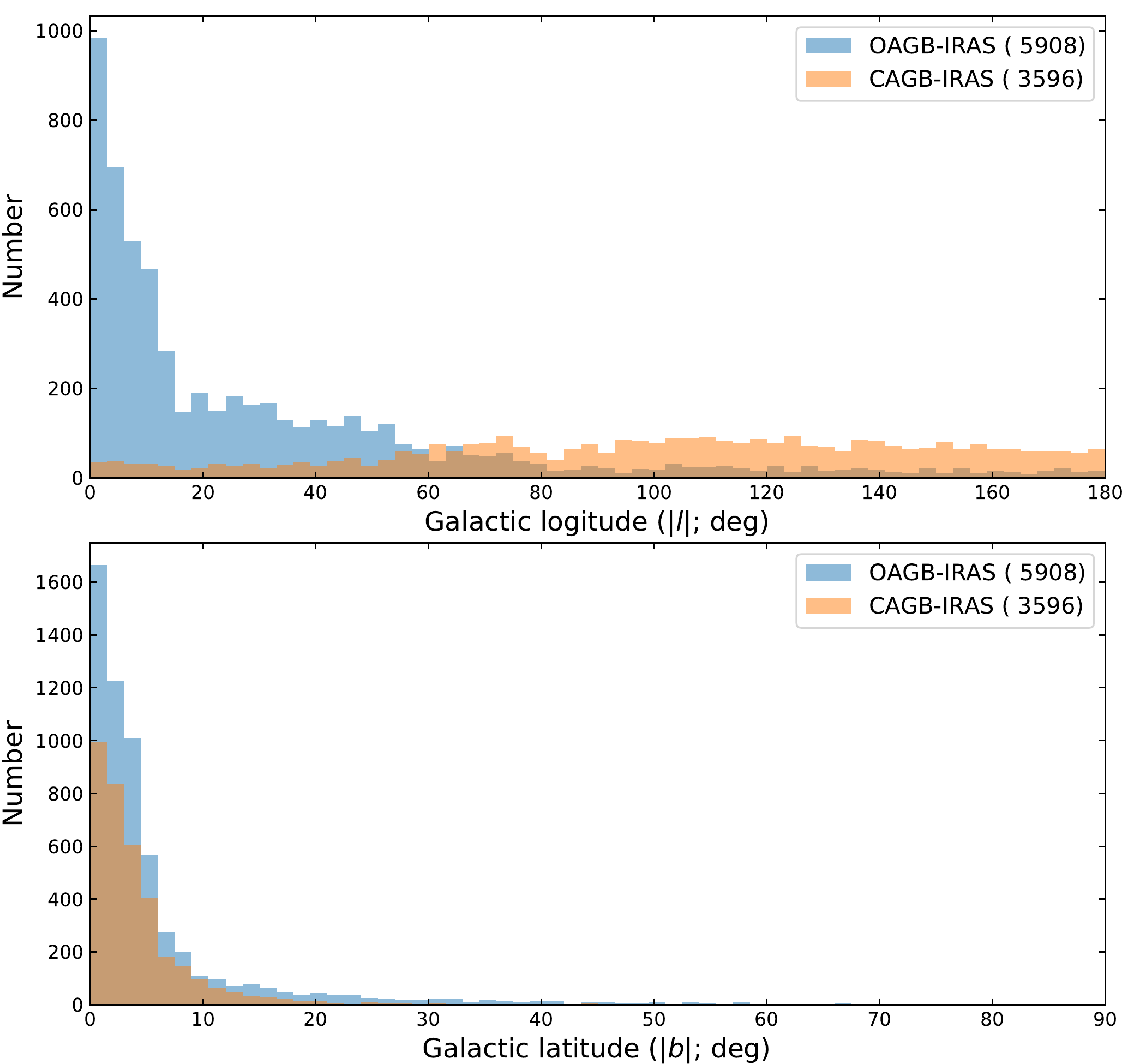}{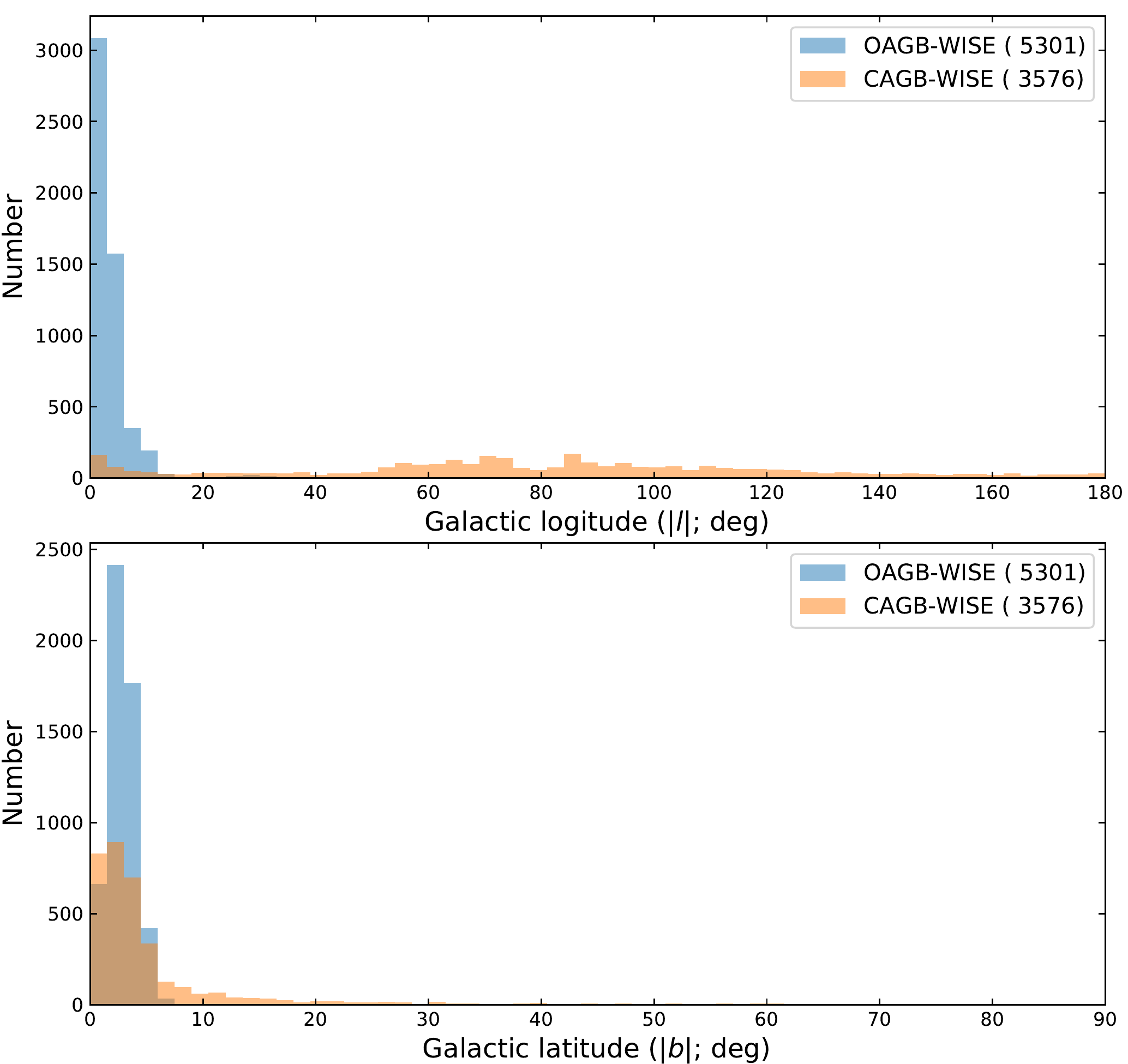}\caption{Number distribution of the Galactic longitude and latitude for OAGB and CAGB stars in AGB-IRAS and AGB-WISE catalogs.
The number in parentheses denotes the number of objects.}
\label{f14}
\end{figure*}

\begin{figure*}
\centering
\smallplottwo{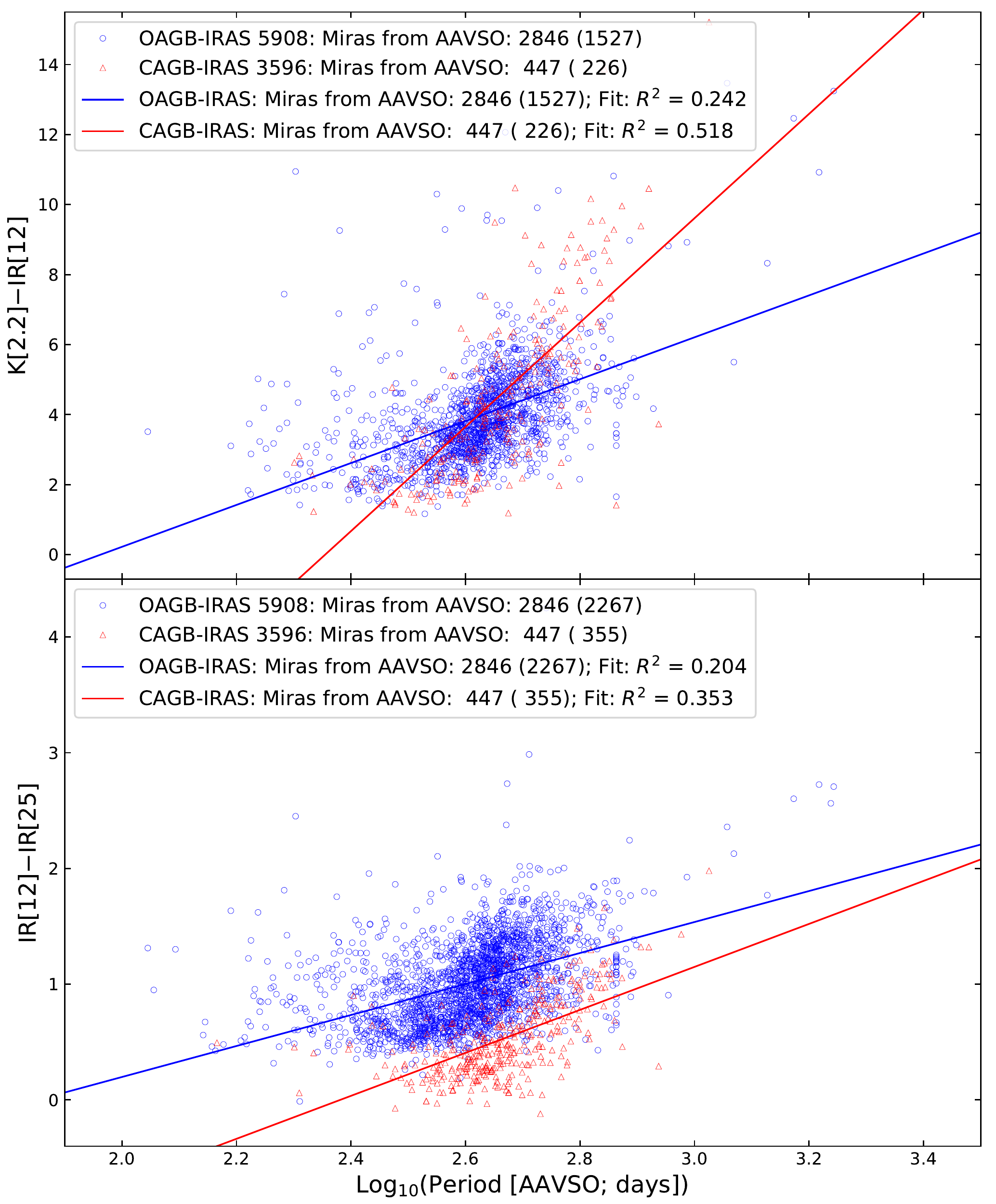}{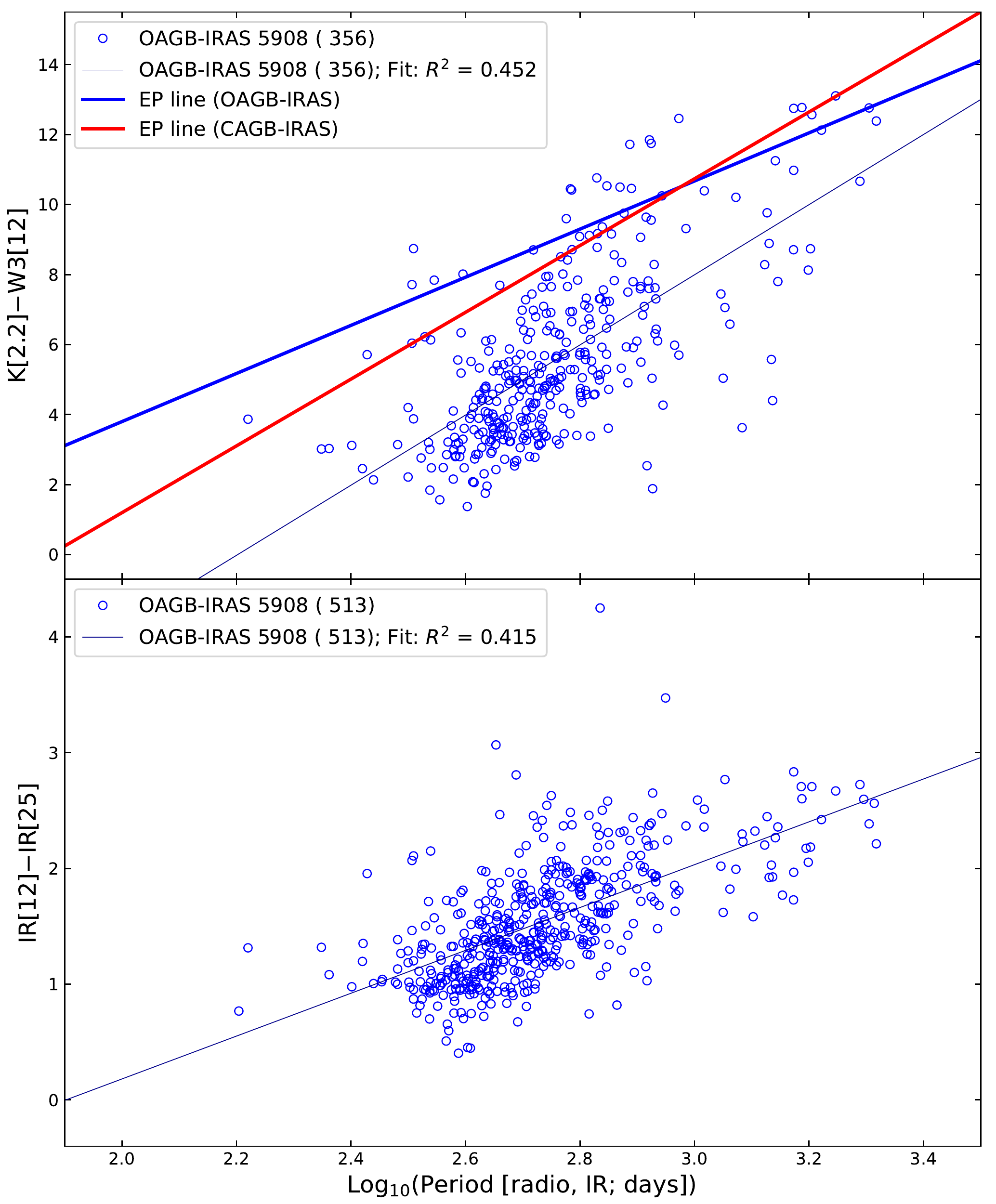}\caption{Left panels show period-color relations for AGB-IRAS objects known as Miras.
Right panels show period-color relations for 522 OAGB-IRAS objects with periods measured at NIR, MIR, and radio bands.
See Section~\ref{sec:pul}.
Note that the period-color relation in the upper-right panel is used to find an expected period (EP) in Section~\ref{sec:neo-nmc}.}
\label{f15}
\end{figure*}

\begin{figure*}
\centering
\xsmallplotfour{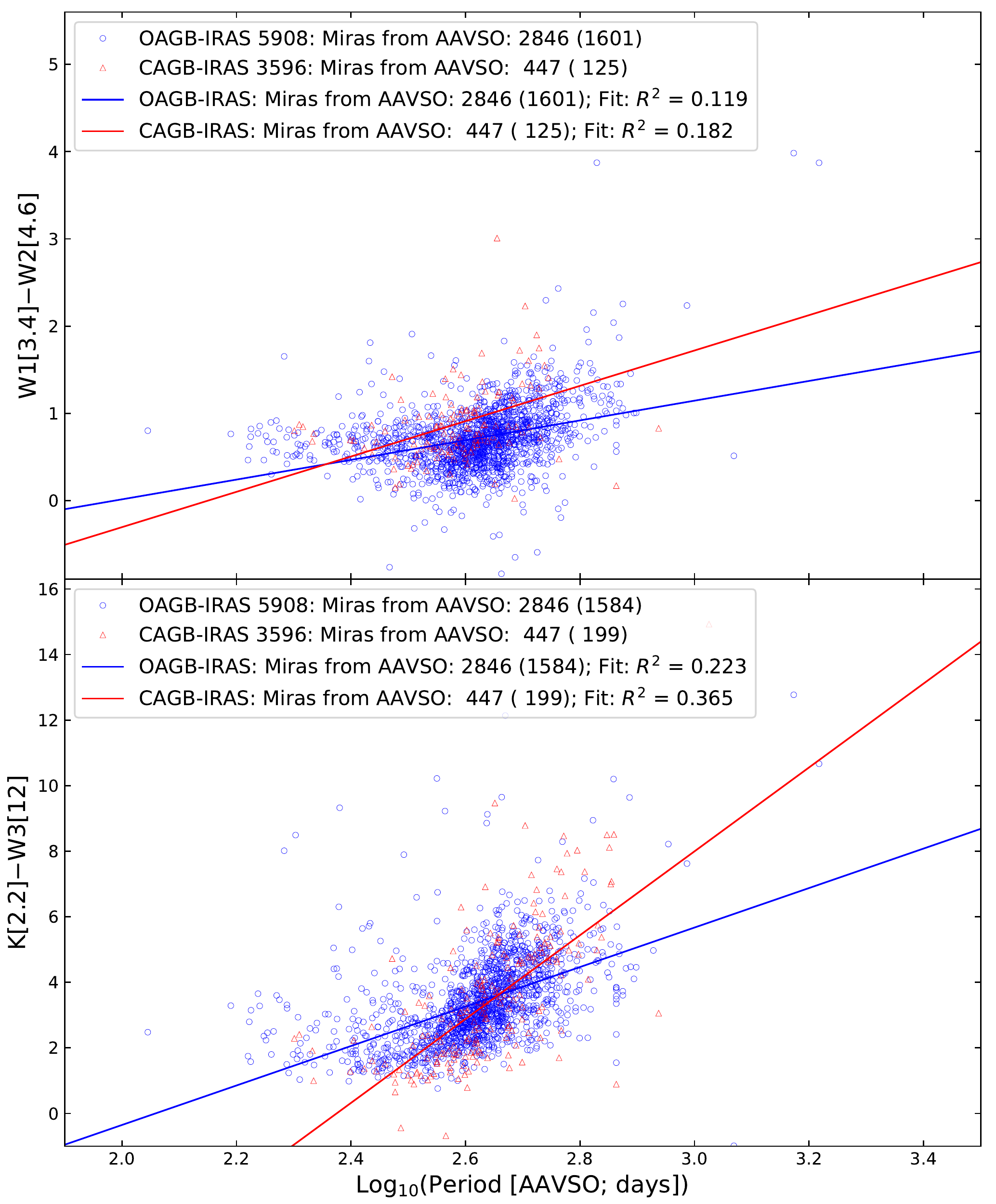}{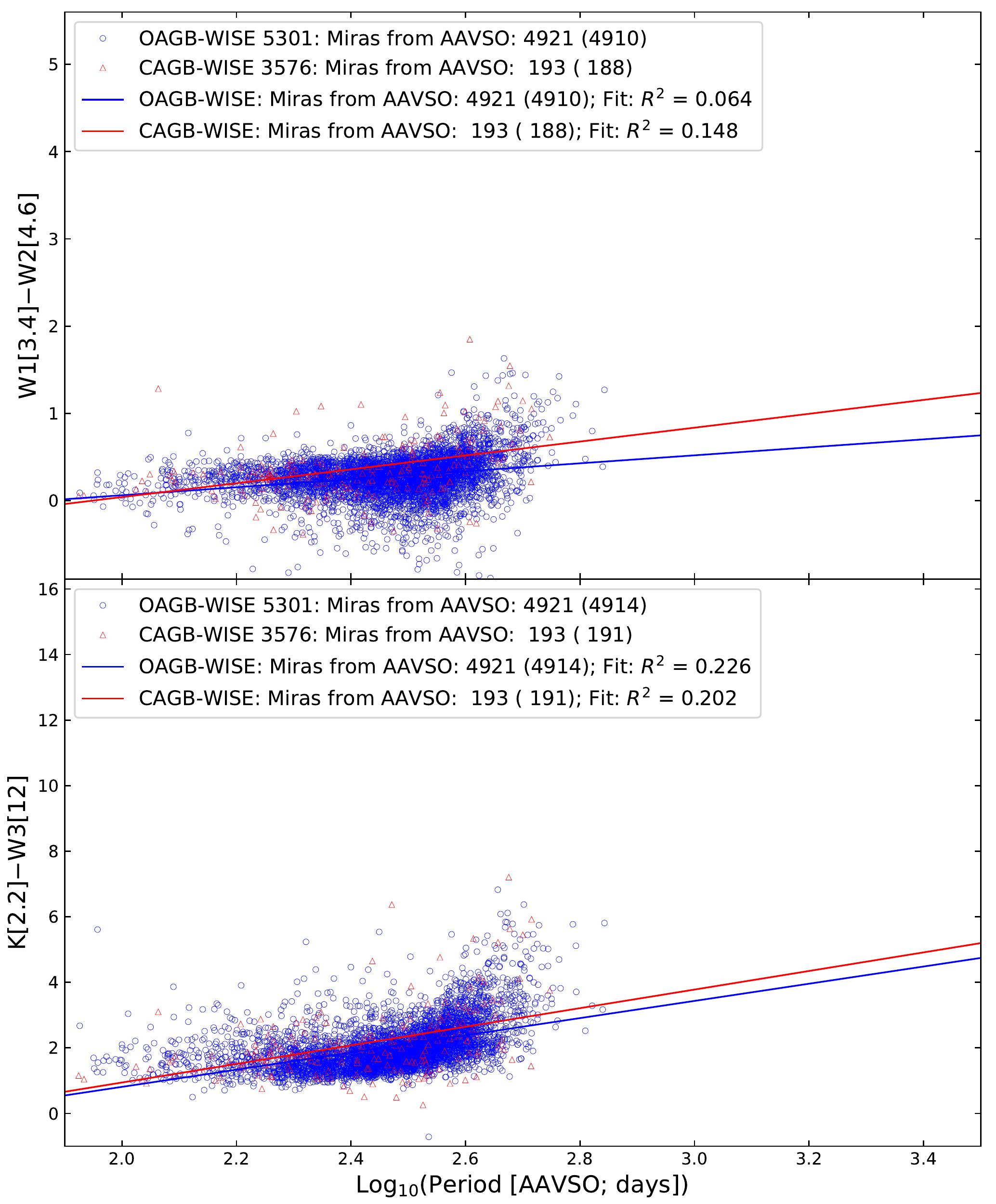}{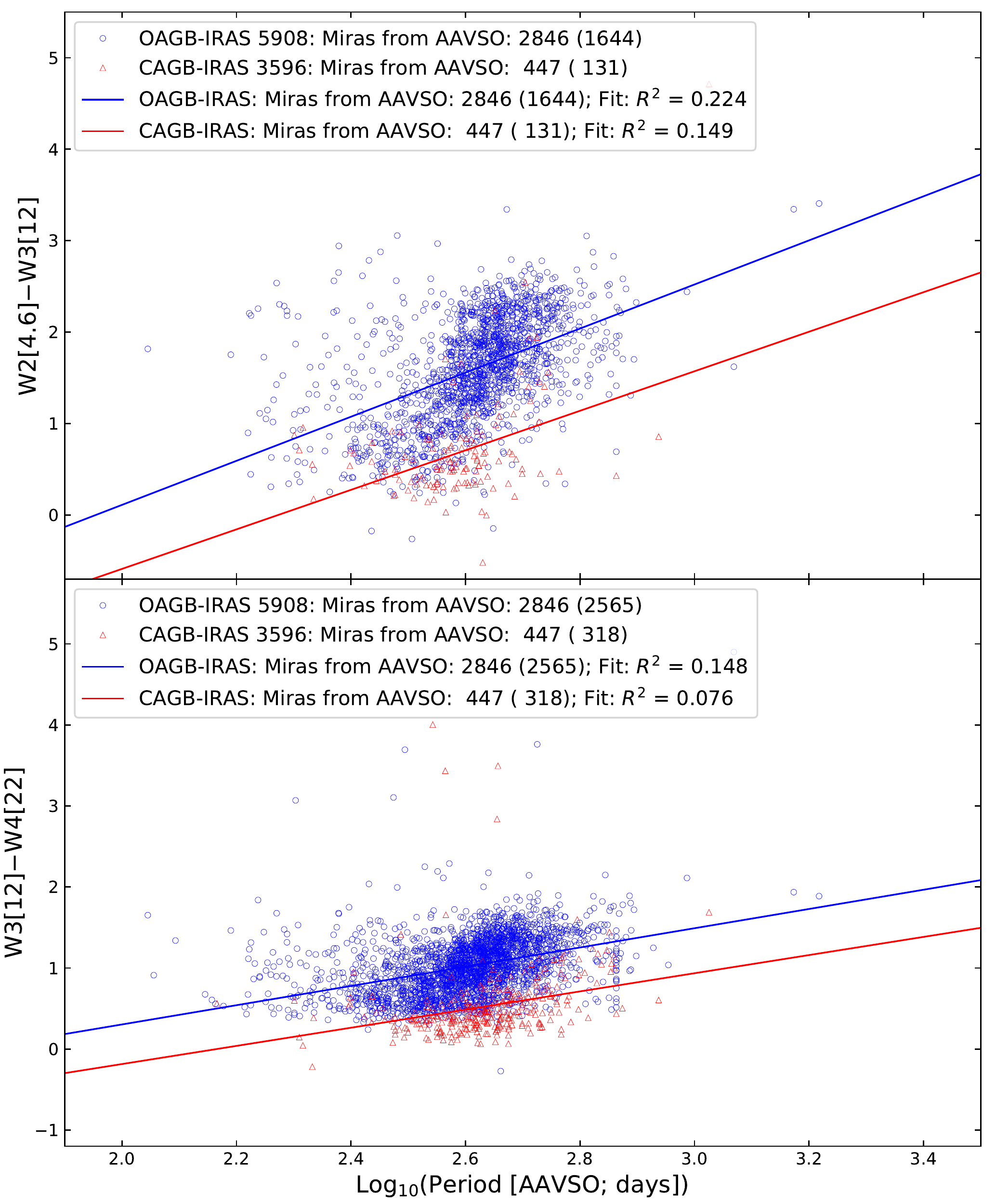}{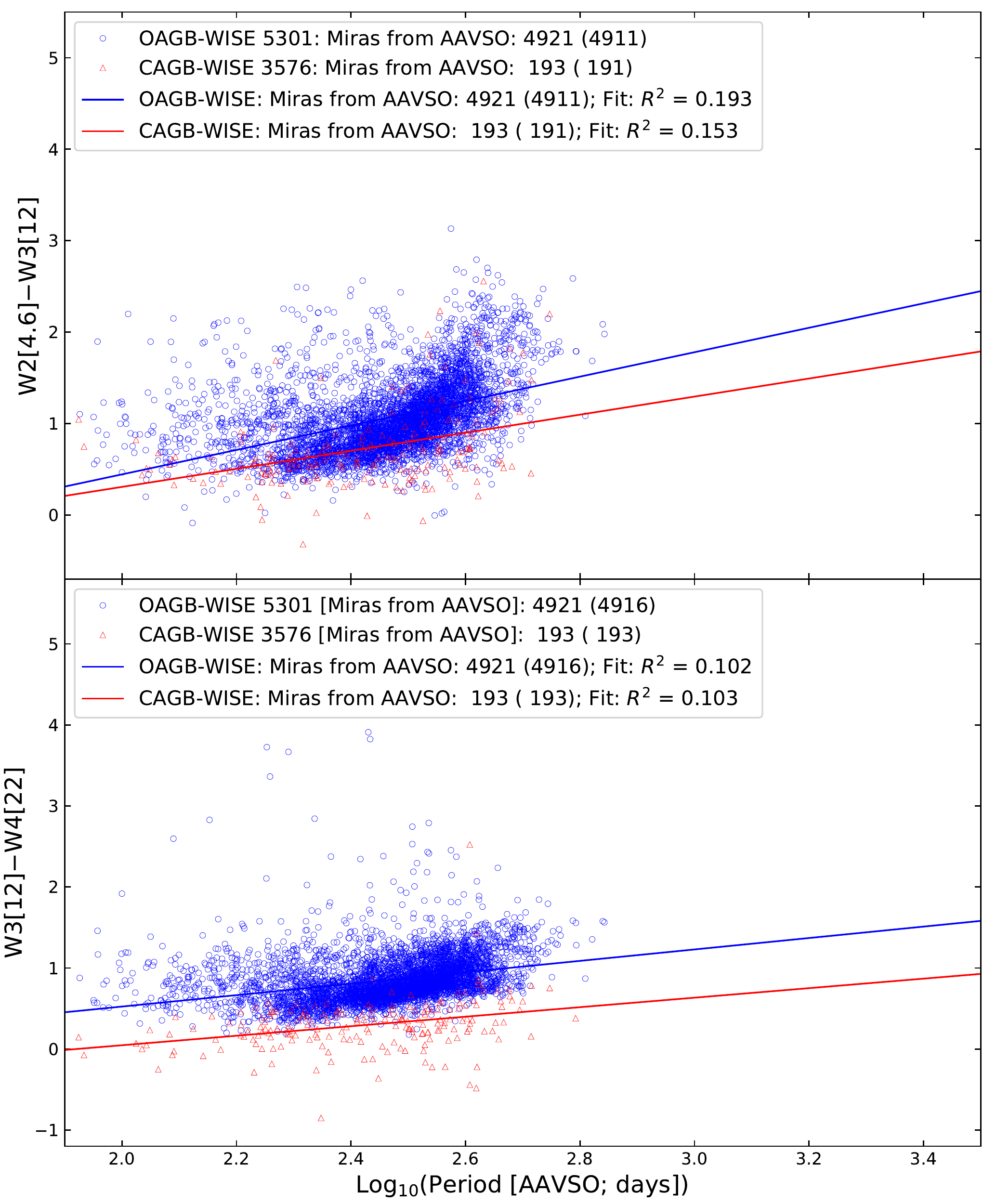}\caption{Period-color relations for AGB-IRAS and AGB-WISE objects known as Miras (AAVSO).
For each class, the number of objects is shown. The number in parentheses denotes the number of the plotted objects with good-quality observed data.
See Section~\ref{sec:pul}.}
\label{f16}
\end{figure*}

\begin{figure*}
\centering
\smallplotfour{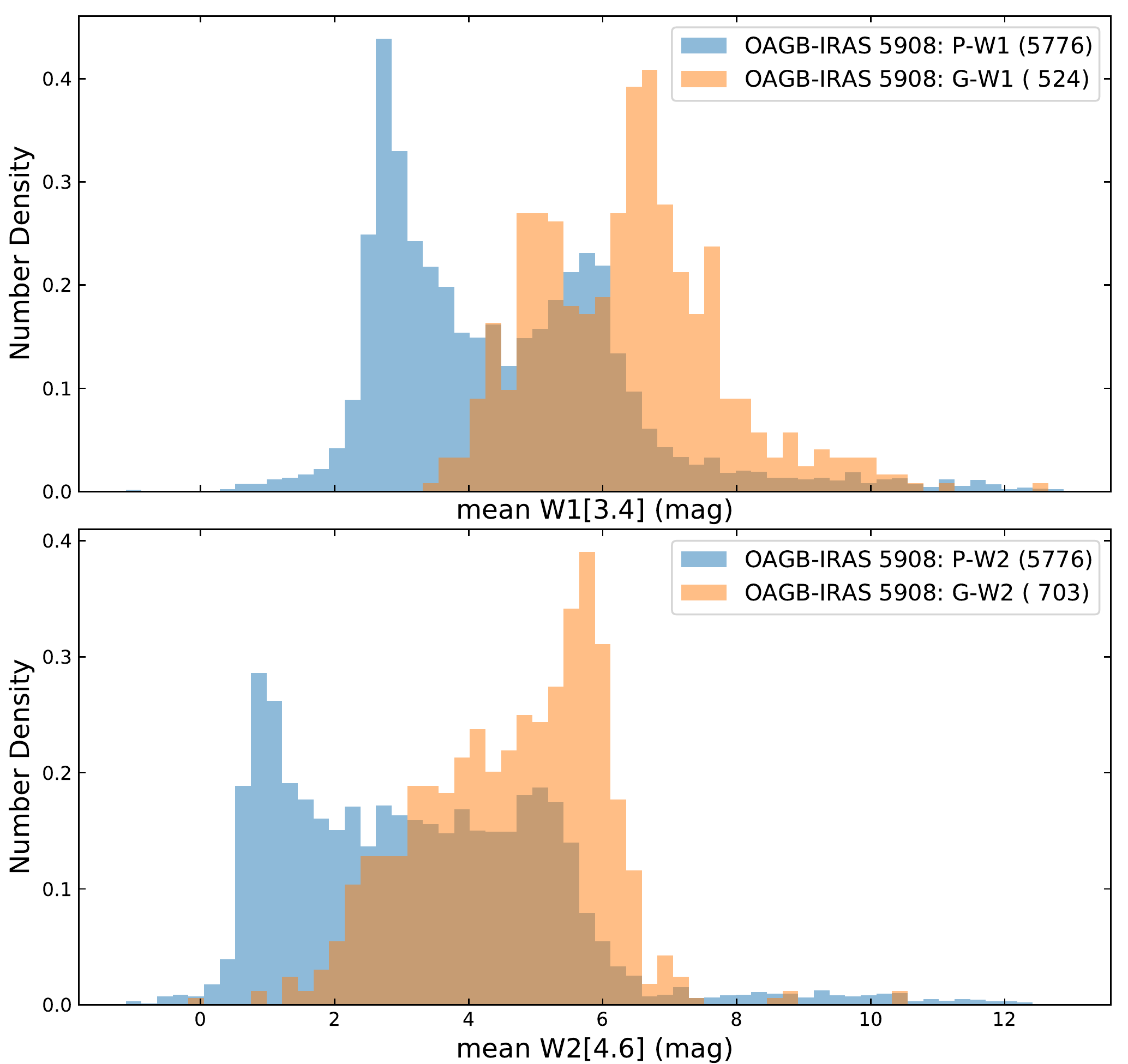}{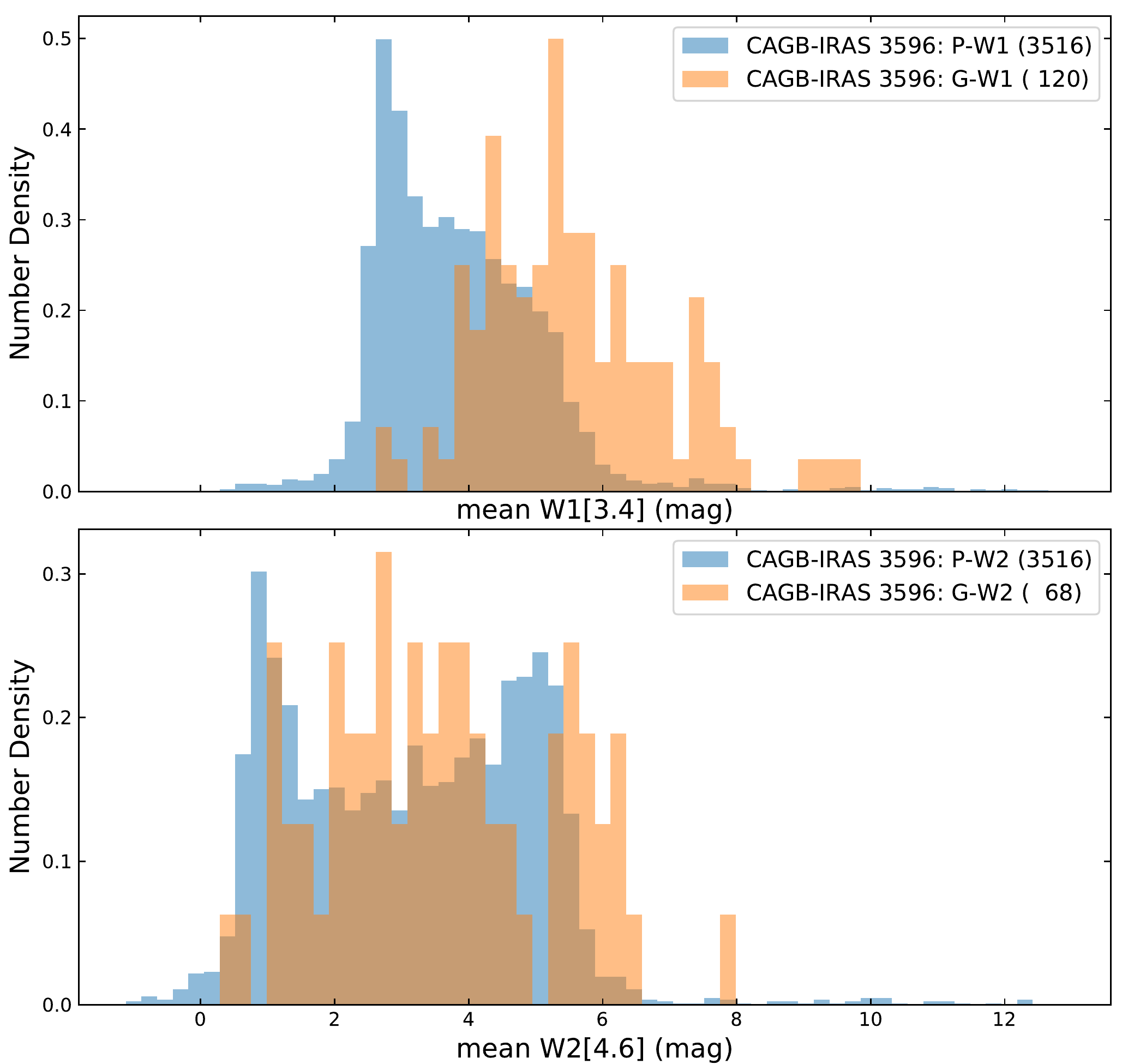}{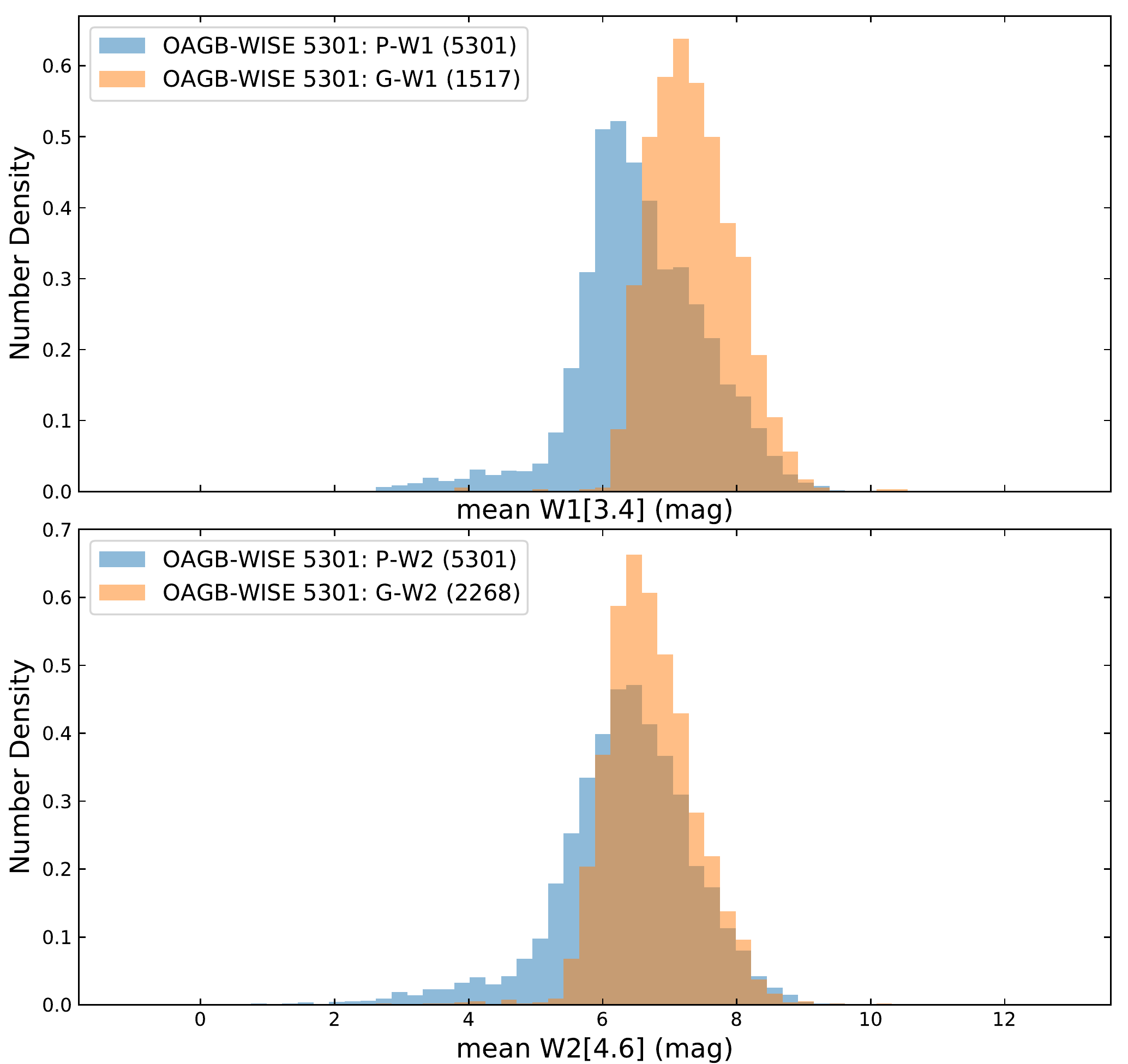}{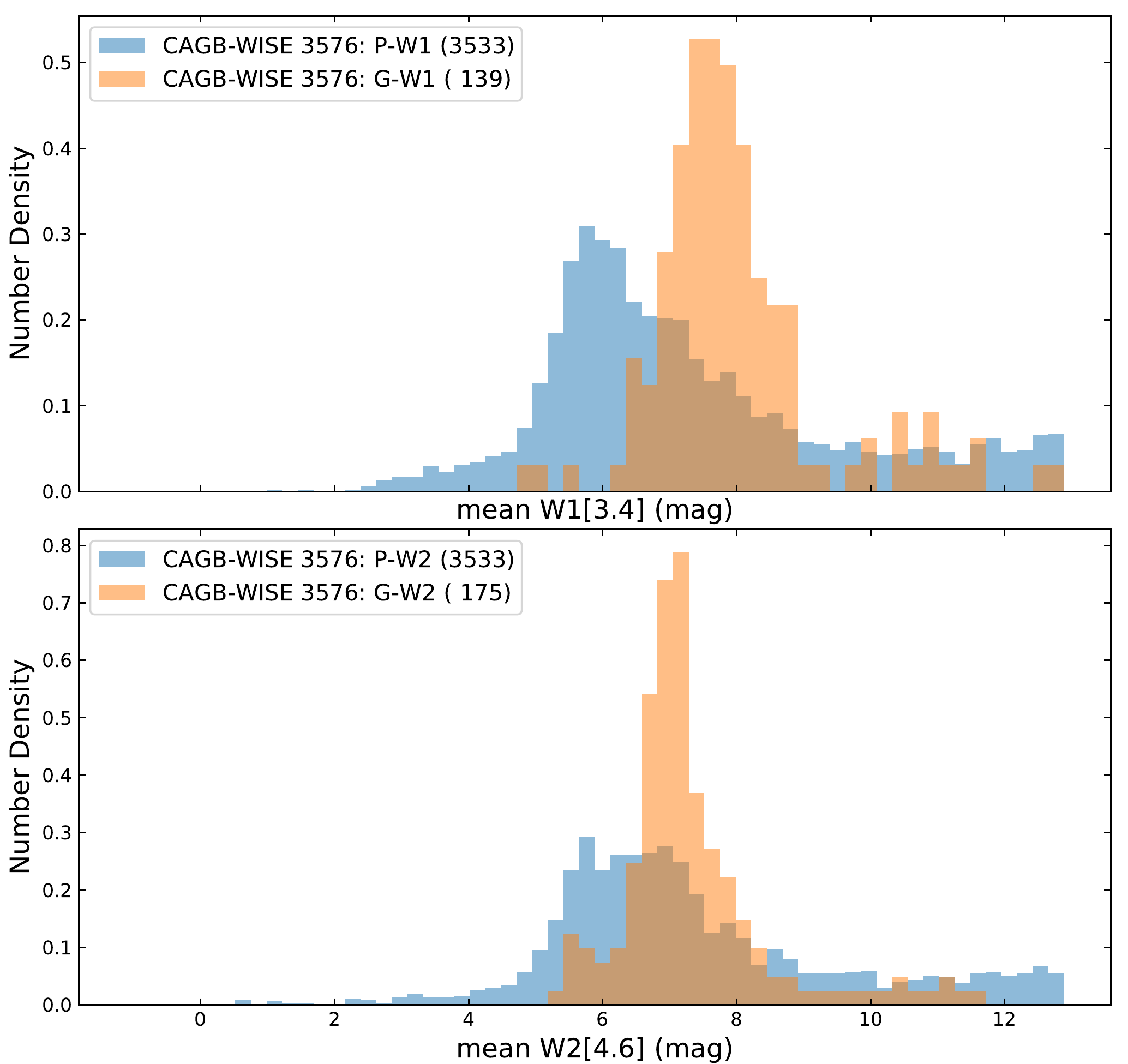}\caption{Number density distributions of magnitudes of the objects, for which the periods are obtained from the WISE light curves.
P-W1 (or P-W2) denotes objects whose periods are obtained from the WISE W1 (or W2) light curves.
G-W1 (or G-W2) denotes objects whose good-quality periods are obtained from the WISE W1 (or W2) light curves.
See Section~\ref{sec:neo} and Table~\ref{tab:tab8}.}
\label{f17}
\end{figure*}

\section{Number distributions of IR colors\label{sec:cnumber}}

We may compare the number distribution of observed IR colors for different
classes (or subgroups) of AGB stars to find differences in the IR properties.
We may also compare the number distributions with the theoretical model colors.

\subsection{OAGB and CAGB stars\label{sec:OCAGB}}

Figure~\ref{f9} shows number density distributions of observed IR magnitudes at
W2[4.6] and W3[12] bands and colors for AGB stars in the AGB-IRAS and AGB-WISE
catalogs. We choose two IR colors: W1[3.4]$-$W2[4.6] and K[2.2]$-$W3[12] (see
Figure~\ref{f6}).

Generally, AGB-IRAS objects are brighter at MIR bands and show redder IR colors
than AGB-WISE objects. For both IR colors (W1[3.4]$-$W2[4.6] and
K[2.2]$-$W3[12]), CAGB stars are generally bluer than OAGB stars. We also find
that AGB-WISE objects are more concentrated toward the bluer colors and the
number decreases with the redder colors, whereas numbers for AGB-IRAS objects
are more dispersed into redder colors. In general, AGB-IRAS objects look to be
more evolved (or more massive) stars with thicker dust envelopes than AGB-WISE
objects.

\subsection{OH and SiO Maser sources among OAGB stars\label{sec:maser}}

A major portion of OAGB stars are SiO, H$_2$O, OH maser sources (e.g.,
\citealt{ks2012}). We use the lists of SiO maser sources compiled by
\citet{ks2012} and new SiO maser sources detected from the Galactic bulge
(\citealt{messineo2018}; \citealt{wu2018}; \citealt{stroh2019}; see
Section~\ref{sec:galb}). We have also compiled SiO maser sources detected by
the Korean VLBI Network single-dish telescopes (\citealt{kim2010};
\citealt{cho2012}; \citealt{kim2013}; \citealt{yoon2014}; \citealt{cho2017}).
There are 2527 and 388 SiO maser sources in OAGB-IRAS and OAGB-WISE catalogs,
respectively (see Tables~\ref{tab:tab1} and \ref{tab:tab2}). In the OAGB-WISE
catalog, all 388 SiO maser sources are in OW-ME and OW-ST.

OH/IR stars are generally considered to be more massive OAGB stars with thicker
dust envelopes and higher mass-loss rates (\citealt{engels1983};
\citealt{blommaert2018}). \citet{chen2001} presented a list of 1065 OH/IR stars
in our Galaxy. The list has been corrected and updated (\citealt{sk2011};
\citealt{ks2012}). We also use the catalog of OH/IR stars presented by
\citet{engels2015}. There are 2079 OH/IR stars in OAGB-IRAS (1988 in OI-SH, 17
in OI-UR, 3 in OI-JB, 1 in OI-ME, 20 in OI-ST, 18 in OI-OG, 1 in OI-WU, and 31
in OI-AM; see Table~\ref{tab:tab1}), but there is only one known OH/IR star (OH
17.434-0.077) in OAGB-WISE (OW-ME) yet.

Figure~\ref{f10} shows number distributions of IR colors (K[2.2]$-$IR[12] and
IR[12]$-$IR[25]). Most OH maser sources are in the range of large dust optical
depths (or mass-loss rates). However, most SiO maser sources are in the range
of moderate dust optical depths (or mass-loss rates) for both colors.

We mark the theoretical model colors on Figures~\ref{f10} and \ref{f11}. For
OAGB stars, the dust shell (silicate; $T_c$=1000 K) model colors for typical
LMOA stars ($\tau_{10}$=0.1 and 1) and HMOA stars ($\tau_{10}$=7) are indicated
(see Section~\ref{sec:models}).

Figure~\ref{f11} shows number distributions of IR colors (K[2.2]$-$W3[12] and
W2[4.6]$-$W3[12]) for SiO maser sources and undetected (or unobserved) sources
among all of the OAGB-IRAS and OAGB-WISE sample stars. Again, most SiO maser
sources in OAGB-IRAS are in the range of moderate dust optical depths (or
mass-loss rates) for both colors. Because the OAGB-WISE class lacks in objects
with thick dust envelopes (or redder IR colors), SiO maser sources look to be
in the range of relatively large dust optical depths for the IR colors. Note
that SiO maser observations have not been performed yet for a major portion of
OAGB-WISE objects, which are OGLE3 Miras in the Galactic bulge (OW-OG; see
Table~\ref{tab:tab2}).

\subsection{Visual carbon stars\label{sec:c-carbon}}

Unlike other subgroups of CAGB stars (in CI-SH, CI-UR, CI-OG, and CW-OG; see
Tables~\ref{tab:tab1}-\ref{tab:tab2}), most objects in CI-GC and CW-GC (from
\citealt{alksnis2001}) are believed to be visual carbon stars in the AGB phase.

To find the differences of the IR colors of visual carbon stars from the colors
of other groups that contain a mixture of infrared carbon stars and visual
carbon stars, we compare the histograms in Figure~\ref{f12}.

We mark the theoretical model colors on Figure~\ref{f12}. For CAGB stars, the
dust shell (AMC; $T_c$=1000 K) model colors for the thin dust shell
($\tau_{10}$=0.1) and thick dust shell ($\tau_{10}$=1) model colors are
indicated (see Section~\ref{sec:models}).

The left panels of Figure~\ref{f12} show number density distributions of
observed IR colors (IR[12]$-$IR[25], IR[25]$-$IR[60], K[2.2]$-$AK[9], and
K[2.2]-MA[8.3]) for CAGB-IRAS objects. Compared with others, visual carbon
stars (in CI-GC) show bluer IR[12]$-$IR[25], K[2.2]$-$AK[9], and
K[2.2]$-$MA[8.3]) colors and but redder IR[25]$-$IR[60] colors. This would be
because visual carbon stars show excessive flux at 60 $\mu$m due to the remnant
of an earlier phase when they were OAGB stars (see Section~\ref{sec:iras-2cd}).

The right panels of Figure~\ref{f12} show number density distributions of
observed IR colors (K[2.2]$-$W3[12] and W3[12]$-$W4[22]) for CAGB-IRAS and
CAGB-WISE objects. For K[2.2]$-$W3[12], non CI-GC objects are in the wide
ranges of large dust optical depths, whereas CI-GC objects are in narrow
ranges. For CAGB-WISE objects, the difference gets smaller because most of the
non CW-GC objects (in CW-OG) are likely to CAGB stars with thin dust shells.

For W3[12]$-$W4[22], the number density distributions for CAGB-IRAS objects are
similar to those for IR[12]$-$IR[25]. For CAGB-WISE objects, some CW-GC objects
show redder colors. This would be because visual carbon stars with detached
dust shell ($T_c$ $<$ 500 K; remnant of an earlier phase when the stars were
OAGB stars) may show redder W3[12]$-$W4[22] (see Section~\ref{sec:wise}). Note
that the theoretical model colors on Figure~\ref{f12} are for hot dust shell
($T_c$=1000 K), which are not be applicable to detached dust shell models.

\section{Spacial distribution of AGB stars\label{sec:spacial}}

Figure~\ref{f13} shows spacial distributions of AGB stars (AGB-IRAS and
AGB-WISE) in Galactic coordinates. Figure~\ref{f14} shows number distribution
of the Galactic longitude and latitude for OAGB and CAGB stars in AGB-IRAS and
AGB-WISE catalogs. In the bulge component, there are more OAGB stars than CAGB
stars. The lack of OAGB-WISE objects at the Galactic center in the lower-left
panel looks to be due to a selection effect of the sample stars.

The histograms for different Galactic latitudes are similar for both OAGB and
CAGB stars. All AGB stars are concentrated toward the Galactic disk.

We find that OAGB stars are more concentrated toward the Galactic center and
the number decreases with the Galactic longitude, while CAGB stars are
distributed more uniformly from the center to large Galactic longitudes.
\citet{ishihara2011} also found that OAGB stars are concentrated toward the
Galactic center and that the density decreases with Galactocentric distance,
whereas CAGB stars show a relatively uniform distribution within about 8 kpc of
Sun.

\section{Period-color relations for known Mira variables\label{sec:pul}}

It is generally believed that more evolved (or more massive) AGB stars would
show the longer pulsation periods, larger pulsation amplitudes, higher
mass-loss rates, thicker dust envelopes, and redder IR colors (e.g.,
\citealt{debeck2010}; \citealt{sk2013}). Studying IR properties of all types of
LPVs in Magellanic clouds, \citet{suh2020} showed that only Mira variables,
among all types of LPVs, show a clear period-color relation (PCR): Miras with
longer pulsation periods generally show redder IR colors. This would be because
Mira variables are usually oscillating in the fundamental mode and occupy a
single sequence in the period-luminosity diagram (\citealt{swu13b}; see
Section~\ref{sec:galb}).

Because most of the known pulsation periods in AAVSO were obtained in optical
observations and the longest wavelength band used by OGLE3 observations was the
$I$ band (0.8 $\mu$m), most Mira variables whose pulsation periods listed in
AAVSO are early phase AGB stars with thin dust envelopes.

The left panels of Figure~\ref{f15} show IR[12]$-$IR[25] and K[2.2]$-$IR[12]
colors versus pulsation periods for Miras in the AGB-IRAS catalog (see
Table~\ref{tab:tab1}). For both IR colors, CAGB stars show larger coefficients
of determination ($R^2$), which mean higher strength of the relationship. We
find that the Mira variables show fairly strong PCRs.

\citet{jimenez2021} investigated variability properties of the Arecibo sample
of OH/IR stars and presented periods for 348 Arecibo sources obtained from
observations at NIR bands. All of those sources are in OI-SH except for one
object, IRAS 18551+0323, which is a CAGB star (in CI-SH with IRAS LRS type C;
this object could be a composite object).

The pulsation periods measured at MIR or radio bands for more evolved or
massive AGB stars with thick dust envelopes are available only for a small
number of AGB stars. We have compiled pulsation periods of 522 OAGB stars (495
OH/IR stars; 214 AAVSO Miras) measured at NIR, MIR, or radio bands presented by
\citet{chen2001}, \citet{ks2010b}, \citet{urago2020}, and \citet{jimenez2021}.
The right panels of Figure~\ref{f15} show PCRs using IR[12]-IR[25] and
K[2.2]$-$W3[12] colors versus the pulsation periods measured at NIR, MIR, or
radio bands for the 522 OAGB stars.

Figure~\ref{f16} shows PCRs using using K[2.2]$-$W3[12], W1[3.4]$-$W2[4.6],
W2[4.6]$-$W3[12] and W3[12]$-$W4[22] colors for Miras in AAVSO. The left and
right panels show the PCRs for Miras in AGB-IRAS and AGB-WISE, respectively.

Though there are large scatters, we find that the PCRs for Miras in the sample
stars show a noticeable trend: Miras with longer pulsation periods generally
show redder IR colors.

\begin{table*}
\caption{Numbers of objects for which good-quality variation parameters are obtained only from the WISE light curves\label{tab:tab8}}
\centering
\begin{tabular}{llllllllll}
\hline \hline
Subgroup$^1$ &Number & G-W1$^2$ &\multicolumn{2}{c}{GM-W1$^3$} &GN-W1$^6$ & G-W2$^2$  &\multicolumn{2}{c}{GM-W2$^3$}  &GN-W2$^6$ \\
       &      &            &GM-W1A$^4$ & GM-W1B$^5$ &  &   & GM-W2A$^4$ & GM-W2B$^5$ &      \\
\hline
OI-SH  & 3826 & 392 (99)    & 16 (2)    & 6         & 370 (97)  & 414 (55)   & 28 (2)     & 14          & 372 (53)\\
OI-UR  & 37   & 13 (3)      & 2 (2)     & 1         & 10 (1)    & 10 (3)     & 3 (2)      & 0           & 7 (1)\\
OI-JB  & 5    & 5 (5)       & 0         & 0         & 5 (5)     & 4 (1)      & 0          & 0           & 4 (1)\\
OI-ME  & 127  & 7 (3)       & 0         & 0         & 7 (3)     & 12 (3)     & 0          & 0           & 12 (3)\\
OI-ST  & 673  & 31 (3)      & 1         & 0         & 30 (3)    & 49 (2)     & 5          & 0           & 44 (2)\\
OI-OG  & 1057 & 76 (12)     & 37 (7)    & 39 (5)    & 0         & 211 (47)   & 105 (24)   & 106 (23)    & 0\\
OI-WU  & 19   & 0           & 0         & 0         & 0         & 0          & 0          & 0           & 0\\
OI-AM  & 163  & 0           & 0         & 0         & 0         & 3 (0)      & 1          & 2           & 0\\
OI-all & 5908 & 524 (125)   & 55 (11)   & 47 (5)    & 422 (109) & 703 (111)  & 140 (28)   & 124 (23)    & 439 (60)\\
\hline						
CI-SH  & 1165 & 91 (29)     & 4 (2)     & 2         & 85 (27)   & 50 (6)     & 1 (1)      & 2           & 47 (5)\\
CI-UR  & 5    & 2           & 0         & 0         & 2         & 0          & 0          & 0           & 0\\
CI-GC  & 2417 & 27 (2)      & 1         & 3 (1)     & 23 (1)    & 17 (3)     & 0          & 1 (1)       & 16 (2)\\
CI-OG  & 9    & 0           & 0         & 0         & 0         & 1          & 1          & 0           & 0\\
CI-all & 3596 & 120 (31)    & 5 (2)     & 5 (1)     & 110 (28)  & 68 (9)     & 2 (1)      & 3 (1)       & 63 (7)\\
\hline						
OW-ME  & 157  & 7 (1)       & 0         & 0         & 7 (1)     & 8 (3)      & 0          & 0           & 8 (3) \\
OW-ST  & 231  & 25 (5)      & 0         & 0         & 25 (5)    & 19 (3)     & 0          & 0           & 19 (3) \\
OW-OG  & 4913 & 1485 (370)  & 761 (212) & 722 (158) & 0         & 2241 (774) & 1100 (417) & 1139 (357)  & 0  \\
OW-all & 5301 & 1517 (376)  & 761 (212) & 722 (158) & 34$^7$ (6)  & 2268 (780) & 1100 (417) & 1139 (357)  & 29$^7$ (6) \\
\hline						
CW-GC  & 3417 & 76 (13)     & 7 (2)     & 5 (3)     & 66 (8)    & 119 (24)   & 6 (3)      & 9 (4)       & 104 (17) \\
CW-OG  & 159  & 63 (20)     & 30 (8)    & 31 (12)   & 0         & 56 (19)    & 28 (10)    & 28 (9)      & 0 \\
CW-all & 3576 & 139 (33)    & 37 (10)   & 36 (15)   & 66 (8)    & 175 (43)   & 34 (13)    & 37 (13)     & 104 (17)\\
\hline
\end{tabular}
\begin{flushleft}
\scriptsize
$^1$See Tables~\ref{tab:tab1} and \ref{tab:tab2}.
$^2$objects with more than 100 observed points for which good-quality ($R^2$ $>$ 0.6) variation parameters (periods and amplitudes) at the WISE W1[3.4] or W2[4.6] band were obtained;
the number in parentheses denotes the number of objects with excellent ($R^2$ $>$ 0.8) quality (See Sections~\ref{sec:neo}).
$^3$G-W1 or G-W2 objects that are known as Mira variables with periods from AAVSO.
$^4$the first peak of the Lomb-Scargle power is the nearest from the known Mira period.
$^5$the second (or up to fourth) peak is the nearest from the known Mira period.
$^6$G-W1 or G-W2 objects with unknown periods or those objects known as non-Mira variables with periods from AAVSO.
$^7$two of them are SRA variables in AAVSO (OGLE3 Miras).
\end{flushleft}
\end{table*}

\section{Finding IR variations of AGB stars from WISE data\label{sec:neo}}

Various observational data obtained in the last 50 yr are available for
studying variability of AGB stars. There are large amounts of photometric data
at visual and NIR bands but the data at MIR bands are available only for a
limited number of objects. For AGB stars with thick dust envelopes, the
variability can be more properly investigated from the observations at MIR
bands (e.g., \citealt{engels1983}; \citealt{ks2010b}).

To study variability of AGB stars at W1[3.4] and W2[4.6] bands during the last
12 yr, we use the ALLWISE multiepoch photometry table obtained in 2009-2010 and
the NEOWISE-R data (2021 data release) which give us the photometry data for 14
epochs, two in every year between 2014 and 2020.

We try to find Mira-like variations from the WISE light curves of the sample
stars using a simple sinusoidal light curve model with periods longer than 50
days (the shortest period of OGLE3 bulge Miras is 78.31 days). In this work, we
use the Lomb-Scargle periodogram which is a commonly used statistical algorithm
for detecting and characterizing periodic signals in unevenly spaced
observations (e.g., \citealt{zechmeister2009}; \citealt{vanderPlas2018}). The
Lomb-Scargle Periodograms are computed using the implementations in AstroPy
(\url{https://docs.astropy.org/en/stable/timeseries/lombscargle.html#}). We use
the AstroPy computing option of autopower using the 'chi2' method, which
utilizes the fact that the Lomb-Scargle periodogram at each frequency is
equivalent to the least-squares fit of a sinusoid to the data. The advantage of
the 'chi2' method is that it allows extensions of the periodogram to multiple
Fourier terms.

For each object in the sample, we have generated the light curves using the
WISE data and produced the Lomb-Scargle periodograms. But the WISE data for a
major part of the sample of AGB stars (mostly bright stars) are either
saturated or show scatters too large to provide meaningful variation
parameters. Therefore, we need to choose the objects with good-quality
variation parameters for which the deviations of the observed points from the
derived sinusoidal model light curves are smaller. To find objects with good
quality variation parameters at W1[3.4] and W2[4.6] bands (G-W1 and G-W2
objects), we choose the objects with more than 100 observed points and the
coefficients of determination ($R^2$) to fit the sinusoidal model to the
observations is larger than 0.6, for which the Lomb-Scargle power is also
stronger. Table~\ref{tab:tab8} summarizes the results for the sample stars.

Figure~\ref{f17} compares the number density distributions of mean magnitudes
for all objects whose WISE light curves are available (P-W1 and P-W2 objects)
and those for the objects with good-quality parameters (G-W1 and G-W2).
Generally, the objects with good-quality variation parameters are less bright.

Using the WISE data, we have obtained good-quality variation parameters for
3710 objects (G-W1 or G-W2; 885 OAGB-IRAS, 141 CAGB-IRAS, 2468 OAGB-WISE, and
216 CAGB-WISE objects) in the catalog (see Table~\ref{tab:tab8}).

There are 2810 objects known as Miras with periods from AAVSO (GM-W1 or GM-W2;
284 OAGB-IRAS, 13 CAGB-IRAS, 2429 OAGB-WISE, and 84 CAGB-WISE objects). For
about half of the objects (GM-W1A or GM-W2A objects), the obtained primary
periods from the WISE data are similar to the period in AAVSO. For another half
(GM-W1B or GM-W2B objects), the obtained primary periods from WISE data are
different from the periods in AAVSO (see Section~\ref{sec:neo-mc}).

And there are 656 objects with unknown periods (GN-W1 or GN-W2; 441 OAGB-IRAS,
97 CAGB-IRAS, 37 OAGB-WISE, and 81 CAGB-WISE objects) and 244 objects known as
non-Mira variables with periods from AAVSO (GN-W1 or GN-W2; 160 OAGB-IRAS, 31
CAGB-IRAS, 2 OAGB-WISE, and 51 CAGB-WISE objects).

\begin{figure*}
\centering
\largeplot{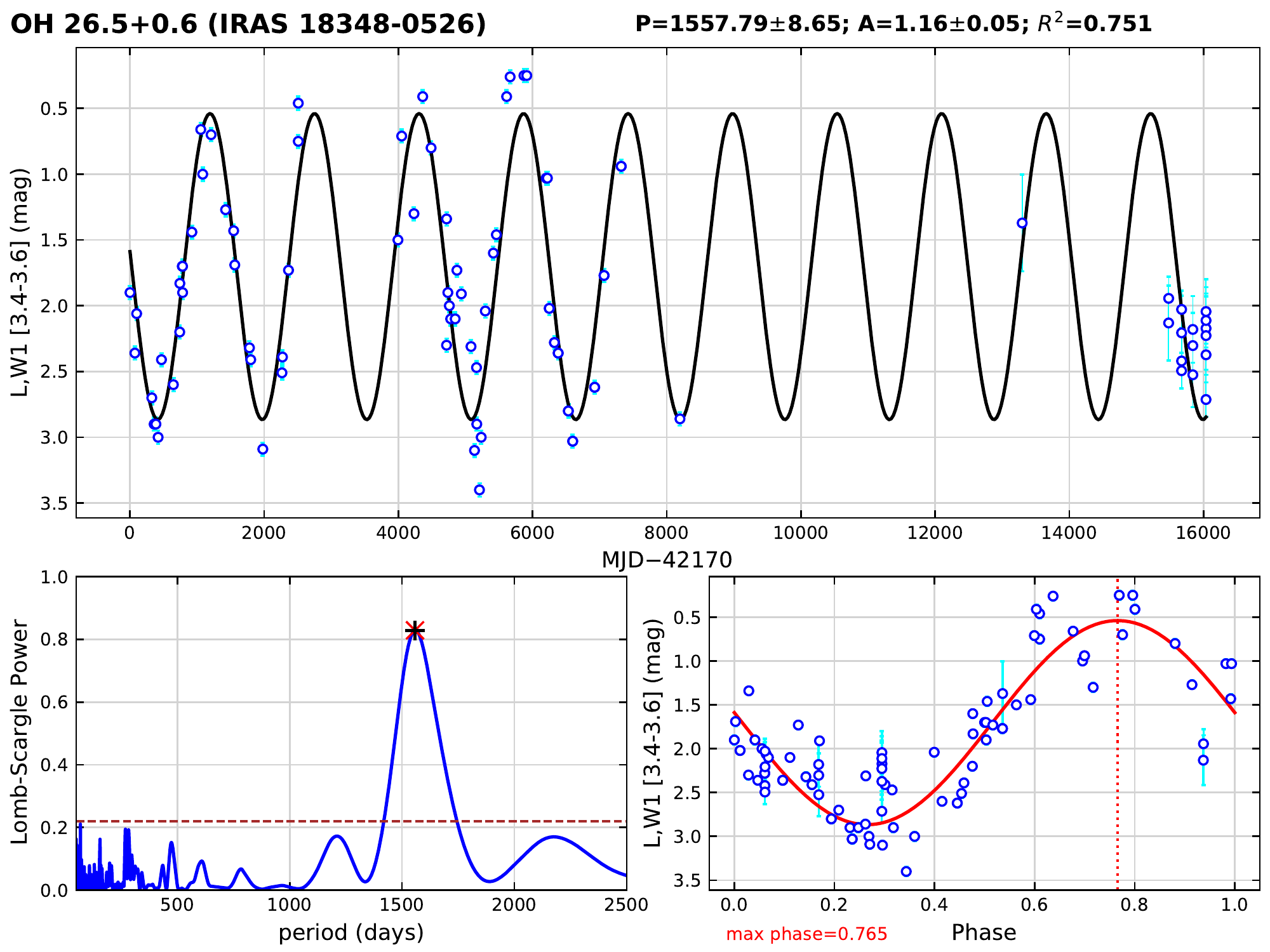}\caption{The combined light curve and Lomb-Scargle periodogram for an OH/IR star OH 26.5+0.6 (in OI-SH) using
the $L$[3.4-3.6] band data acquired in 1974-2003 and the WISE W1[3.4] band data acquired in the last 12 yr.
In the Lomb-Scargle periodogram, the dashed brown horizontal line indicates the periodogram level corresponding to a maximum peak false alarm probability of 1 \%.
Refer to \citet{ks2010a} for details of the $L$ band data. See Section~\ref{sec:neo-m}.
} \label{f18}
\end{figure*}

\begin{figure*}
\centering
\smallplottwo{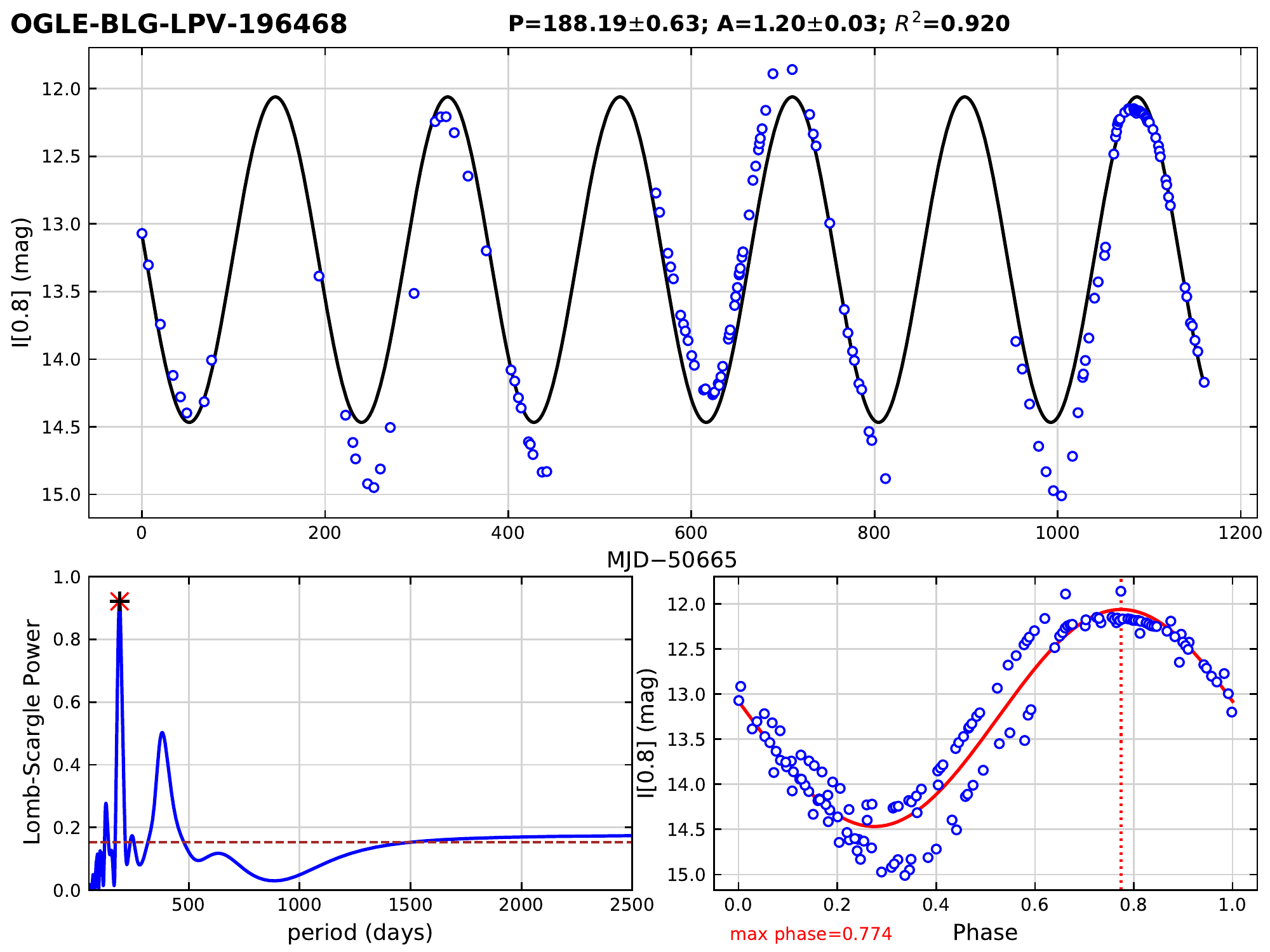}{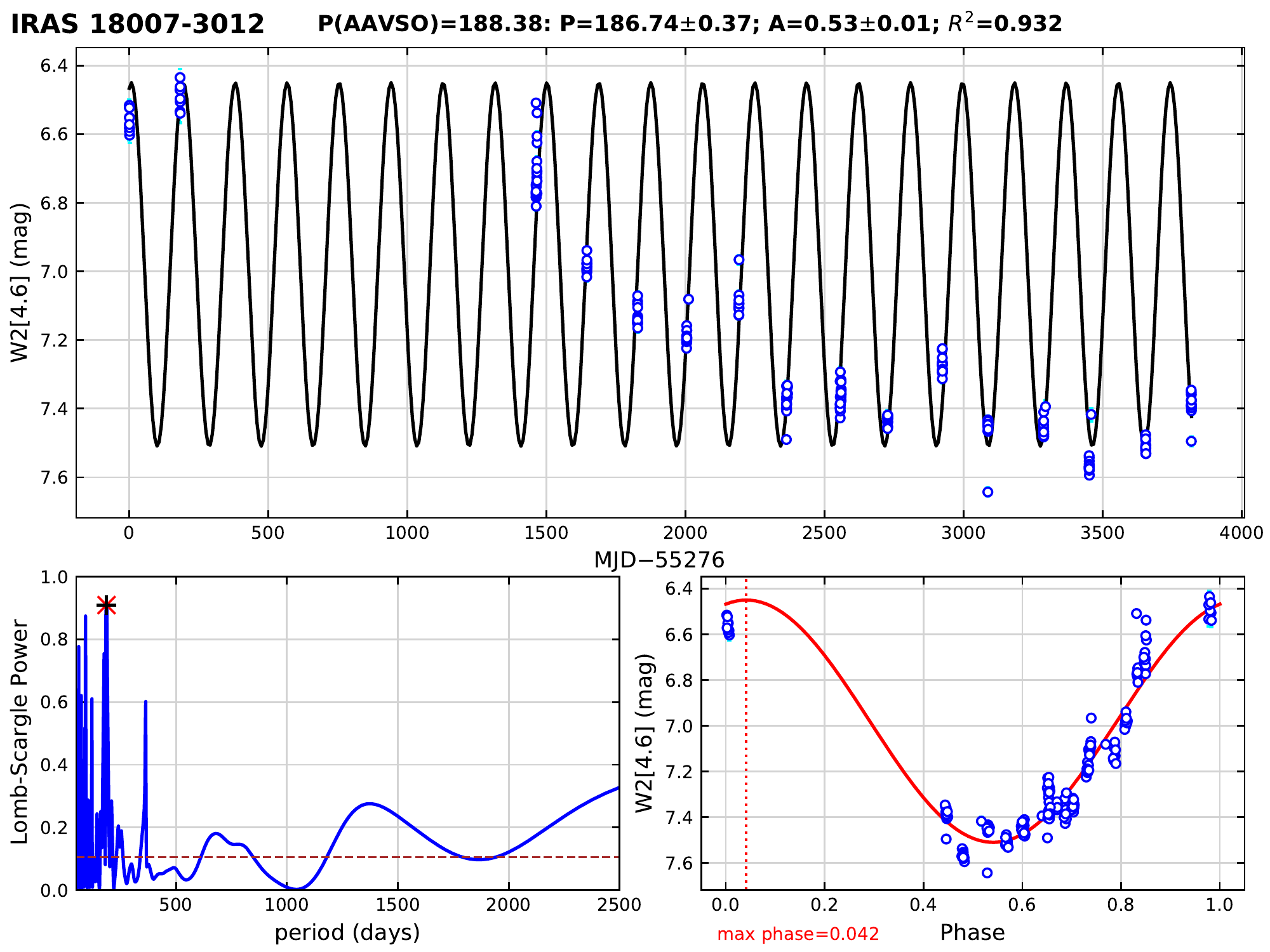}
\caption{Light curves and Lomb-Scargle periodograms for an OAGB-IRAS (OI-OG) object IRAS 18007-3012 (OGLE-BLG-LPV-196468) using the OGLE3 ($I$ band) and WISE (W2 band) data.
\citet{sus13a} obtained a period of 188.38 days. See Section~\ref{sec:neo-m}.}
\label{f19}
\end{figure*}

\begin{figure*}
\centering
\smallplotsix{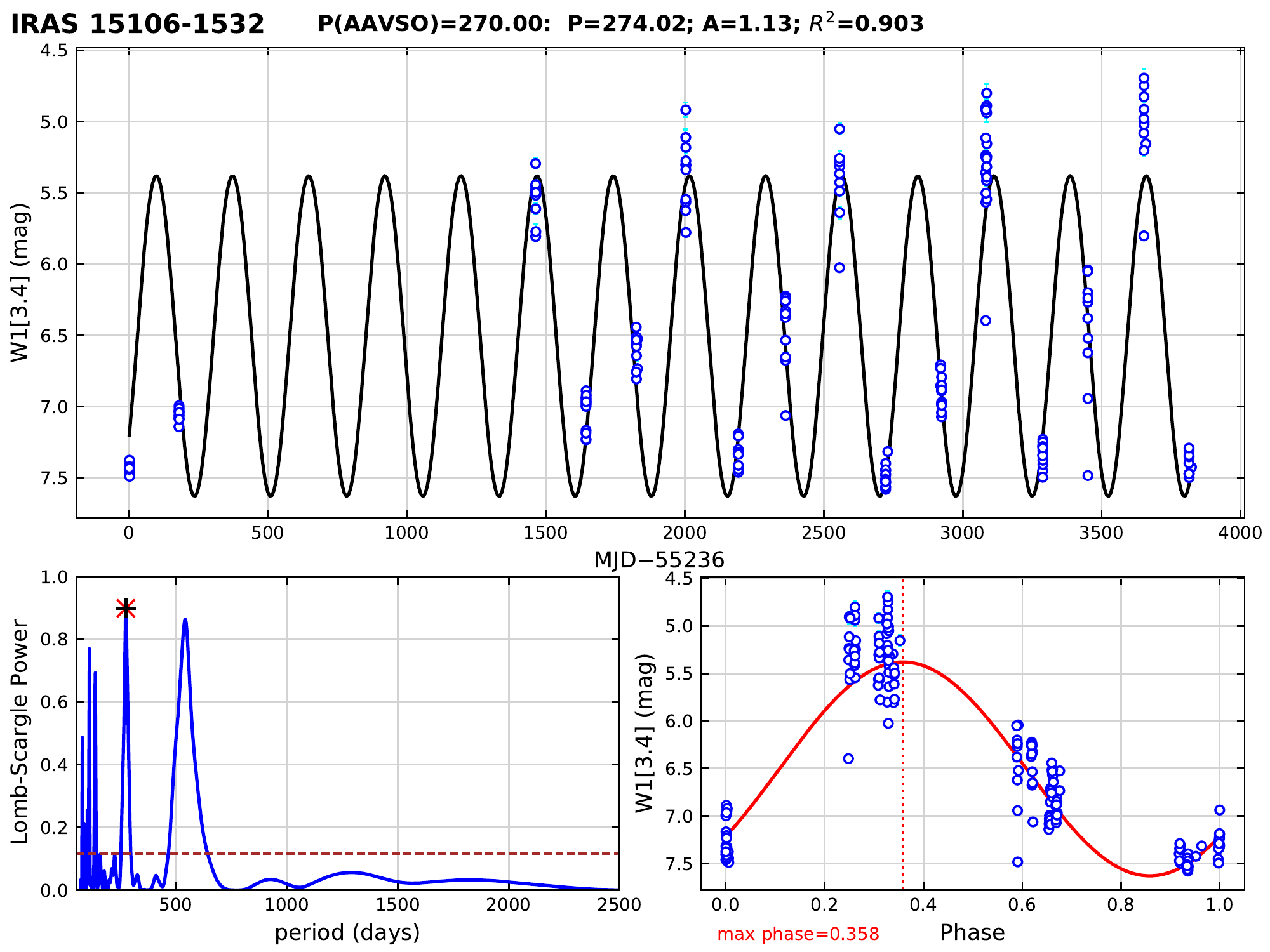}{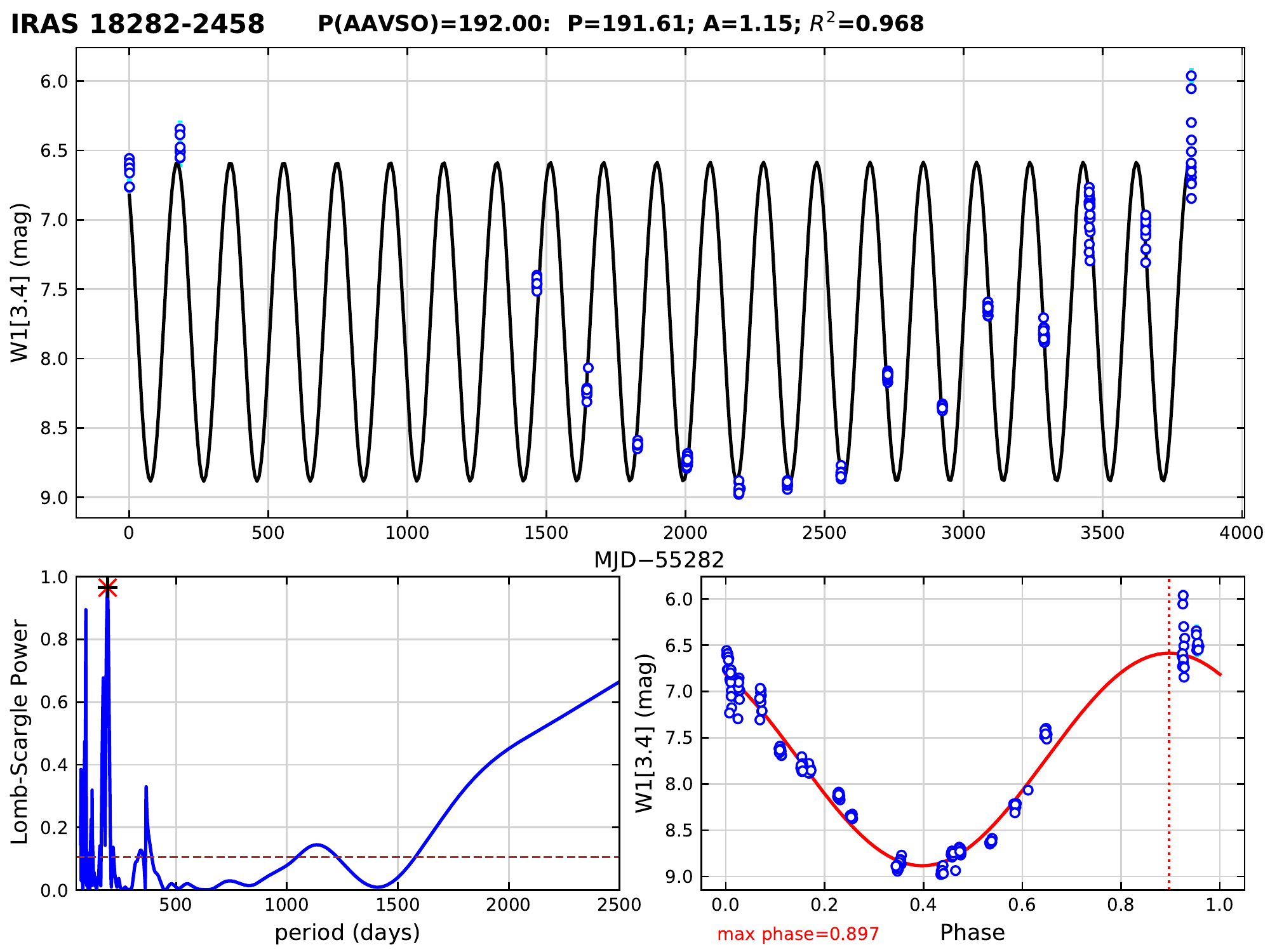}{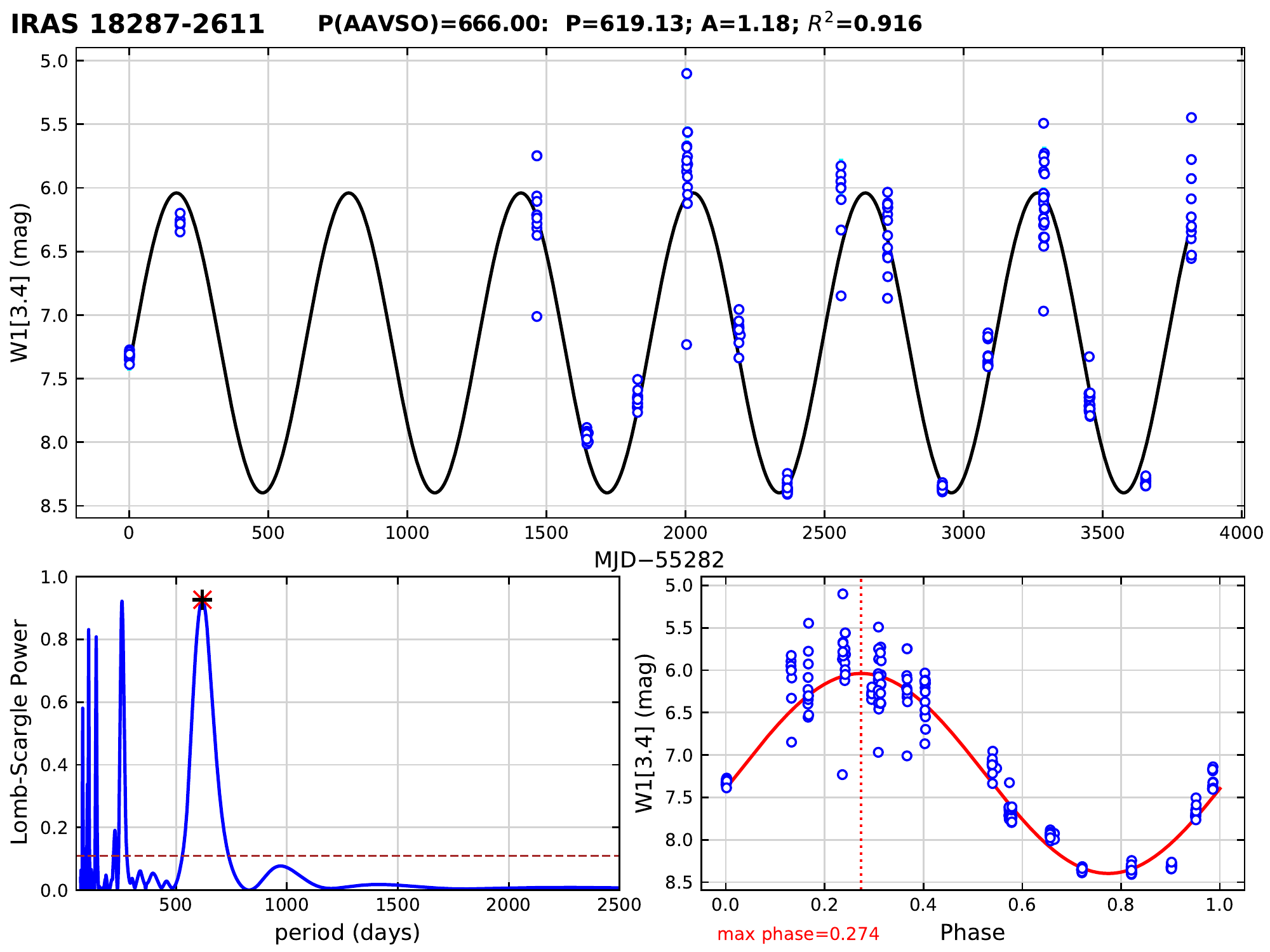}{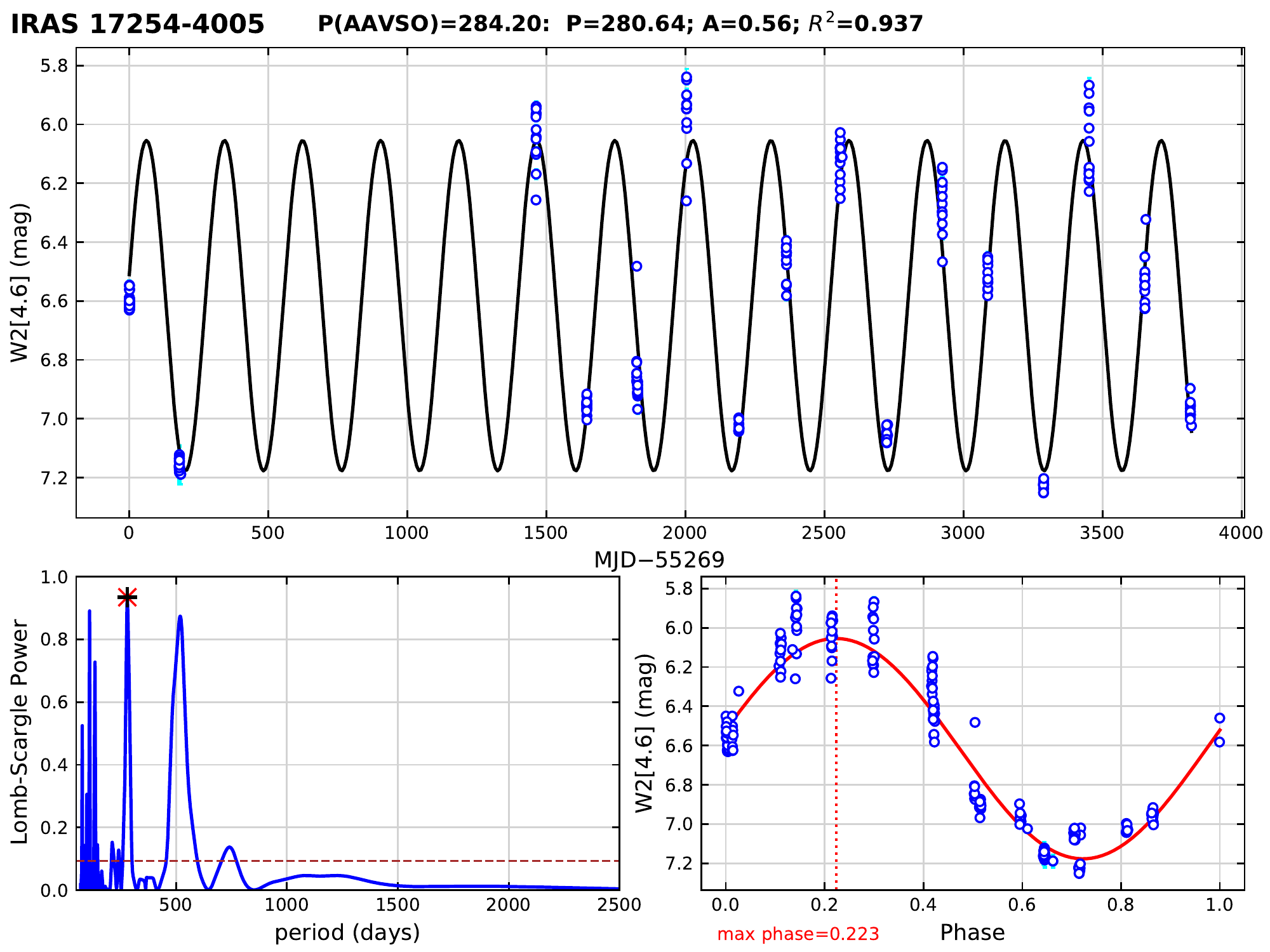}{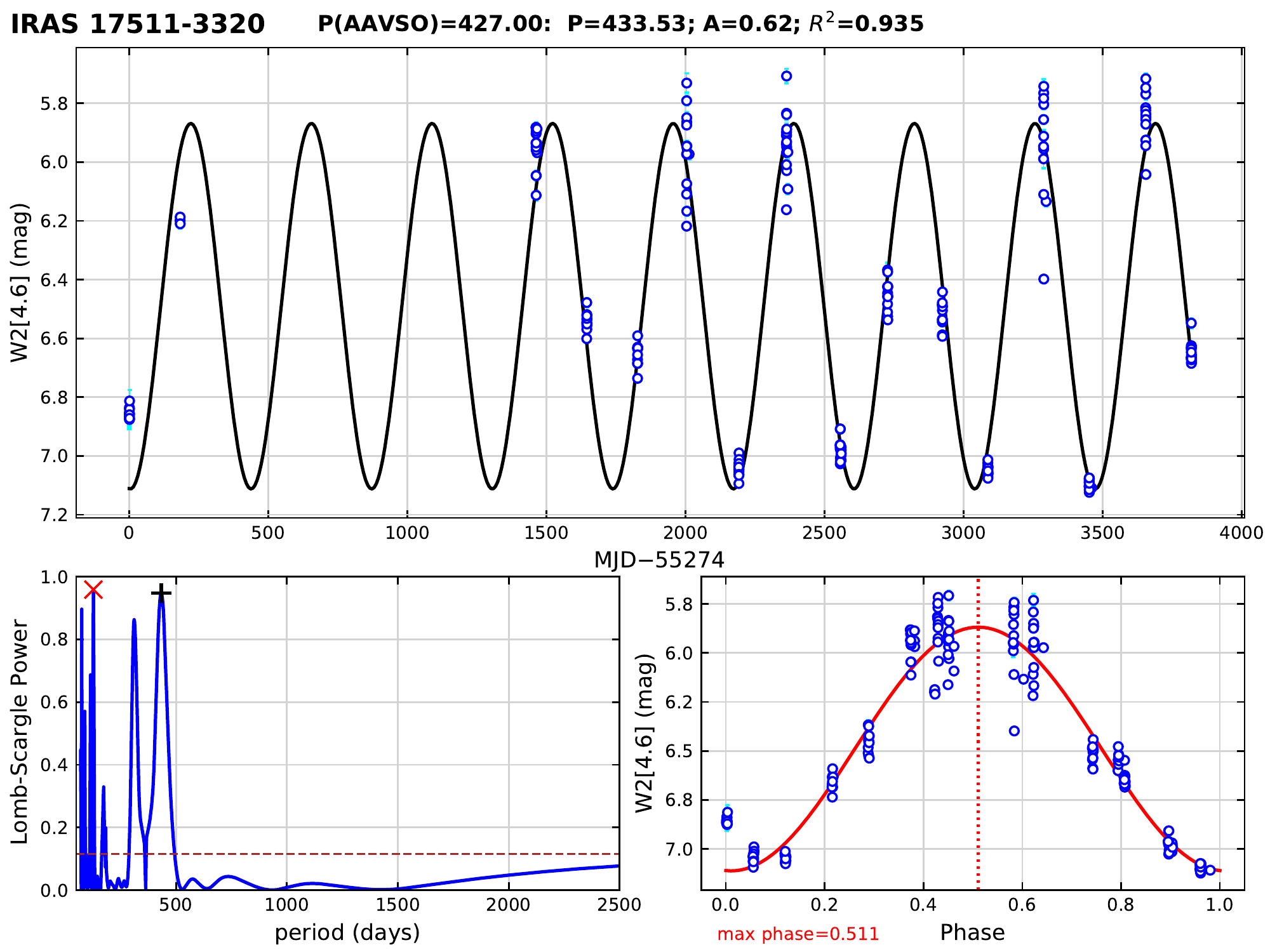}{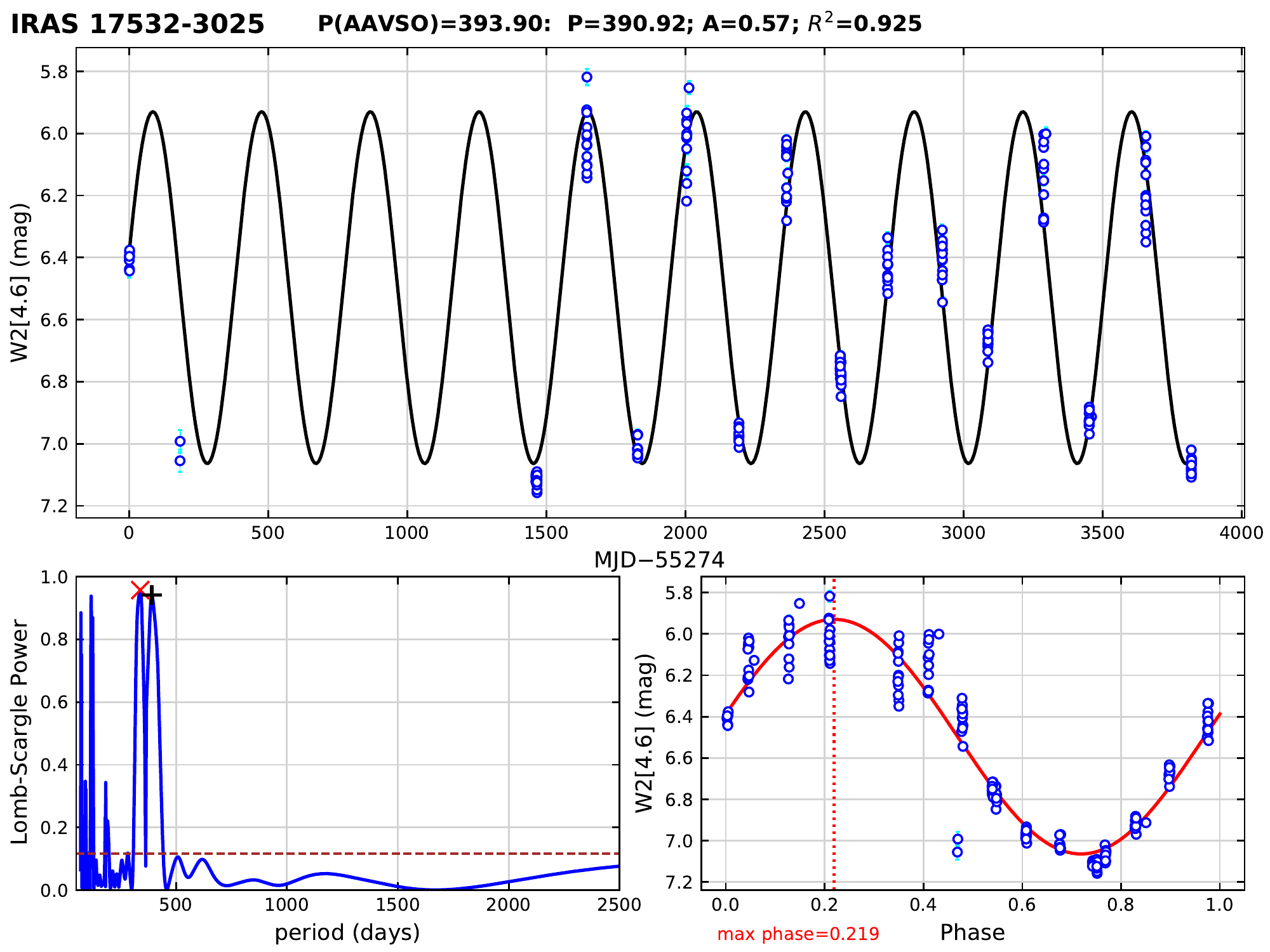}
\caption{WISE light curves and Lomb-Scargle periodograms for six OAGB-IRAS objects known as Miras with periods.
In the Lomb-Scargle periodogram, the red X and black cross marks indicate the primary and selected peaks, respectively
and the red dashed brown horizontal line indicates the periodogram level corresponding to a maximum peak false alarm probability of 1 \%.
Upper two panels show OI-UR objects whose periods from \citet{urago2020} are 268 and 394 days, respectively.
Middle two panels show OI-SH and OI-OG objects (see Table~\ref{tab:tab1}).
For the two objects (OI-OG objects) in lower panels, the second peak of the Lomb-Scargle power is selected for the period (GM-W2B objects).
See Section~\ref{sec:neo-m}.}
\label{f20}
\end{figure*}

\begin{figure*}
\centering
\smallplotsix{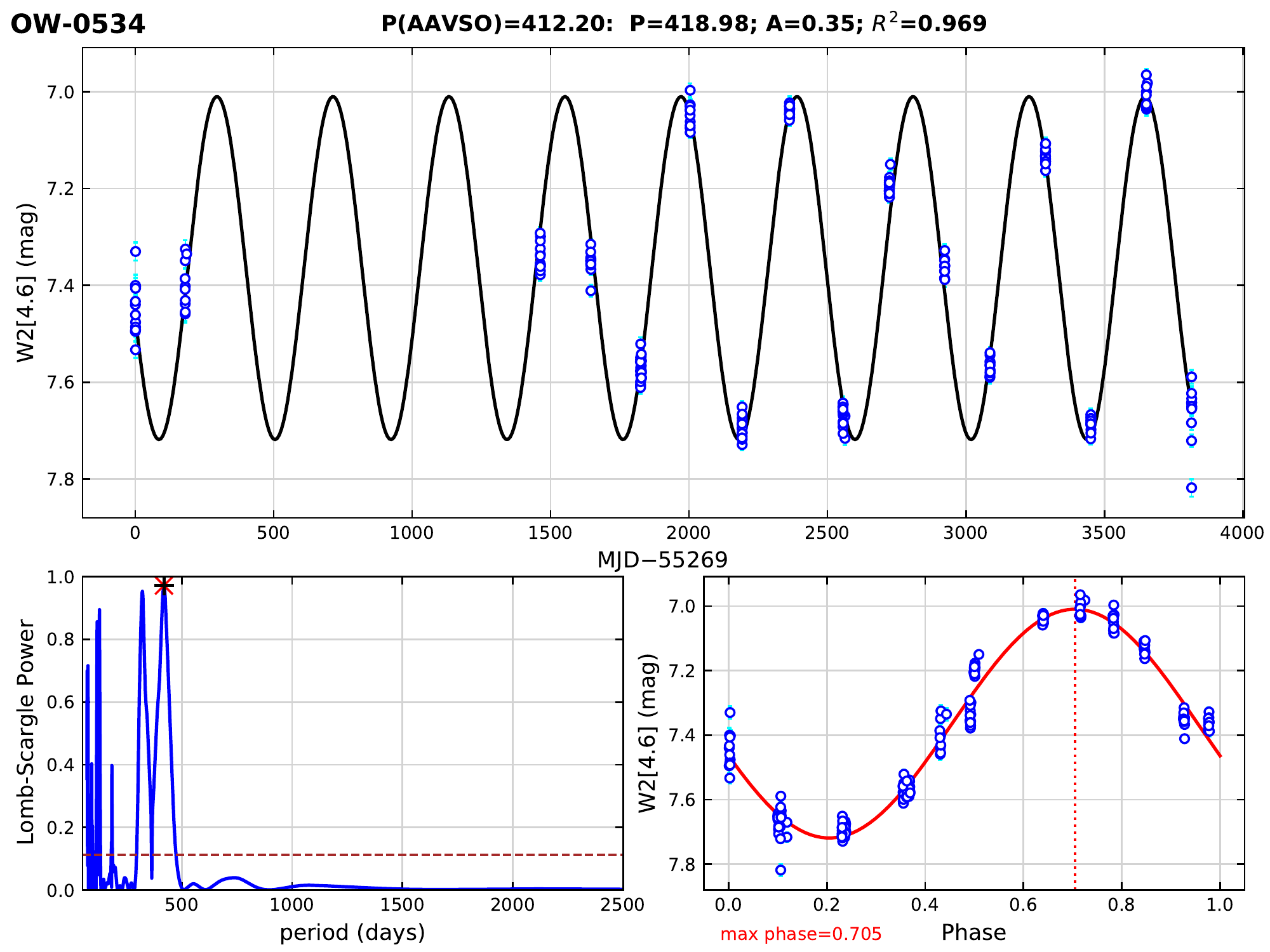}{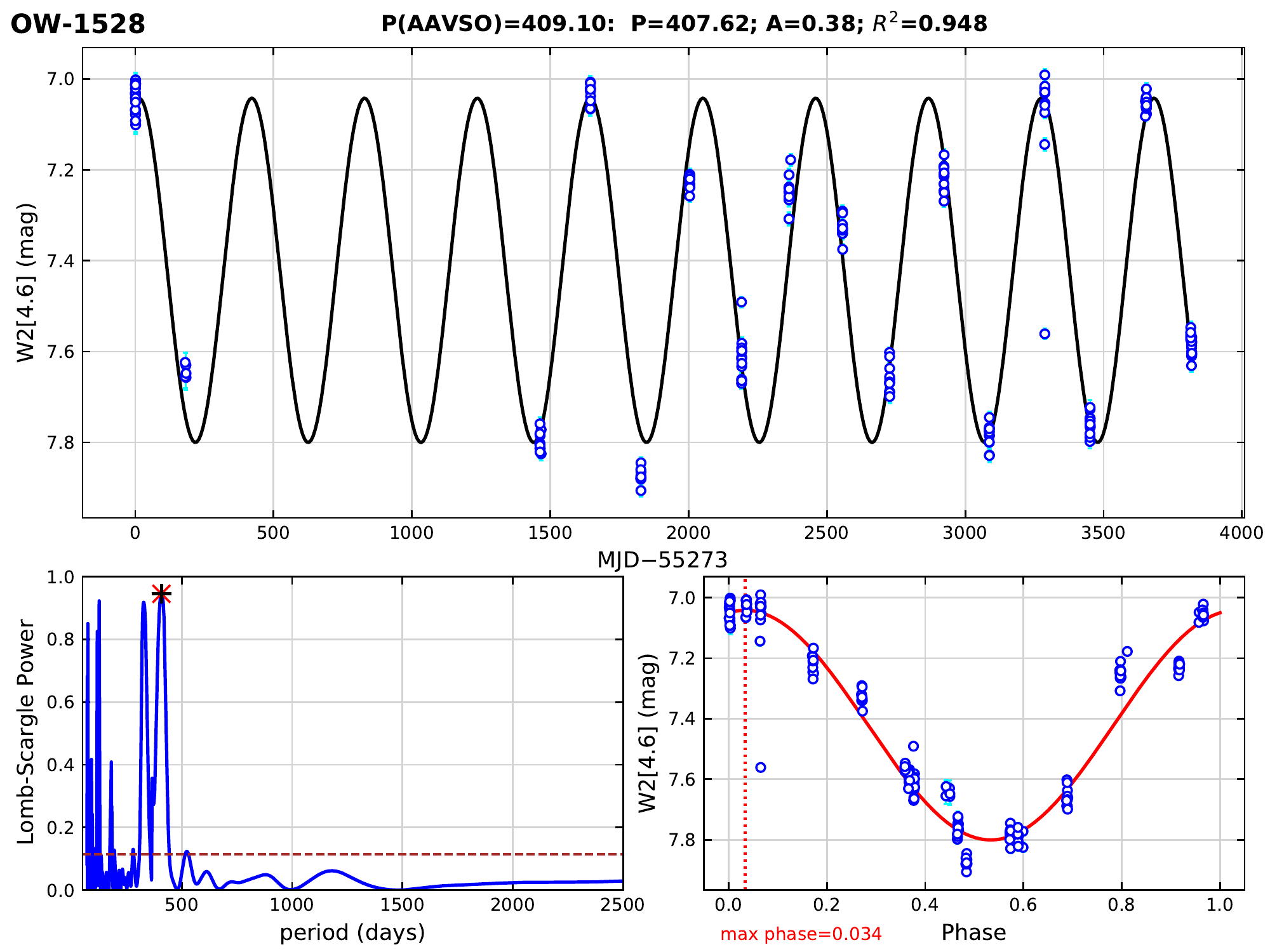}{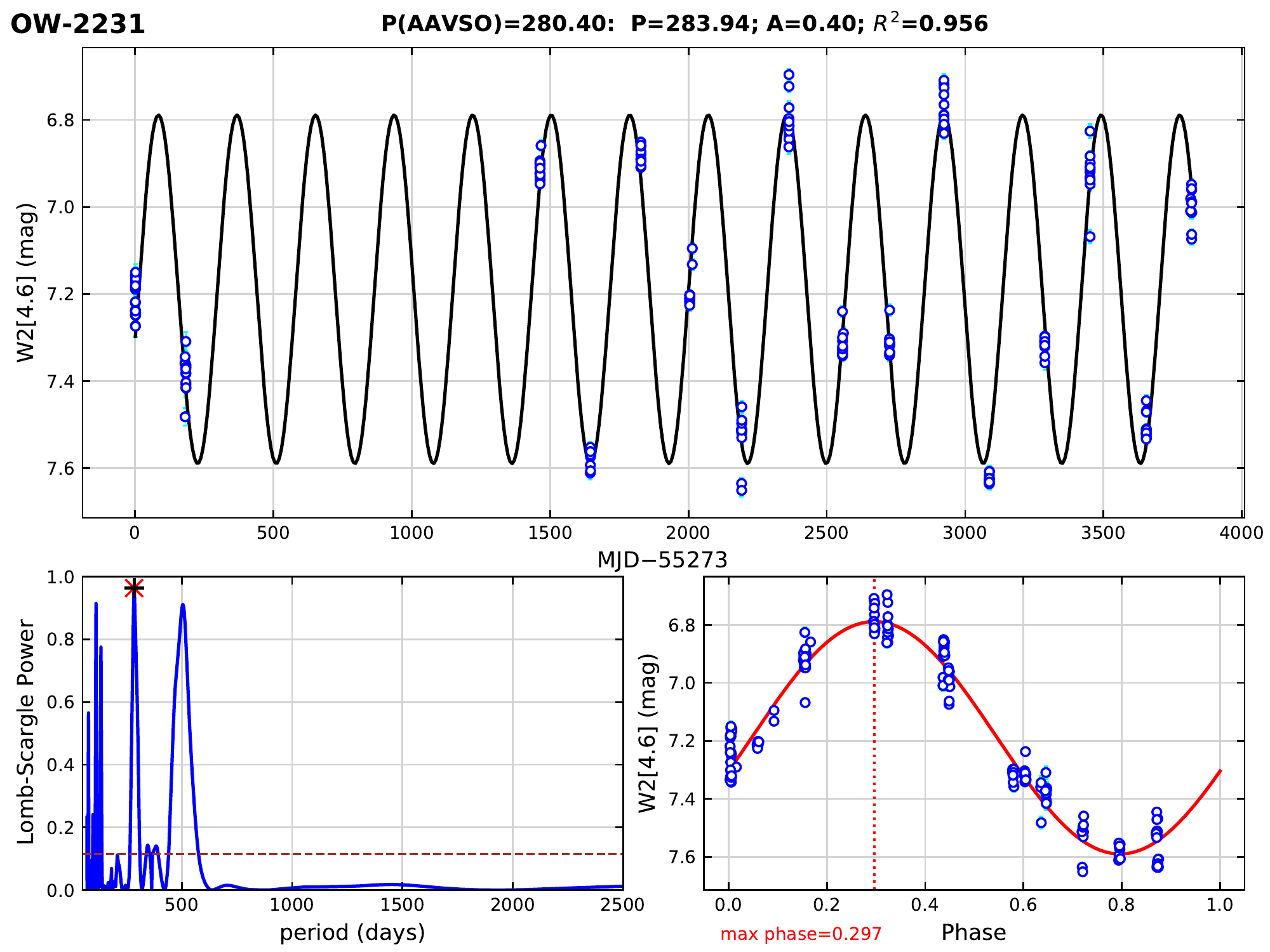}{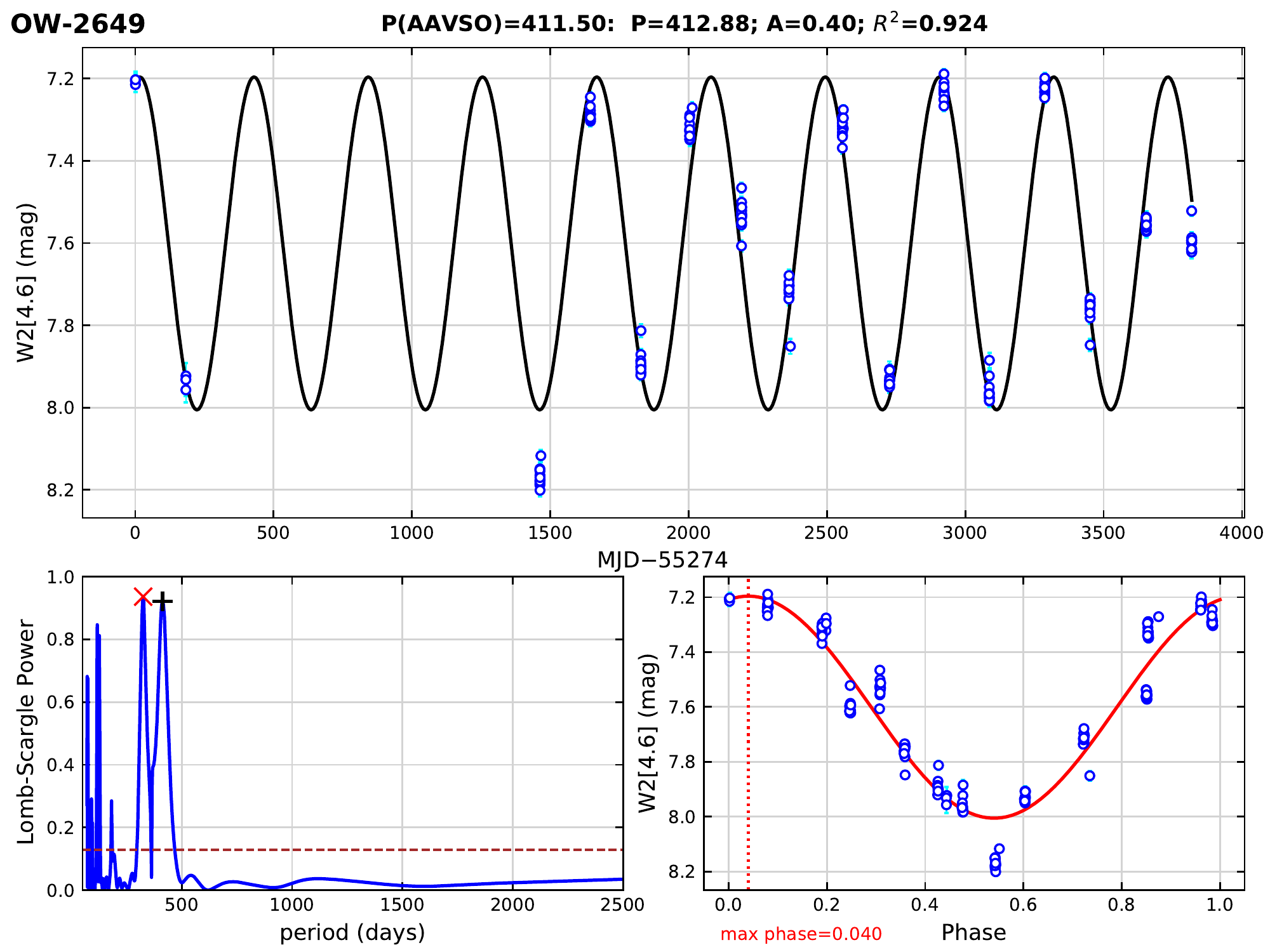}{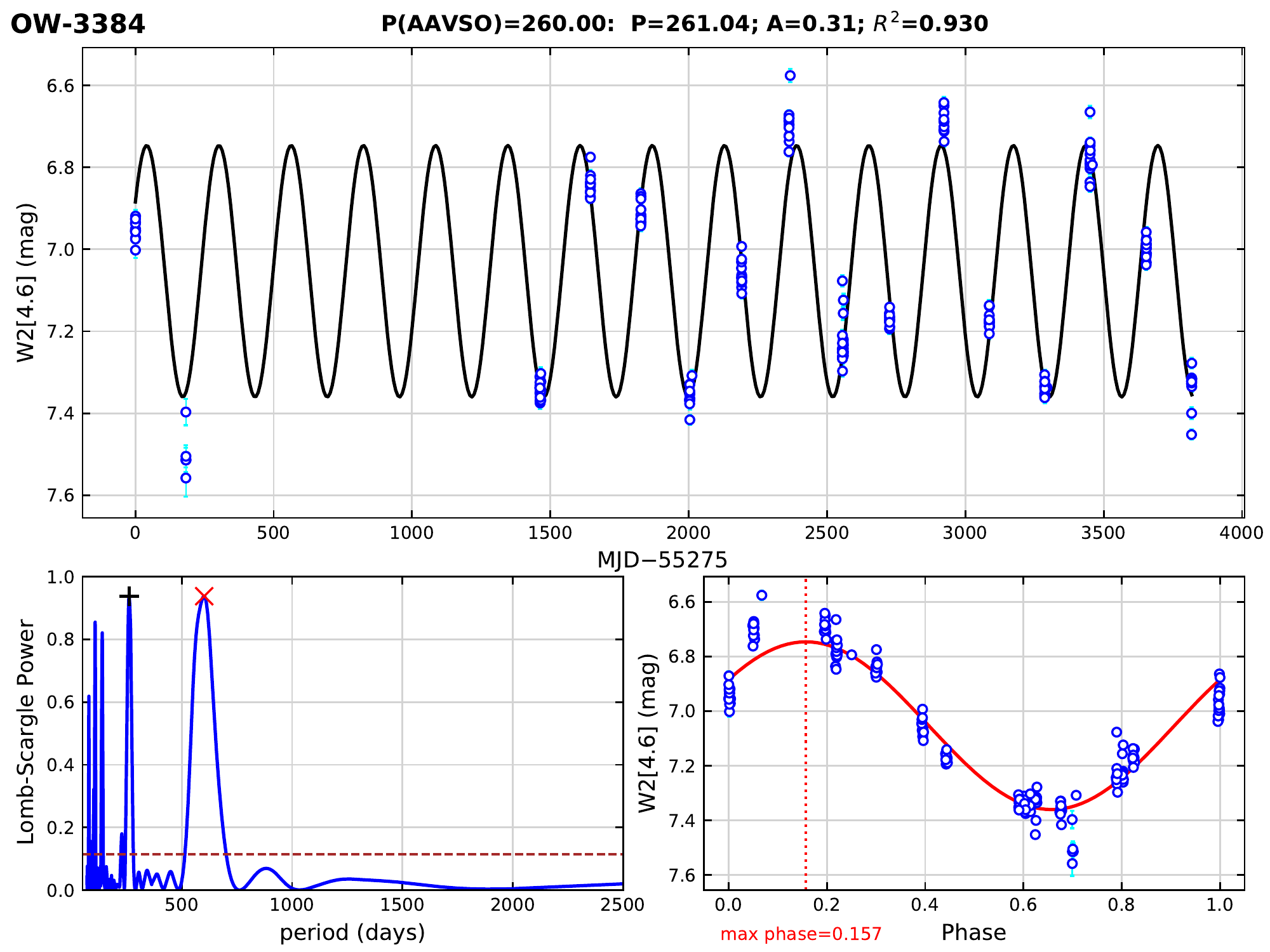}{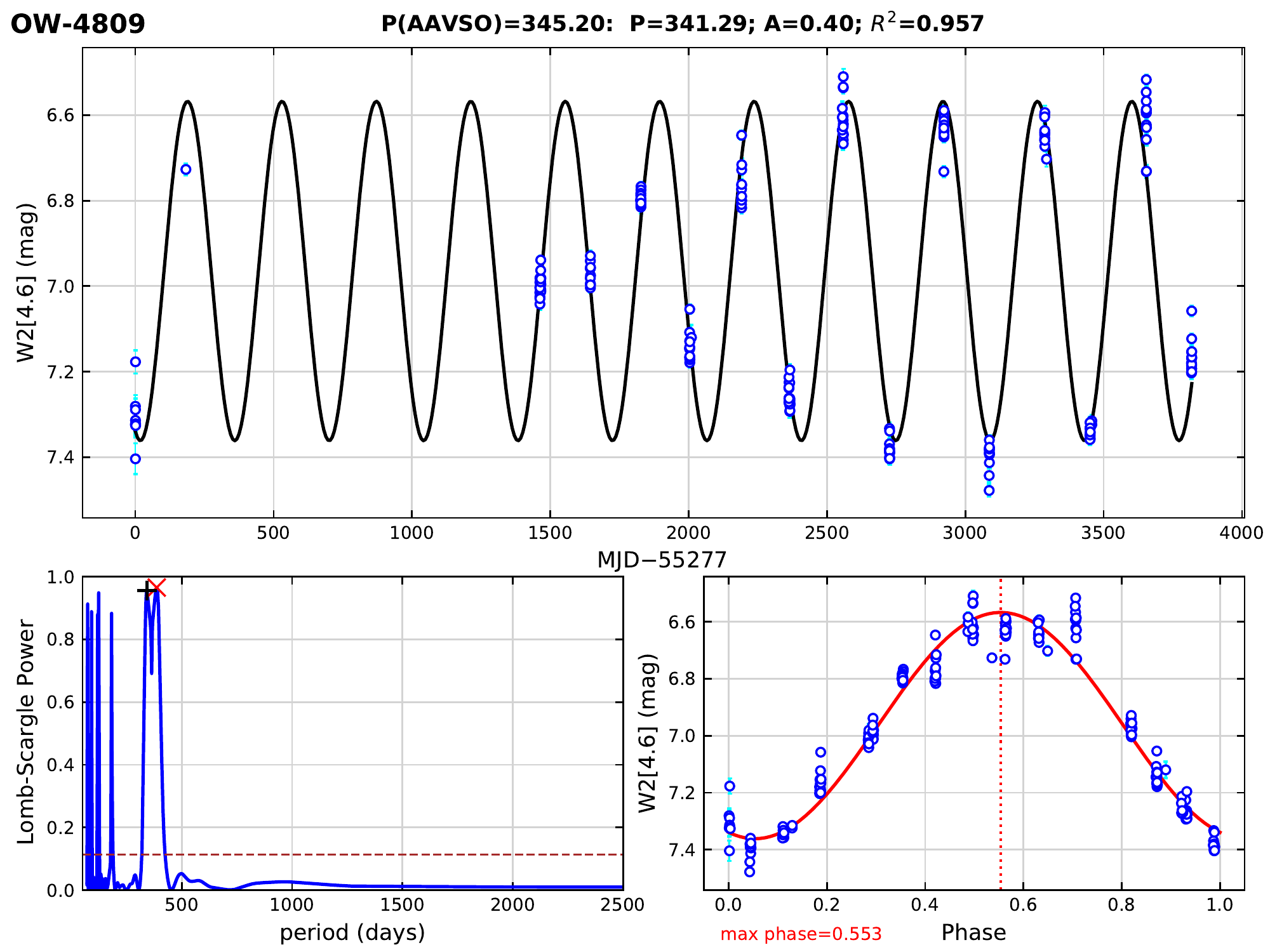}
\caption{WISE light curves and Lomb-Scargle periodograms for six OAGB-WISE (OW-OG) objects known as Miras with periods.
The object name is denoted by the OAGB-WISE identifier (OW-N; see Table~\ref{tab:tab11}).
In the Lomb-Scargle periodogram, the red X and black cross marks indicate the primary and selected peaks, respectively
and the red dashed brown horizontal line indicates the periodogram level corresponding to a maximum peak false alarm probability of 1 \%.
For three objects in lower panels, the second peak of the Lomb-Scargle power is selected for the period.
See Section~\ref{sec:neo-m}.}
\label{f21}
\end{figure*}

\begin{figure*}
\centering
\smallplotsix{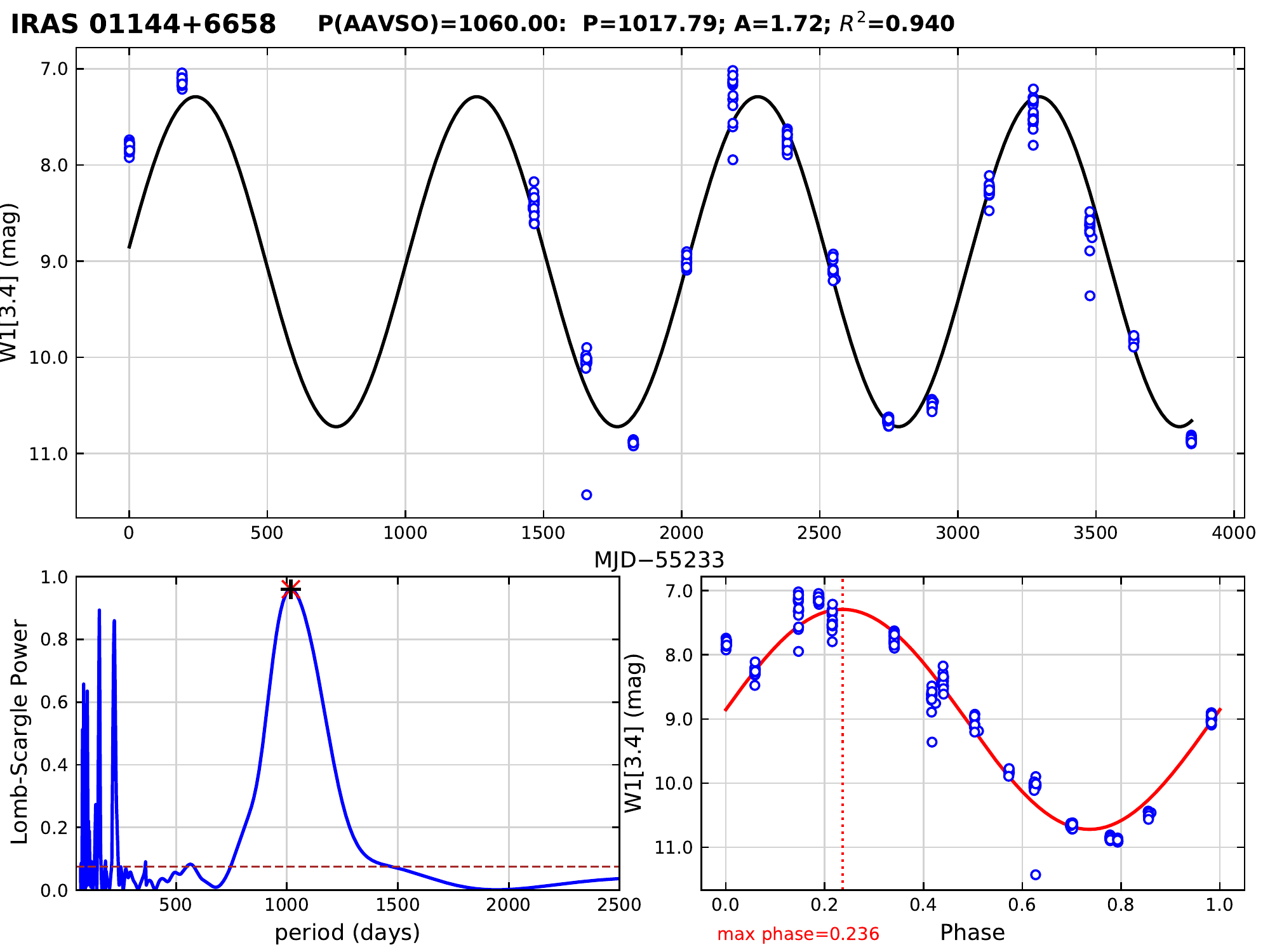}{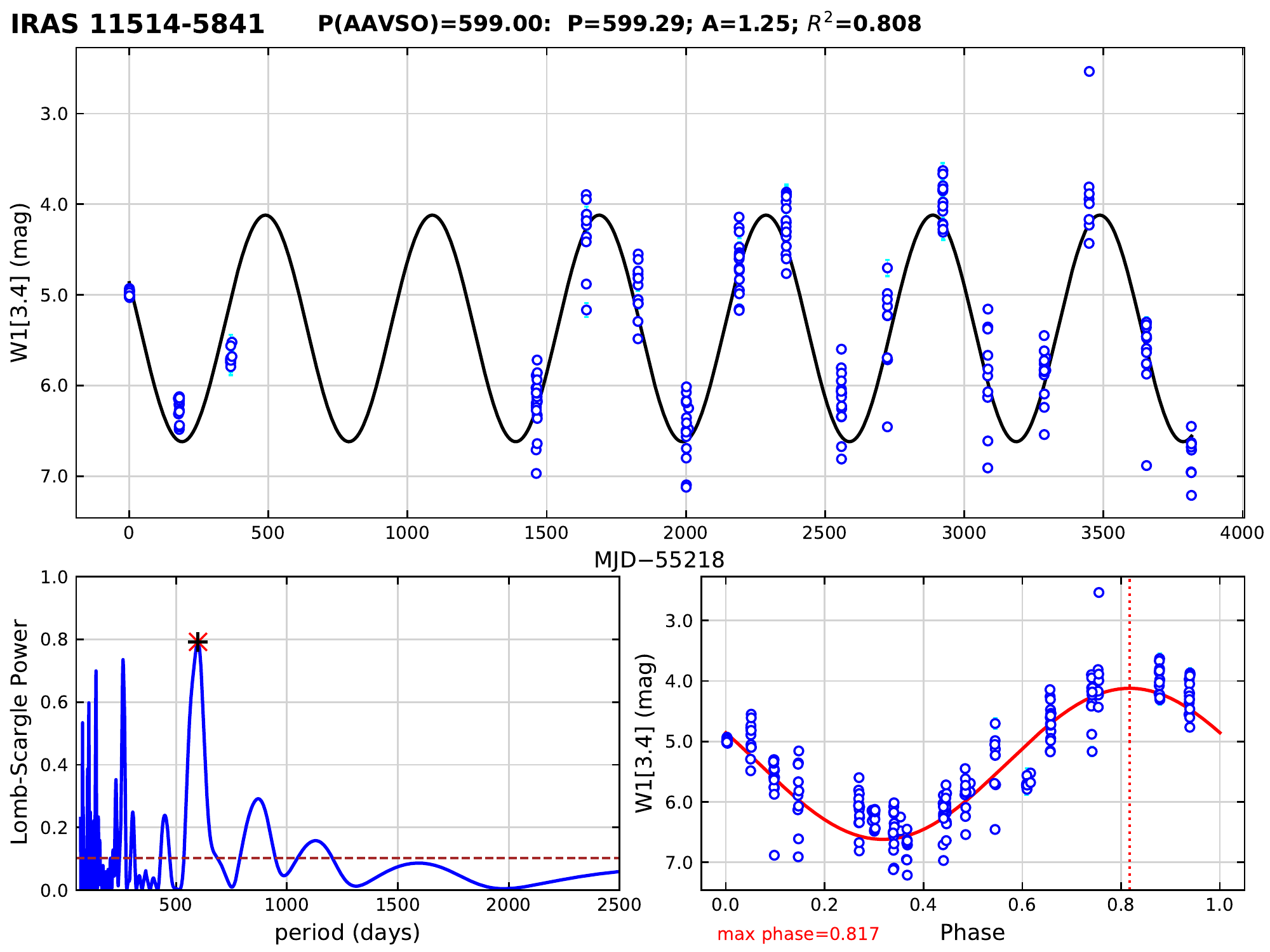}{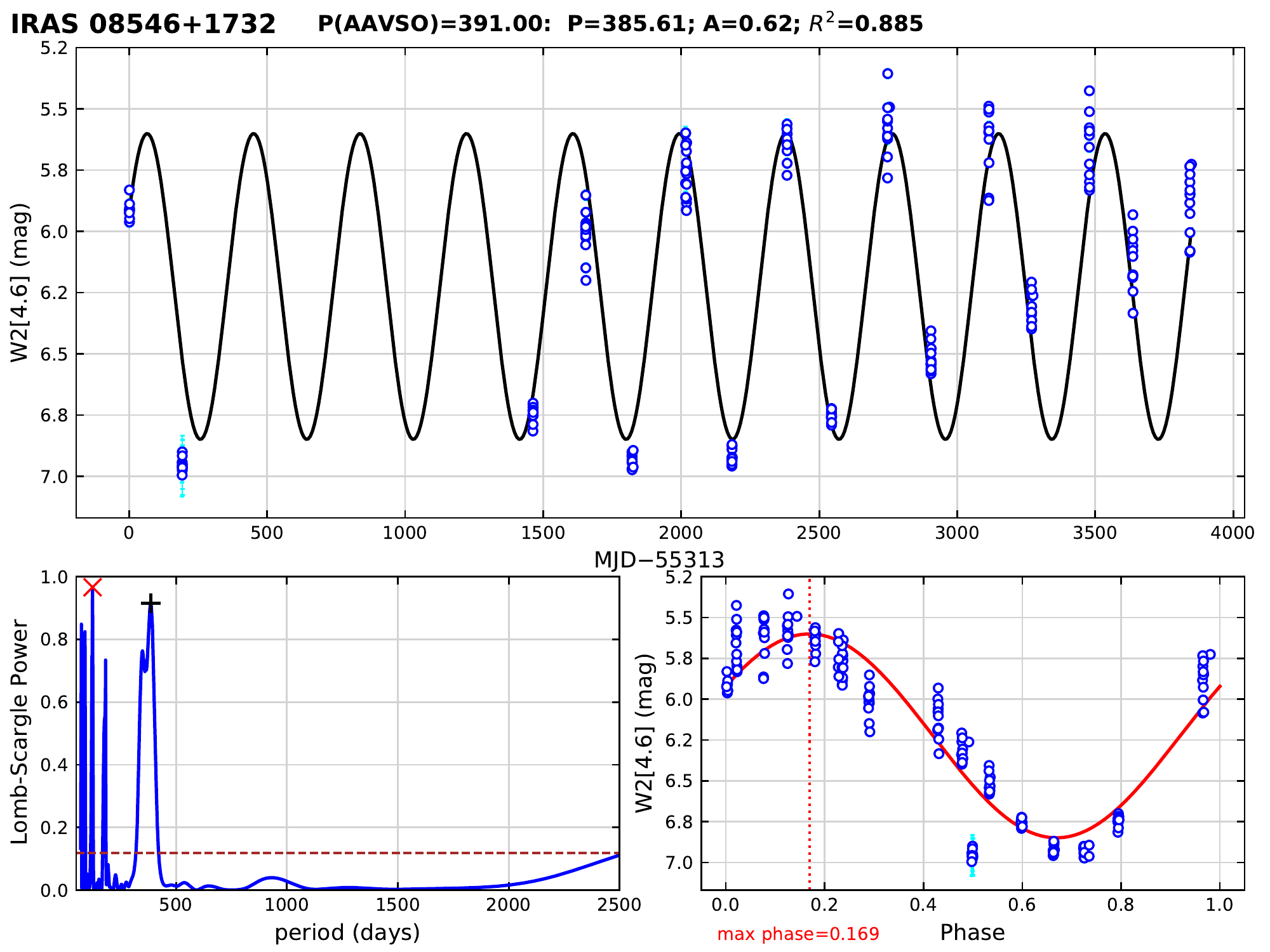}{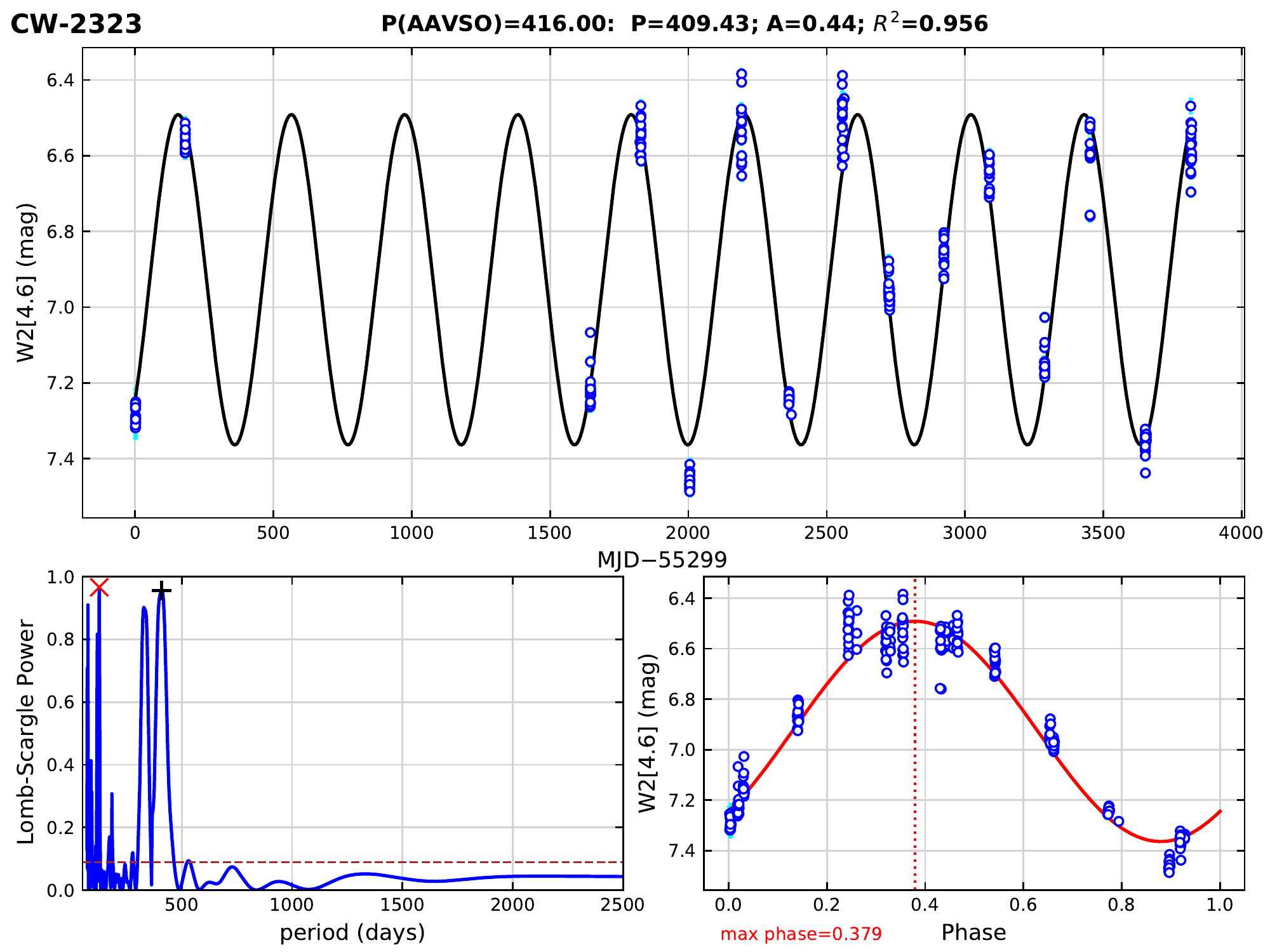}{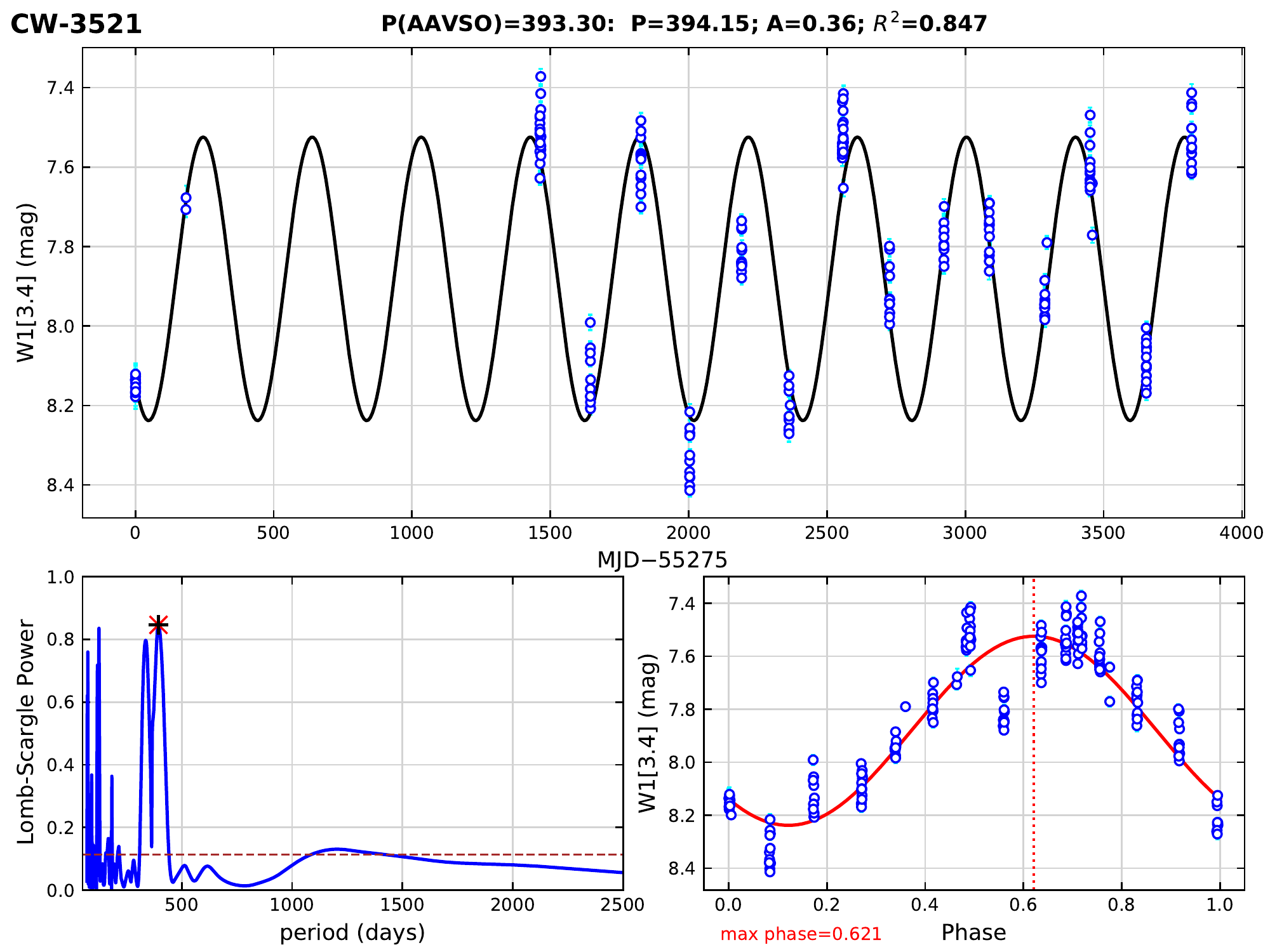}{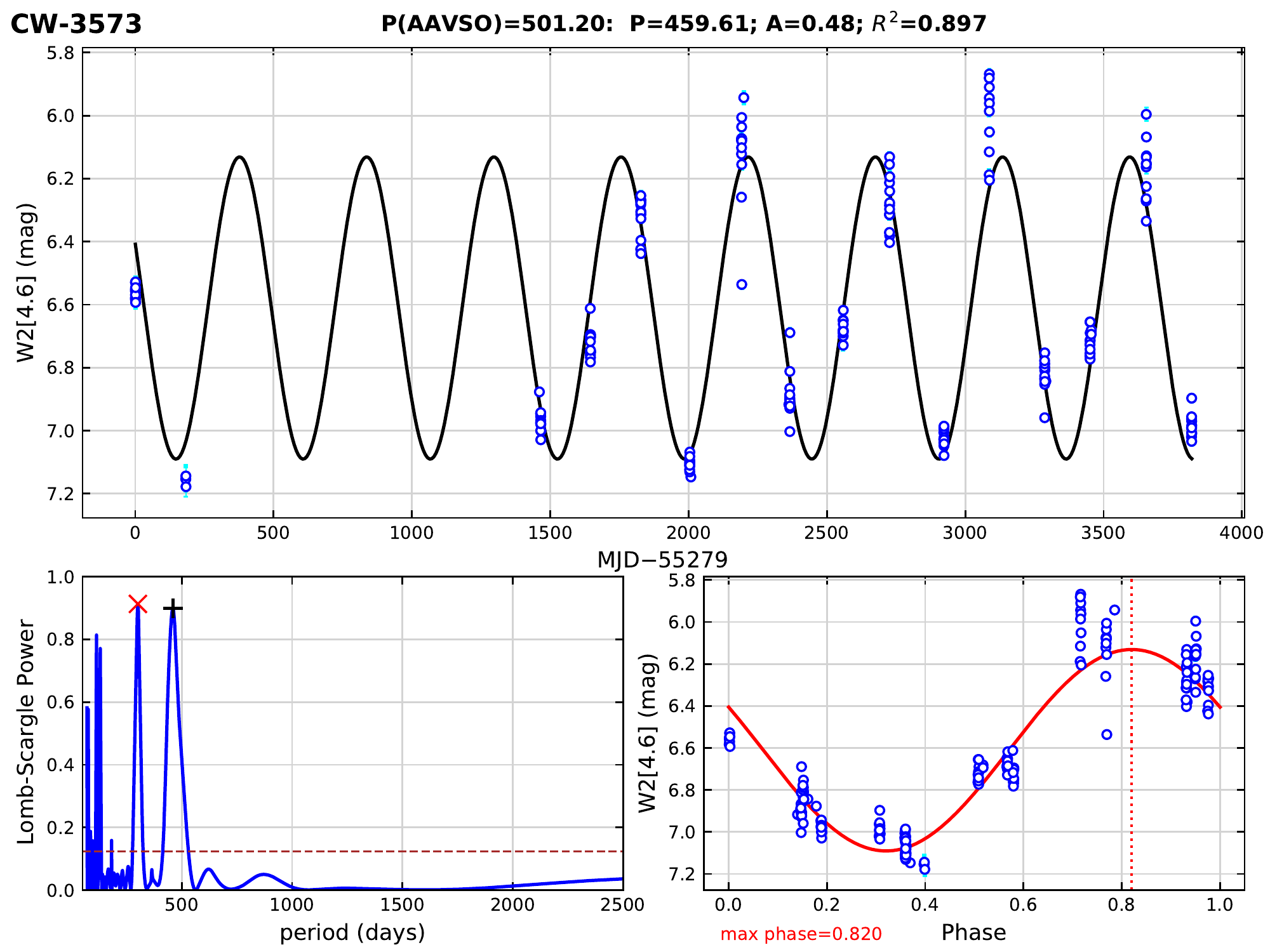}
\caption{WISE light curves and Lomb-Scargle periodograms for three CAGB-IRAS objects (objects in upper panels: CI-SH; IRAS 08546+1732: CI-GC) and
three CAGB-WISE objects (CW-2323: CW-GC; objects in lower panels: CW-OG) known as Miras with periods.
For CAGB-WISE, the object name is denoted by the CAGB-WISE identifier (CW-N; see Table~\ref{tab:tab12}).
In the Lomb-Scargle periodogram, the red X and black cross marks indicate the primary and selected peaks, respectively
and the red dashed brown horizontal line indicates the periodogram level corresponding to a maximum peak false alarm probability of 1 \%.
Note that the second peak of the Lomb-Scargle power is selected for the periods of the three objects. See Section~\ref{sec:neo-m}.}
\label{f22}
\end{figure*}

\begin{figure*}
\centering
\smallplotfour{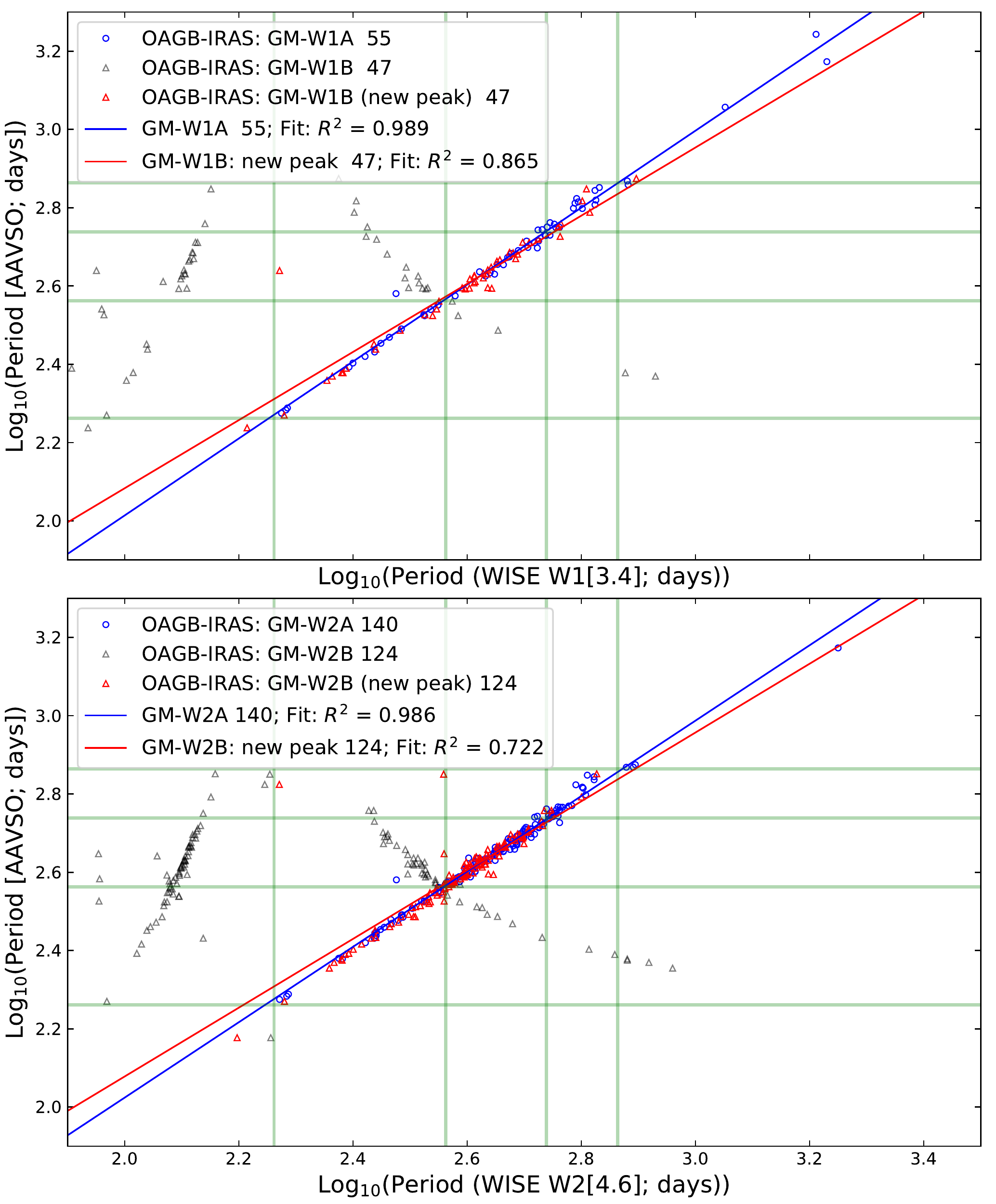}{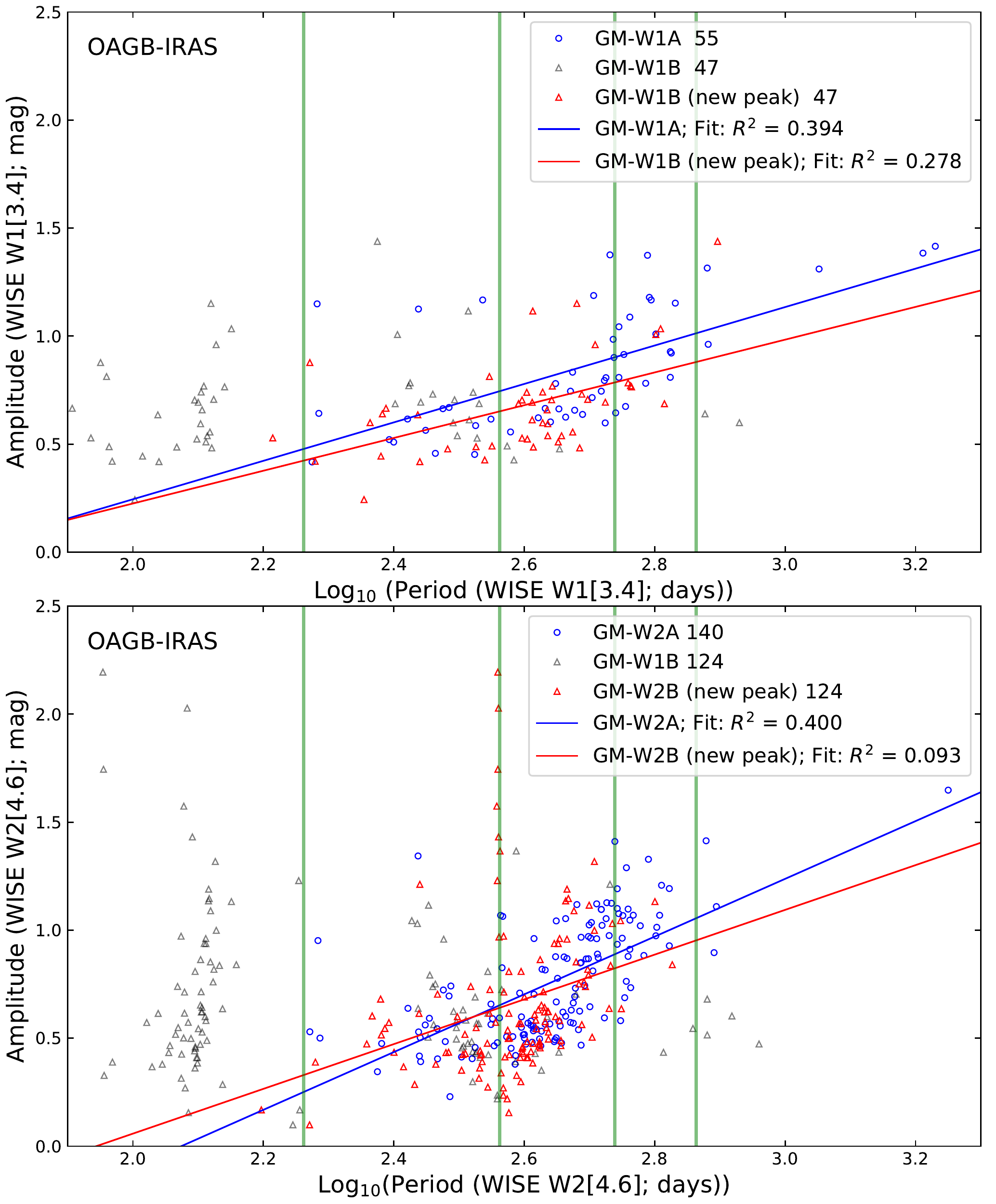}{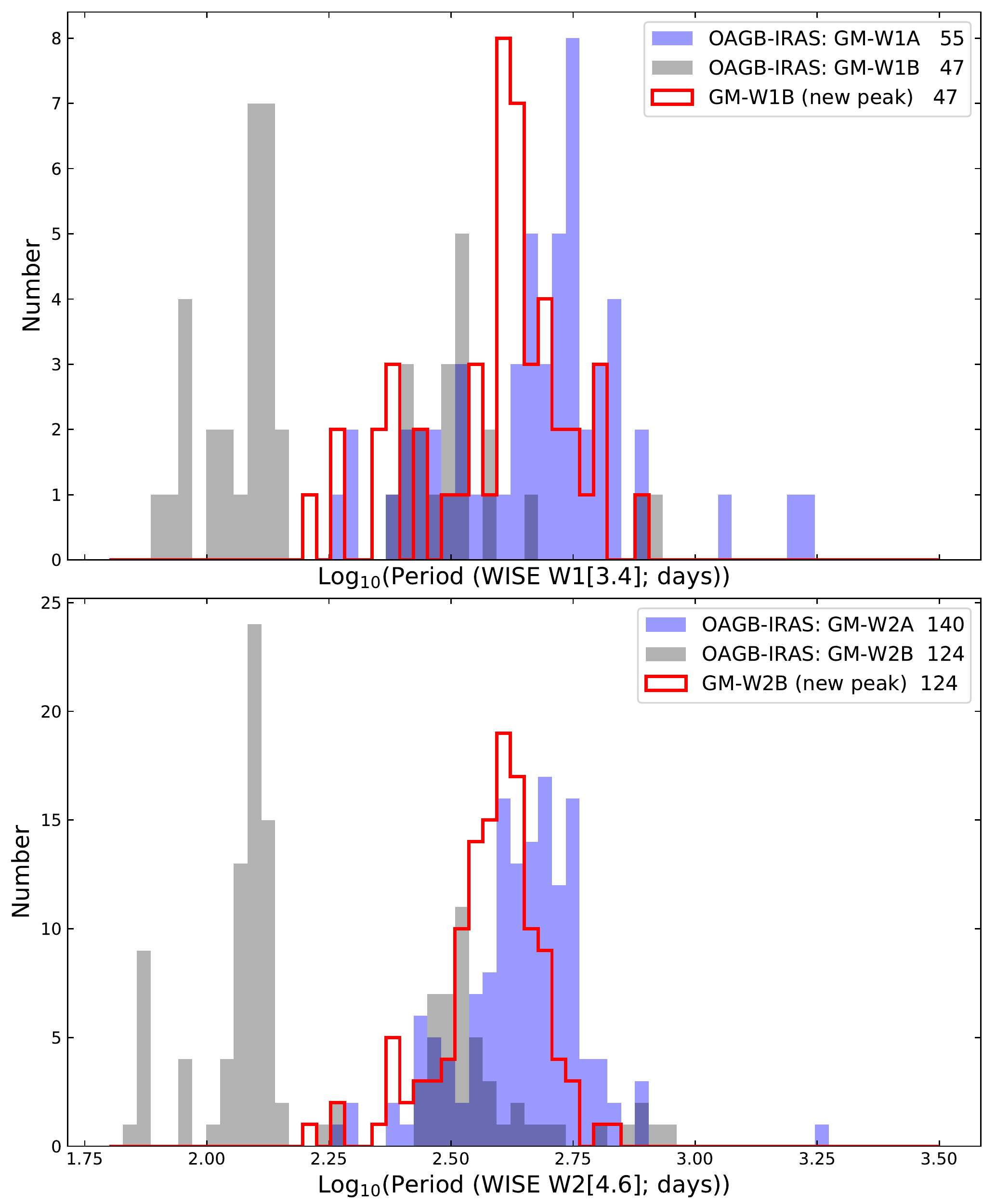}{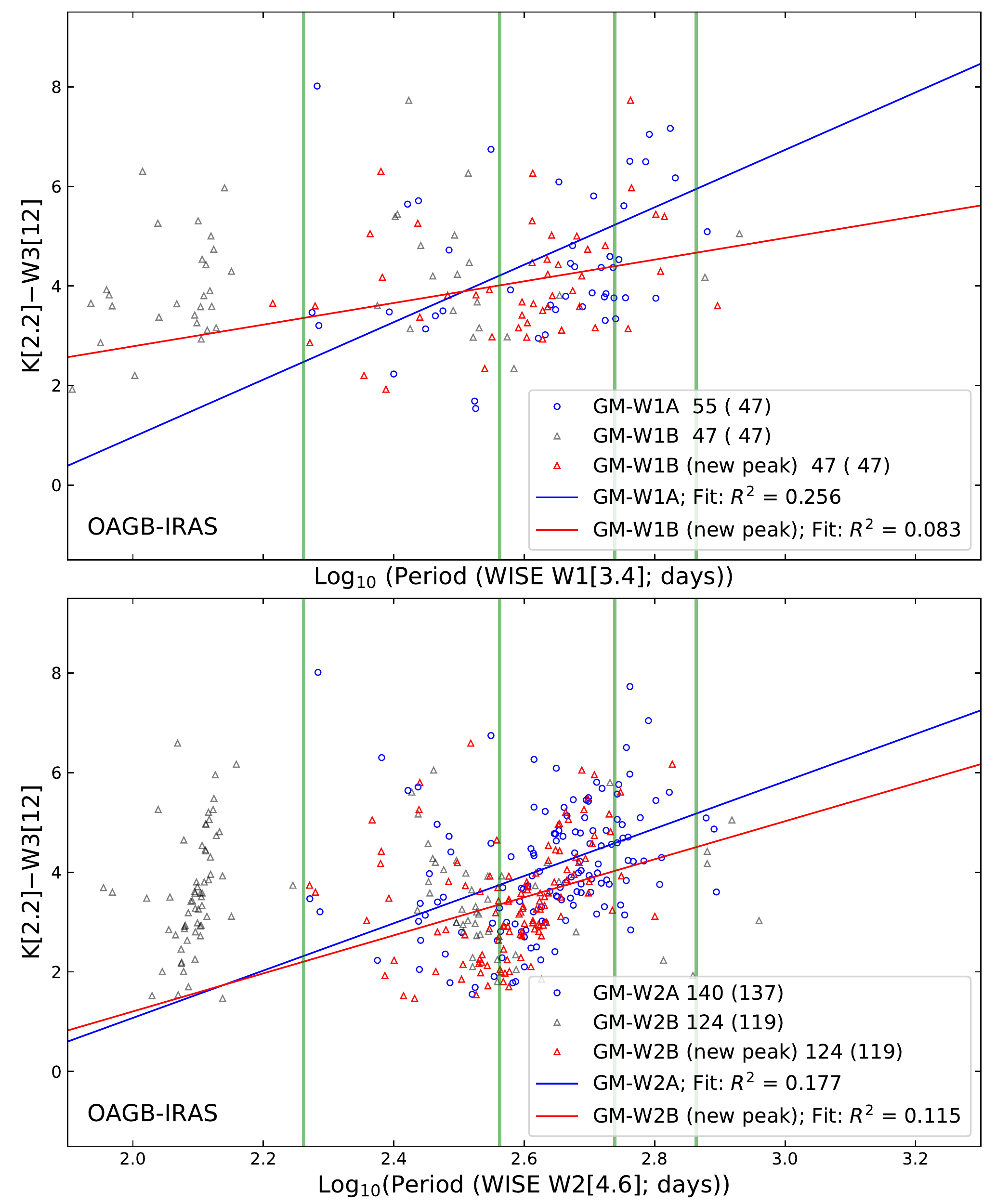}
\caption{Properties of variation for OAGB-IRAS objects known as Miras with periods.
The left panels show comparisons of the periods from AAVSO and the periods obtained from the WISE light curves.
The green vertical (and horizontal) lines indicate the multiples of the interval of WISE observations (6 months).
The right panels shows the period-amplitude and period-color relations.
See Section~\ref{sec:neo-mc}.} \label{f23}
\end{figure*}

\begin{figure*}
\centering \smallplotfour{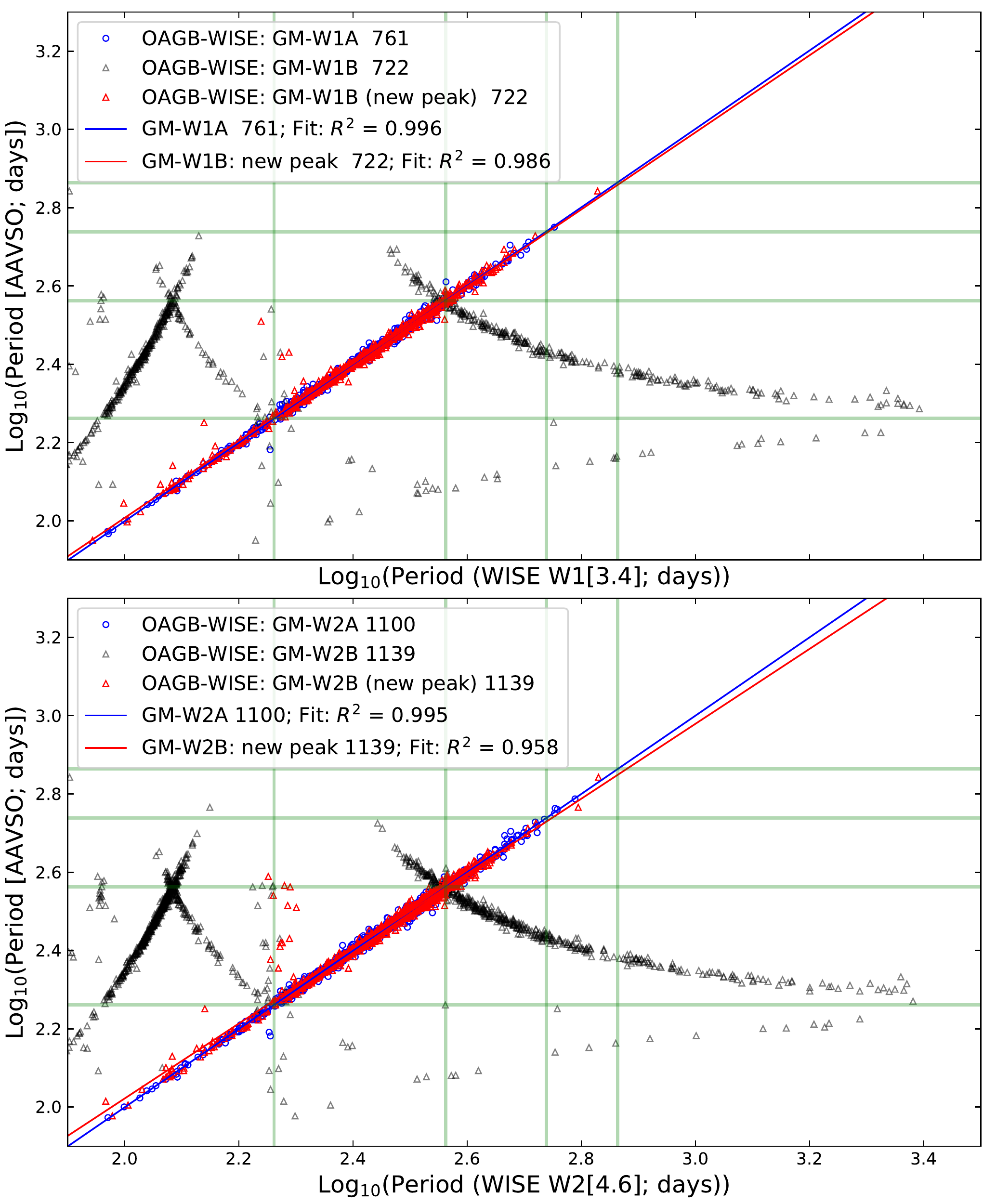}{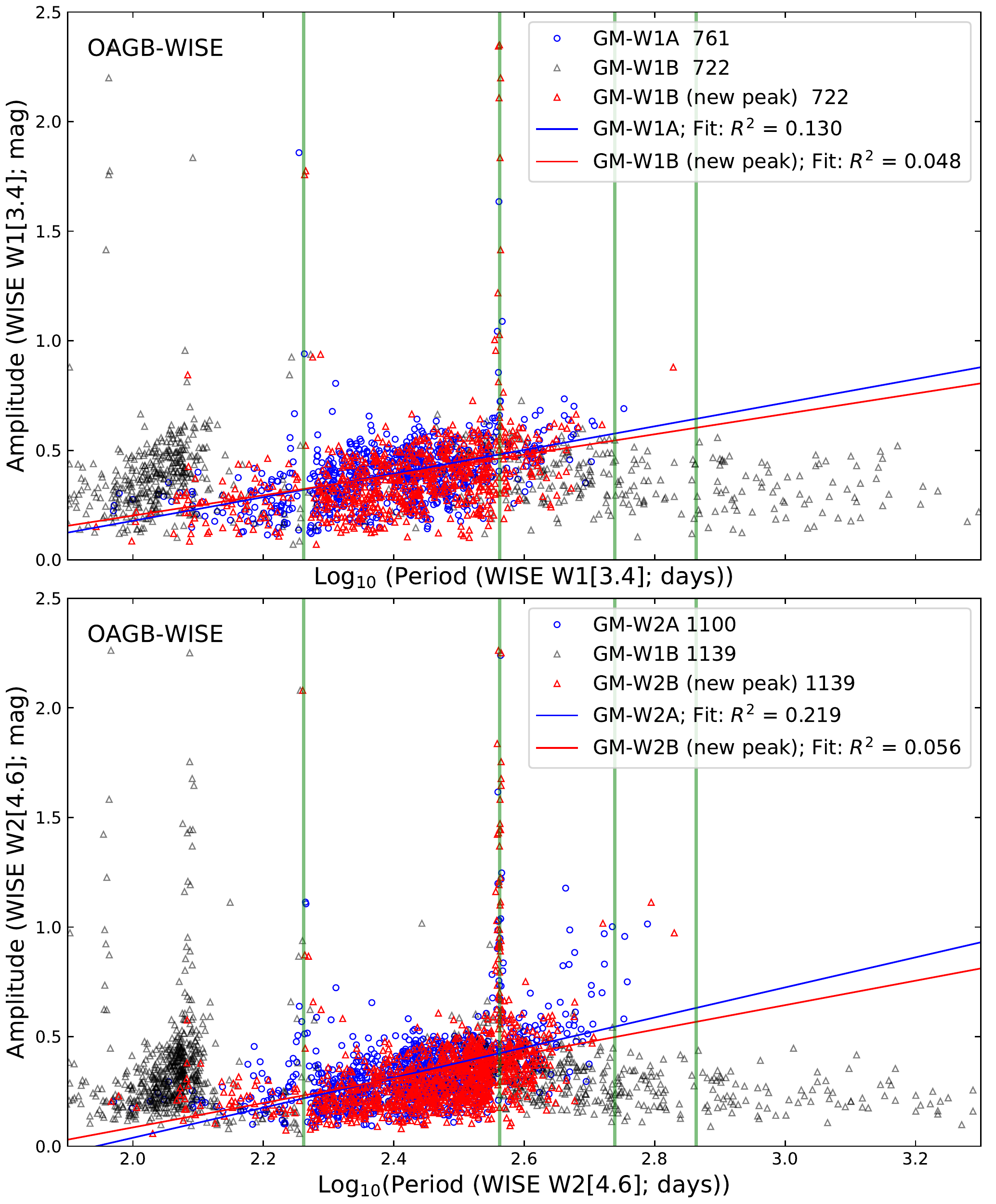}{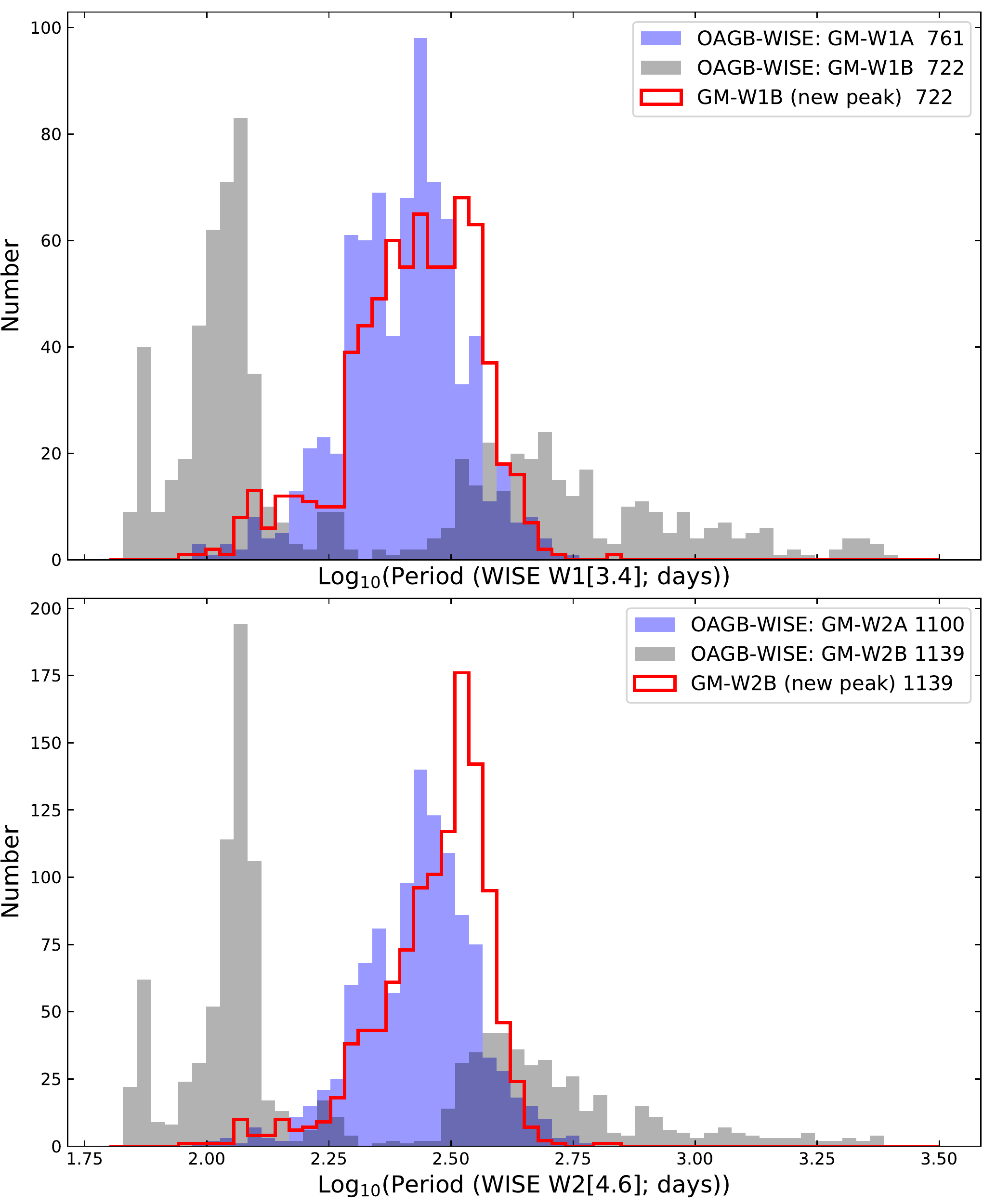}{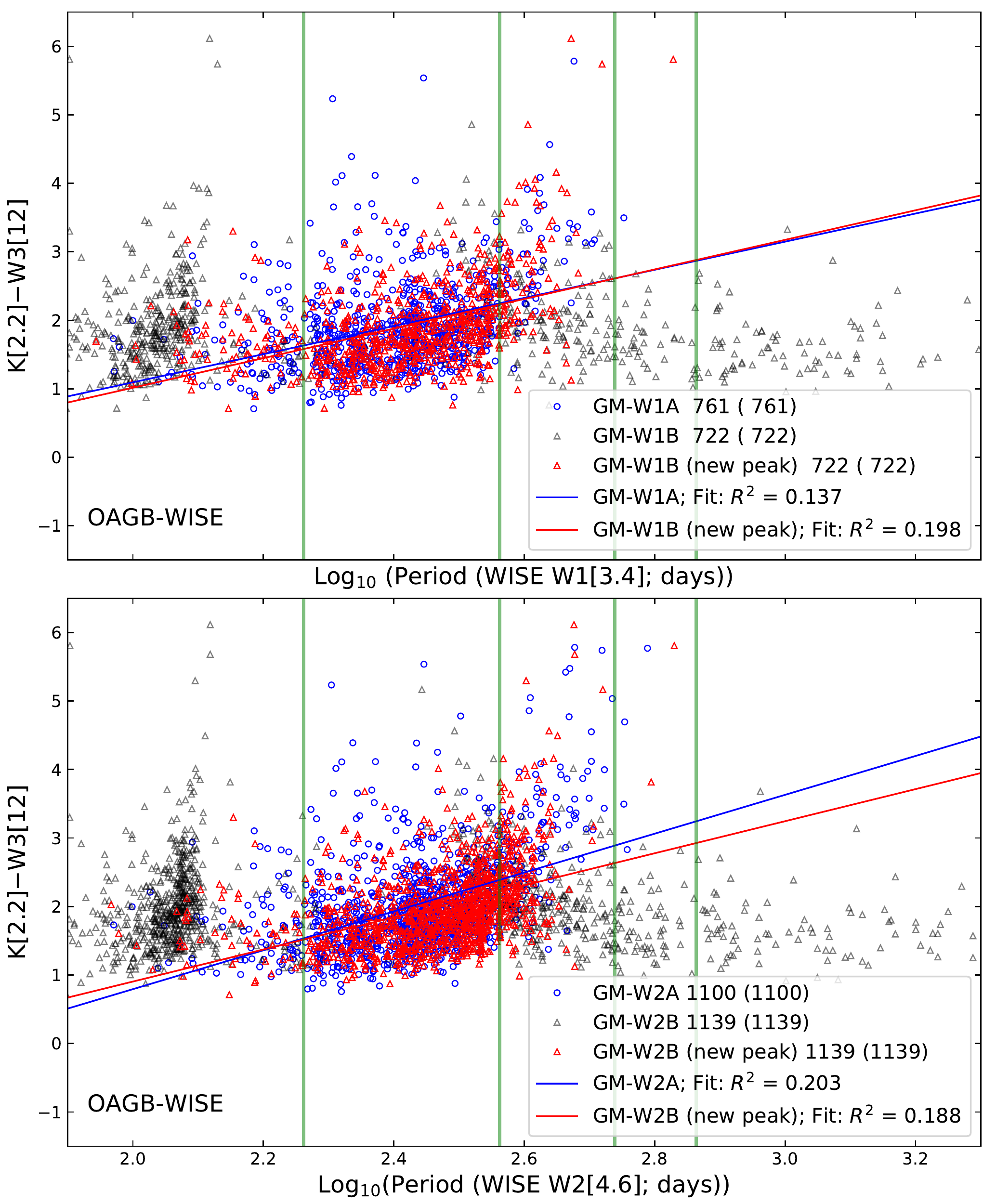}
\caption{Properties of variation for OAGB-WISE objects known as Miras with periods.
The left panels show comparisons of the periods from AAVSO and the periods obtained from the WISE light curves.
The green vertical (and horizontal) lines indicate the multiples of the interval of WISE observations (6 months).
The right panels shows the period-amplitude and period-color relations.
See Section~\ref{sec:neo-mc}.} \label{f24}
\end{figure*}

\begin{figure*}
\centering
\smallplotfour{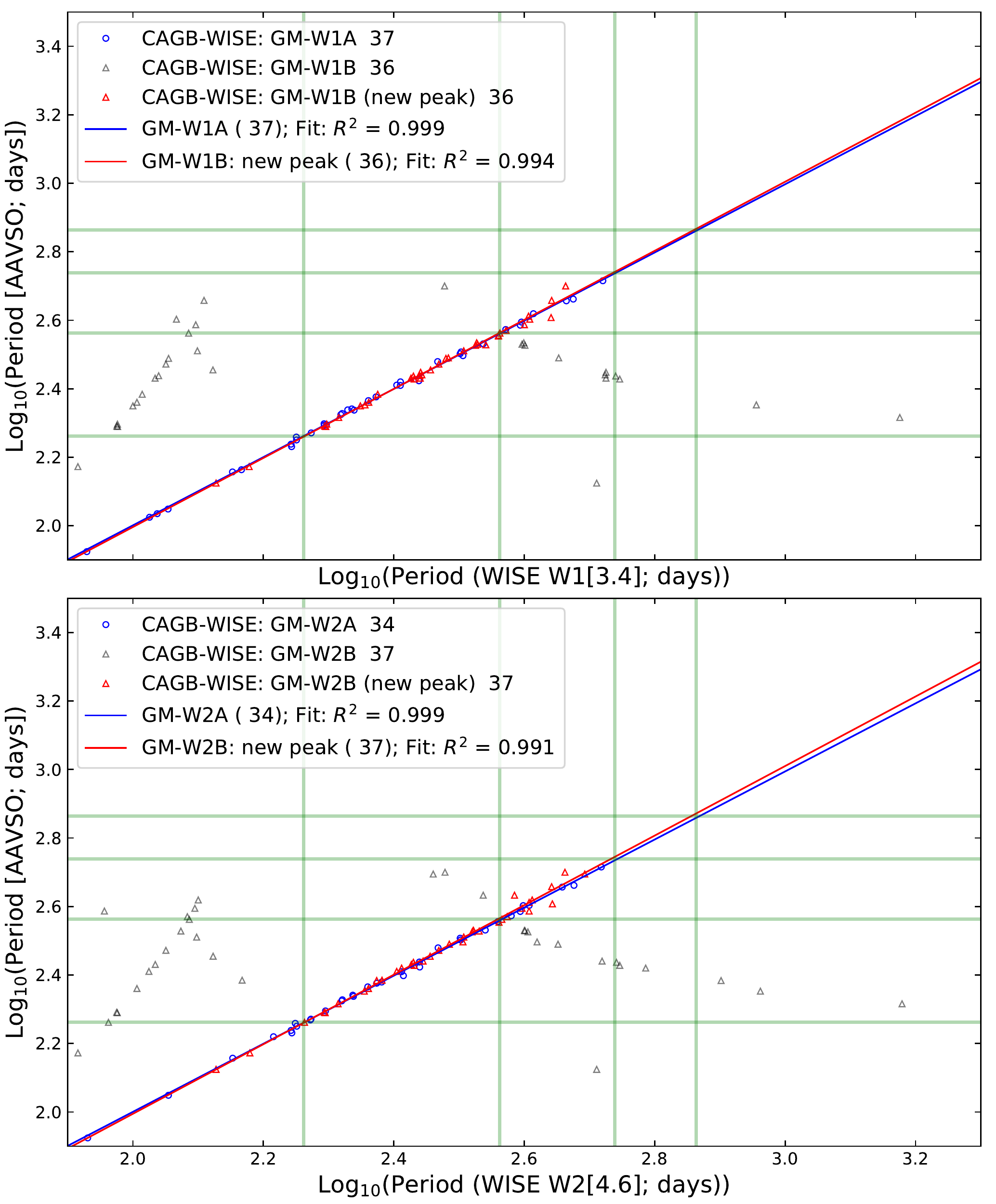}{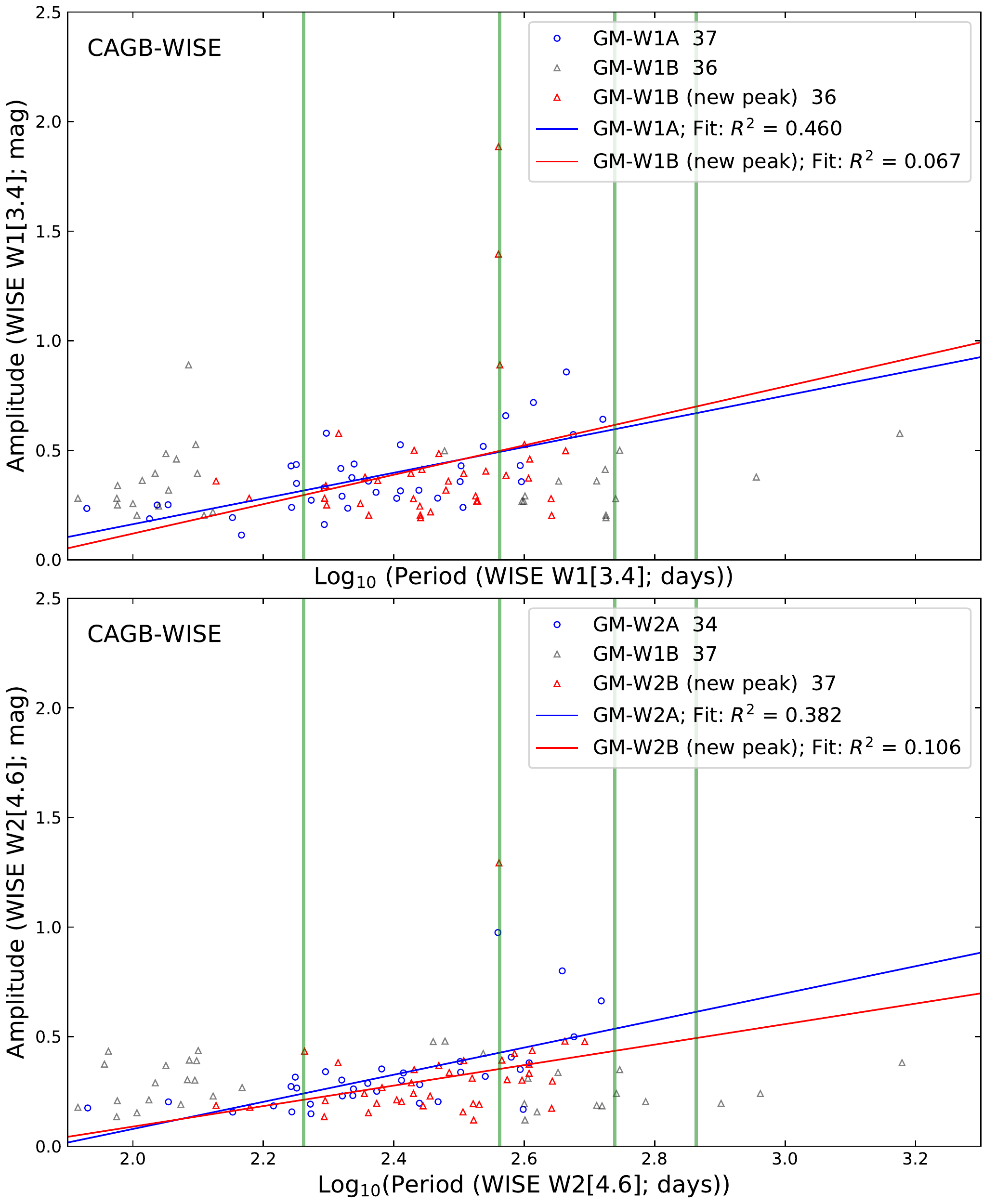}{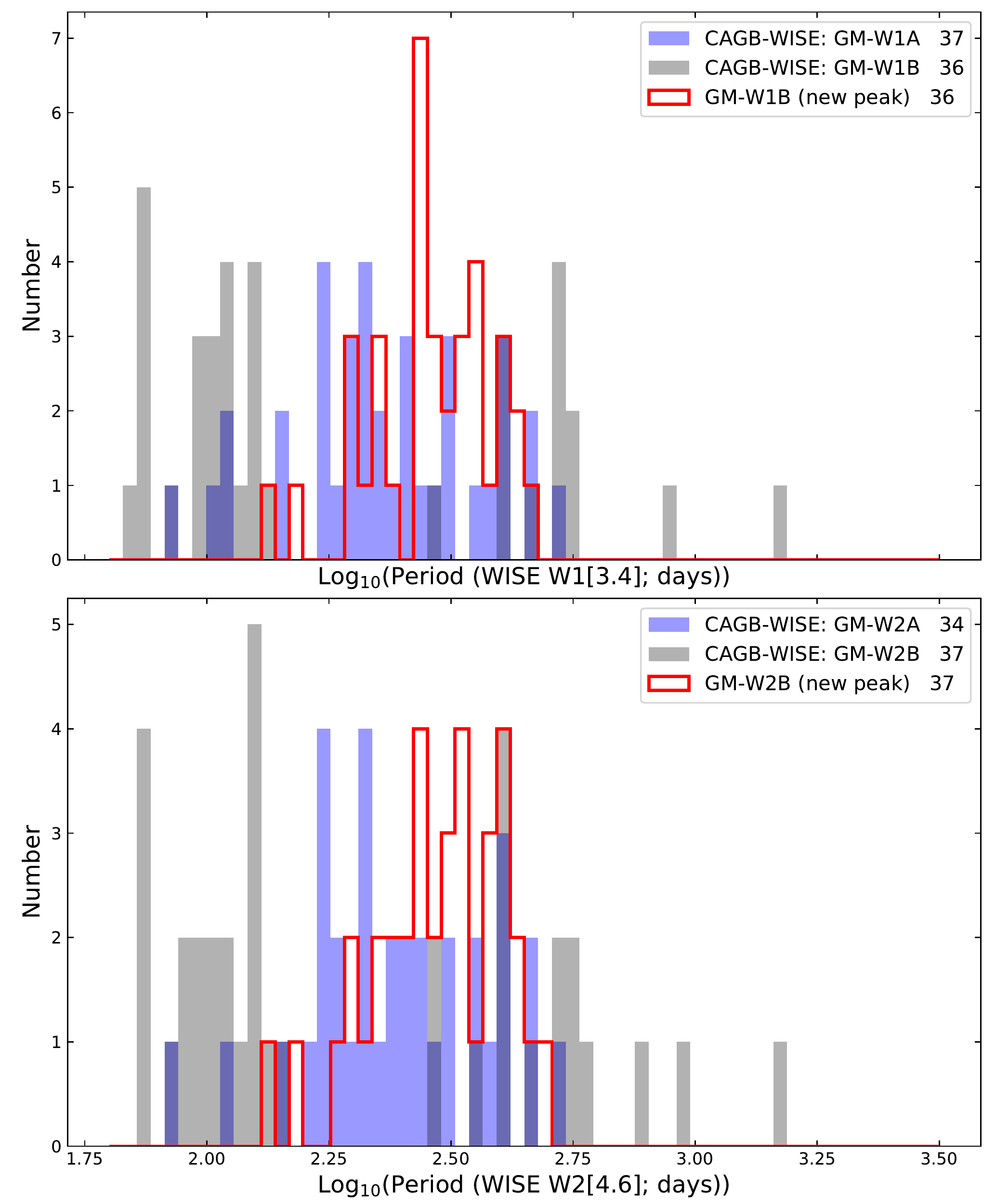}{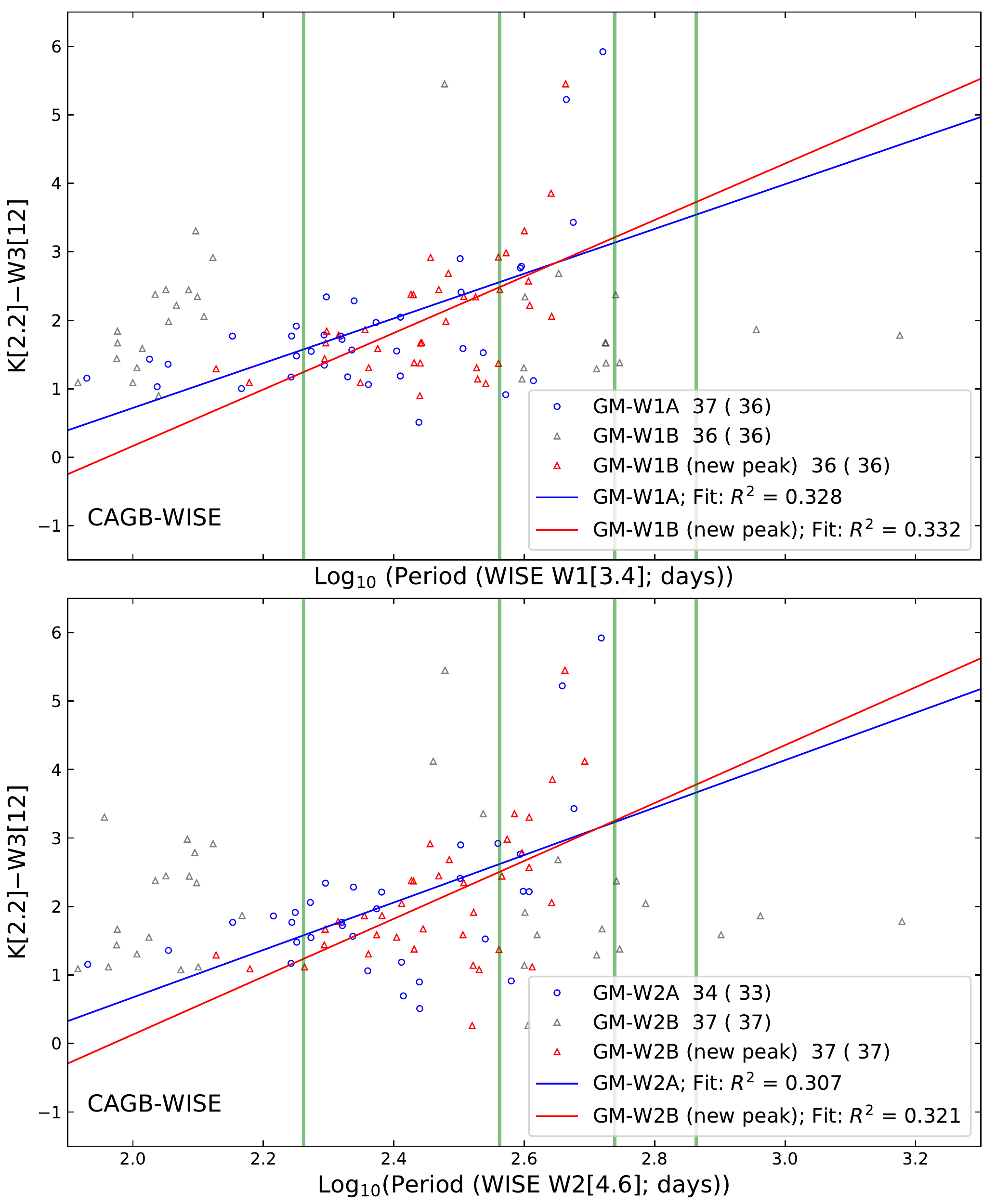}
\caption{Properties of variation for CAGB-WISE objects known as Miras with periods.
The left panels show comparisons of the periods from AAVSO and the periods obtained from the WISE light curves.
The green vertical (and horizontal) lines indicate the multiples of the interval of WISE observations (6 months).
The right panels shows the period-amplitude and period-color relations.
See Section~\ref{sec:neo-mc}.} \label{f25}
\end{figure*}

\begin{figure}
\centering
\smallplot{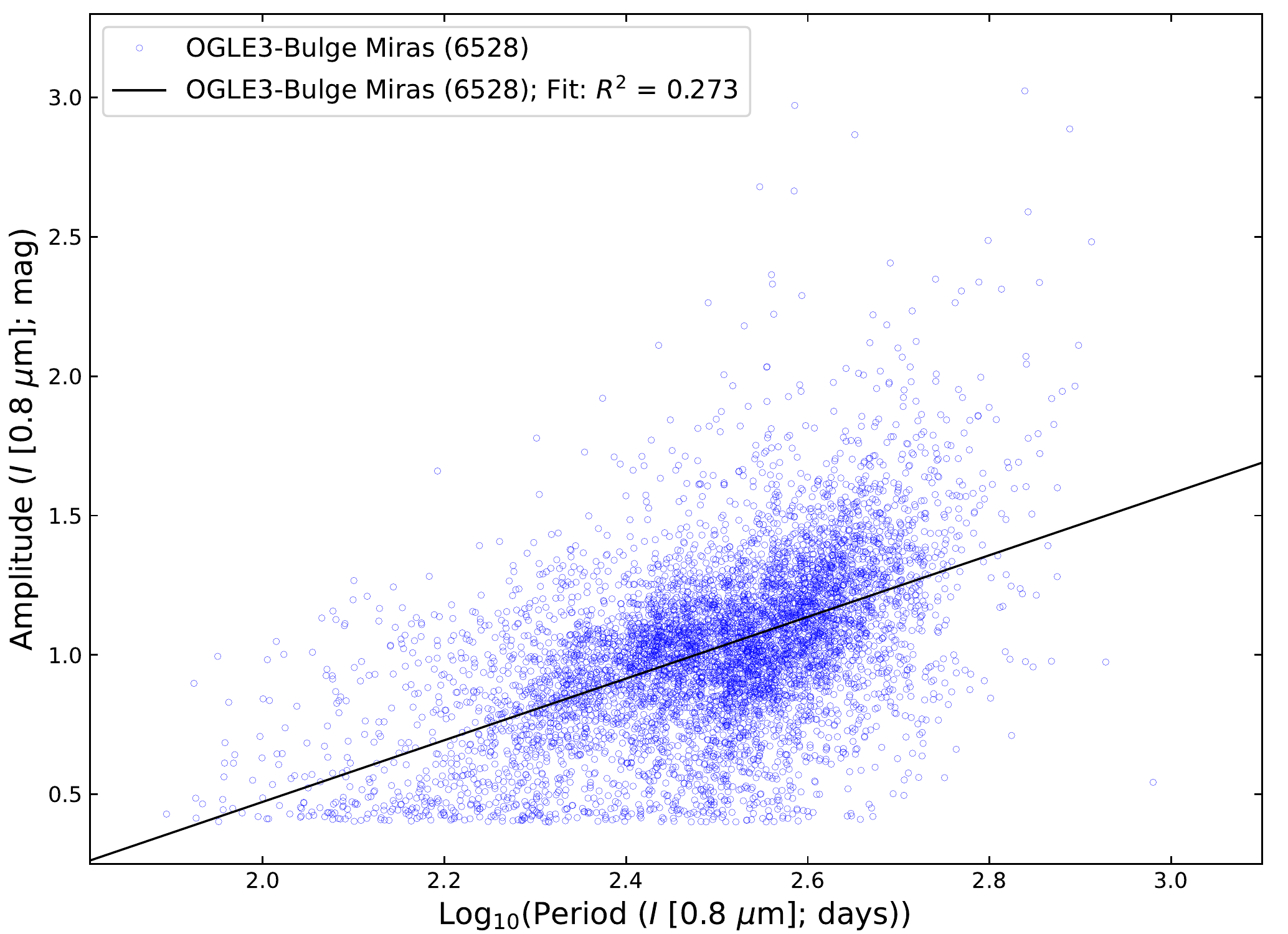}\caption{The period-amplitude relation for all Miras in the Galactic bulge
obtained from OGLE3 $I$ (0.8 $\mu$m) band observations (\citealt{sus13a}).}
\label{f26}
\end{figure}

\subsection{WISE light curves for known Mira Variables\label{sec:neo-m}}

\citet{ks2010a} and \citet{ks2010b} analyzed $L$ and $M$ band photometric data
for 12 OH/IR stars acquired between the 1970s and 2004 and found their periods
and amplitudes. Because $L$[3.4-3.6] and $M$[4.8-4.9] band wavelengths are near
WISE W1[3.4] and W2[4.6] band wavelengths, we have analyzed the combined light
curves. But the WISE data for most of these bright AGB stars are either
saturated or show too large scatters. We could plot a meaningful combined light
curve for only one source (OH 26.5+0.6).

Figure~\ref{f18} shows the light curve for the OH/IR star OH 26.5+0.6 (in
OI-SH) using the data at the $L$[3.4-3.6] band acquired in 1974-2003 and the
WISE W1[3.4] band data acquired in the last 12 yr.

Figure~\ref{f19} shows the OGLE3 and WISE light curves for IRAS 18007-3012 (in
OI-OG). The light curves at $I$ (0.8 $\mu$m) and W2[4.6] bands show similar
characteristics, though the amplitude at the W2[4.6] band is smaller. Unlike
the OGLE3 light curve, there are multiple peaks with similar power values in
the Lomb-Scargle periodogram obtained from the WISE light curve because the
WISE data were taken in a regular interval (every six months).

We have obtained pulsation periods from the WISE light curves for 2810 objects
known as Miras with periods from AAVSO (GM-W1 or GM-W2; 284 OAGB-IRAS, 13
CAGB-IRAS, 2429 OAGB-WISE, and 84 CAGB-WISE objects; see Table~\ref{tab:tab8}).

Figure~\ref{f20} shows the light curves for six OAGB-IRAS objects known as
Miras. The periods obtained from WISE data are similar to the ones in AAVSO,
but there are multiple peaks with similar Lomb-Scargle power values. For the
two objects in lower panels (GM-W2B objects), though the period from the
primary peak of the Lomb-Scargle power is different from the AAVSO period, the
period from the second peak is very similar to the AAVSO period and produce a
similar fit to the observations.

Figure~\ref{f21} shows the light curves for six OAGB-WISE objects known as
Miras. The obtained primary period from WISE data is similar to the ones in
AAVSO for three objects (GM-W2A objects). For the other three objects (GM-W2B
objects), the period from the second peak of Lomb-Scargle power is similar to
the AAVSO period.

Figure~\ref{f22} shows the light curves for for three CAGB-IRAS objects
(objects in upper panels: CI-SH; IRAS 08546+1732: CI-GC) and three CAGB-WISE
objects (CW-2323: CW-GC; objects in lower panels: CW-OG) known as Miras with
periods. For three objects (GM-W2B objects), the period from the second peak of
Lomb-Scargle power is similar to the AAVSO period.

For most Mira variables whose primary periods obtained from the WISE light
curves are different from AAVSO periods (GM-W1B or GM-W2B objects), we can find
a new (second or up to fourth) peak in the Lomb-Scargle power, for which the
new period is similar to the AAVSO period and produces a similar fit (a little
worse fit with a smaller $R^2$ value) to the observations.

Because there are multiple peaks with similar Lomb-Scargle power values, it is
not easy to obtain a precise period using only the WISE data (see
Figures~\ref{f20}-\ref{f22}). The uncertainties in period and amplitude
specified in Figures~\ref{f18} and \ref{f19} are calculated by measuring the
smooth imprecision in the selected peak of the Lomb-Scargle power. However,
when there are multiple peaks with similar Lomb-Scargle power values,
uncertainties expressed in this way cannot be meaningful because the
uncertainties of periods can be more affected by the false peaks
(\citealt{vanderPlas2018}).

\subsection{Comparison between AAVSO periods and new periods from WISE data for known Miras\label{sec:neo-mc}}

If we compare the WISE periods with the AAVSO periods for the objects known as
Miras with known periods from AAVSO (GM-W1 or GM-W2 objects; see
Table~\ref{tab:tab8}), we may check the reliability of the the WISE periods.
For AGB-WISE objects, most of AAVSO periods are from OGLE3 $I$ (0.8 $\mu$m)
band observations (\citealt{sus13a}).

Figures~\ref{f23}-\ref{f25} show the relations between AAVSO periods and the
periods obtained from the WISE light curves for OAGB-IRAS, OAGB-WISE, and
CAGB-WISE objects (known as Miras with known periods from AAVSO), respectively.
The figures also show the histograms of the periods obtained from the WISE
light curves, period-amplitude relations, and period-color relations. For
CAGB-IRAS objects, the numbers of sample stars (GM-W1 or GM-W2; see
Table~\ref{tab:tab8}) are too small to make proper plots.

The upper-left panels of Figures~\ref{f23}-\ref{f25} compare the periods from
AAVSO (most of them are are from OGLE3 $I$ band observations) and the periods
obtained from the WISE light curves for OAGB-IRAS, OAGB-WISE, and CAGB-WISE
objects, respectively. For about a half of the objects (GM-W1A or GM-W2A), the
obtained primary periods from the WISE data are similar to the periods in
AAVSO. For another half (GM-W1B or GM-W2B), the obtained primary periods from
WISE data are different from the periods in AAVSO. The deviations look to occur
more severely when the AAVSO or WISE periods are similar to the interval of the
WISE observations (6 months).

These deviations could be due to the characteristic of the Lomb-Scargle
periodogram with similar multiple peaks, which could be due to the regularity
of the WISE observations (6 months). \citet{vanderPlas2018} compared the true
period and peak Lomb-Scargle period for 1000 simulated periodic light curves
and found that the Lomb-Scargle peak does not always coincide with the true
period, and there is noticeable structure among these failures, which is
similar to the ones in upper-left panels of Figures~\ref{f23}-\ref{f25}. The
noticeable structure is clearer for OAGB-WISE objects because the sample number
is much larger than other classes.

But for most GM-W1B or GM-W2B (see Table~\ref{tab:tab8}) objects, we can find a
new (second or up to fourth) peak in the Lomb-Scargle power values, for which
the new period is similar to the AAVSO period. This would be because the AAVSO
periods can be regarded as true periods and the multiple peaks in the
Lomb-Scargle power values are very similar (see Figures~\ref{f20}-\ref{f22}).
When we select the new peak that is similar to the AAVSO period for the GM-W1B
or GM-W2B objects, the periods obtained from the WISE light curves are very
similar to the AAVSO periods for most objects.

The lower-left panels of Figures~\ref{f23}-\ref{f25} show the histograms of the
periods obtained from the WISE light curves. GM-W1A or GM-W2A objects show a
roughly single peak whereas GM-W1B or GM-W2B objects show multiple peaks. When
we select new peaks for GM-W1B or GM-W2B objects, the histogram shows a roughly
single peak just like GM-W1A or GM-W2A objects. Again, this effect is clearer
for OAGB-WISE objects because the sample number is much larger than other
classes.

On the whole, the periods from the WISE data and AAVSO show good correlations
for all of the sample stars known as Miras (see upper-left panels of
Figures~\ref{f23}-\ref{f25}).

Upper-right panels of Figures~\ref{f23}-\ref{f25} show the period-amplitude
relations. Both the relations for the objects using primary peaks (GM-W1A or
GM-W2A) and for the objects using new peaks (GM-W1B or GM-W2B) look similar to
the one for Miras in OGLE3 bulge (see Figure~\ref{f26}), which show larger
amplitudes. But for some objects, the amplitudes show large deviation from the
general trend, especially when obtained periods are similar to the multiples of
the interval of WISE observations (6 months). Compared with OAGB-WISE objects,
OAGB-IRAS objects show generally larger periods and amplitudes.

Lower-right panels of Figures~\ref{f23}-\ref{f25} show the PCRs for OAGB-IRAS,
OAGB-WISE, and CAGB-WISE objects using the K[2.2]$-$W3[12] color. The objects
show the similar PCRs to those for known Mira variables using AAVSO periods
(see Figure~\ref{f16}).

For most objects that are known as Miras with periods in the catalog, we find
that the new periods from the WISE W1[3.4] and W2[4.6] light curves are very
similar to the periods in AAVSO. They also show very similar period-amplitude
relations, though the amplitudes from OGLE3 are generally larger.

Though the reliability of the periods obtained from WISE observations has a
weak point due to the regularity of observation (6 months), the new periods
obtained from the WISE W1[3.4] and W2[4.6] light curves, whether they are from
primary peaks of Lomb-Scargle power or not, could be more reliable than AAVSO
periods for some objects depending on the quality of the model fit.

\begin{figure*}
\centering
\smallplotsix{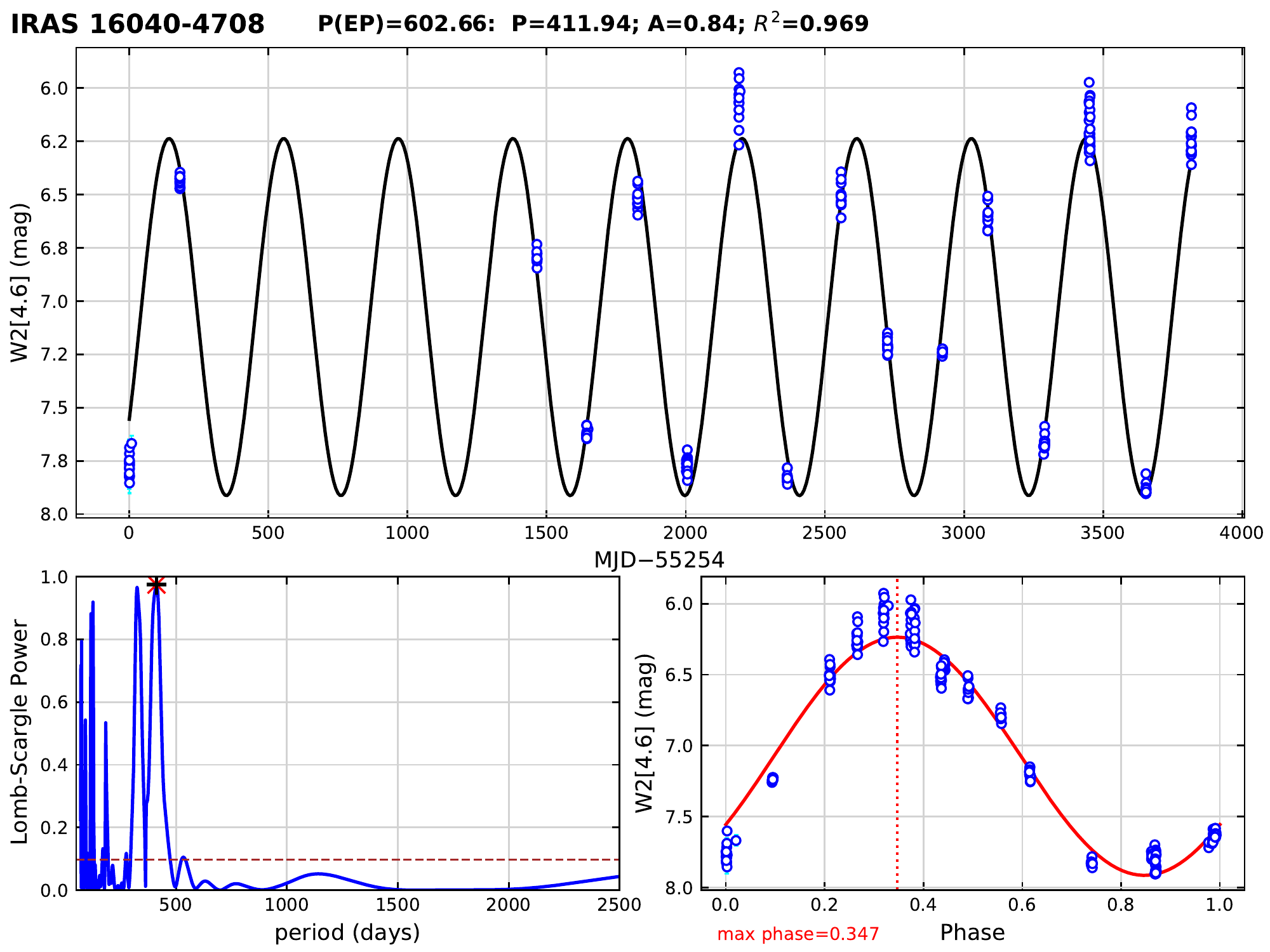}{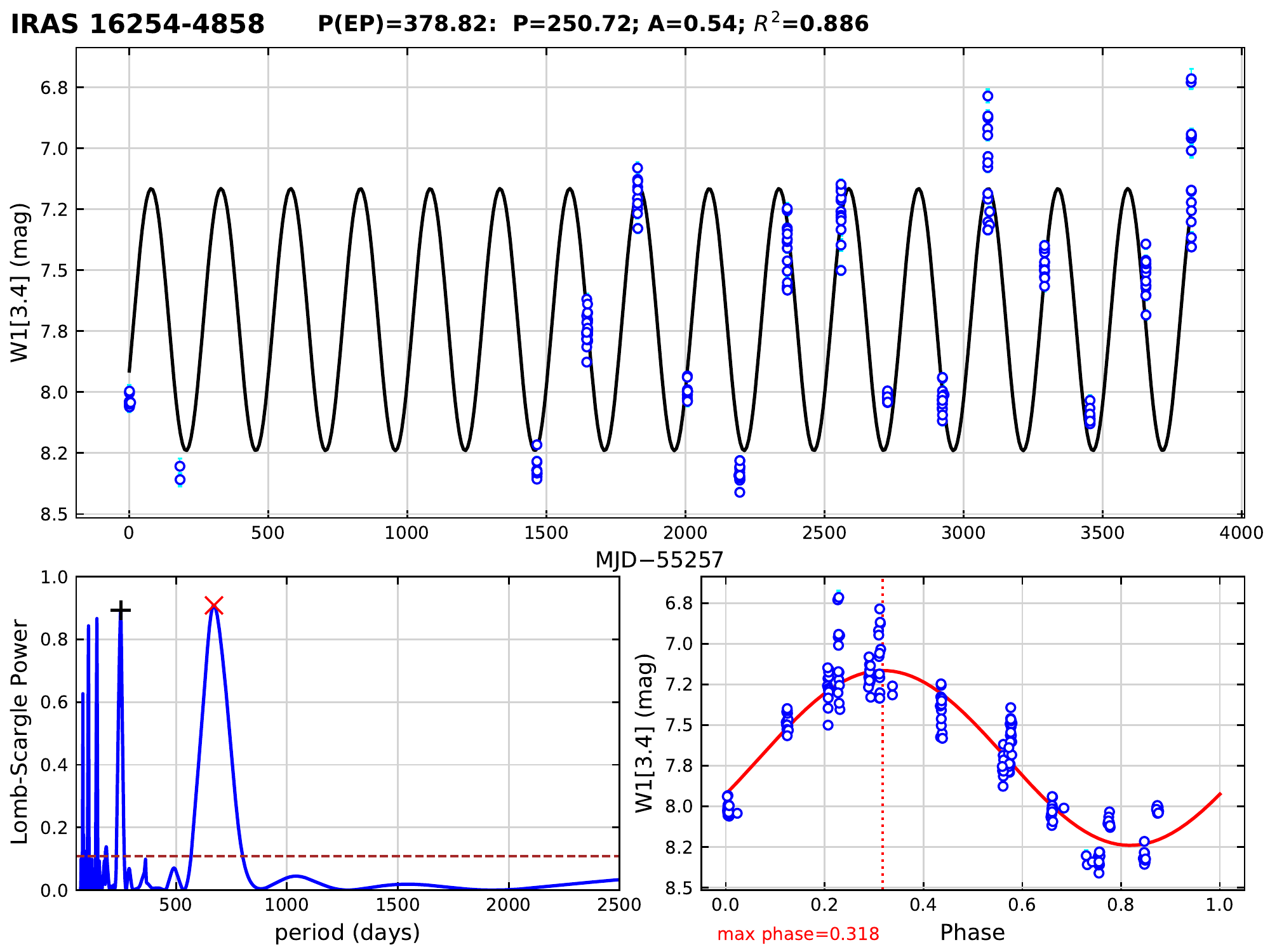}{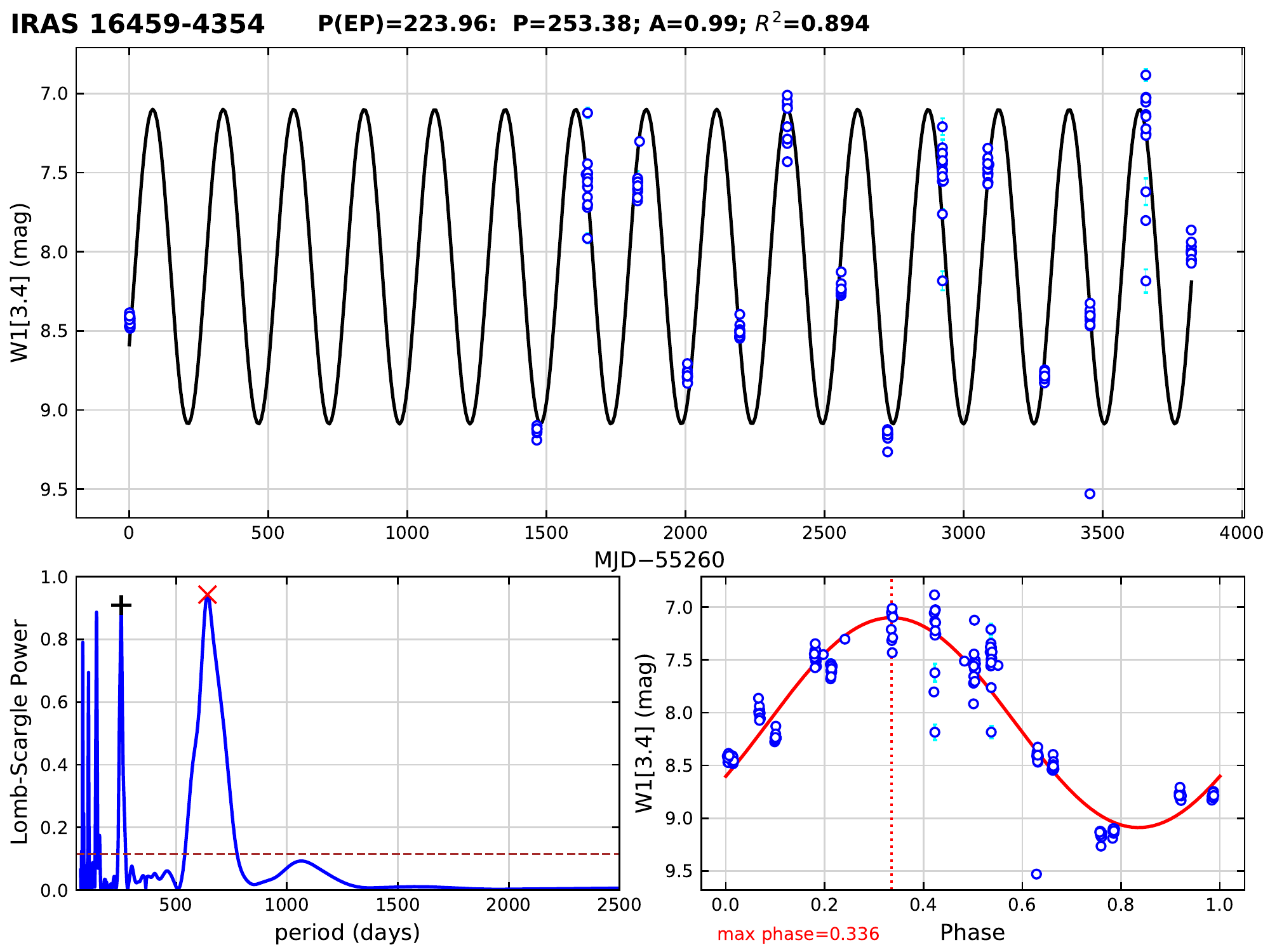}{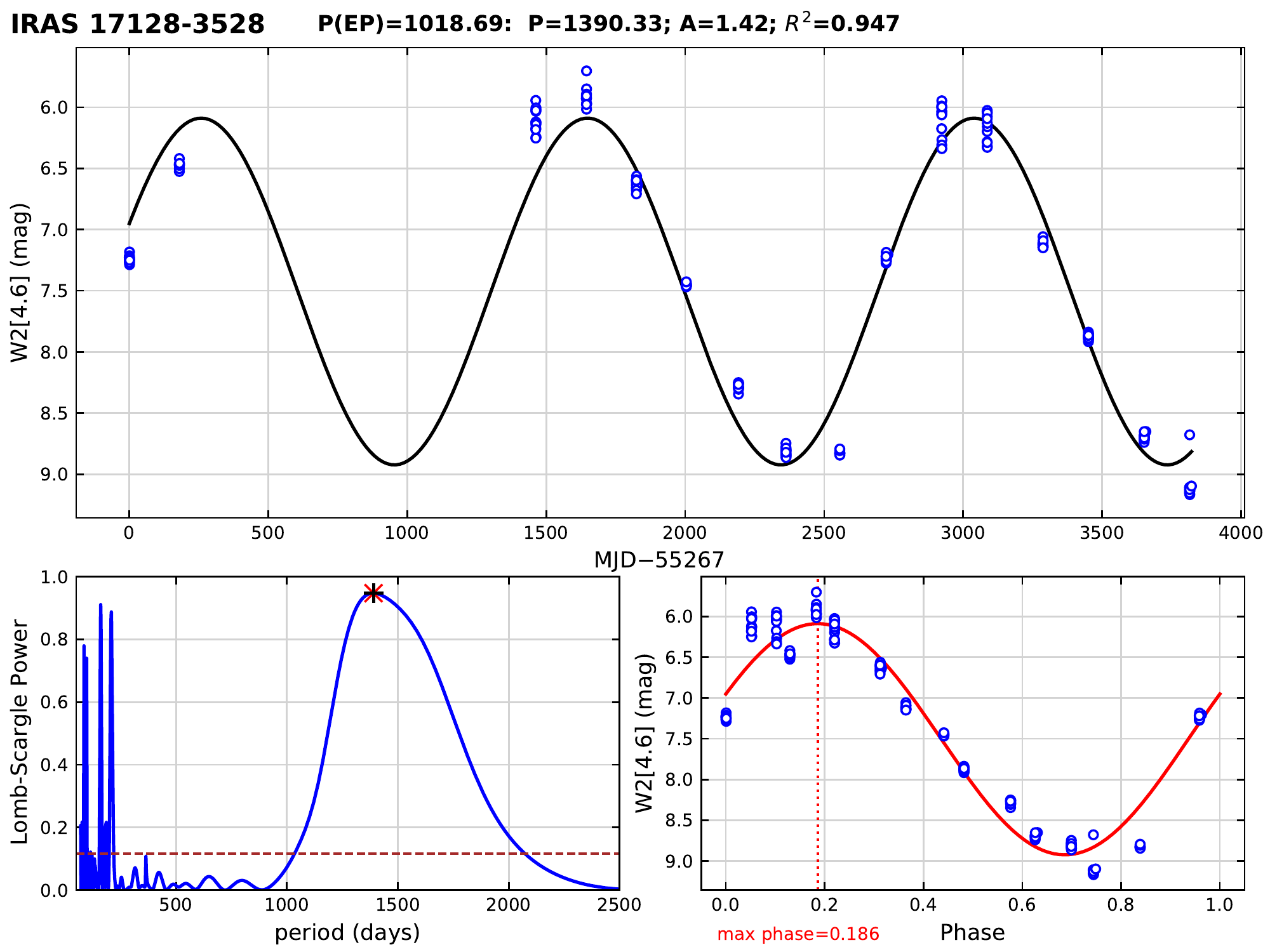}{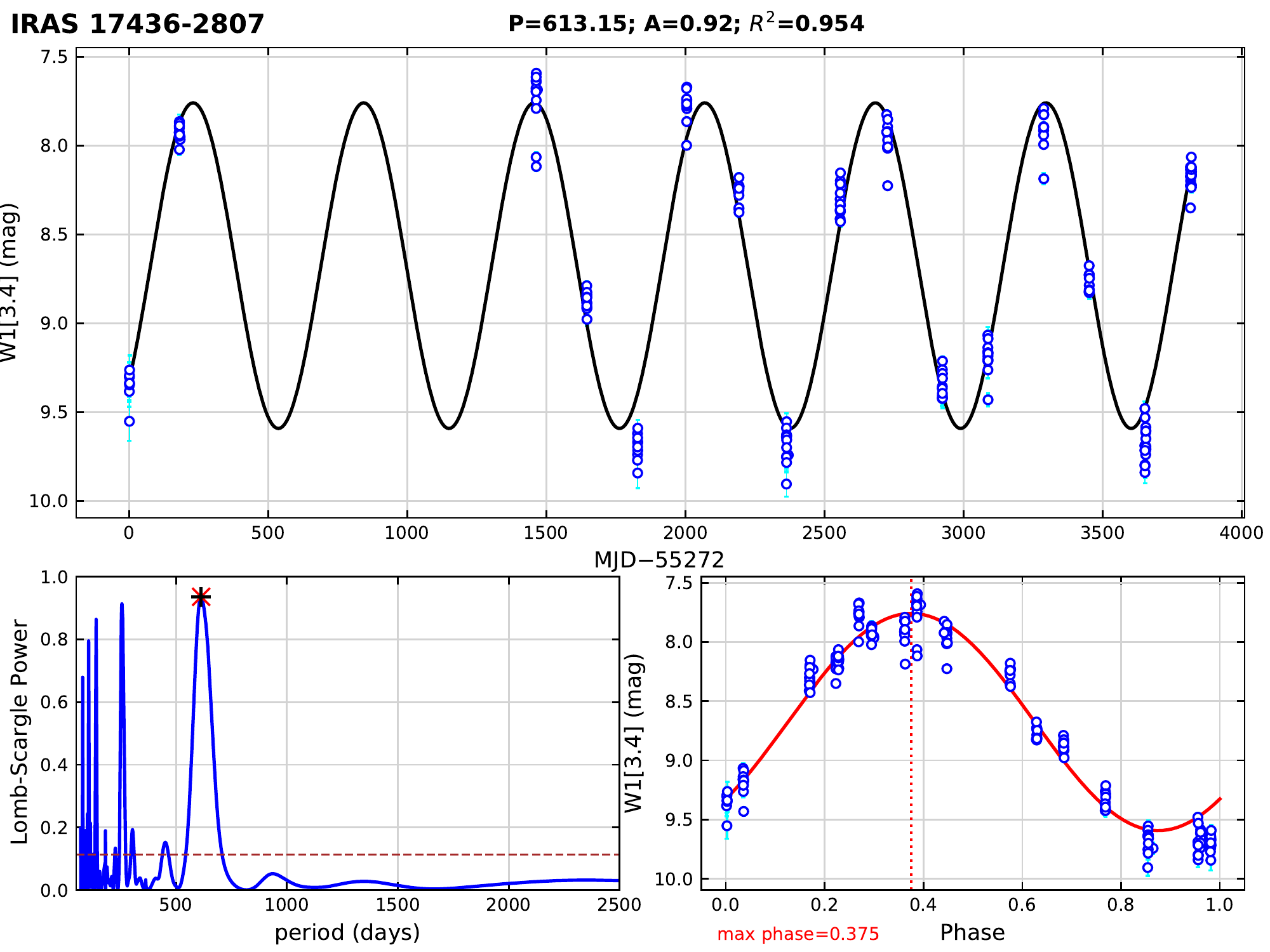}{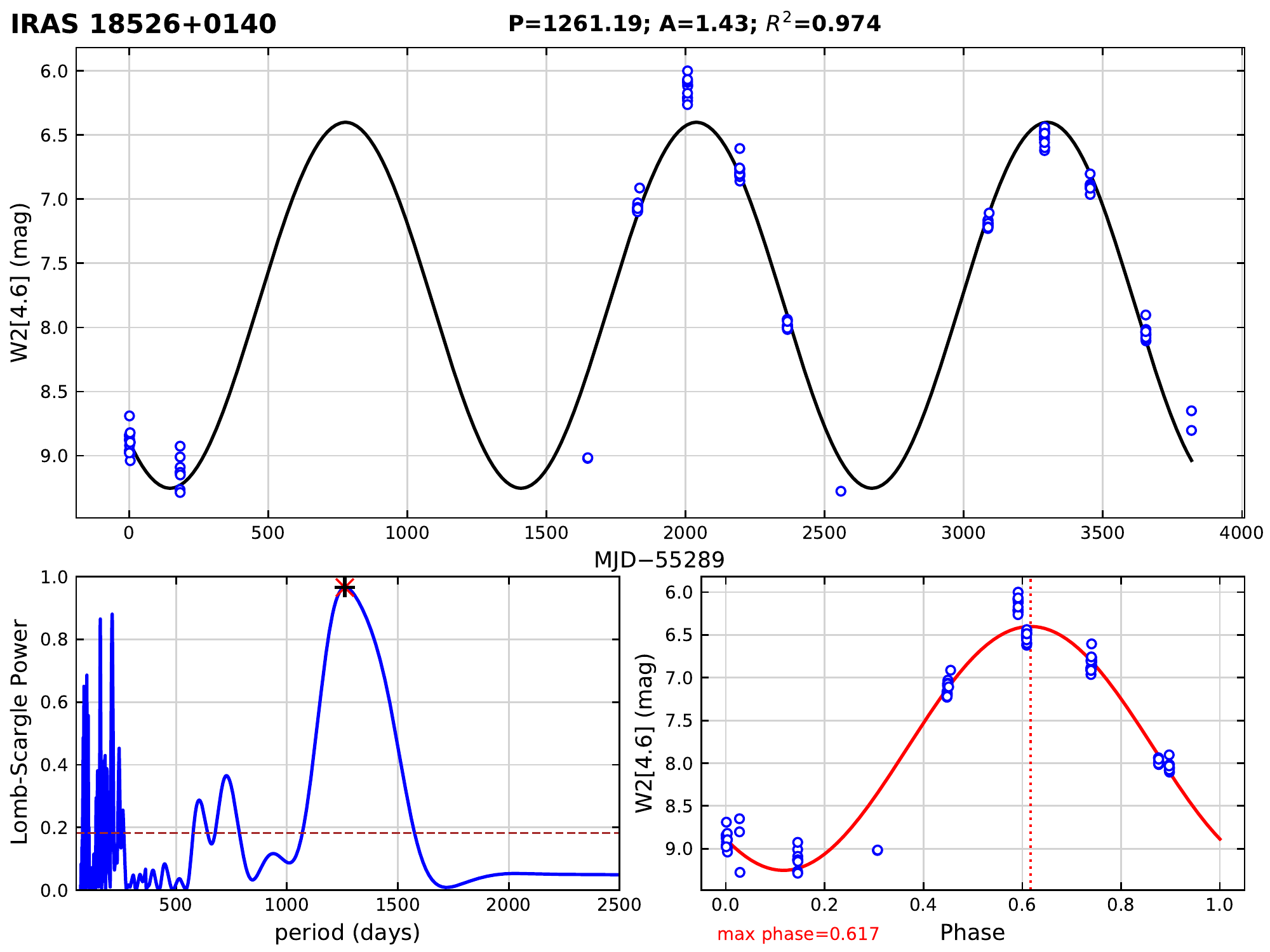}
\caption{WISE light curves and Lomb-Scargle periodograms for OAGB-IRAS objects with unknown periods (IRAS 16254-4858 in OI-ST, IRAS 17128-3528 in OI-JB, and others in OI-SH).
In the Lomb-Scargle periodogram, the red X and black cross marks indicate the primary and selected peaks, respectively
and the red dashed brown horizontal line indicates the periodogram level corresponding to a maximum peak false alarm probability of 1 \%.
See Section~\ref{sec:neo-nm}.
For four objects, P(EP) (the expected period from the IR color K[2.2]$-$W3[12]; see Section~\ref{sec:neo-nmc}) is also shown.
For two objects, the second peak of the Lomb-Scargle power is selected for the period because it is more similar to EP.
} \label{f27}
\end{figure*}

\begin{figure*}
\centering
\smallplotsix{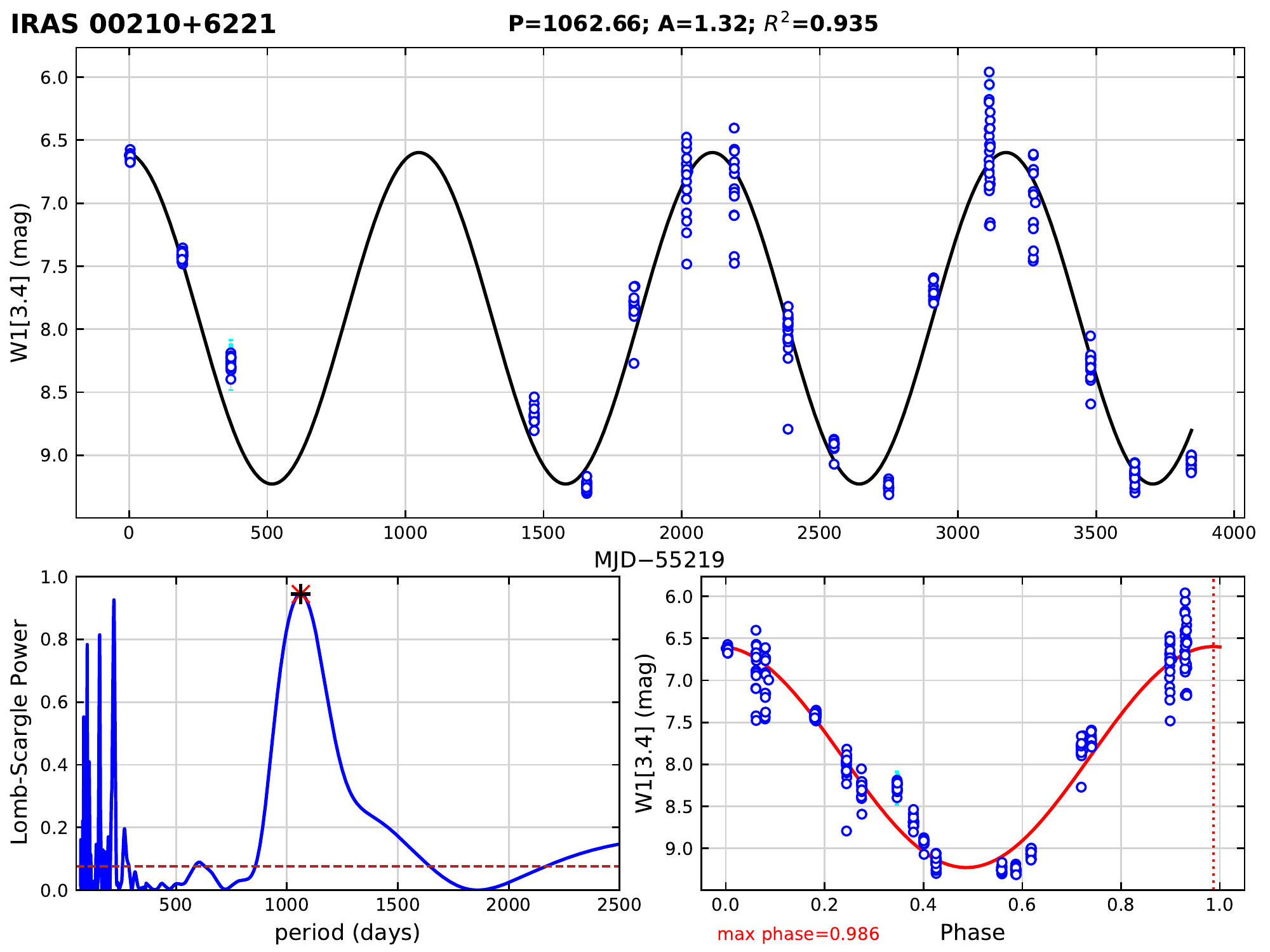}{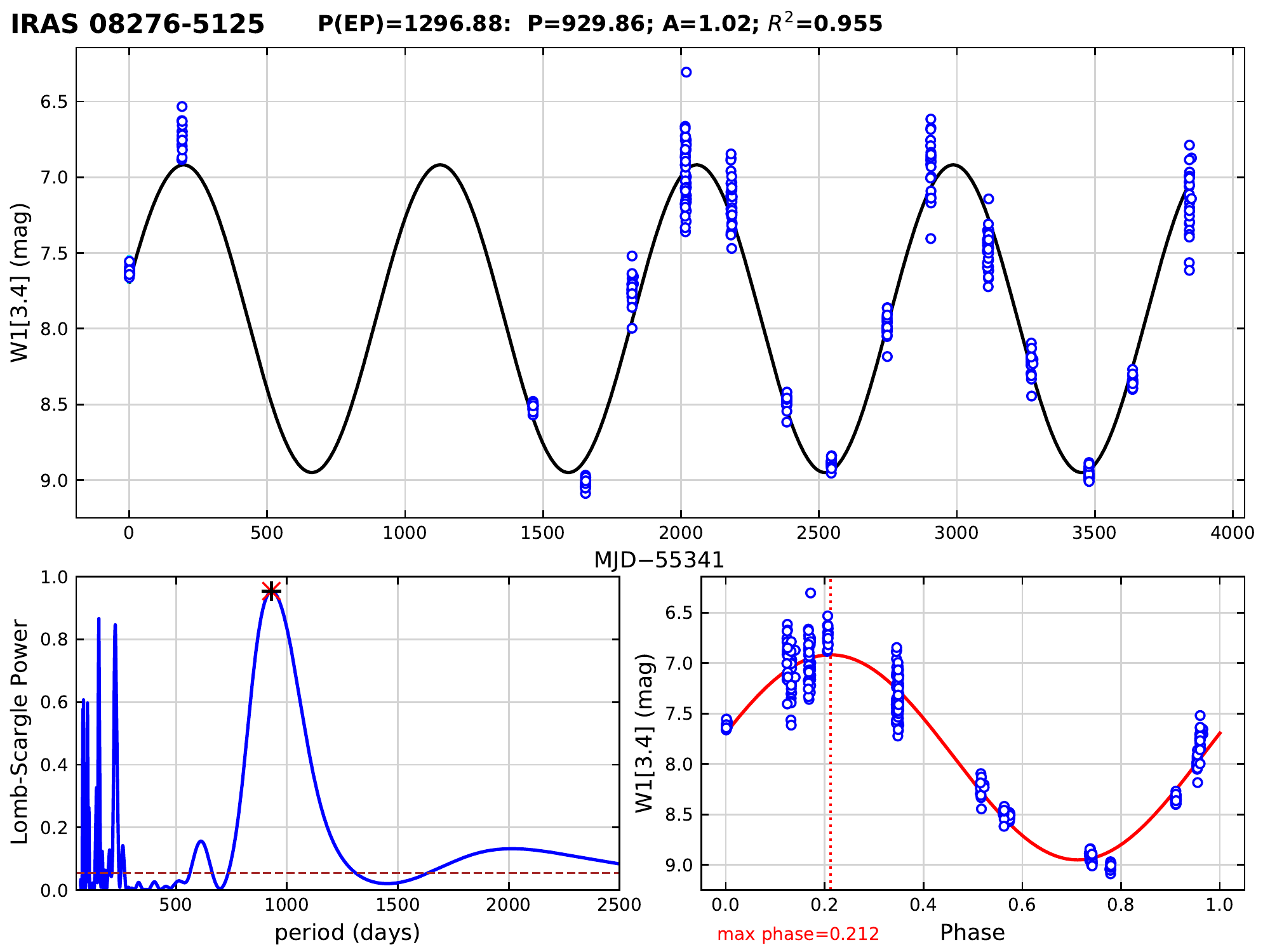}{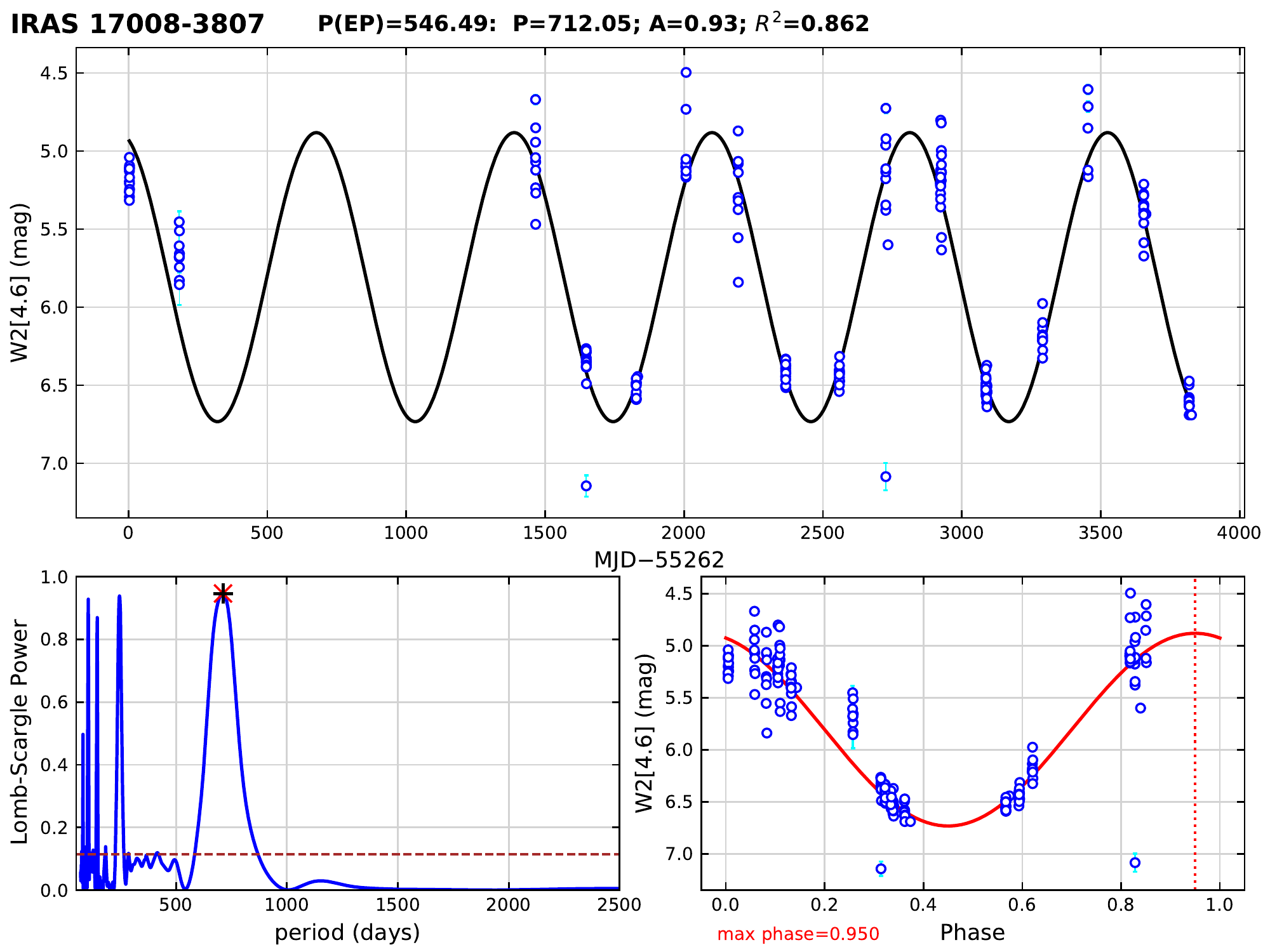}{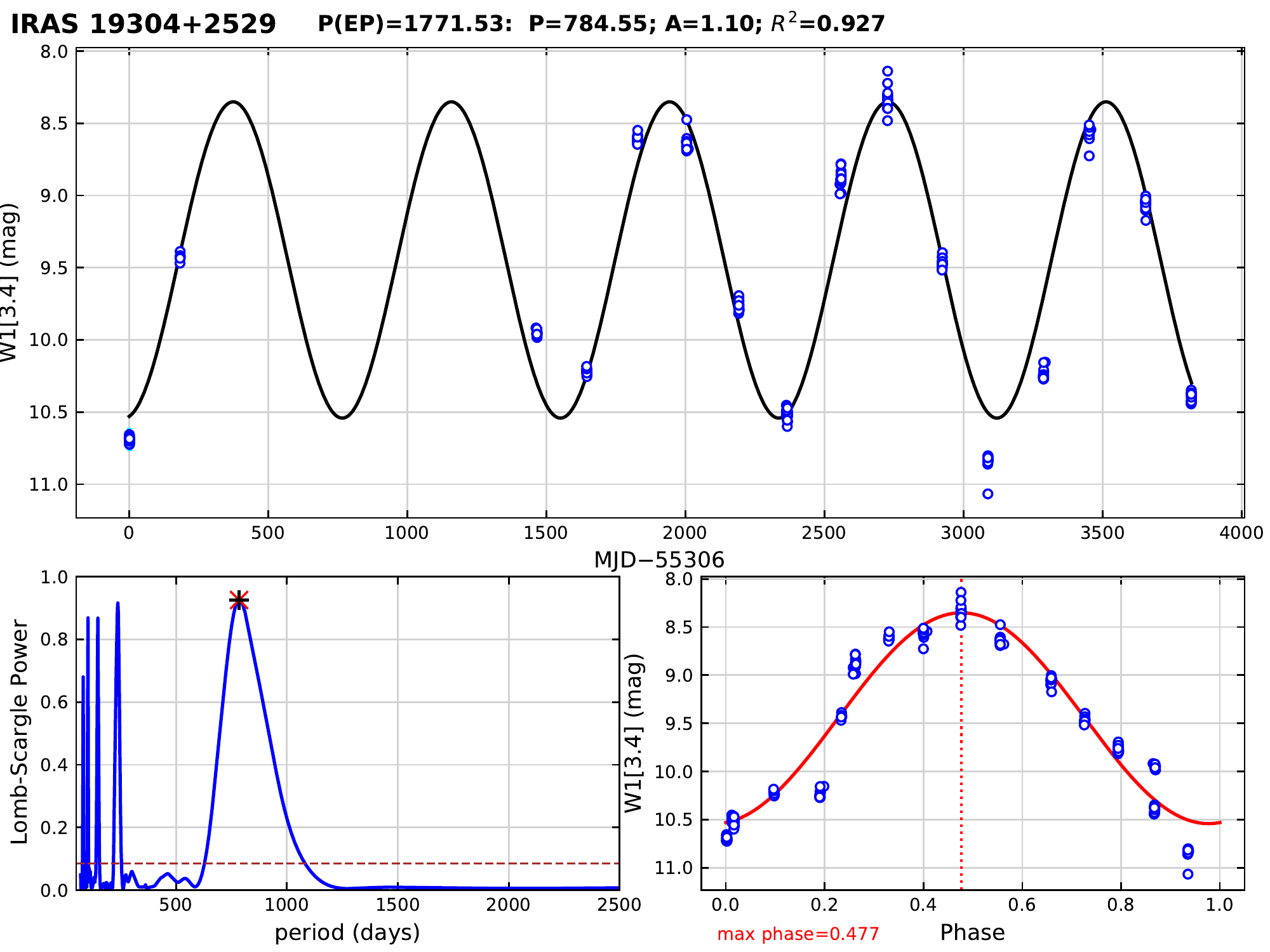}{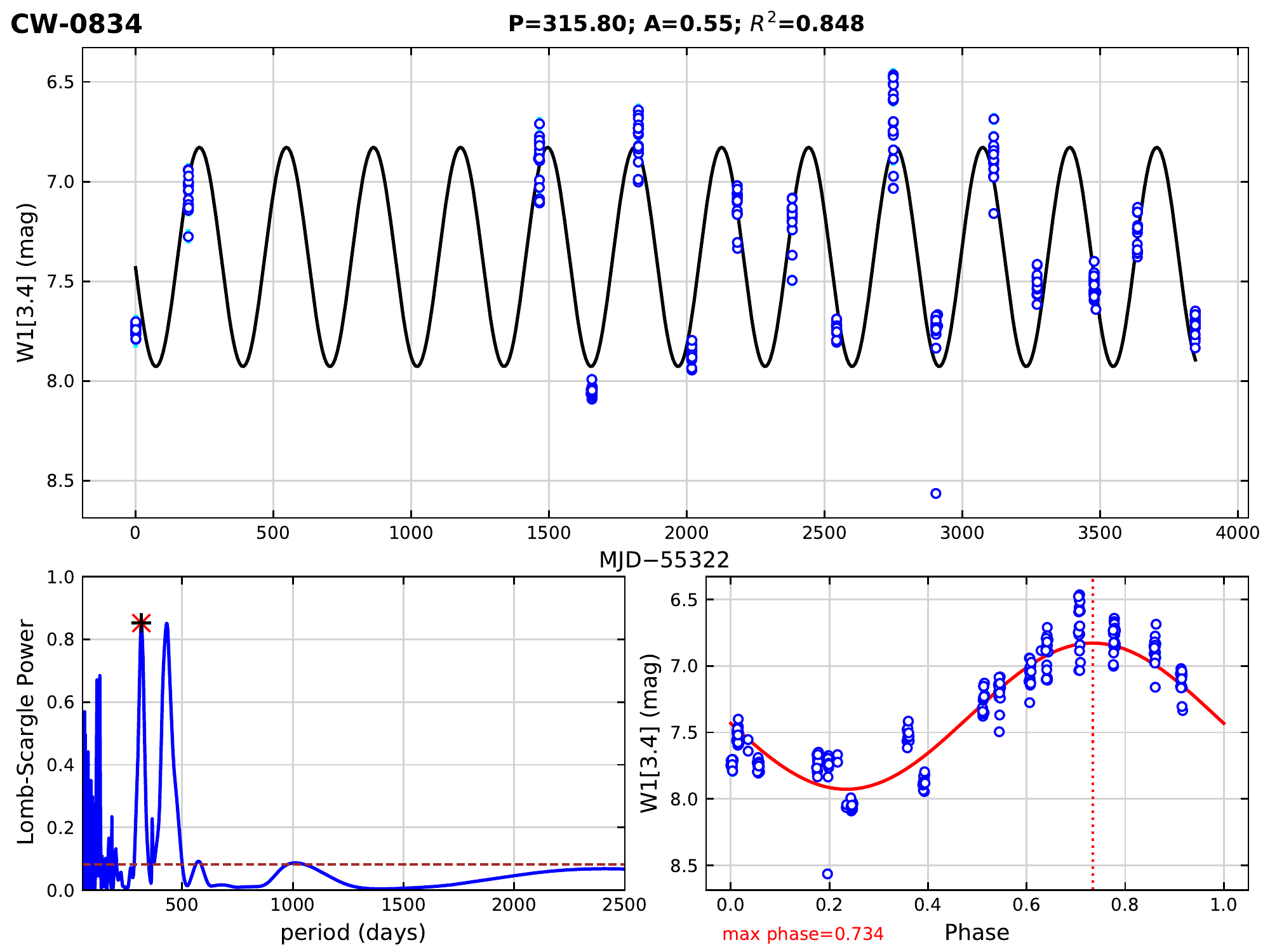}{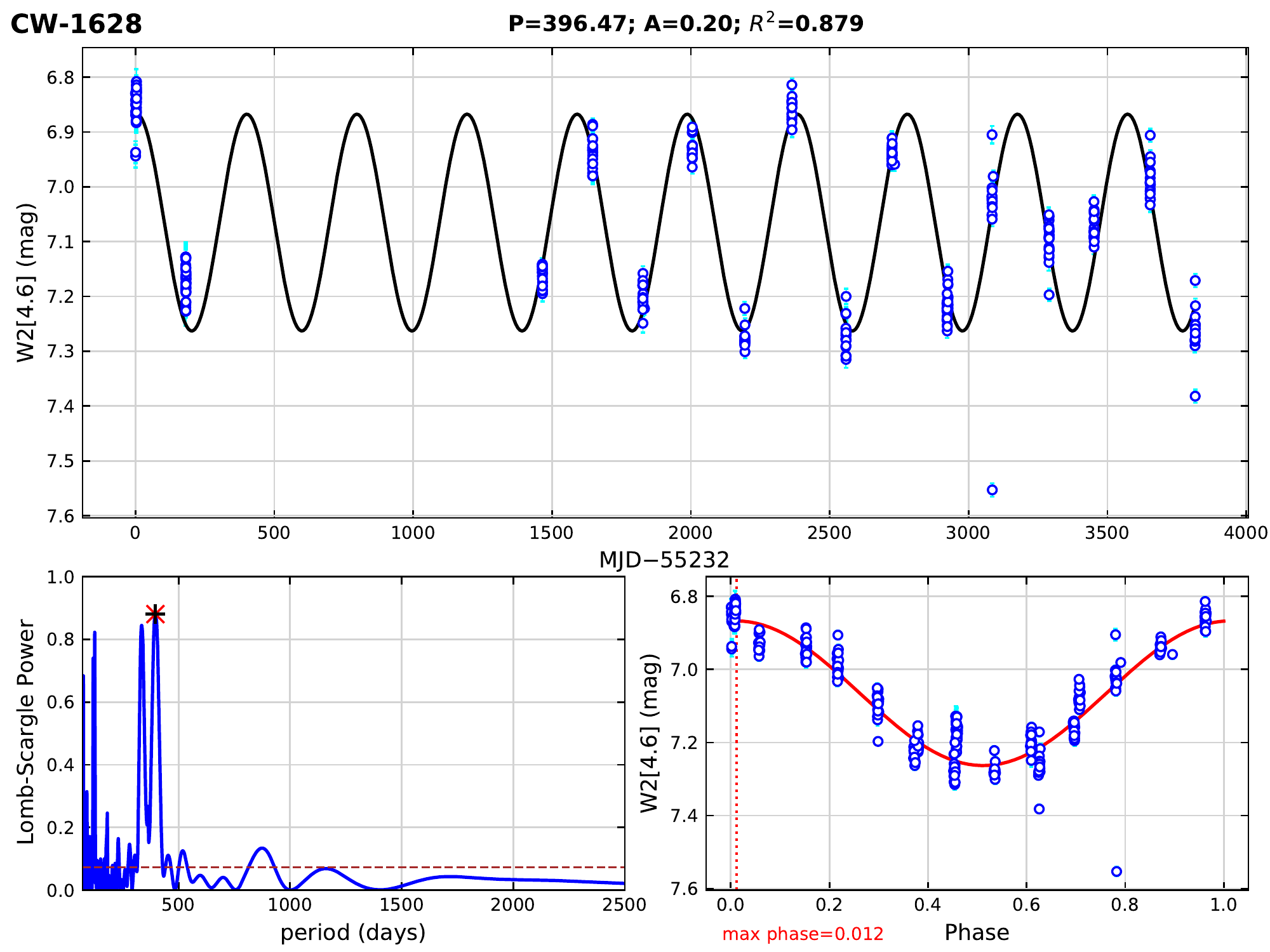}
\caption{WISE light curves and Lomb-Scargle periodograms for four CAGB-IRAS (CI-SH) and two CAGB-WISE (CW-GC) objects with unknown periods.
For CAGB-WISE, the object name is denoted by the CAGB-WISE identifier (CW-N; see Table~\ref{tab:tab12}).
In the Lomb-Scargle periodogram, the red X and black cross marks indicate the primary and selected peaks, respectively
and the red dashed brown horizontal line indicates the periodogram level corresponding to a maximum peak false alarm probability of 1 \%.
See Section~\ref{sec:neo-nm}.
For three objects, P(EP) (the expected period from the IR color K[2.2]$-$W3[12]; see Section~\ref{sec:neo-nmc}) is also shown.
} \label{f28}
\end{figure*}

\begin{figure*}
\centering
\smallplotfour{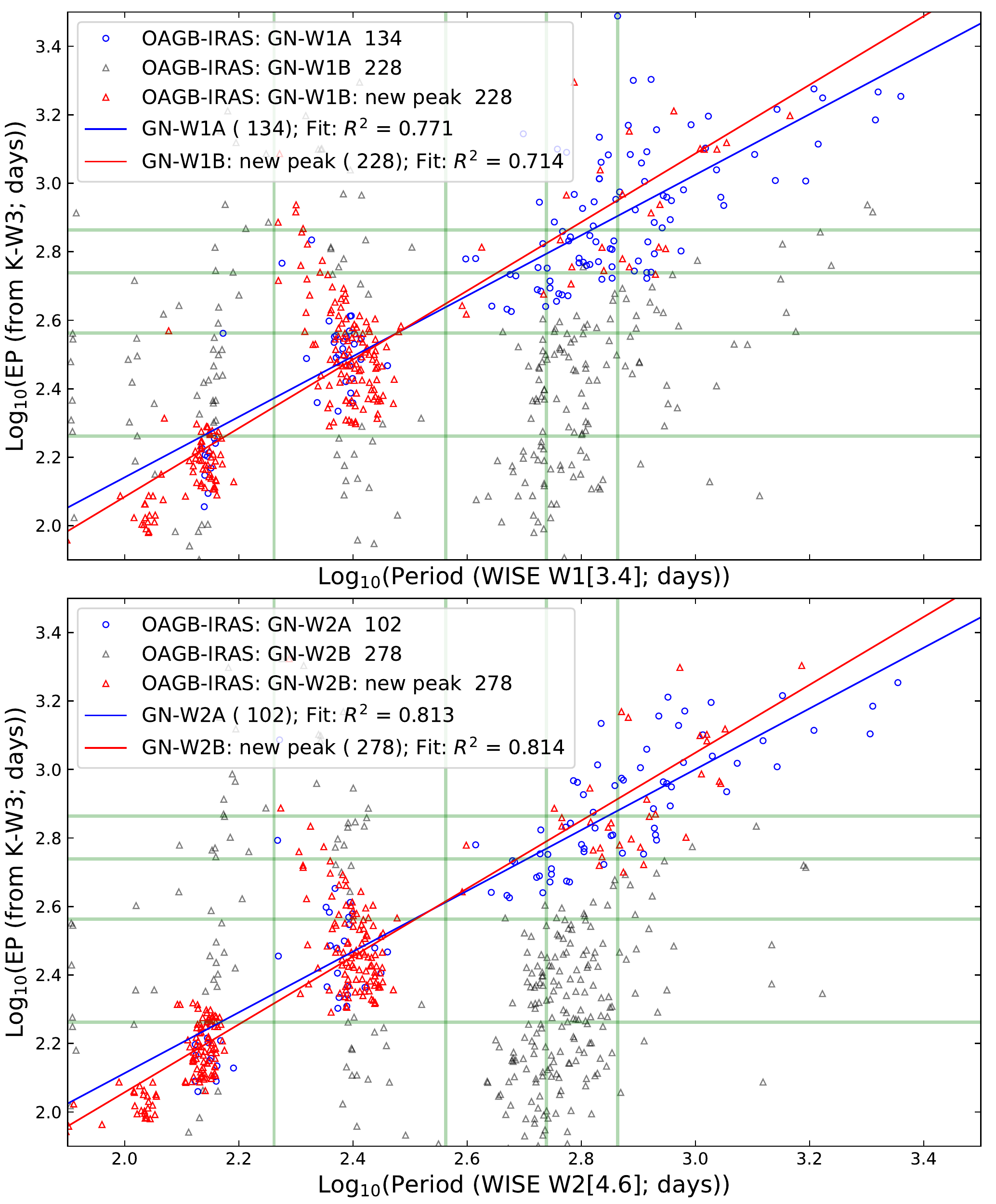}{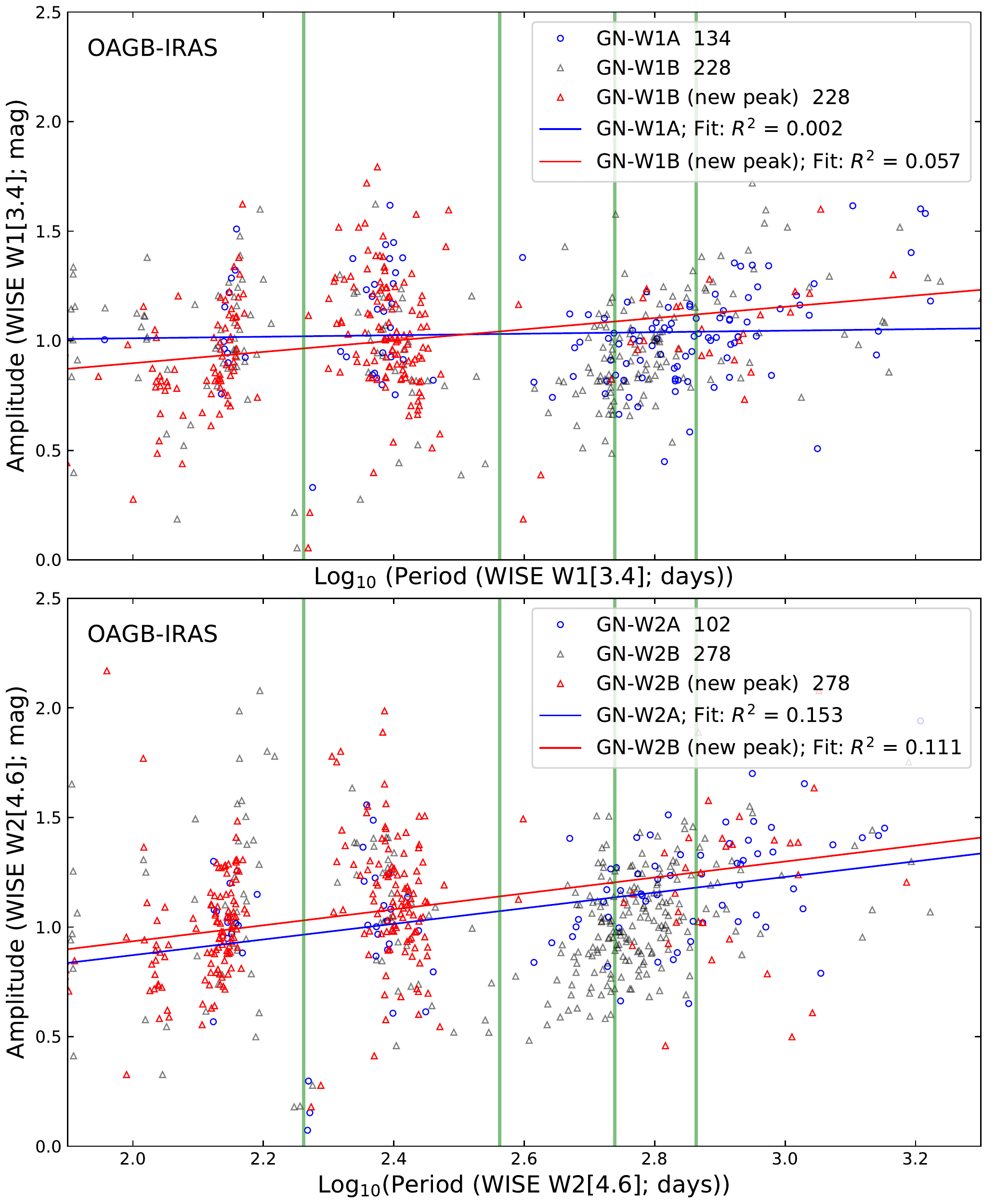}{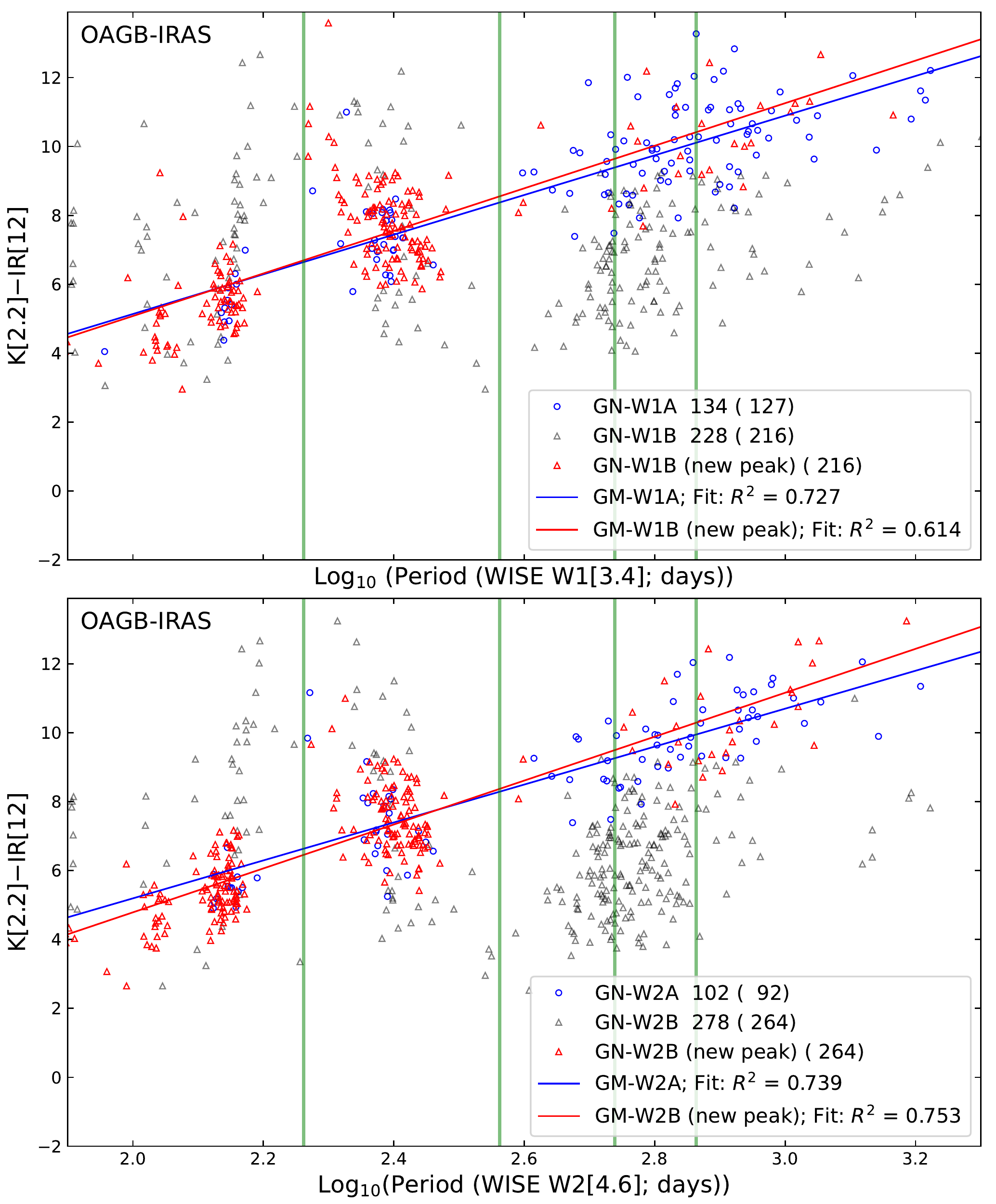}{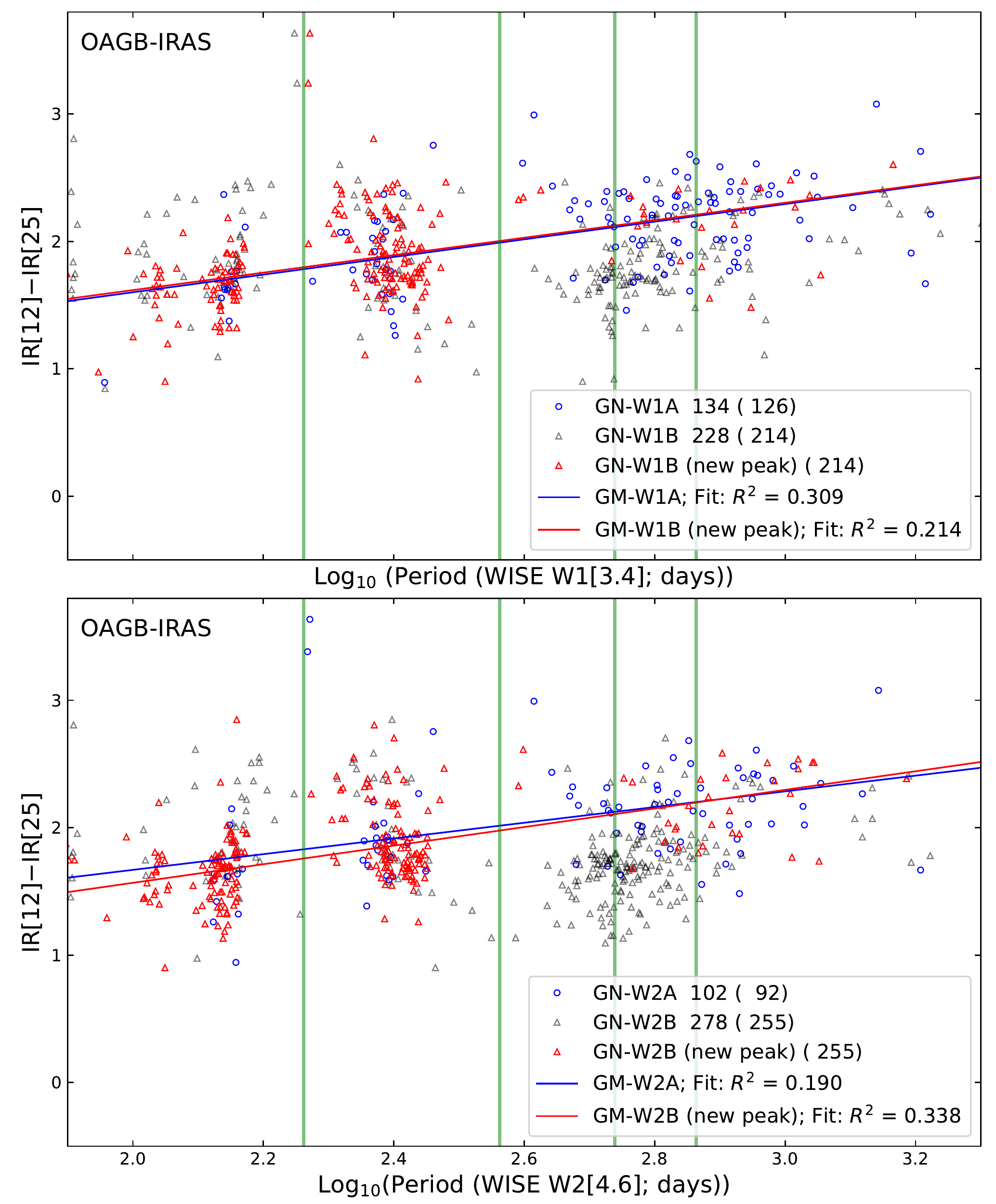}
\caption{For OAGB-IRAS objects with unknown periods or those objects known as non-Mira variables with periods, the upper-left panel compares the expected periods (EPs) obtained from the IR color (K[2.2]$-$W3[12]) with the periods obtained from the WISE light curves.
The upper-right panel show the period-amplitude relation.
The green vertical lines indicate the multiples of the interval of WISE observations (6 months).
The lower panels show the relations between the WISE periods and IR colors (K[2.2]$-$W3[12] and IR[12]$-$IR[25]). See Section~\ref{sec:neo-nmc}.}
\label{f29}
\end{figure*}

\begin{figure*}
\centering
\smallplotfour{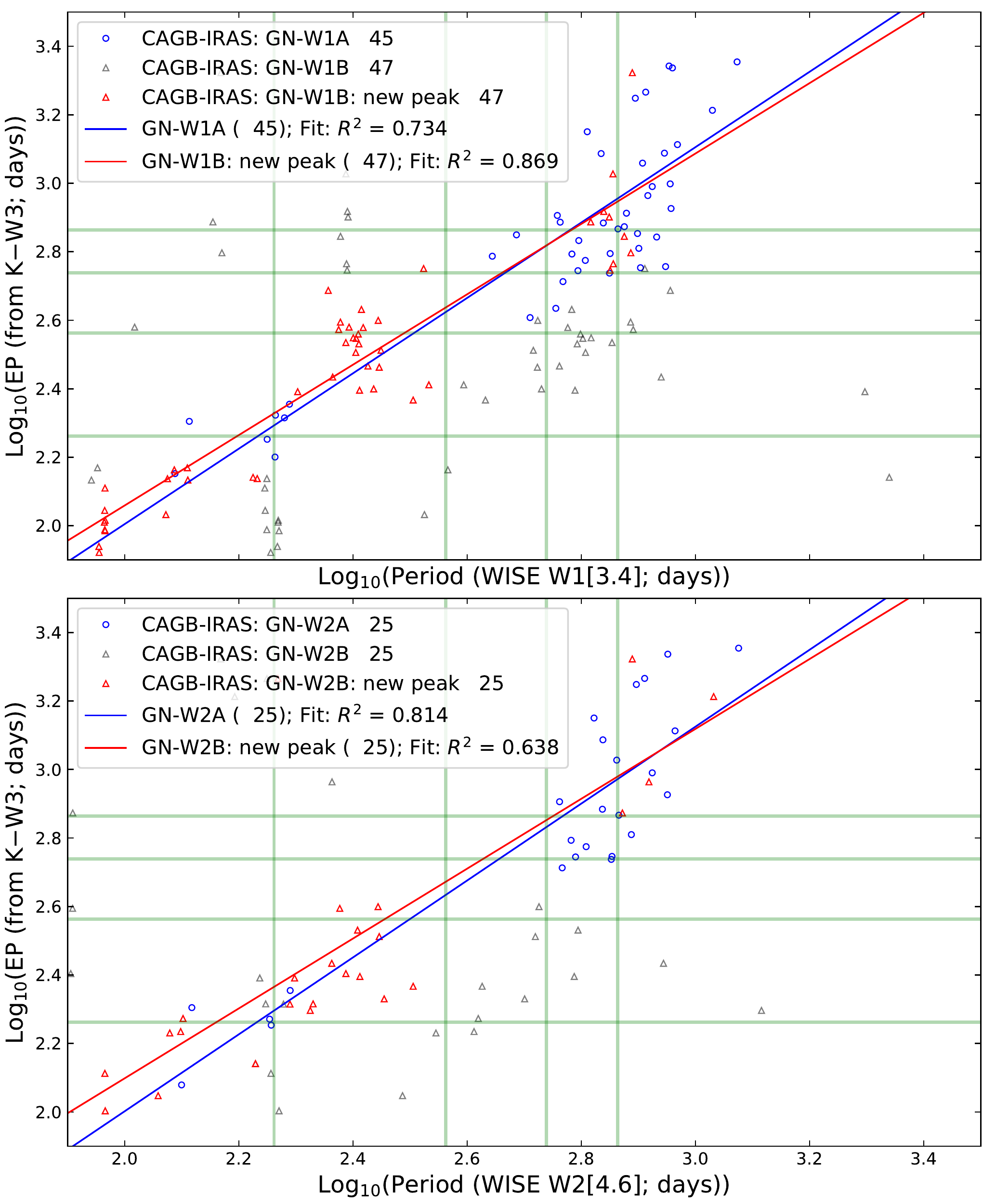}{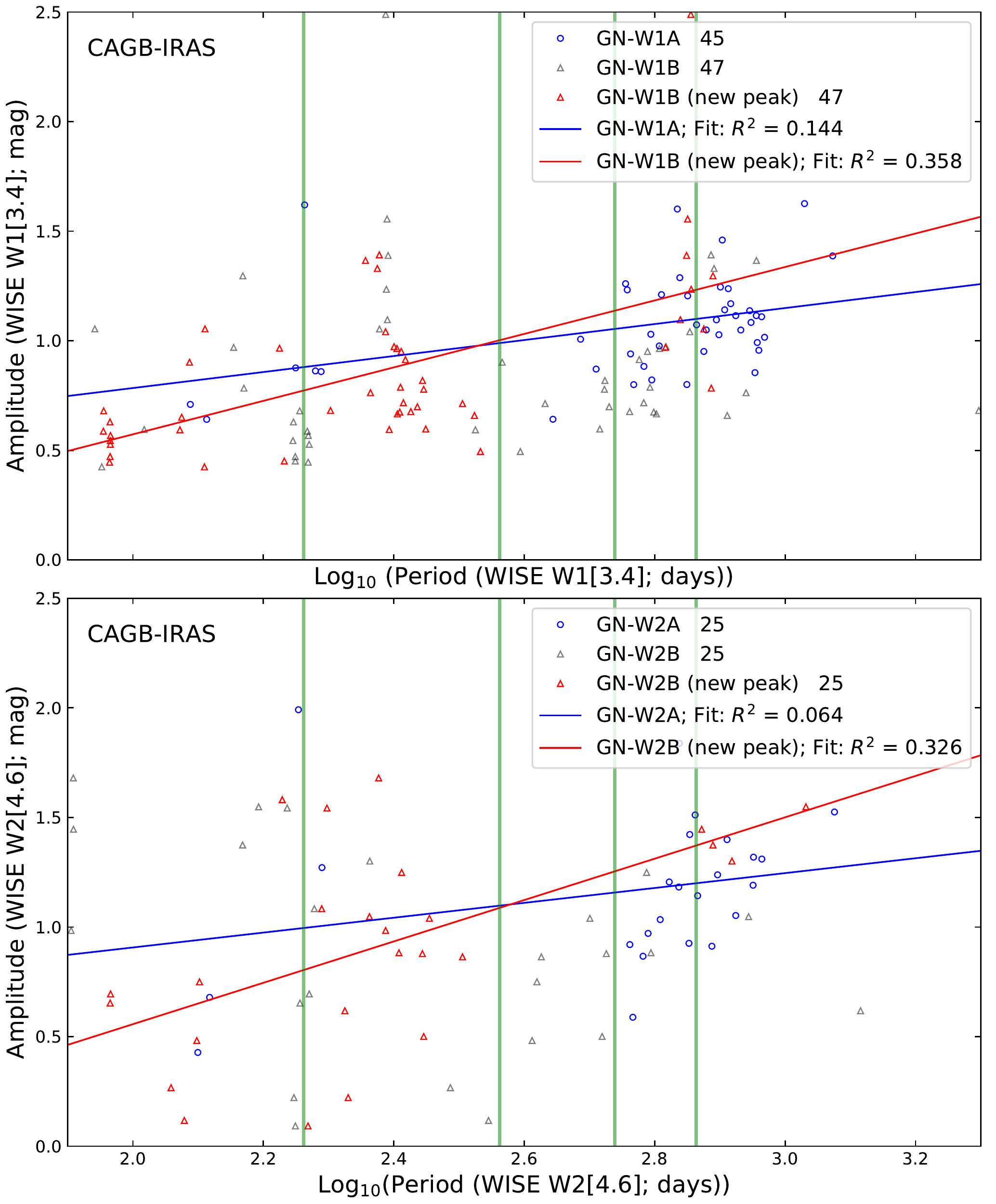}{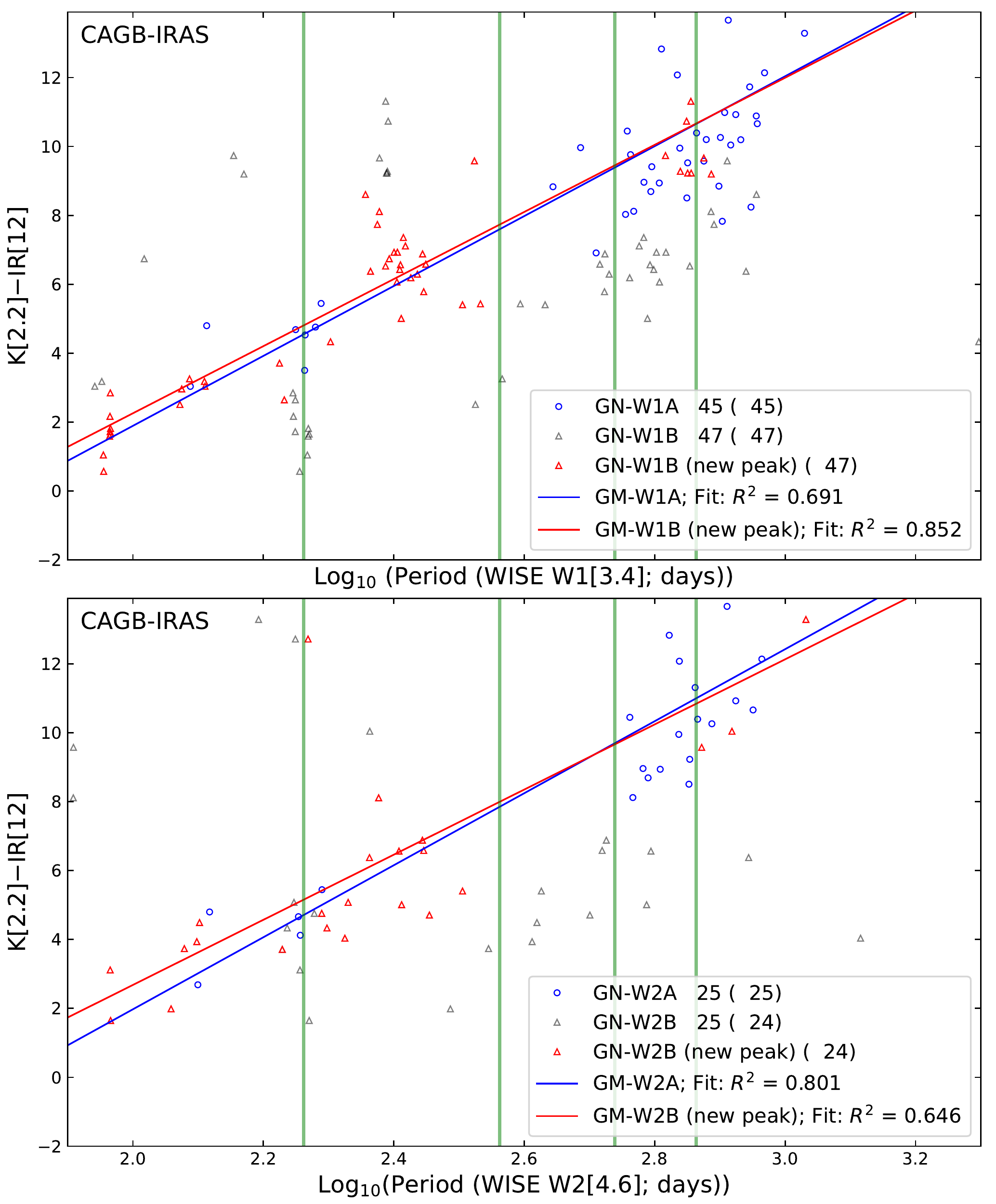}{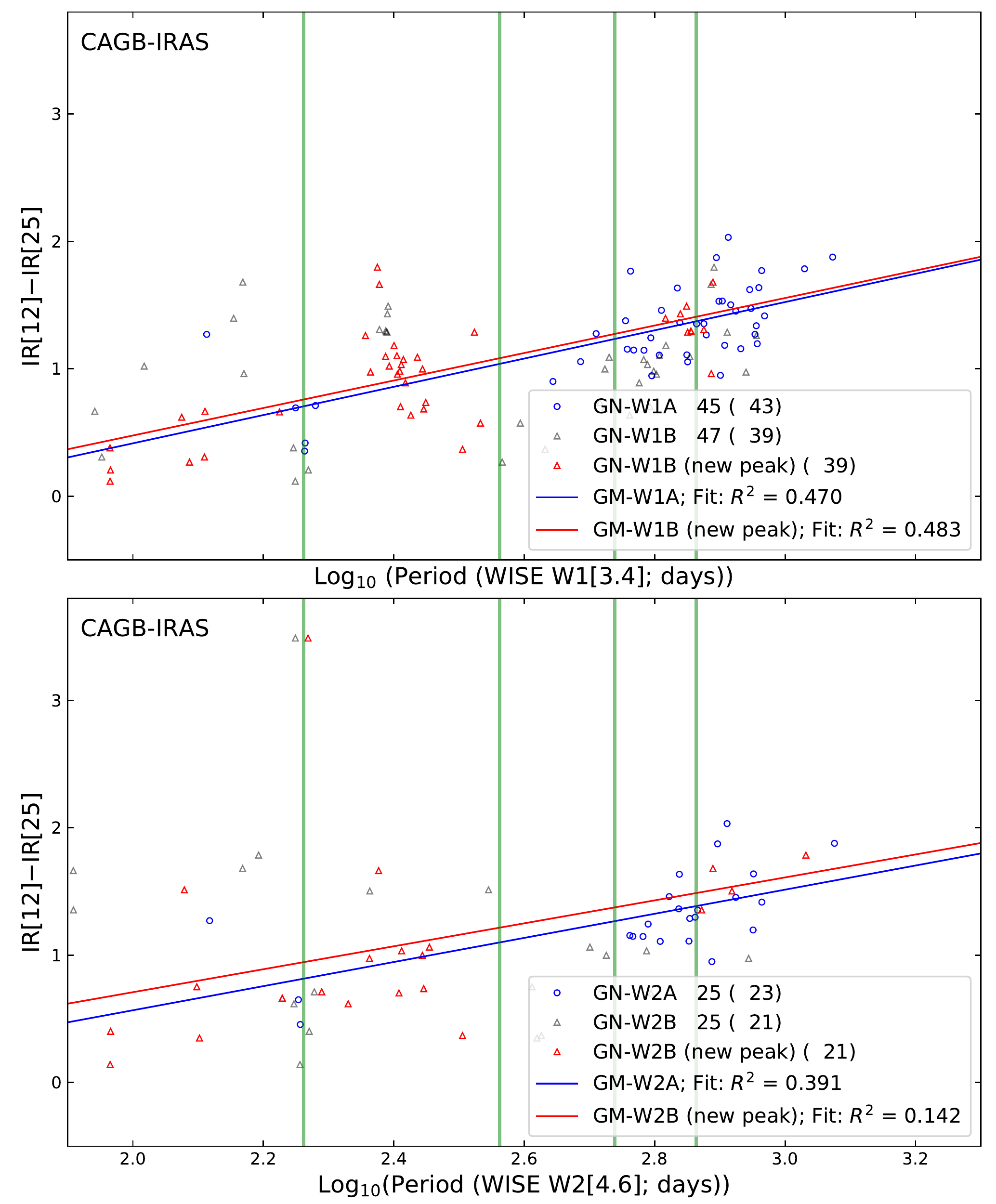}
\caption{For CAGB-IRAS objects with unknown periods or those objects known as non-Mira variables with periods, the upper-left panel compares the expected periods (EPs) obtained from the IR color (K[2.2]$-$W3[12]) with the periods obtained from the WISE light curves.
The upper-right panel show the period-amplitude relation.
The green vertical lines indicate the multiples of the interval of WISE observations (6 months).
The lower panels show the relations between the WISE periods and IR colors (K[2.2]$-$W3[12] and IR[12]$-$IR[25]). See Section~\ref{sec:neo-nmc}.}
\label{f30}
\end{figure*}

\begin{figure*}
\centering
\smallplotsix{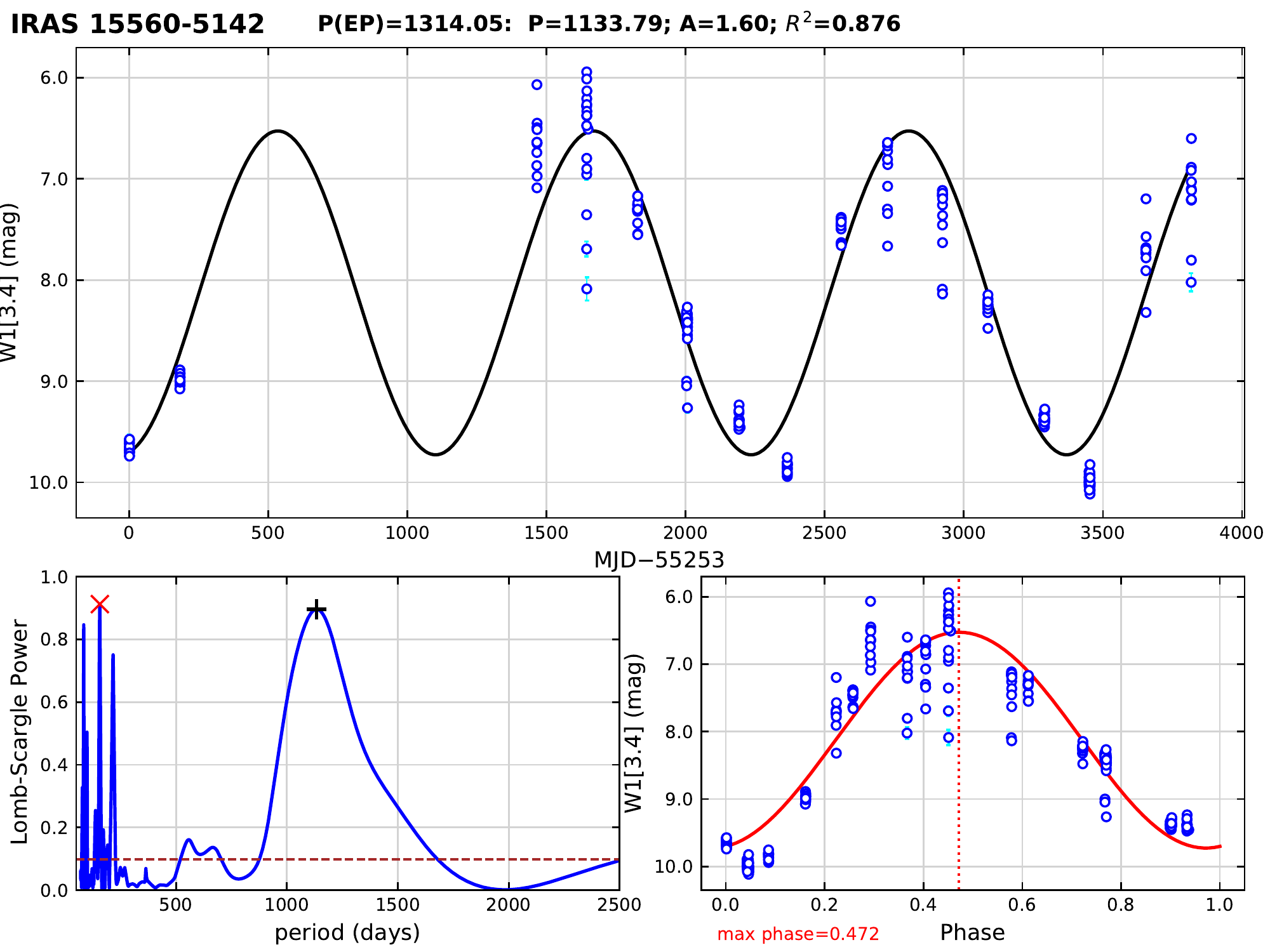}{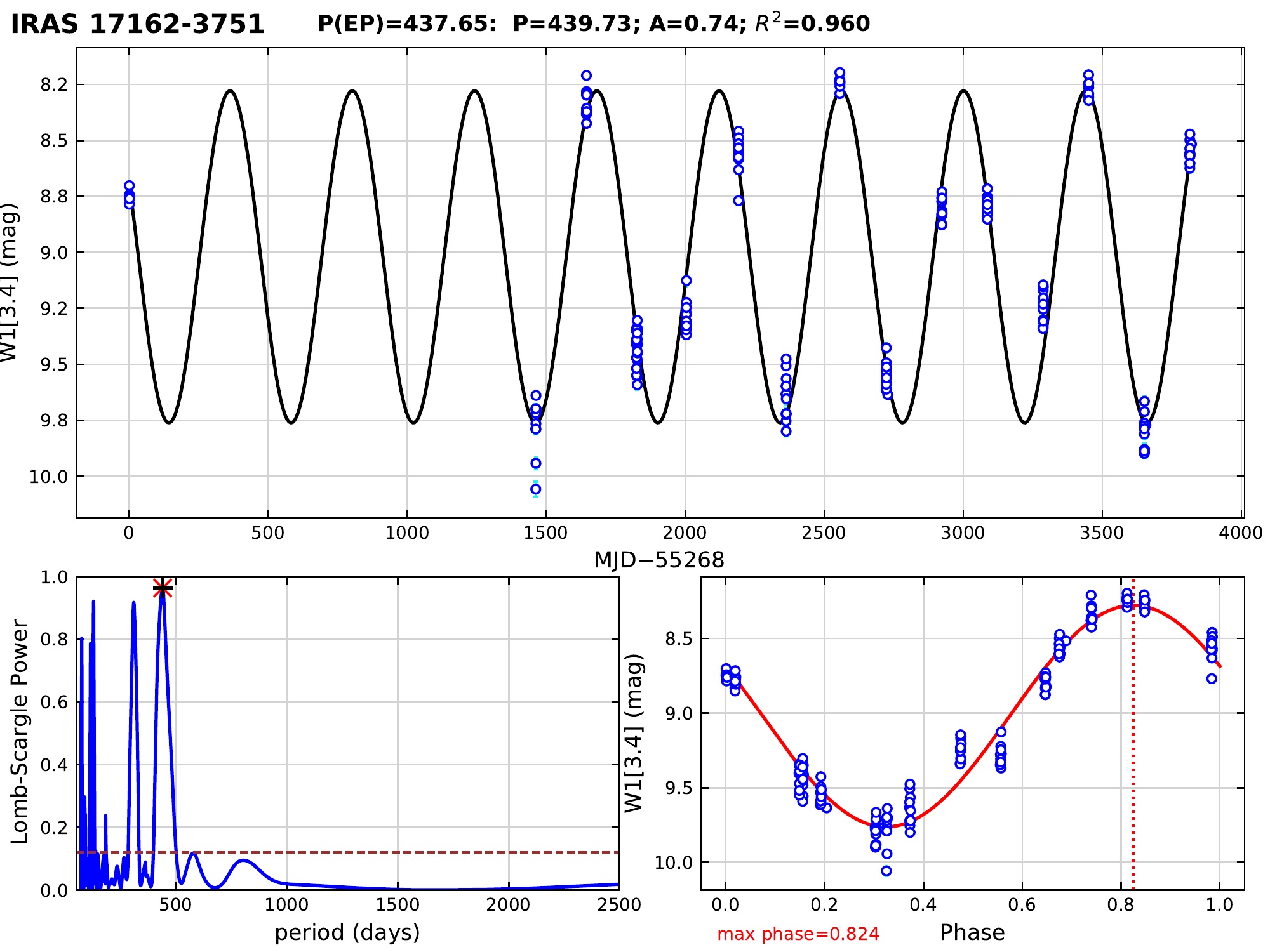}{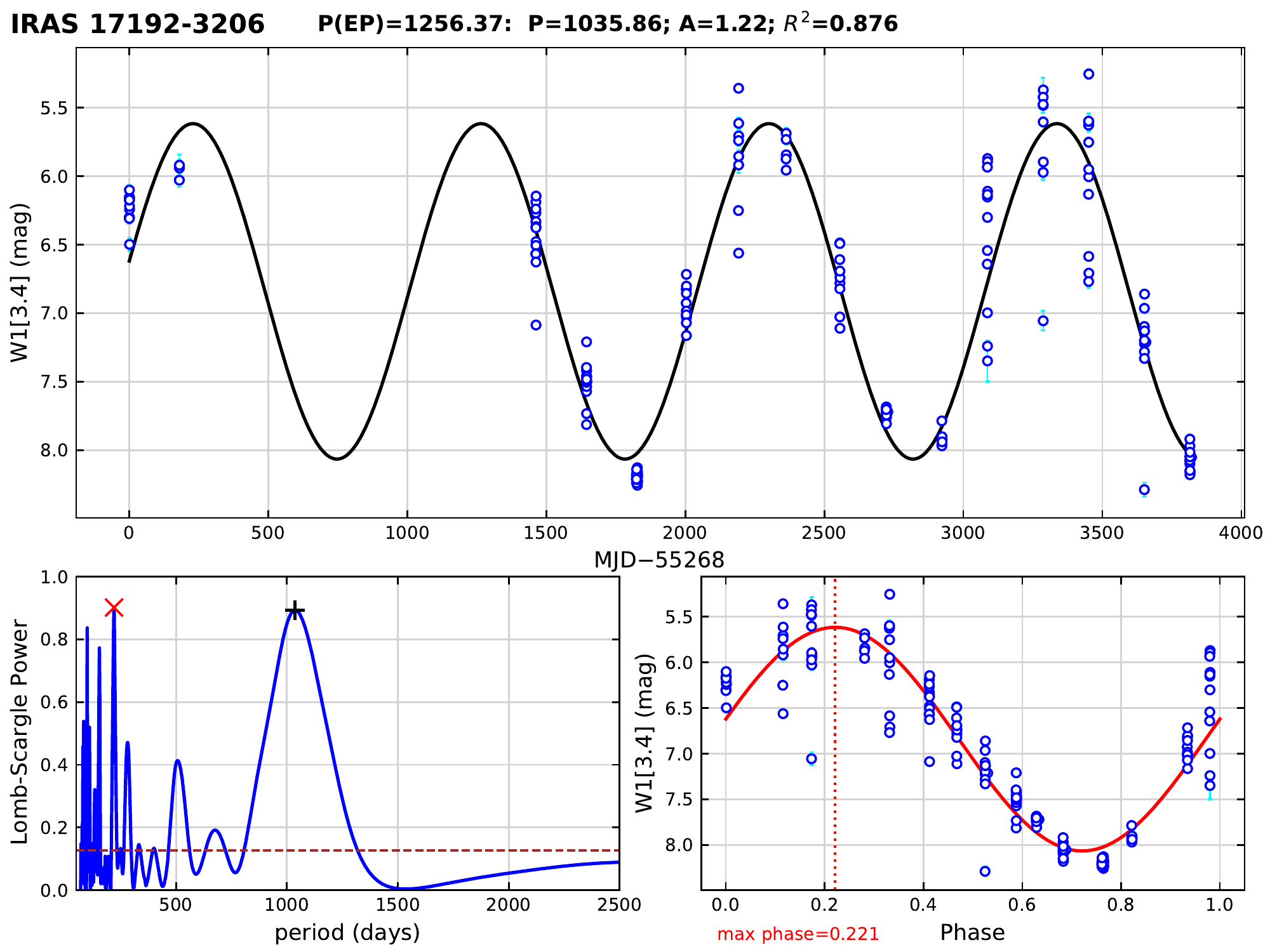}{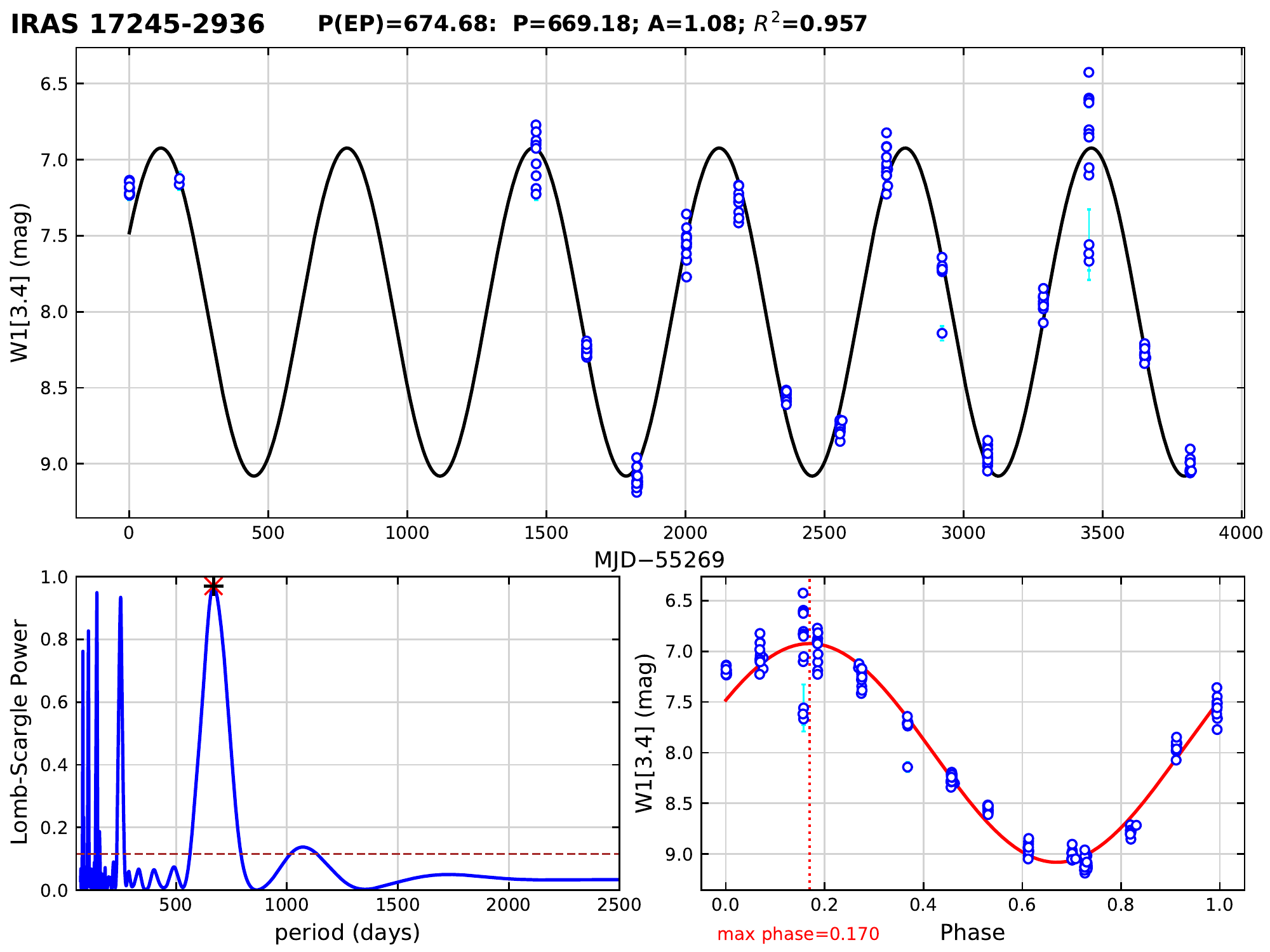}{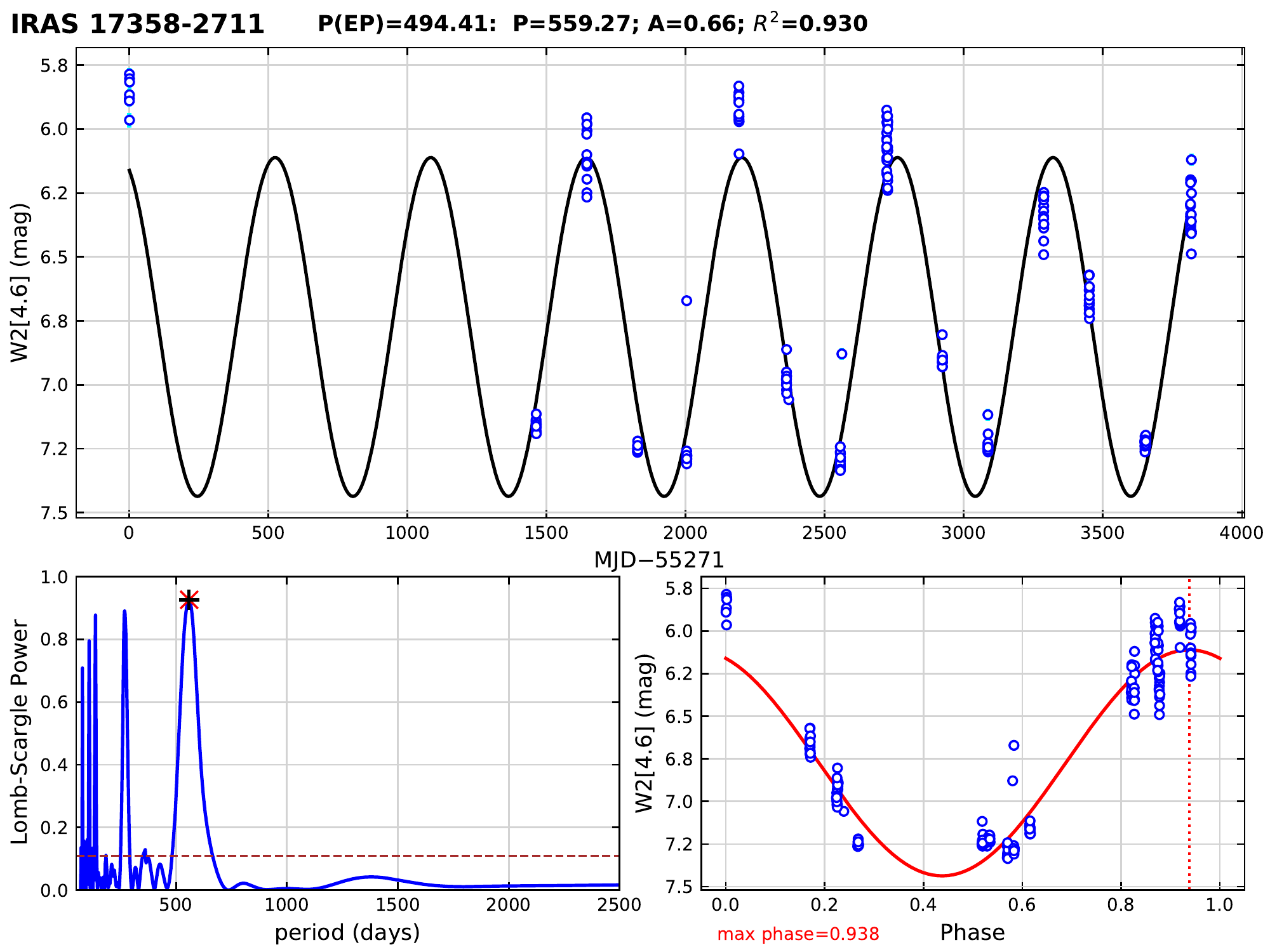}{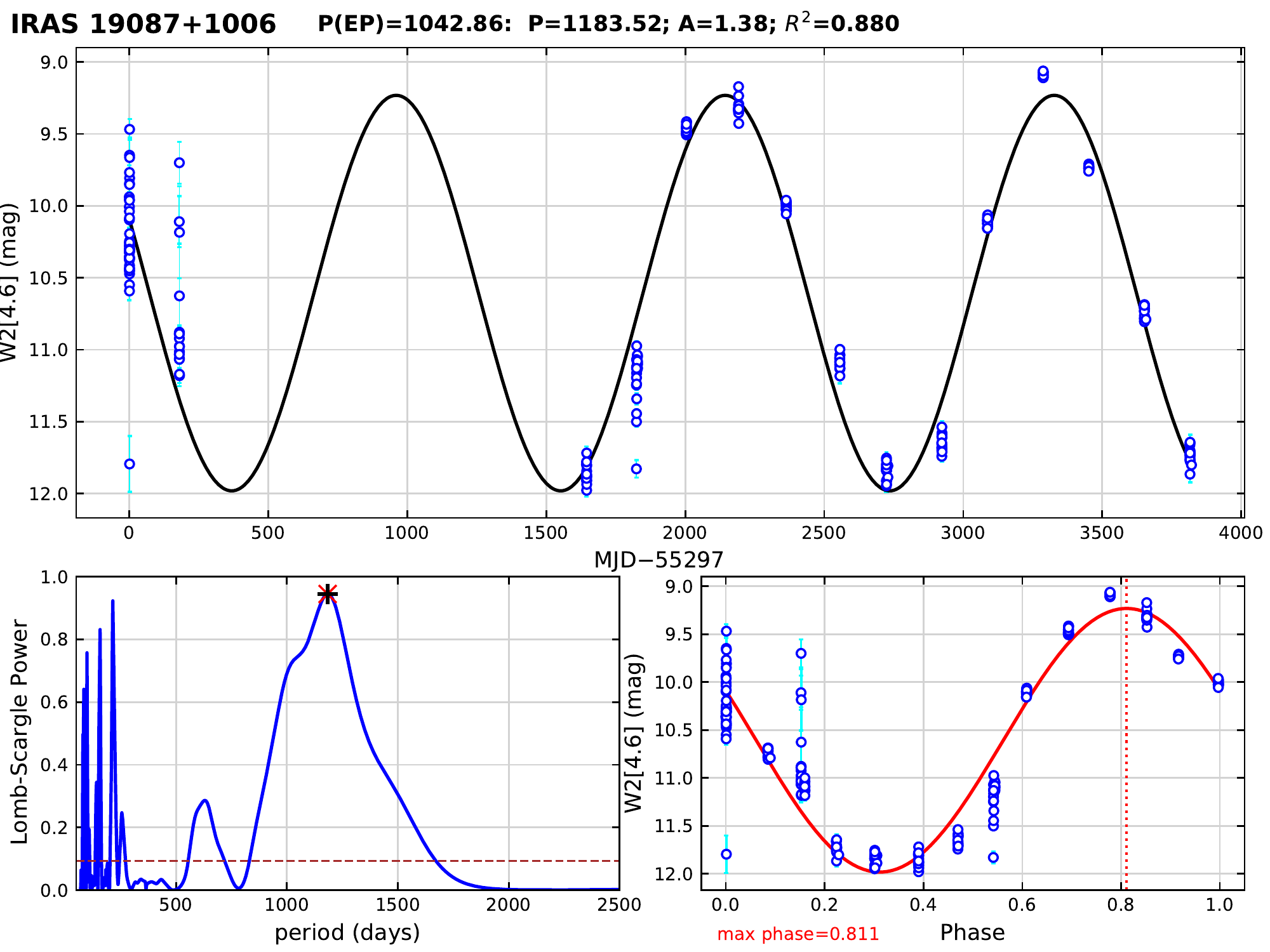}
\caption{WISE light curves and Lomb-Scargle periodograms for six OAGB (in OI-SH) objects with unknown periods.
In the Lomb-Scargle periodogram, the red X and black cross marks indicate the primary and selected peaks, respectively
and the red dashed brown horizontal line indicates the periodogram level corresponding to a maximum peak false alarm probability of 1 \%.
P(EP) is the expected period from the IR color K[2.2]$-$W3[12] (see Section~\ref{sec:neo-nmc}).
For two objects, the second peak of the Lomb-Scargle power is selected for the period because it is more similar to EP.}
\label{f31}
\end{figure*}

\begin{figure*}
\centering
\smallplotsix{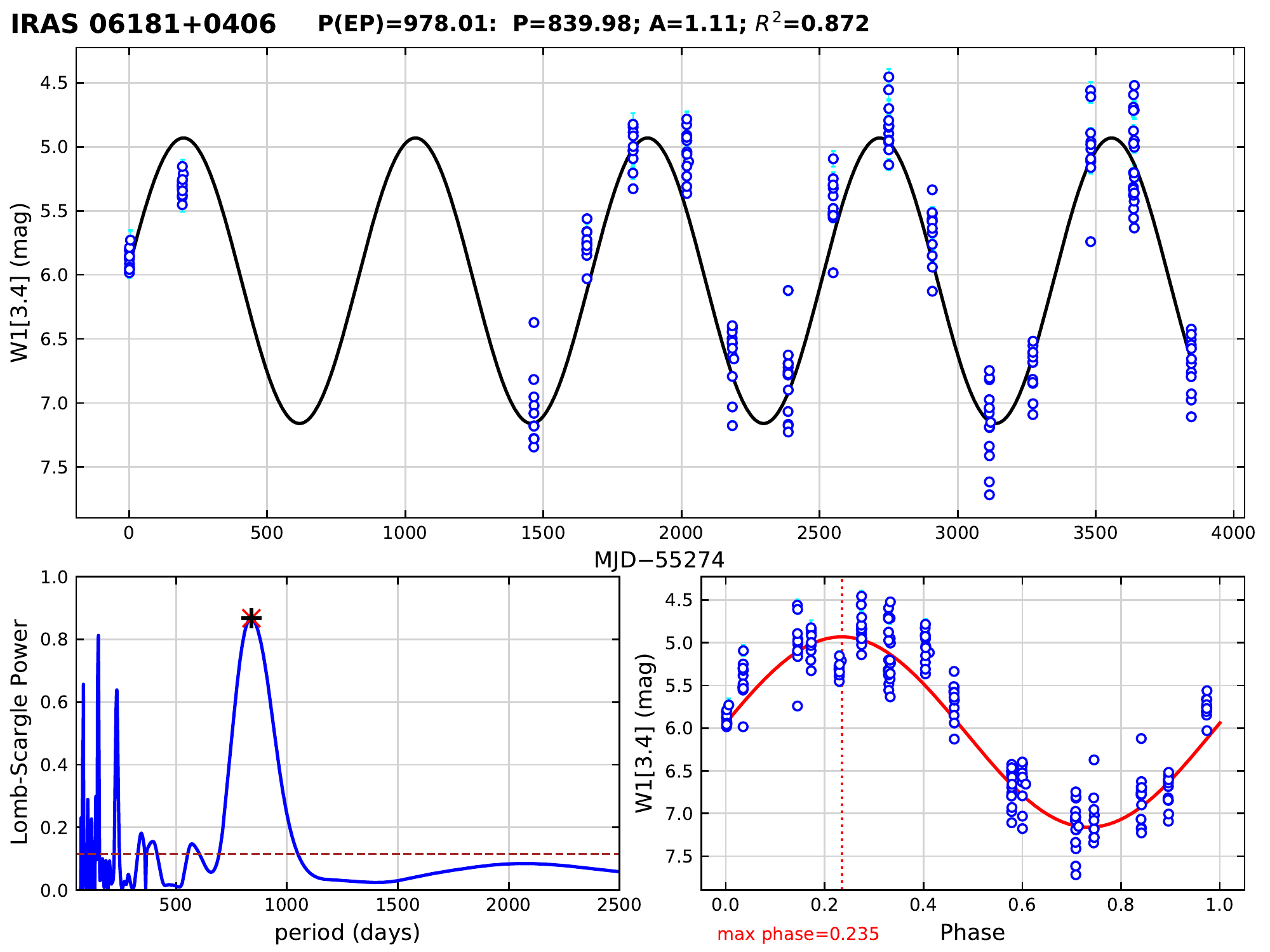}{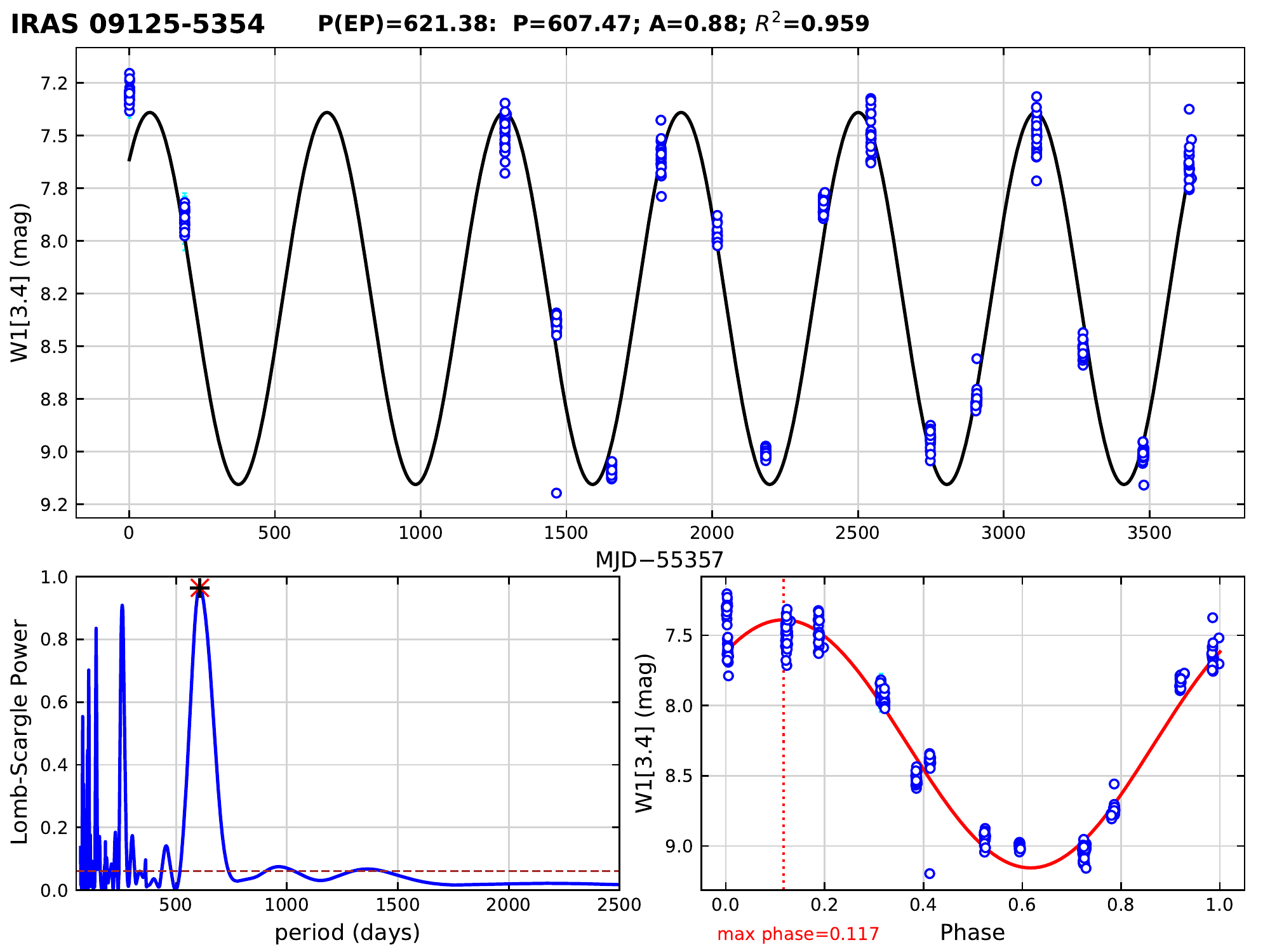}{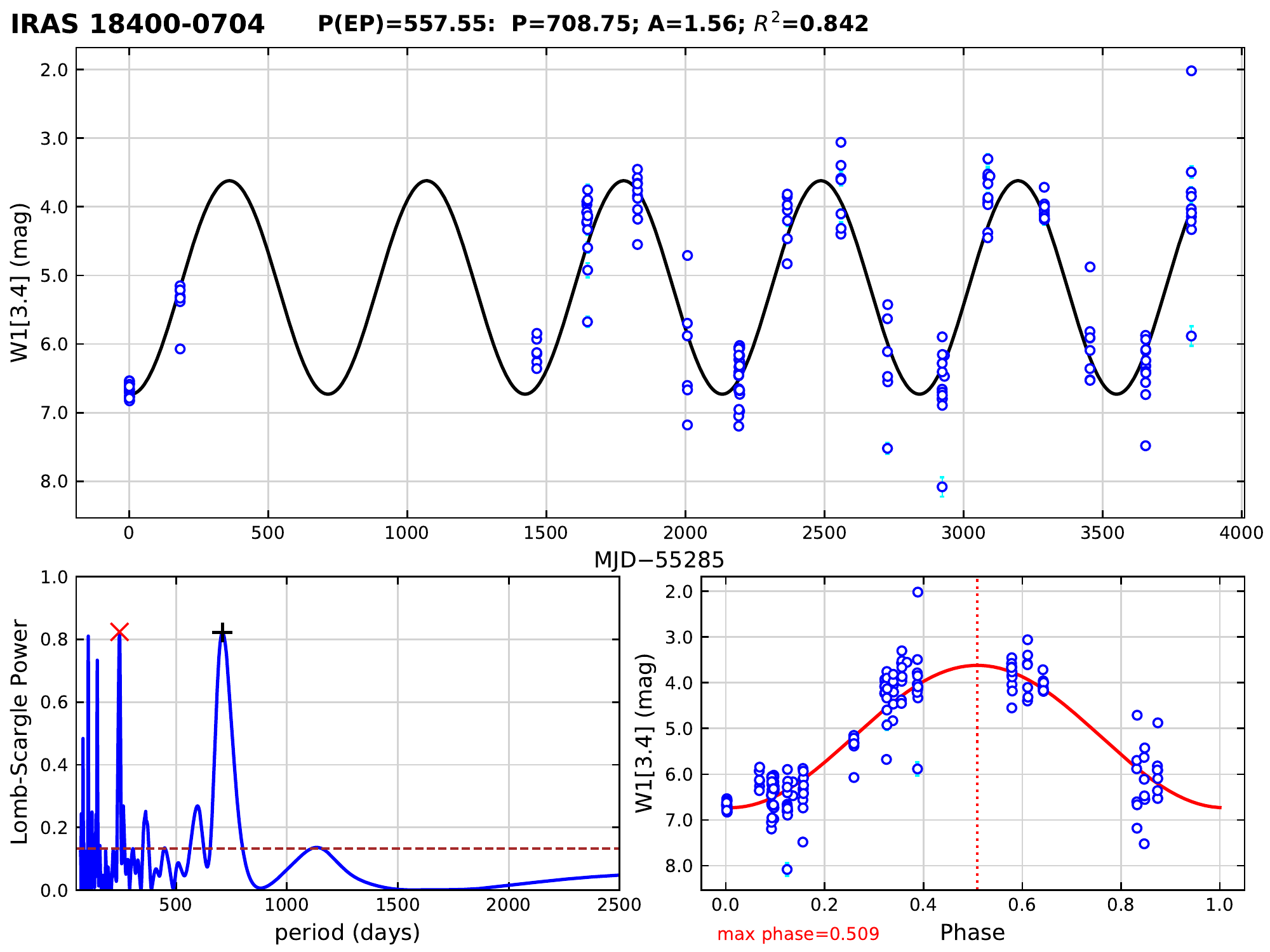}{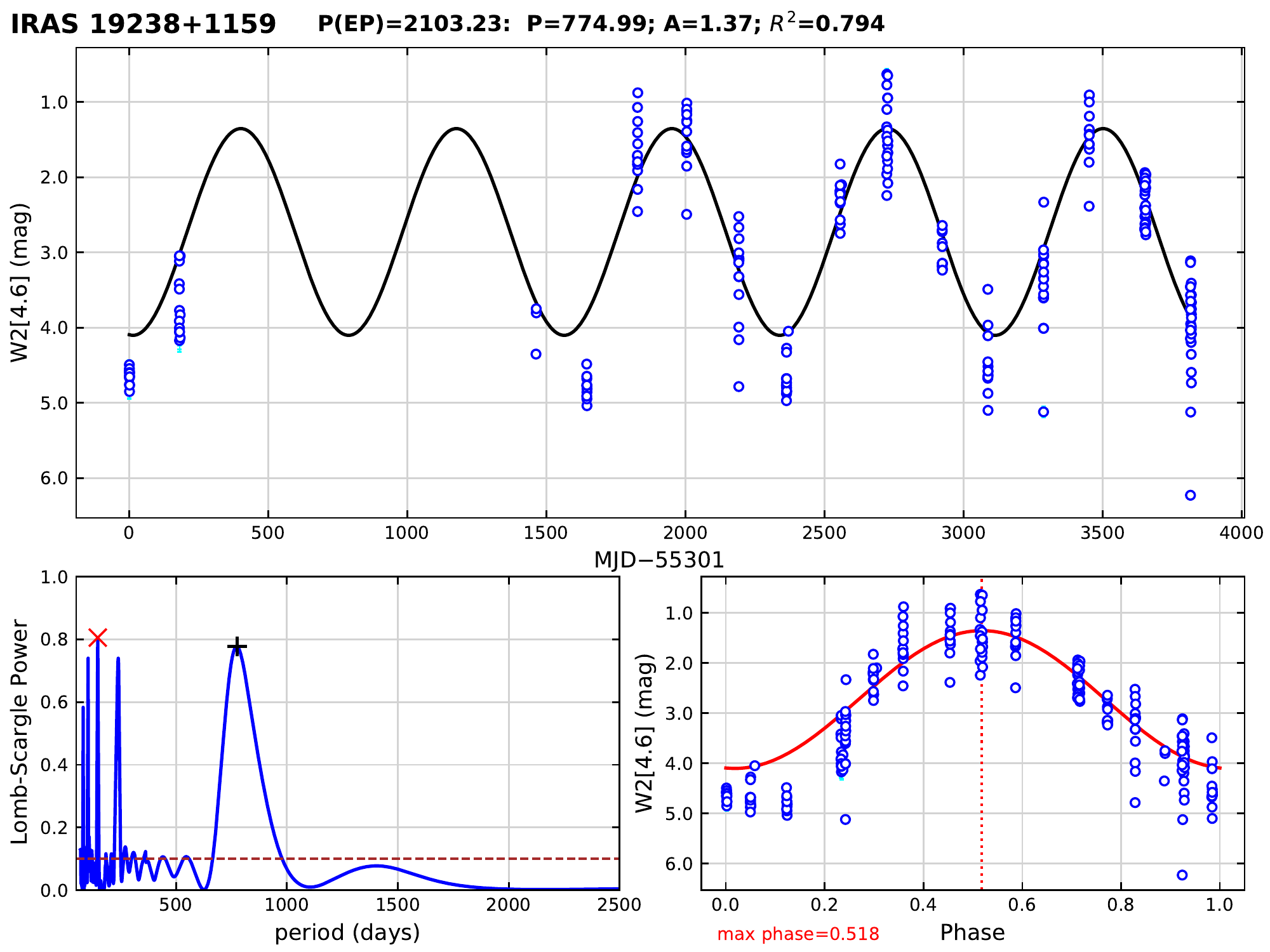}{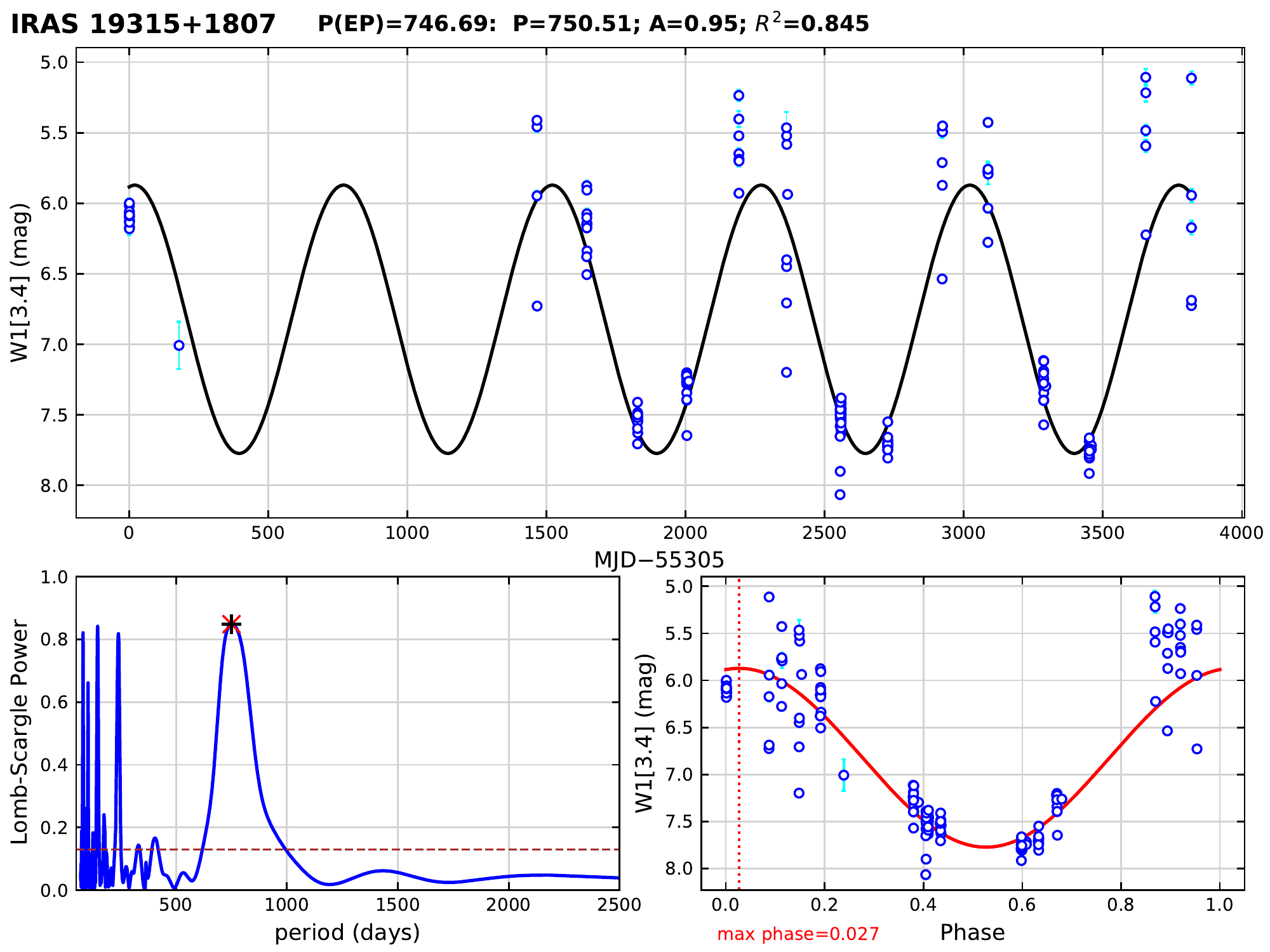}{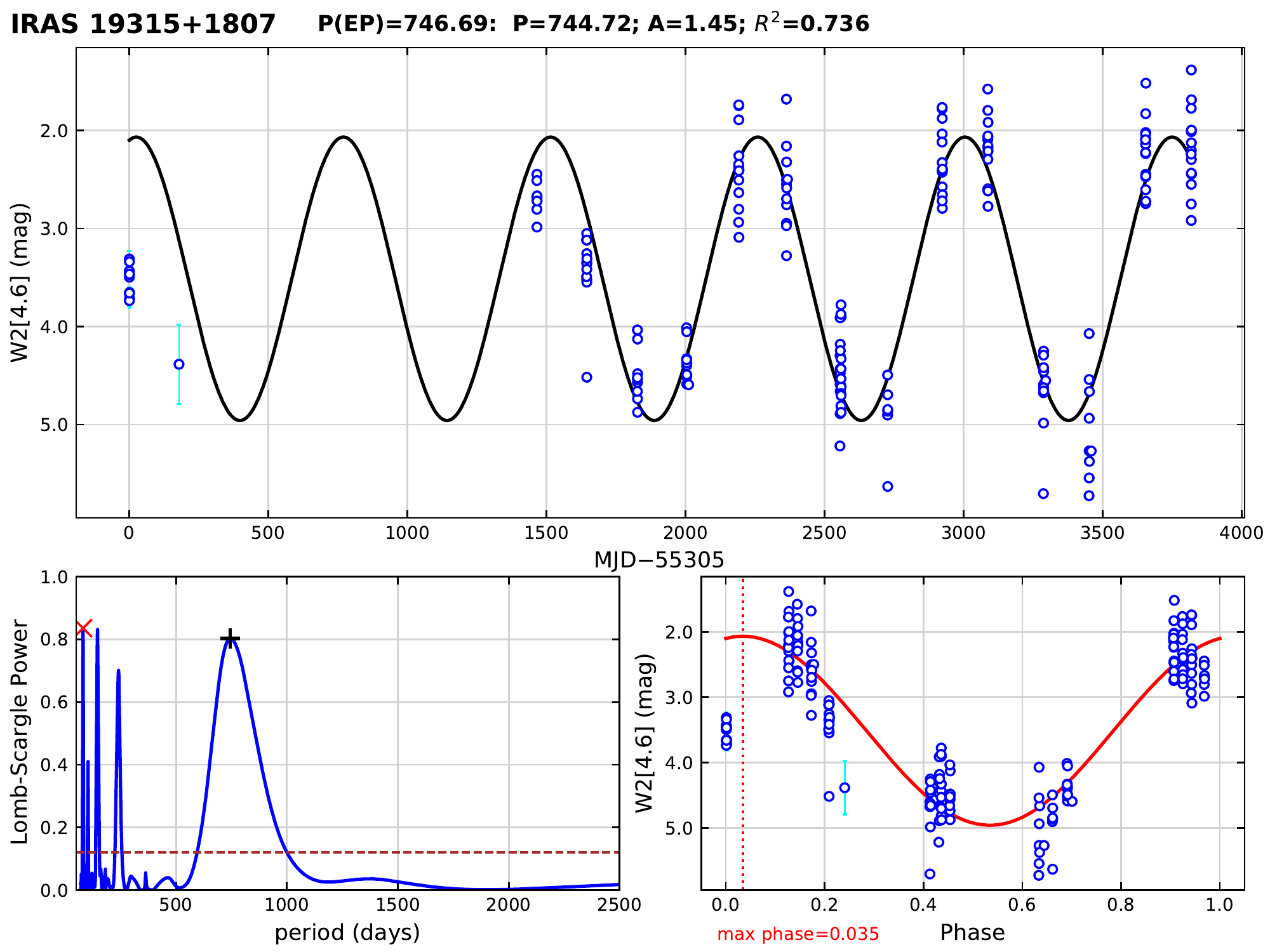}
\caption{WISE light curves and Lomb-Scargle periodograms for six CAGB (in CI-SH) objects with unknown periods.
In the Lomb-Scargle periodogram, the red X and black cross marks indicate the primary and selected peaks, respectively
and the red dashed brown horizontal line indicates the periodogram level corresponding to a maximum peak false alarm probability of 1 \%.
P(EP) is the expected period from the IR color K[2.2]$-$W3[12] (see Section~\ref{sec:neo-nmc}).
For three objects, the second peak of the Lomb-Scargle power is selected for the period because it is more similar to EP.
Note that the primary peaks at W1 and W2 bands are different for IRAS 19315+1807 (see lower two panels).}
\label{f32}
\end{figure*}

\subsection{Candidate objects for new Mira variables\label{sec:neo-nm}}

We have obtained new pulsation periods from the WISE light curves for 656
objects with unknown periods and 244 objects known as non-Mira variables with
periods from AAVSO (GN-W1 or GN-W2 objects; see Table~\ref{tab:tab8}). The 656
objects with unknown periods consist of 441 OAGB-IRAS, 97 CAGB-IRAS, 37
OAGB-WISE, and 81 CAGB-WISE objects. The 244 objects known as non-Mira
variables with periods from AAVSO consist of 160 OAGB-IRAS, 31 CAGB-IRAS, 2
OAGB-WISE, and 51 CAGB-WISE objects. These objects can be candidates for new
Mira variables.

Though it is not be easy to obtain precise periods from the WISE light curves
because the periodograms show multiple peaks with similar Lomb-Scargle power
values, the derived variation parameters would be useful if there are enough
observed points that fit the model well (stronger Lomb-Scargle power values or
larger coefficients of determination for the model fit).

Figure~\ref{f27} shows the WISE light curves and periodograms for the six
OAGB-IRAS objects (OH/IR stars except for IRAS 16254-48580) with unknown
periods, which can be candidate objects for new Miras. IRAS 16254-48580 (in
OI-ST), which is a SiO maser source without any known characteristics, shows a
Mira-like regular pulsation period (251 or 671 days). Though IRAS 17436-2807
(in OI-SH) was suspected to be a HII region (\citealt{walsh1998}), the object
is likely to be a typical OH/IR star because it shows a Mira-like regular
pulsation period (613 days) obtained from the WISE light curves.

Figure~\ref{f28} shows the WISE light curves and periodograms for the four
CAGB-IRAS (CI-SH) and two CAGB-WISE (CW-GC) objects with unknown periods, which
can be candidate objects for new Miras. Note that IRAS 08276-5125 (CI-SH) was
suspected to be a SRV with unknown period (in AAVSO).

We have found new periods from the WISE light curves at W2[4.6] for 104 CW-GC
objects (GN-W2; see Table~\ref{tab:tab8}) from which 33 objects were known to
be SRVs with periods in AAVSO.

\subsection{Properties of the candidate objects for new Miras\label{sec:neo-nmc}}

As we discussed in Section~\ref{sec:neo-mc}, the obtained primary period from
the WISE light curve are different from the true period (AAVSO period) for
about a half of the known Mira variables. Likewise, when we use only primary
periods, the candidate objects for new Miras do not show the period-amplitude
or period-color relations typical for Miras variables (see the upper-left
panels of Figures~\ref{f29}-\ref{f30}).

Generally, Mira variables with longer pulsation periods show redder IR colors
(see Figures~\ref{f15}-\ref{f16} and \ref{f23}-\ref{f25}). Though the PCRs show
large scatter, we may roughly estimate the expected periods from the IR colors
for the new Mira candidates. We use the relation between IR color
(K[2.2]$-$W3[12]) and the period of Miras to estimate the expected period (EP).

We tried to use the PCR from the upper-right panel of Figure~\ref{f15} for OAGB
stars with known periods from radio or IR observations to obtain EP from
K[2.2]$-$W3[12]. However, the PCR from the fit line produced too large EPs.
After some trials with different slopes and intercepts, we find new PCRs (the
blue and red lines in the panel) that produce a better relation between EPs and
WISE periods for OAGB-IRAS objects (see the upper-left panel of
Figure~\ref{f29}) and CAGB-IRAS objects (see the upper-left panel of
Figure~\ref{f30}). Some discrepancies were not avoidable because EP cannot be
regarded as the true period for all objects.

We use the following PCRs: K[2.2]$-$W3[12] =6.87
* $log_{10}$ (EP) - 9.95 (for OAGB-IRAS) and K[2.2]$-$W3[12] = 9.54 * $log_{10}$
(EP) - 17.9 (for CAGB-IRAS); see the upper-right panel of Figure~\ref{f15}. For
a major portion of the best quality periodograms of OAGB-IRAS and CAGB-IRAS
objects (see Figures~\ref{f31} and \ref{f32} for examples), the WISE periods
are similar to EPs when we use these PCRs.

From the GW1-N and GW2-N objects in AGB-IRAS (Table~\ref{tab:tab8}), we select
the objects with good-quality K[2.2]$-$W3[12] colors (OAGB-IRAS: 362 from 422
GW1-N objects and 380 from 439 GW2-N objects; CAGB-IRAS: 92 from 110 GW1-N
objects and 50 from 63 GW2-N objects).

Figures~\ref{f29} and \ref{f30} show the relations between EPs and the WISE
periods, period-amplitude relations, and PCRs for OAGB-IRAS and CAGB-IRAS
objects (with unknown periods or known as non-Mira variables with periods from
AAVSO; GN-W1 and GN-W2 objects in Table~\ref{tab:tab8}), respectively.

The upper-left panels in Figures~\ref{f29} and \ref{f30} compare EPs obtained
from the IR colors (K[2.2]$-$W3[12]) and the periods obtained from the WISE
light curves. There are objects whose primary WISE period is similar to EP
(GN-W1A or GN-W2A) or objects whose new secondary (or up to fourth) period is
similar to EP (GN-W1B or GN-W2B). For the GW1-NB or GW2-NB objects, the new
WISE periods and EPs show a better correlation when we select the new
(secondary or up to fourth) peak of the Lomb-Scargle power that is similar to
EP. But the new correlation is still not as good as the one that uses the AAVSO
period for the target period (see Figures~\ref{f23}-\ref{f25}) and shows larger
deviations. There would be two possible reasons for larger deviations: the
target period (EP from K[2.2]$-$W3[12]) cannot be the true period for all
objects and some objects could not be actual Mira variables. Though this rough
correlation is not as strong as the one between AAVSO and WISE periods, this
comparison would be useful to find general characteristics of Mira variables
for the candidate objects.

The upper-right panels in Figures~\ref{f29} and \ref{f30} show the
period-amplitude relations. And the lower panels in Figures~\ref{f29} and
\ref{f30} show PCRs using K[2.2]$-$IR[12] and IR[12]$-$IR[25] colors.

When we select the new peak, which is similar to EP, for the GN-W1B or GN-W2B
objects, the candidate objects for new Miras show roughly similar
period-amplitude relations and PCRs (see Figures~\ref{f29} and \ref{f30}) to
those for known Mira variables (see Figures~\ref{f15}-\ref{f16} and
\ref{f23}-\ref{f25}).

Figures~\ref{f27} and \ref{f28} show WISE light curves and periodograms for AGB
objects with unknown periods. For related objects with good-quality
K[2.2]$-$W3[12] colors, EP is also shown. We find that the periods obtained
from WISE light curves show some deviations from EPs for these objects.

Figure~\ref{f31} shows WISE light curves and periodograms for six OAGB-IRAS
objects (in OI-SH) with unknown periods, which can be candidate objects for new
Miras. For these objects, we find that the periods obtained from WISE light
curves are similar to EPs obtained from K[2.2]$-$W3[12]. Note that IRAS
17358-2711 can be identified as OGLE-BLG-ECL-054370 (angular distance:
33$\arcsec$) with an E type variable (period: 1.08 days) from AAVSO, but they
could be different objects.

Figure~\ref{f32} shows WISE light curves and periodograms for six CAGB-IRAS
objects (in CI-SH) with unknown periods, which can be candidate objects for new
Miras. Again, we find that the periods obtained from WISE light curves are
similar to EPs obtained from K[2.2]$-$W3[12] for these objects.

\section{The catalog data\label{sec:catalog}}

For OAGB-IRAS objects (see Table~\ref{tab:tab1}), Table~\ref{tab:tab9} lists
the OAGB-IRAS identifier (OI-N), subgroup name, IRAS PSC number, counterparts
of AKARI PSC and BSC, best position (right ascension and declination J2000; RA
and DEC), source of the best position (AKARI PSC, AKARI BSC, or IRAS PSC; see
Section~\ref{sec:agb-c}; BP), ALLWISE counterpart, spectral type (SP), and IRAS
LRS type (LRS) from \citet{kwok1997}; variable type (A-Type) and period
(A-Period) from AAVSO; and information about SiO and OH maser emission (see
Section~\ref{sec:maser}). For CAGB-IRAS objects (see Table~\ref{tab:tab1}),
Table~\ref{tab:tab10} lists the CAGB-IRAS identifier (CI-N) and the same
information except for the information about SiO and OH maser emission.

For OAGB-WISE objects (see Table~\ref{tab:tab2}), Table \ref{tab:tab11} lists
the OAGB-WISE identifier (OW-N), subgroup name, source number from the original
reference (recno/star), ALLWISE source name, ALLWISE source position (W-RA,
W-DEC), counterparts of 2MASS, AKARI PSC, MSX, and variable type (A-Type) and
period (A-Period) from AAVSO. For CAGB-WISE objects (see Table~\ref{tab:tab2}),
Table \ref{tab:tab12} lists the CAGB-WISE identifier (CW-N) and the same
information.

For 3710 objects in the catalog (G-W1 or G-W2; see Table~\ref{tab:tab8}), the
variation parameters obtained from WISE light curves at W1 and W2 bands are
listed in Tables \ref{tab:tab13}-\ref{tab:tab16}. For each obtained period (P),
the amplitude (A) and coefficients of determination ($R^2$) to fit the
sinusoidal model to the observations are also listed.

For OAGB-IRAS objects with variation parameters obtained from WISE light
curves, Table \ref{tab:tab13} lists the OAGB-IRAS identifier (OI-N), variable
type (A-Type) and period (A-Period) from AAVSO, expected period from the IR
color K[2.2]$-$W3[12] (EP; see Section~\ref{sec:neo-nmc}), G-W1(P), G-W1(A),
G-W1($R^2$),	GM-W1(P), GM-W1(A), GM-W1($R^2$), GN-W1(P), GN-W1(A),
GN-W1($R^2$), G-W2(P),	G-W2(A), G-W2($R^2$), GM-W2(P), GM-W2(A), GM-W2($R^2$),
GN-W2(P), GN-W2(A), GN-W2($R^2$), subgroup name, and IRAS PSC number. For
CAGB-IRAS objects with variation parameters obtained from WISE light curves,
Table \ref{tab:tab14} lists the CAGB-IRAS identifier (CI-N) and the same
information.

For OAGB-WISE objects with variation parameters obtained from WISE light
curves, Table \ref{tab:tab15} lists the OAGB-WISE identifier (OW-N), variable
type (A-Type) and period (A-Period) from AAVSO, G-W1(P), G-W1(A), G-W1($R^2$),	
GM-W1(P), GM-W1(A), GM-W1($R^2$), G-W2(P), G-W2(A), G-W2($R^2$), GM-W2(P),
GM-W2(A), and GM-W2($R^2$). For CAGB-WISE objects with variation parameters
obtained from WISE light curves, Table \ref{tab:tab16} lists the CAGB-WISE
identifier (CW-N) and the same information.

Note that G-W1(P) or G-W2(P) listed in Tables \ref{tab:tab13}-\ref{tab:tab16}
are primary periods. However, GM-W1(P) and GN-W1(P) (or GM-W2(P) and GN-W2(P))
are selected periods, which can be different from the primary periods (see
Sections~\ref{sec:neo-mc} and \ref{sec:neo-nmc}).

\section{summary\label{sec:sum}}

We have presented a new catalog of 11,209 OAGB stars and 7172 CAGB stars in our
Galaxy identifying more AGB stars in the bulge component and considering more
visual carbon stars. For each object, we have cross-identified the IRAS, AKARI,
MSX, WISE, 2MASS, and AAVSO counterparts.

We have presented the new catalog in two parts: one (AGB-IRAS) is based on the
IRAS PSC for brighter or more isolated objects, the other one (AGB-WISE) is
based on the ALLWISE source catalog for less bright objects or small objects in
crowded regions.

We have performed radiative transfer model calculations for AGB stars using
various parameters of central stars and spherically symmetric dust shells.

We have presented various IR 2CDs for the sample stars. We have compared the
various sequences of theoretical dust shell models at increasing dust optical
depth with the observations of AGB stars on the IR 2CDs. We find that the
theoretical dust shell models can roughly explain the observations of AGB stars
on the various IR 2CDs.

We have compared number distributions of observed IR magnitudes and colors for
AGB stars in the AGB-IRAS and AGB-WISE catalogs. Most AGB-IRAS objects are
brighter at MIR bands and they show redder IR colors than AGB-WISE objects. In
general, AGB-IRAS objects look to be more evolved (or massive) stars with
thicker dust envelopes than AGB-WISE objects.

We have investigated the IR properties of SiO and OH maser emission sources in
the OAGB-IRAS and OAGB-WISE catalogs. Almost all known OH maser sources are in
the OAGB-IRAS catalog. We have found that most OH maser sources are in the
range of large dust optical depths (or mass-loss rates). On the other hand,
most SiO maser sources in the OAGB-IRAS catalog are in the range of moderate
dust optical depths (or mass-loss rates) for various IR colors. For a major
portion of the OAGB stars in the OAGB-WISE catalog, OH or SiO maser
observations have not been performed yet.

We have compared the IR properties of visual carbon stars from those of
infrared carbon stars. Generally, visual carbon stars show bluer colors than
infrared carbon stars because the dust shell optical depths for visual carbon
stars are smaller. But some visual carbon stars show redder colors at MIR bands
using longer wavelengths, which would be due to detached circumstellar dust
shells that are remnants of an earlier phase when the stars were OAGB stars.

We have investigated number distribution of the Galactic longitude and latitude
for AGB stars in the AGB-IRAS and AGB-WISE catalogs. We have found that OAGB
stars are more concentrated toward the Galactic center and the number decreases
with the Galactic longitude, while CAGB stars are distributed more uniformly
from the center to large Galactic longitudes. The histograms for different
Galactic latitudes are similar for both OAGB and CAGB stars. All AGB stars are
concentrated toward the Galactic disk.

For known Mira variables in the sample stars, we have investigated the
period-color relations and found that objects with longer pulsation periods
generally show redder colors.

We have investigated infrared variability of the sample stars using the WISE
photometric data in the last 12 yr: the ALLWISE multiepoch data that were
acquired between 2009 and 2010 and the NEOWISE-R 2021 data release that were
acquired from 2013 until the end of 2020.

We have tried to find Mira-like variations from the WISE light curves of the
sample stars using a simple sinusoidal light curve model. Using the WISE data
at W1 and W2 bands, we have generated the light curves and computed the
Lomb-Scargle periodograms for all of the sample stars and found good-quality
variation parameters for 3710 objects in the catalog, for which periods were
either known or unknown in previous works.

We have obtained pulsation periods from the WISE light curves for 2810 objects,
which are known to be Miras with periods from AAVSO. For about a half of the
objects, the obtained primary periods from the WISE data are very similar to
AAVSO periods. For another half of objects whose primary periods from WISE data
are different from AAVSO periods, we can find a new (second or up to fourth)
peak in the Lomb-Scargle power values, for which the new period is similar to
the AAVSO period. This would be because the AAVSO periods can be regarded as
true periods and the multiple peaks in the Lomb-Scargle power values are
similar, which would be due to the regularity of the WISE observations (6
months). On the whole, the periods from the WISE data and AAVSO showed very
good correlations for all of the sample stars known as Miras.

We have obtained pulsation periods from the WISE light curves for 656 objects
with unknown periods and 244 objects known as non-Mira variables with periods
from AAVSO. We have found that a major portion of these objects could be
candidate objects for new Mira variables because they show similar
period-amplitude and period-color relations to those of known Mira variables.

If we perform new photometric observations at $L$ or $M$ band at different
pulsation phases of AGB stars, the WISE data would be more useful to find
precise periods.

The catalog data are presented in Section \ref{sec:catalog} (and Tables
\ref{tab:tab9}-\ref{tab:tab16}). The data will also be accessible through the
author's webpage \url{http://web.chungbuk.ac.kr/~kwsuh/agb.htm}.


\begin{acknowledgments}
I thank the anonymous referee for helpful comments and constructive
suggestions. This work was supported by the National Research Foundation of
Korea (NRF) grant funded by the Korea government (MSIT; Ministry of Science and
ICT) (No. NRF-2017R1A2B4002328). This research has made use of the VizieR
catalogue access tool, CDS, Strasbourg, France. This research has made use of
the NASA/ IPAC Infrared Science Archive, which is operated by the Jet
Propulsion Laboratory, California Institute of Technology, under contract with
the National Aeronautics and Space Administration.
\end{acknowledgments}

\software{For a major part of the computations and figures in this paper, we
have used Python codes. We have used the Numpy (\citealt{van der Walt2011}),
Pandas (\citealt{McKinney2010}), AstroPy (\citealt{astropy2013}), and
Matplotlib (\citealt{hunter2007}) packages.}

\bibliographystyle{aasjournal}
\bibliography{pms-R1}



\begin{longrotatetable}
\begin{table*}
\centering
\scriptsize
\caption{OAGB-IRAS objects (16 columns; 5908 rows)$^0$\label{tab:tab9}}
\begin{tabular}{llllllllllllll}
\hline \hline
OI-N$^1$	&	Subgroup	&	IRAS PSC	&	AKARI PSC	&	AKARI BSC	&	RA$^2$	& DEC$^2$	& BP$^3$	&	ALLWISE	&	SP$^4$	& LRS$^5$	&	A-Name$^6$	&	A-Type$^7$	&	A-Period$^8$	\\
\hline
1	&	OI-SH	&	00007+5524	&	0003214+554051	&	0003210+554052	&	0.83958	&	55.68099	&	akari-p	&	J000321.45+554052.1	&	M7e:                          	&	E	&	Y Cas	&	M	&	417	\\
2	&	OI-SH	&	00017+3949	&	0004200+400635	&		&	1.08372	&	40.10991	&	akari-p	&	J000420.08+400635.9	&	M5.5e                         	&	F	&	SV And	&	M	&	313	\\
3	&	OI-SH	&	00042+4248	&	0006527+430502	&	0006526+430501	&	1.71972	&	43.08394	&	akari-p	&	J000652.77+430502.2	&	M10                           	&	E	&	KU And	&	M	&	720	\\
4	&	OI-SH	&	00050-2546	&	0007362-252940	&	0007369-252943	&	1.90107	&	-25.49451	&	akari-p	&	J000736.26-252939.9	&	M6e                           	&	E	&	SY Scl	&	M	&	411	\\
5	&	OI-SH	&	00060-3929	&	0008373-391304	&		&	2.15578	&	-39.21801	&	akari-p	&	J000837.35-391304.8	&	M5.5e:                        	&	F	&	V Scl	&	M	&	296.1	\\
6	&	OI-SH	&	00075+5435	&	0010092+545233	&		&	2.5386	&	54.87602	&	akari-p	&	J001009.14+545234.3	&	M6e                           	&	F	&	TT Cas	&	M	&	396	\\
7	&	OI-SH	&	00127+5437	&	0015248+545421	&	0015248+545426	&	3.85362	&	54.90609	&	akari-p	&	J001524.85+545422.1	&	                              	&	E	&		&		&		\\
8	&	OI-SH	&	00128-3219	&	0015223-320243	&	0015218-320239	&	3.84295	&	-32.04537	&	akari-p	&	J001522.28-320243.0	&	M6-e                          	&	F	&	S Scl	&	M	&	367	\\
9	&	OI-SH	&	00138+6544	&	0016365+660109	&	0016362+660113	&	4.15224	&	66.01937	&	akari-p	&	J001636.49+660110.5	&		&		&	NSV 15060	&	M	&	330	\\
10	&	OI-SH	&	00170+6542	&	0019512+655929	&	0019516+655930	&	4.9637	&	65.99148	&	akari-p	&	J001951.28+655930.5	&	(OH)                          	&	E	&		&		&		\\
11	&	OI-SH	&	00176+6931	&	0020255+694756	&	0020251+694759	&	5.10634	&	69.79903	&	akari-p	&	J002025.52+694757.2	&	                              	&	F	&	NSVS J0020257+694759	&	M	&	425	\\
12	&	OI-SH	&	00193-4033	&	0021474-401715	&	0021475-401713	&	5.44763	&	-40.28772	&	akari-p	&	J002147.41-401715.5	&	M9                            	&	E	&	BE Phe	&	M:	&		\\
13	&	OI-SH	&	00205+5530	&	0023142+554732	&	0023135+554735	&	5.80928	&	55.79243	&	akari-p	&	J002314.29+554733.1	&	M7.5e                         	&	F	&	T Cas	&	M	&	440	\\
14	&	OI-SH	&	00207-6156	&	0023077-614017	&		&	5.78216	&	-61.67143	&	akari-p	&	J002307.67-614016.9	&		&		&	S Tuc	&	M	&	242.4	\\
15	&	OI-SH	&	00222+6952	&	0025101+700851	&	0025087+700901	&	6.29223	&	70.14756	&	akari-p	&	J002510.03+700851.9	&	M6                            	&	F	&	NQ Cep	&	M	&	454	\\
16	&	OI-SH	&	00245-0652	&	0027064-063616	&	0027064-063610	&	6.77682	&	-6.60471	&	akari-p	&	J002706.42-063617.0	&	M7                            	&	E	&	UY Cet	&	SRB	&	440	\\
17	&	OI-SH	&	00309-7956	&	0032404-794021	&		&	8.16842	&	-79.67251	&	akari-p	&	J003240.38-794021.0	&		&		&	W Hyi	&	M	&	280	\\
18	&	OI-SH	&	00336+6744	&	0036365+680119	&	0036372+680125	&	9.1524	&	68.02211	&	akari-p	&	J003636.61+680119.8	&		&		&	V0861 Cas	&	M	&	210	\\
19	&	OI-SH	&	00340+6251	&	0036594+630800	&	0036593+630758	&	9.24777	&	63.13344	&	akari-p	&	J003659.43+630801.9	&	M6                            	&	E	&	TY Cas	&	M	&	645	\\
20	&	OI-SH	&	00347+8004	&	0038228+802125	&		&	9.59516	&	80.35698	&	akari-p	&	J003822.82+802124.9	&	M5e:                          	&	F	&	Y Cep	&	M	&	332.57	\\
21	&	OI-SH	&	00381-8018	&	0039391-800204	&		&	9.91314	&	-80.03467	&	akari-p	&	J003938.99-800204.5	&	                              	&	E	&	X Hyi	&	M	&	304	\\
22	&	OI-SH	&	00420+7533	&	0045280+755022	&	0045264+755022	&	11.36698	&	75.83947	&	akari-p	&	J004528.05+755022.1	&	                              	&	E	&	NSVS J0045283+755024	&	M	&	599	\\
23	&	OI-SH	&	00428+6854	&	0046002+691053	&	0045594+691055	&	11.50094	&	69.18145	&	akari-p	&	J004600.17+691053.8	&	M8:                           	&	E	&	V0524 Cas	&	M	&	490	\\
24	&	OI-SH	&	00445+3224	&	0047189+324108	&	0047191+324108	&	11.82894	&	32.68564	&	akari-p	&	J004718.91+324108.1	&	S6/2-e                        	&	F	&	RW And	&	M	&	430	\\
25	&	OI-SH	&	00453+5317	&	0048100+533400	&		&	12.04174	&	53.56691	&	akari-p	&	J004809.99+533401.2	&	M8                            	&	F	&	V0414 Cas	&	M	&	195	\\
26	&	OI-SH	&	00459+6749	&	0049069+680547	&		&	12.27877	&	68.09647	&	akari-p	&	J004906.85+680547.3	&	M8                            	&	F	&	V0865 Cas	&	M	&	180	\\
27	&	OI-SH	&	00479+4614	&	0050433+463029	&	0050437+463029	&	12.68047	&	46.5083	&	akari-p	&	J005043.29+463030.5	&	                              	&	E	&	V0415 And	&	SRB	&	460	\\
28	&	OI-SH	&	00498+4708	&	0052427+472455	&	0052425+472502	&	13.17824	&	47.41547	&	akari-p	&	J005242.78+472456.2	&	M6Se                          	&	E	&	RV Cas	&	M	&	331.68	\\
29	&	OI-SH	&	00503+6445	&	0053250+650156	&	0053252+650201	&	13.35421	&	65.03224	&	akari-p	&	J005325.18+650155.9	&	M8/9                          	&	E	&	NSV 15193	&	M	&	507	\\
30	&	OI-SH	&	00534+6031	&	0056283+604710	&		&	14.11806	&	60.78614	&	akari-p	&	J005628.33+604709.3	&		&		&	V0867 Cas	&	M	&	412	\\
-	&	-	&	-	&	-	&	-	&	-	&	-	&	-	&	-	&	-	&	-	&	-	&	-	&	-	\\
5899	&	OI-AM	&	20499+4657	&	2051366+470909	&	2051365+470907	&	312.90288	&	47.15266	&	akari-p	&	J205136.63+470909.1	&	                              	&	E	&	ZTF J205136.64+470909.3	&	M	&	457.3509671	\\
5900	&	OI-AM	&	21112+3150	&	2113182+320326	&		&	318.32608	&	32.05743	&	akari-p	&	J211318.23+320327.2	&	                              	&	E	&	V0472 Cyg	&	M	&	297	\\
5901	&	OI-AM	&	21444+4752	&	2146199+480648	&	2146219+480652	&	326.5831	&	48.1135	&	akari-p	&	J214619.97+480650.0	&	M7:                           	&	E	&	LP Cyg	&	M	&	419	\\
5902	&	OI-AM	&	21543+5605	&	2156008+561927	&		&	329.00362	&	56.32422	&	akari-p	&	J215600.90+561927.5	&	M9                            	&	E	&	V0720 Cep	&	M	&	305	\\
5903	&	OI-AM	&	22000+5643	&	2201477+565810	&	2201479+565811	&	330.44911	&	56.96954	&	akari-p	&	J220147.74+565810.2	&	M7                            	&	E	&	YY Cep	&	M	&	526.08	\\
5904	&	OI-AM	&	22049+4813	&	2206516+482755	&	2206523+482755	&	331.71531	&	48.46548	&	akari-p	&	J220651.71+482756.2	&	M7                            	&	E	&	AP Lac	&	M	&	524	\\
5905	&	OI-AM	&	22124+7315	&	2213218+732958	&	2213215+732958	&	333.34096	&	73.4995	&	akari-p	&	J221321.64+732958.3	&	M7                            	&	E	&	NSVS J2213225+733019	&	M	&	429	\\
5906	&	OI-AM	&	22450+5829	&	2247042+584513	&		&	341.76759	&	58.75388	&	akari-p	&	J224704.07+584514.2	&	M7-8                          	&	E	&	Mis V1170	&	M	&	730	\\
5907	&	OI-AM	&	23548-6539	&	2357263-652304	&		&	359.35966	&	-65.38461	&	akari-p	&	J235726.38-652304.9	&	M5e                           	&	E	&	R Tuc	&	M	&	286.06	\\
5908	&	OI-AM	&	23561+6037	&	2358382+605342	&	2358370+605336	&	359.65948	&	60.89513	&	akari-p	&	J235838.26+605342.5	&	                              	&	E	&	EU Cas	&	M	&	447	\\
\hline
\end{tabular}
\begin{flushleft}
$^0$Only 14 columns are shown in this example table. In the data file, there are two more columns: OH and SiO maser detection (see Section~\ref{sec:maser}).
$^1$The OAGB-IRAS identifier (see Table~\ref{tab:tab1}).
$^2$The best position (right ascension and declination J2000).
$^3$The source of the best position (AKARI PSC, AKARI BSC, or IRAS PSC; see Section~\ref{sec:agb-c}).
$^4$The spectral type from \citet{kwok1997}.
$^5$The IRAS LRS type from \citet{kwok1997}.
$^6$The object name from AAVSO.
$^7$The variable type from AAVSO.
$^8$The period from AAVSO.
\end{flushleft}
\end{table*}
\end{longrotatetable}

\begin{longrotatetable}
\begin{table*}
\centering
\scriptsize
\caption{CAGB-IRAS objects (14 columns; 3596 rows)\label{tab:tab10}}
\begin{tabular}{llllllllllllll}
\hline \hline
CI-N$^1$	&	Subgroup	&	IRAS PSC	&	AKARI PSC	&	AKARI BSC	&	RA$^2$	& DEC$^2$	& BP$^3$	&	ALLWISE	&	SP$^4$	& LRS$^5$	&	A-Name$^6$	&	A-Type$^7$	&	A-Period$^8$ \\
\hline
1	&	CI-SH	&	00020+4316	&	0004364+433304	&		&	1.15175	&	43.55133	&	akari-p	&	J000436.41+433304.6	&	C6,4	&	S	&	SU And	&	LC	&		\\
2	&	CI-SH	&	00036+6947	&	0006142+700402	&	0006131+700409	&	1.55954	&	70.06723	&	akari-p	&	J000614.21+700402.1	&	C	&	C	&	OR Cep	&	M	&	355.32	\\
3	&	CI-SH	&	00050+7357	&	0007432+741411	&		&	1.93001	&	74.23647	&	akari-p	&	J000743.08+741411.5	&		&	C	&	NSVS J0007434+741409	&	L	&	476	\\
4	&	CI-SH	&	00084-1851	&	0010579-183422	&		&	2.74137	&	-18.57287	&	akari-p	&	J001057.93-183422.7	&	M4III	&	C	&	AC Cet	&	LB	&		\\
5	&	CI-SH	&	00172+4425	&	0019540+444234	&	0019539+444230	&	4.97506	&	44.70953	&	akari-p	&	J001954.00+444233.8	&	C4,5J	&	C	&	VX And	&	SRA	&	375	\\
6	&	CI-SH	&	00210+6221	&	0023508+623811	&	0023507+623813	&	5.96206	&	62.6364	&	akari-p	&	J002350.87+623810.9	&	(CO)	&	U	&		&		&		\\
7	&	CI-SH	&	00247+6922	&	0027411+693851	&	0027405+693857	&	6.92134	&	69.64762	&	akari-p	&	J002741.04+693851.0	&	(SiC)	&	C	&	V0668 Cas	&	M	&	650	\\
8	&	CI-SH	&	00248+3518	&	0027316+353514	&	0027315+353510	&	6.88205	&	35.58736	&	akari-p	&	J002731.69+353514.2	&	C5,4	&	C	&	AQ And	&	SRB	&	169	\\
9	&	CI-SH	&	00422+5310	&	0045070+532647	&	0045069+532652	&	11.27958	&	53.44656	&	akari-p	&	J004507.07+532647.6	&	(SiC)	&	C	&	V0720 Cas	&	M	&	431	\\
10	&	CI-SH	&	00523+6812	&	0055340+682854	&		&	13.89167	&	68.48174	&	akari-p	&	J005533.95+682855.2	&	C6-,4	&	C	&	V0880 Cas	&	SR	&	377	\\
11	&	CI-SH	&	00535+5923	&	0056330+593944	&	0056329+593951	&	14.13752	&	59.66235	&	akari-p	&	J005633.03+593943.9	&	C	&	C	&	V0721 Cas	&	M:	&		\\
12	&	CI-SH	&	00596+6135	&	0102436+615142	&	0102448+615128	&	15.68168	&	61.8618	&	akari-p	&	J010243.58+615143.4	&	C4,3	&	S	&	HO Cas	&	LB	&		\\
13	&	CI-SH	&	01022+6542	&	0105272+655900	&		&	16.36364	&	65.98334	&	akari-p	&	J010527.41+655859.7	&	M7	&	C	&	V0888 Cas	&	M	&	341	\\
14	&	CI-SH	&	01080+5327	&	0111035+534339	&	0111035+534338	&	17.76461	&	53.72774	&	akari-p	&	J011103.46+534339.9	&	C4,3e	&	C	&	HV Cas	&	M	&	527	\\
15	&	CI-SH	&	01105+6241	&	0113444+625736	&	0113450+625735	&	18.4352	&	62.96007	&	akari-p	&	J011344.54+625735.9	&	C6,3:e	&	C	&	NSV 438	&	M:	&	280	\\
16	&	CI-SH	&	01133+2530	&	0116050+254609	&	0116047+254604	&	19.02099	&	25.76919	&	akari-p	&	J011605.03+254609.7	&	C7,2	&	F	&	Z Psc	&	SRB	&	155.8	\\
17	&	CI-SH	&	01142+6306	&	0117335+632205	&	0117337+632211	&	19.38993	&	63.36806	&	akari-p	&	J011733.56+632205.1	&	(SiC)	&	C	&		&		&		\\
18	&	CI-SH	&	01144+6658	&	0117515+671352	&	0117507+671357	&	19.4647	&	67.23136	&	akari-p	&	J011751.34+671353.2	&	(HCN)	&	U	&	V0829 Cas	&	M	&	1060	\\
19	&	CI-SH	&	01156+6237	&	0118539+625258	&	0118534+625306	&	19.7246	&	62.88294	&	akari-p	&	J011853.84+625259.4	&	C	&	C	&	NSVS J0118537+625300	&	L	&	279	\\
20	&	CI-SH	&	01215+6430	&	0124580+644613	&	0124583+644609	&	21.24174	&	64.77053	&	akari-p	&	J012458.25+644613.2	&	C	&	C	&	NSVS 1695145	&	M:	&	258	\\
21	&	CI-SH	&	01246-3248	&	0126580-323235	&	0126581-323237	&	21.74177	&	-32.54324	&	akari-p	&	J012658.05-323236.0	&	C6,4	&	C	&	R Scl	&	SRB	&	370	\\
22	&	CI-SH	&	01324+4907	&	0135288+492242	&		&	23.87005	&	49.37836	&	akari-p	&	J013528.78+492242.0	&	(SiC)	&	C	&	Dauban V257	&	VAR	&		\\
23	&	CI-SH	&	01327+6503	&	0136165+651839	&	0136159+651849	&	24.06899	&	65.3109	&	akari-p	&	J013616.59+651839.8	&		&	C	&	MGAB-V1402	&	M	&		\\
24	&	CI-SH	&	01348+5543	&	0138056+555814	&		&	24.52342	&	55.97078	&	akari-p	&	J013805.62+555815.2	&		&		&	PT Cas	&	M	&	300	\\
25	&	CI-SH	&	01411+7104	&	0145087+711915	&		&	26.28656	&	71.32098	&	akari-p	&	J014508.73+711916.1	&	C	&	F	&	NSVS J0145096+711919	&	L	&	730	\\
26	&	CI-SH	&	01443+6417	&	0147555+643255	&	0147553+643256	&	26.98151	&	64.54869	&	akari-p	&	J014755.72+643256.1	&	C	&	C	&	NSVS J0147560+643254	&	L	&	476	\\
27	&	CI-SH	&	01531+5900	&	0156381+591533	&	0156374+591528	&	29.15886	&	59.25942	&	akari-p	&	J015638.14+591533.9	&	C5,4e	&	C	&	X Cas	&	M	&	415	\\
28	&	CI-SH	&	01551+5458	&	0158294+551258	&	0158294+551303	&	29.62284	&	55.21626	&	akari-p	&	J015829.47+551258.8	&	Ce	&	F	&	V0437 Per	&	M	&	470	\\
29	&	CI-SH	&	01580+5803	&	0201281+581814	&		&	30.36713	&	58.30403	&	akari-p	&	J020128.16+581814.0	&	Ce	&	C	&	V0666 Cas	&	M	&	432.1	\\
30	&	CI-SH	&	02152+2822	&	0218061+283646	&	0218062+283645	&	34.52542	&	28.61286	&	akari-p	&	J021806.05+283645.3	&	(CO)	&	C	&	YY Tri	&	M	&	624	\\
-	&	-	&	-	&	-	&	-	&	-	&	-	&	-	&	-	&	-	&	-	&	-	&	-	&	-	\\
3587	&	CI-GC	&	23573+5642	&	2359542+565846	&		&	359.97602	&	56.97963	&	akari-p	&	J235954.25+565846.6	&		&		&	V0533 Cas	&	SRA	&	305	\\
3588	&	CI-OG	&	17398-2146	&	1742510-214729	&	1742510-214728	&	265.71265	&	-21.79161	&	akari-p	&	J174250.99-214728.7	&		&		&	OGLE-BLG-LPV-020858	&	M	&	469.6	\\
3589	&	CI-OG	&	17468-3320	&	1750084-332118	&		&	267.53534	&	-33.35513	&	akari-p	&	J175008.37-332118.3	&		&		&	OGLE-BLG-LPV-059872	&	M	&	400.4	\\
3590	&	CI-OG	&	17490-3414	&	1752190-341509	&		&	268.07937	&	-34.25274	&	akari-p	&	J175219.08-341510.2	&		&		&	OGLE-BLG-LPV-085234	&	M	&	555.8	\\
3591	&	CI-OG	&	17514-3354	&		&		&	268.682037	&	-33.916191	&	iras	&	J175443.43-335507.5	&		&		&	OGLE-BLG-LPV-114752	&	M	&	448	\\
3592	&	CI-OG	&	17552-2814	&	1758219-281454	&		&	269.59127	&	-28.24834	&	akari-p	&	J175821.90-281452.7	&	(PN)	&	I	&	OGLE-BLG-LPV-149402	&	ZAND+M	&	416.2	\\
3593	&	CI-OG	&	17570-3056	&	1800175-305623	&		&	270.07295	&	-30.93992	&	akari-p	&	J180017.57-305624.1	&		&		&	OGLE-BLG-LPV-169921	&	M	&	449.7	\\
3594	&	CI-OG	&	18031-3229	&	1806270-322914	&		&	271.61265	&	-32.48727	&	akari-p	&	J180627.05-322913.8	&		&		&	OGLE-BLG-LPV-209822	&	M	&	516.9	\\
3595	&	CI-OG	&	18039-3411	&	1807152-341125	&		&	271.81348	&	-34.19048	&	akari-p	&	J180715.23-341125.1	&		&		&	OGLE-BLG-LPV-212597	&	M	&	378.8	\\
3596	&	CI-OG	&	18204-2133	&	1823282-213129	&		&	275.86784	&	-21.52491	&	akari-p	&	J182328.27-213130.6	&		&		&	OGLE-BLG-LPV-231922	&	M	&	513.8	\\
\hline
\end{tabular}
\begin{flushleft}
$^1$The CAGB-IRAS identifier (see Table~\ref{tab:tab1}).
$^2$The best position (right ascension and declination J2000).
$^3$The source of the best position (AKARI PSC, AKARI BSC, or IRAS PSC; see Section~\ref{sec:agb-c}).
$^4$The spectral type from \citet{kwok1997}.
$^5$The IRAS LRS type from \citet{kwok1997}.
$^6$The object name from AAVSO.
$^7$The variable type from AAVSO.
$^8$The period from AAVSO.
\end{flushleft}
\end{table*}
\end{longrotatetable}

\begin{longrotatetable}
\begin{table*}
\centering
\scriptsize
\caption{OAGB-WISE objects (12 columns; 5301 rows)\label{tab:tab11}}
\begin{tabular}{llllllllllll}
\hline \hline
OW-N$^1$	&	Subgroup	&	recno/star	&	ALLWISE	&	W-RA$^2$	&	W-DEC$^2$	&	2MASS	&	AKARI PSC	&	MSX	&	A-Name	&	A-Type	&	A-Period	\\
\hline
1	&	OW-ME	&	1	&	J173140.97-320355.7	&	262.9207368	&	-32.0654792	&	17314097-3203559	&	1731409-320355	&	G355.7512+00.8670	&		&		&		\\
2	&	OW-ME	&	3	&	J173707.28-312131.2	&	264.2803606	&	-31.3586915	&	17370728-3121312	&	1737072-312132	&	G356.9721+00.2780	&		&		&		\\
3	&	OW-ME	&	4	&	J173729.36-311716.8	&	264.3723546	&	-31.2880029	&	17372934-3117166	&	1737292-311717	&	G357.0739+00.2497	&		&		&		\\
4	&	OW-ME	&	6	&	J173813.13-293941.4	&	264.5547352	&	-29.6615113	&	17381248-2939385	&	1738125-293938	&	G358.5318+00.9902	&		&		&		\\
5	&	OW-ME	&	7	&	J173817.04-294231.9	&	264.5710302	&	-29.7088827	&	17381707-2942324	&	1738170-294231	&	G358.4999+00.9505	&	GDS\_J1738177-294232	&	VAR	&		\\
6	&	OW-ME	&	9	&	J173832.49-312042.5	&	264.6353944	&	-31.3451427	&	17383250-3120427	&	1738324-312043	&	G357.1459+00.0294	&		&		&		\\
7	&	OW-ME	&	12	&	J174057.23-294531.4	&	265.238484	&	-29.7587303	&	17405722-2945314	&	1740571-294532	&	G358.7654+00.4335	&		&		&		\\
8	&	OW-ME	&	16	&	J174136.83-292930.9	&	265.4034951	&	-29.4919261	&	17413685-2929309	&		&	G359.0679+00.4525	&		&		&		\\
9	&	OW-ME	&	17	&	J174137.39-293205.7	&	265.4058107	&	-29.5349427	&	17413740-2932057	&		&	G359.0325+00.4283	&		&		&		\\
10	&	OW-ME	&	18	&	J174204.33-295846.1	&	265.5180671	&	-29.9794904	&	17420435-2958463	&	1742042-295846	&	G358.7064+00.1100	&		&		&		\\
11	&	OW-ME	&	19	&	J174206.85-281832.4	&	265.5285433	&	-28.3090071	&	17420685-2818323	&	1742068-281831	&	G000.1309+00.9838	&		&		&		\\
12	&	OW-ME	&	20	&	J174223.29-293935.3	&	265.5970442	&	-29.6598205	&	17422328-2939355	&	1742232-293936	&	G359.0141+00.2210	&		&		&		\\
13	&	OW-ME	&	21	&	J174232.89-294126.0	&	265.6370525	&	-29.6905646	&	17423291-2941251	&	1742325-294112	&	G359.0087+00.1783	&		&		&		\\
14	&	OW-ME	&	22	&	J174232.47-294110.6	&	265.6352975	&	-29.6862809	&	17423247-2941107	&	1742325-294112	&	G359.0087+00.1783	&		&		&		\\
15	&	OW-ME	&	24	&	J174309.81-292403.1	&	265.7909036	&	-29.4008852	&	17430981-2924033	&	1743097-292403	&	G359.3228+00.2140	&		&		&		\\
16	&	OW-ME	&	25	&	J174323.47-285350.1	&	265.8477983	&	-28.8972522	&	17432345-2853503	&	1743234-285349	&	G359.7779+00.4361	&		&		&		\\
17	&	OW-ME	&	26	&	J174325.25-294528.6	&	265.8552409	&	-29.7579569	&	17432525-2945285	&	1743251-294529	&	G359.0484-00.0212	&		&		&		\\
18	&	OW-ME	&	27	&	J174332.70-291539.2	&	265.8862747	&	-29.2608935	&	17433271-2915393	&	1743326-291539	&	G359.4858+00.2165	&		&		&		\\
19	&	OW-ME	&	28	&	J174333.12-295133.1	&	265.8880322	&	-29.8592045	&	17433312-2951331	&	1743330-295134	&	G358.9772-00.0984	&		&		&		\\
20	&	OW-ME	&	29	&	J174334.83-294029.2	&	265.8951536	&	-29.6747897	&	17433479-2940304	&	1743346-294031	&	G359.1369-00.0067	&		&		&		\\
21	&	OW-ME	&	30	&	J174335.12-292447.3	&	265.896337	&	-29.4131541	&	17433512-2924472	&		&	G359.3607+00.1297	&		&		&		\\
22	&	OW-ME	&	32	&	J174341.73-290118.9	&	265.9239097	&	-29.0219439	&	17434176-2901190	&	1743417-290118	&	G359.7068+00.3140	&		&		&		\\
23	&	OW-ME	&	33	&	J174349.53-290319.9	&	265.9564094	&	-29.0555301	&	17434954-2903194	&		&	G359.6931+00.2721	&		&		&		\\
24	&	OW-ME	&	34	&	J174349.98-290121.4	&	265.9582628	&	-29.0226238	&	17434997-2901213	&		&	G359.7219+00.2882	&		&		&		\\
25	&	OW-ME	&	35	&	J174351.10-290028.8	&	265.9629185	&	-29.0080118	&	17435106-2900292	&		&	G359.7368+00.2924	&		&		&		\\
26	&	OW-ME	&	36	&	J174354.00-284128.5	&	265.975037	&	-28.6912772	&	17435399-2841285	&		&	G000.0118+00.4492	&		&		&		\\
27	&	OW-ME	&	37	&	J174355.30-285649.9	&	265.9804464	&	-28.9471972	&	17435531-2856498	&		&	G359.7964+00.3112	&		&		&		\\
28	&	OW-ME	&	38	&	J174357.27-293146.6	&	265.988648	&	-29.5296325	&	17435729-2931465	&		&	G359.3035-00.0007	&		&		&		\\
29	&	OW-ME	&	39	&	J174358.02-293052.2	&	265.99179	&	-29.5145069	&	17435804-2930520	&		&	G359.3182+00.0059	&		&		&		\\
30	&	OW-ME	&	40	&	J174359.29-291951.5	&	265.997061	&	-29.3309929	&	17435926-2919517	&		&	G359.4766+00.0982	&		&		&		\\
-	&	-	&	-	&	-	&	-	&	-	&	-	&	-	&	-	&	-	&	-	&	-	\\
5292	&	OW-OG	&	232164	&	J182407.05-212750.1	&	276.0294118	&	-21.4639172	&	18240706-2127500	&	1824072-212748	&	G010.7840-03.9197	&	OGLE-BLG-LPV-232164	&	M	&	305.1	\\
5293	&	OW-OG	&	232216	&	J182415.74-214049.6	&	276.0656208	&	-21.6804648	&	18241570-2140498	&	1824159-214049	&	G010.6070-04.0495	&	OGLE-BLG-LPV-232216	&	M	&	325.8	\\
5294	&	OW-OG	&	232256	&	J182421.96-212222.4	&	276.0915081	&	-21.372902	&	18242197-2122224	&	1824221-212222	&	G010.8916-03.9294	&	OGLE-BLG-LPV-232256	&	M	&	306.8	\\
5295	&	OW-OG	&	232307	&	J182431.29-220041.6	&	276.1304043	&	-22.0115722	&	18243129-2200414	&	1824312-220041	&	G010.3404-04.2554	&	OGLE-BLG-LPV-232307	&	M	&	339.4	\\
5296	&	OW-OG	&	232329	&	J182458.95-252439.4	&	276.2456492	&	-25.4109592	&	18245895-2524394	&		&		&	OGLE-BLG-LPV-232329	&	M	&	229.2	\\
5297	&	OW-OG	&	232339	&	J182508.99-253350.7	&	276.2874842	&	-25.5641061	&	18250899-2533505	&	1825089-253350	&		&	OGLE-BLG-LPV-232339	&	M	&	197.57	\\
5298	&	OW-OG	&	232340	&	J182510.62-255542.0	&	276.2942638	&	-25.9283545	&	18251062-2555419	&	1825106-255541	&		&	OGLE-BLG-LPV-232340	&	M	&	263.2	\\
5299	&	OW-OG	&	232350	&	J182547.17-251527.2	&	276.4465735	&	-25.2575831	&	18254717-2515270	&	1825472-251527	&		&	OGLE-BLG-LPV-232350	&	M	&	298.8	\\
5300	&	OW-OG	&	232364	&	J182640.02-251436.3	&	276.6667693	&	-25.2434178	&	18264001-2514363	&	1826400-251435	&		&	OGLE-BLG-LPV-232364	&	M	&	199.55	\\
5301	&	OW-OG	&	232397	&	J183121.84-242406.1	&	277.8410132	&	-24.4017158	&	18312183-2424060	&	1831218-242406	&		&	OGLE-BLG-LPV-232397	&	M	&	303.8	\\
\hline
\end{tabular}
\begin{flushleft}
$^1$The OAGB-WISE identifier (see Table~\ref{tab:tab2}).
$^2$The ALLWISE source position (right ascension and declination J2000).
See Section~\ref{sec:catalog}.
\end{flushleft}
\end{table*}
\end{longrotatetable}

\begin{longrotatetable}
\begin{table*}
\centering
\scriptsize
\caption{CAGB-WISE objects (12 columns; 3576 rows)\label{tab:tab12}}
\begin{tabular}{llllllllllll}
\hline \hline
CW-N$^1$	&	Subgroup	&	recno/star	&	ALLWISE	&	W-RA$^2$	&	W-DEC$^2$	&	2MASS	&	AKARI PSC	&	MSX	&	A-Name	&	A-Type	&	A-Period	\\
\hline
1	&	CW-GC	&	1	&	J000010.87+642554.6	&	0.0453247	&	64.4318379	&	00001089+6425546	&		&		&		&		&		\\
2	&	CW-GC	&	3	&	J000052.97+565807.3	&	0.2207225	&	56.9687165	&	00005295+5658072	&		&		&		&		&		\\
3	&	CW-GC	&	6	&	J000129.81+645507.8	&	0.3742141	&	64.9188449	&	00012982+6455077	&	0001298+645508	&	G117.6688+02.5591	&	ZTF J000129.84+645507.9	&	SR	&	253.5952999	\\
4	&	CW-GC	&	8	&	J000203.10+625859.4	&	0.5129277	&	62.983188	&	00020311+6258593	&	0002032+625858	&	G117.3577+00.6467	&	NSV 26200	&		&		\\
5	&	CW-GC	&	9	&	J000207.99-024913.1	&	0.5333304	&	-2.8203301	&	00020802-0249121	&	0002080-024912	&		&		&		&		\\
6	&	CW-GC	&	11	&	J000223.06+560231.9	&	0.5961116	&	56.0422064	&	00022306+5602319	&		&		&		&		&		\\
7	&	CW-GC	&	13	&	J000300.86+303823.4	&	0.7536194	&	30.6398393	&	00030086+3038233	&		&		&	2MASS J00030086+3038233	&	L	&		\\
8	&	CW-GC	&	14	&	J000322.05+084752.4	&	0.8419097	&	8.7978933	&	null	&		&		&		&		&		\\
9	&	CW-GC	&	15	&	J000341.96+594413.0	&	0.9248438	&	59.736952	&	00034196+5944130	&	0003419+594412	&		&	NSV 15007	&	M:	&	259	\\
10	&	CW-GC	&	19	&	J000524.99+011003.8	&	1.3541477	&	1.1677398	&	00052499+0110039	&		&		&		&		&		\\
11	&	CW-GC	&	21	&	J000544.48+653057.1	&	1.4353489	&	65.515888	&	00054449+6530569	&	0005444+653056	&	G118.2161+03.0626	&	ZTF J000544.52+653057.2	&	SR	&	121.6376906	\\
12	&	CW-GC	&	22	&	J000545.12+592948.0	&	1.4380045	&	59.4966744	&	00054511+5929479	&	0005452+592948	&	G117.1563-02.8616	&	ZTF J000545.13+592948.0	&	SR	&	274.403834	\\
13	&	CW-GC	&	30	&	J000911.97+650126.3	&	2.2999015	&	65.0239937	&	00091198+6501262	&	0009120+650124	&		&		&		&		\\
14	&	CW-GC	&	31	&	J001109.46+642925.5	&	2.7894529	&	64.4904191	&	00110946+6429257	&	0011093+642925	&	G118.6098+01.9573	&	NSVS 1612314	&	LPV	&	236.7535227	\\
15	&	CW-GC	&	32	&	J001135.45+631830.8	&	2.897721	&	63.3085687	&	00113544+6318308	&	0011353+631830	&	G118.4749+00.7823	&	ZTF J001135.47+631830.7	&	SR	&	310.7410887	\\
16	&	CW-GC	&	33	&	J001204.70+613249.6	&	3.0195937	&	61.5471352	&	00120469+6132496	&	0012046+613249	&	G118.2606-00.9672	&	Mis V1359	&	SR:	&		\\
17	&	CW-GC	&	36	&	J001310.30+593635.2	&	3.2929478	&	59.6097837	&	00131045+5936326	&		&	G118.1186-02.9019	&		&		&		\\
18	&	CW-GC	&	37	&	J001316.63+630200.3	&	3.3193147	&	63.0334316	&	00131664+6302003	&	0013163+630159	&	G118.6208+00.4810	&	ZTF J001316.65+630200.2	&	SR	&	255.4633061	\\
19	&	CW-GC	&	38	&	J001322.15-304045.1	&	3.3423166	&	-30.6792219	&	null	&		&		&		&		&		\\
20	&	CW-GC	&	41	&	J001346.40+681730.2	&	3.4433605	&	68.2917322	&	00134645+6817301	&	0013463+681730	&		&	V0966 Cep	&	SR:	&	130	\\
21	&	CW-GC	&	43	&	J001431.09+605531.0	&	3.629571	&	60.9252917	&	00143110+6055310	&	0014310+605530	&	G118.4594-01.6258	&	ZTF J001431.10+605531.1	&	SR	&	152.8841528	\\
22	&	CW-GC	&	44	&	J001444.57+574812.7	&	3.6857147	&	57.8035551	&	00144456+5748128	&		&		&		&		&		\\
23	&	CW-GC	&	46	&	J001501.62+600821.8	&	3.7567599	&	60.1394082	&	00150169+6008222	&		&	G118.4225-02.4038	&		&		&		\\
24	&	CW-GC	&	47	&	J001553.66+605328.6	&	3.9736087	&	60.8912972	&	00155366+6053286	&		&		&	NSV 107	&		&		\\
25	&	CW-GC	&	50	&	J001718.41+105203.1	&	4.3267474	&	10.8675438	&	00171842+1052032	&		&		&		&		&		\\
26	&	CW-GC	&	52	&	J001751.22+630406.1	&	4.4634252	&	63.0683664	&	00175122+6304060	&	0017511+630406	&	G119.1401+00.4449	&	ZTF J001751.23+630406.2	&	SR	&	103.9275228	\\
27	&	CW-GC	&	53	&	J001814.60+594601.9	&	4.5608417	&	59.7672053	&	00181456+5946018	&		&		&		&		&		\\
28	&	CW-GC	&	57	&	J002006.57+615919.6	&	5.027391	&	61.9887875	&	00200656+6159196	&		&		&	ZTF J002006.58+615919.7	&	SR	&	209.421136	\\
29	&	CW-GC	&	58	&	J002021.60+011206.7	&	5.0900124	&	1.201865	&	00202160+0112068	&		&		&		&		&		\\
30	&	CW-GC	&	60	&	J002134.74+591448.6	&	5.3947506	&	59.246851	&	00213483+5914483	&		&		&		&		&		\\
-	&	-	&	-	&	-	&	-	&	-	&	-	&	-	&	-	&	-	&	-	&	-	\\
3567	&	CW-OG	&	224006	&	J181108.13-261611.8	&	272.7839129	&	-26.2699694	&	18110815-2616119	&		&	G005.1362-03.5662	&	OGLE-BLG-LPV-224006	&	M	&	222.7	\\
3568	&	CW-OG	&	224160	&	J181113.68-251656.4	&	272.8070377	&	-25.28234	&	18111369-2516563	&		&	G006.0256-03.1142	&	OGLE-BLG-LPV-224160	&	M	&	111.86	\\
3569	&	CW-OG	&	226511	&	J181316.97-253135.1	&	273.3207119	&	-25.5264335	&	18131697-2531351	&	1813169-253136	&	G006.0216-03.6336	&	OGLE-BLG-LPV-226511	&	RCB:	&		\\
3570	&	CW-OG	&	226979	&	J181346.26-253415.6	&	273.4427696	&	-25.5710236	&	18134625-2534156	&		&		&	OGLE-BLG-LPV-226979	&	M	&	231.9	\\
3571	&	CW-OG	&	227901	&	J181447.50-282438.8	&	273.6979355	&	-28.4107823	&	18144749-2824389	&	1814474-282438	&		&	OGLE-BLG-LPV-227901	&	M	&	359.7	\\
3572	&	CW-OG	&	228903	&	J181602.35-240937.0	&	274.0098284	&	-24.1603045	&	18160237-2409370	&	1816023-240936	&	G007.5149-03.5473	&	OGLE-BLG-LPV-228903	&	M	&	293.6	\\
3573	&	CW-OG	&	230266	&	J181744.12-252836.3	&	274.433864	&	-25.4767681	&	18174413-2528361	&	1817441-252837	&	G006.5396-04.4973	&	OGLE-BLG-LPV-230266	&	M	&	501.2	\\
3574	&	CW-OG	&	230608	&	J181944.05-222311.2	&	274.9335824	&	-22.3864556	&	18194444-2223039	&	1819443-222304	&	G009.4924-03.4501	&	OGLE-BLG-LPV-230608	&	M	&	366.9	\\
3575	&	CW-OG	&	231818	&	J182313.69-252653.8	&	275.8070795	&	-25.4482779	&	18231371-2526540	&		&		&	OGLE-BLG-LPV-231818	&	M	&	355.8	\\
3576	&	CW-OG	&	231979	&	J182335.93-262852.2	&	275.8997362	&	-26.4811782	&	18233594-2628521	&		&		&	OGLE-BLG-LPV-231979	&	M	&	263.1	\\
\hline
\end{tabular}
\begin{flushleft}
$^1$The CAGB-WISE identifier (see Table~\ref{tab:tab2}).
$^2$The ALLWISE source position (right ascension and declination J2000).
See Section~\ref{sec:catalog}.
\end{flushleft}
\end{table*}
\end{longrotatetable}

\begin{longrotatetable}
\begin{table*}
\centering
\scriptsize
\caption{OAGB-IRAS objects with obtained variation parameters from WISE light curves$^0$ (24 columns; 885 rows)\label{tab:tab13}}
\begin{tabular}{llllllllllllllll}
\hline \hline
OI-N$^1$	&	A-Type	&	A-Period	& EP	&	G-W1(P)	&	G-W1(A)	&	G-W1($R^2$)	&	GM-W1(P)	&	GM-W1(A)	&	GM-W1($R^2$)	&	GN-W1(P)	&	GN-W1(A)	&	GN-W1($R^2$)	&	G-W2(P)	&	G-W2(A)	&	G-W2($R^2$)	\\
\hline
181	&		&		&	880.56 	&	532.78 	&	0.816 	&	0.731 	&		&		&		&	532.78 	&	0.816 	&	0.731 	&		&		&		\\
183	&	L:	&		&	70.48 	&	59.55 	&	0.829 	&	0.677 	&		&		&		&	88.53 	&	0.837 	&	0.615 	&	125.44 	&	0.839 	&	0.700 	\\
264	&		&		&	316.24 	&	546.50 	&	0.763 	&	0.726 	&		&		&		&	273.61 	&	0.725 	&	0.629 	&		&		&		\\
295	&		&		&	338.04 	&	1233.38 	&	0.986 	&	0.639 	&		&		&		&	213.70 	&	1.030 	&	0.504 	&		&		&		\\
392	&		&		&	439.42 	&	124.46 	&	1.114 	&	0.937 	&		&		&		&	390.09 	&	1.165 	&	0.902 	&	124.38 	&	1.134 	&	0.940 	\\
523	&	VAR	&		&	720.22 	&	1654.50 	&	1.340 	&	0.601 	&		&		&		&	204.20 	&	1.288 	&	0.496 	&		&		&		\\
533	&		&		&	115.72 	&	559.05 	&	0.987 	&	0.628 	&		&		&		&	108.34 	&	0.789 	&	0.335 	&	137.05 	&	1.023 	&	0.643 	\\
555	&	M	&	383	&		&		&		&		&		&		&		&		&		&		&	90.34 	&	1.368 	&	0.787 	\\
560	&		&		&	1576.20 	&	207.57 	&	1.315 	&	0.764 	&		&		&		&	1464.18 	&	1.302 	&	0.779 	&		&		&		\\
579	&		&		&	132.85 	&	543.11 	&	1.213 	&	0.808 	&		&		&		&	136.26 	&	1.199 	&	0.771 	&	550.95 	&	1.318 	&	0.627 	\\
611	&		&		&	58.82 	&		&		&		&		&		&		&		&		&		&	111.02 	&	0.423 	&	0.698 	\\
642	&		&		&	428.80 	&	467.42 	&	1.123 	&	0.766 	&		&		&		&	467.42 	&	1.123 	&	0.766 	&	467.42 	&	1.404 	&	0.660 	\\
653	&		&		&	865.06 	&	2001.50 	&	1.214 	&	0.687 	&		&		&		&	199.64 	&	1.193 	&	0.628 	&		&		&		\\
656	&		&		&	341.11 	&		&		&		&		&		&		&		&		&		&	607.19 	&	1.159 	&	0.606 	\\
657	&	VAR	&		&	621.10 	&		&		&		&		&		&		&		&		&		&	185.25 	&	0.073 	&	0.624 	\\
666	&		&		&	554.24 	&	781.39 	&	1.213 	&	0.775 	&		&		&		&	781.39 	&	1.213 	&	0.775 	&		&		&		\\
688	&		&		&	452.11 	&	570.87 	&	0.896 	&	0.620 	&		&		&		&	570.87 	&	0.896 	&	0.620 	&		&		&		\\
704	&	VAR	&		&	664.36 	&	1420.69 	&	1.050 	&	0.632 	&		&		&		&	208.97 	&	1.084 	&	0.622 	&		&		&		\\
709	&		&		&	359.41 	&	234.67 	&	1.257 	&	0.726 	&		&		&		&	234.67 	&	1.257 	&	0.726 	&		&		&		\\
731	&		&		&		&	952.69 	&	0.868 	&	0.860 	&		&		&		&		&		&		&	962.30 	&	0.946 	&	0.800 	\\
773	&		&		&		&	835.88 	&	1.361 	&	0.736 	&		&		&		&		&		&		&		&		&		\\
796	&		&		&	125.16 	&		&		&		&		&		&		&		&		&		&	644.07 	&	0.959 	&	0.623 	\\
802	&		&		&	282.57 	&		&		&		&		&		&		&		&		&		&	549.50 	&	0.977 	&	0.648 	\\
803	&		&		&	419.00 	&	878.19 	&	1.498 	&	0.710 	&		&		&		&	229.32 	&	1.414 	&	0.565 	&		&		&		\\
806	&		&		&	1151.96 	&	683.28 	&	0.883 	&	0.745 	&		&		&		&	683.28 	&	0.883 	&	0.745 	&		&		&		\\
819	&		&		&		&	821.49 	&	1.427 	&	0.692 	&		&		&		&		&		&		&		&		&		\\
821	&	VAR	&		&	263.99 	&		&		&		&		&		&		&		&		&		&	505.85 	&	0.902 	&	0.608 	\\
827	&	VAR:	&		&	1793.77 	&	2290.05 	&	1.148 	&	0.845 	&		&		&		&	2290.05 	&	1.148 	&	0.845 	&	2262.90 	&	1.241 	&	0.750 	\\
833	&		&		&	455.15 	&	716.64 	&	0.935 	&	0.611 	&		&		&		&	243.98 	&	0.788 	&	0.615 	&		&		&		\\
835	&		&		&	436.78 	&	546.37 	&	1.035 	&	0.694 	&		&		&		&	546.37 	&	1.035 	&	0.694 	&	540.18 	&	1.265 	&	0.627 	\\
-	&	-	&	-	&	-	&	-	&	-	&	-	&	-	&	-	&	-	&	-	&	-	&	-	&	-	&	-	&	-	\\
5699	&	M	&	301.3	&		&		&		&		&		&		&		&		&		&		&	292.65 	&	0.544 	&	0.708 	\\
5702	&	M	&	331.2	&		&		&		&		&		&		&		&		&		&		&	71.90 	&	0.444 	&	0.840 	\\
5703	&	M	&	278	&		&		&		&		&		&		&		&		&		&		&	274.15 	&	0.529 	&	0.757 	\\
5711	&	M	&	362.9	&		&		&		&		&		&		&		&		&		&		&	73.05 	&	0.528 	&	0.754 	\\
5716	&	M	&	310	&		&		&		&		&		&		&		&		&		&		&	306.27 	&	0.229 	&	0.616 	\\
5725	&	M	&	445.6	&		&		&		&		&		&		&		&		&		&		&	444.64 	&	0.877 	&	0.603 	\\
5726	&	M	&	390.9	&		&		&		&		&		&		&		&		&		&		&	124.67 	&	0.467 	&	0.670 	\\
5752	&	M	&	575	&		&		&		&		&		&		&		&		&		&		&	570.73 	&	0.763 	&	0.616 	\\
5799	&	M	&	444	&		&		&		&		&		&		&		&		&		&		&	89.94 	&	0.991 	&	0.782 	\\
5861	&	M	&	398.8247036	&		&		&		&		&		&		&		&		&		&		&	124.54 	&	0.842 	&	0.641 	\\
\hline
\end{tabular}
\begin{flushleft}
$^0$Only 16 columns are shown in this example table. In the data file, there are eight more columns:
GM-W2(P), GM-W2(A), GM-W2($R^2$), GN-W2(P), GN-W2(A), GN-W2($R^2$), Subgroup, and IRAS PSC.
$^1$The OAGB-IRAS identifier (see Tables~\ref{tab:tab1} and \ref{tab:tab9}). See Section~\ref{sec:catalog}.
\end{flushleft}
\end{table*}
\end{longrotatetable}

\begin{longrotatetable}
\begin{table*}
\centering
\scriptsize
\caption{CAGB-IRAS objects with obtained variation parameters from WISE light curves$^0$ (24 columns; 141 rows)\label{tab:tab14}}
\begin{tabular}{llllllllllllllll}
\hline \hline
CI-N$^1$	&	A-Type	&	A-Period	&	EP	&	G-W1(P)	&	G-W1(A)	&	G-W1($R^2$)	&	GM-W1(P)	&	GM-W1(A)	&	GM-W1($R^2$)	&	GN-W1(P)	&	GN-W1(A)	&	GN-W1($R^2$)	&	G-W2(P)	&	G-W2(A)	&	G-W2($R^2$)	\\
\hline
6	&		&		&		&	1062.66 	&	1.315 	&	0.935 	&		&		&		&		&		&		&	1068.57 	&	1.730 	&	0.671 	\\
18	&	M	&	1060	&		&	1017.79 	&	1.715 	&	0.940 	&	1017.79 	&	1.715 	&	0.940 	&		&		&		&	1001.88 	&	2.180 	&	0.895 	\\
34	&	M	&		&		&	798.05 	&	1.243 	&	0.796 	&		&		&		&		&		&		&		&		&		\\
35	&		&		&	1224.47 	&	882.29 	&	1.137 	&	0.780 	&		&		&		&	882.29 	&	1.137 	&	0.780 	&		&		&		\\
58	&		&		&	713.21 	&	791.47 	&	1.028 	&	0.606 	&		&		&		&	791.47 	&	1.028 	&	0.606 	&		&		&		\\
90	&	EW	&	0.5546584	&	626.50 	&	147.96 	&	0.753 	&	0.702 	&		&		&		&	770.42 	&	0.784 	&	0.732 	&		&		&		\\
91	&		&		&	362.81 	&	628.79 	&	0.749 	&	0.716 	&		&		&		&	256.52 	&	0.675 	&	0.678 	&		&		&		\\
106	&	M:	&		&	137.15 	&	51.55 	&	0.760 	&	0.698 	&		&		&		&	118.80 	&	0.651 	&	0.552 	&		&		&		\\
111	&		&		&	339.67 	&	620.59 	&	0.828 	&	0.764 	&		&		&		&	257.16 	&	0.788 	&	0.638 	&	622.59 	&	1.014 	&	0.605 	\\
120	&	SR	&	89.1313574	&	563.64 	&	815.10 	&	1.004 	&	0.663 	&		&		&		&	333.90 	&	0.659 	&	-0.030 	&		&		&		\\
192	&		&		&	2199.82 	&	898.91 	&	0.855 	&	0.696 	&		&		&		&	898.91 	&	0.855 	&	0.696 	&		&		&		\\
210	&		&		&	978.01 	&	839.98 	&	1.115 	&	0.872 	&		&		&		&	839.98 	&	1.115 	&	0.872 	&	839.98 	&	1.053 	&	0.743 	\\
230	&		&		&	706.87 	&	485.67 	&	1.007 	&	0.833 	&		&		&		&	485.67 	&	1.007 	&	0.833 	&		&		&		\\
233	&	M	&	455.3437419	&		&	419.91 	&	0.907 	&	0.670 	&	419.91 	&	0.907 	&	0.670 	&		&		&		&		&		&		\\
246	&		&		&	397.66 	&	529.24 	&	0.886 	&	0.670 	&		&		&		&	278.01 	&	0.819 	&	0.615 	&	531.91 	&	0.912 	&	0.614 	\\
251	&		&		&		&	837.44 	&	1.135 	&	0.895 	&		&		&		&		&		&		&	816.15 	&	1.132 	&	0.603 	\\
277	&	SR	&	339.9025322	&	232.80 	&	428.51 	&	0.703 	&	0.704 	&		&		&		&	320.13 	&	0.713 	&	0.624 	&	422.86 	&	0.904 	&	0.702 	\\
280	&	ELL|ROT	&	1.083	&	696.54 	&	855.04 	&	1.049 	&	0.818 	&		&		&		&	855.04 	&	1.049 	&	0.818 	&		&		&		\\
283	&		&		&	257.89 	&	392.57 	&	0.458 	&	0.605 	&		&		&		&	341.06 	&	0.495 	&	0.658 	&		&		&		\\
292	&	M	&	443.003349	&		&	179.95 	&	2.096 	&	0.659 	&	255.25 	&	0.660 	&	0.220 	&		&		&		&		&		&		\\
313	&		&		&	595.25 	&	641.20 	&	0.976 	&	0.883 	&		&		&		&	641.20 	&	0.976 	&	0.883 	&	643.35 	&	1.034 	&	0.783 	\\
317	&		&		&	290.27 	&	528.43 	&	0.820 	&	0.694 	&		&		&		&	279.14 	&	0.779 	&	0.622 	&		&		&		\\
338	&		&		&	325.23 	&	519.87 	&	0.644 	&	0.740 	&		&		&		&	281.19 	&	0.597 	&	0.692 	&	524.12 	&	0.550 	&	0.768 	\\
348	&	VAR	&		&	379.01 	&	597.37 	&	0.908 	&	0.704 	&		&		&		&	261.67 	&	0.914 	&	0.650 	&		&		&		\\
373	&		&		&	352.36 	&	634.81 	&	0.654 	&	0.684 	&		&		&		&	254.39 	&	0.666 	&	0.514 	&		&		&		\\
380	&	EW	&	0.3634808	&	765.67 	&	689.44 	&	1.288 	&	0.867 	&		&		&		&	689.44 	&	1.288 	&	0.867 	&	686.98 	&	1.183 	&	0.612 	\\
396	&		&		&	135.96 	&	87.41 	&	0.907 	&	0.641 	&		&		&		&	128.97 	&	1.054 	&	0.638 	&		&		&		\\
406	&		&		&	292.60 	&	577.60 	&	0.668 	&	0.605 	&		&		&		&	266.73 	&	0.677 	&	0.585 	&		&		&		\\
407	&		&		&	380.11 	&	104.06 	&	1.378 	&	0.644 	&		&		&		&	247.19 	&	0.595 	&	0.594 	&		&		&		\\
-	&	-	&	-	&	-	&	-	&	-	&	-	&	-	&	-	&	-	&	-	&	-	&	-	&	-	&	-	&	-	\\
2984	&	SR	&	409	&	147.59 	&	60.08 	&	0.703 	&	0.625 	&		&		&		&	128.72 	&	0.425 	&	0.228 	&		&		&		\\
3052	&	SRB	&	175.4	&		&		&		&		&		&		&		&		&		&		&	177.87 	&	0.775 	&	0.609 	\\
3089	&	SR	&	858	&	128.80 	&	175.93 	&	0.641 	&	0.650 	&		&		&		&	92.32 	&	0.544 	&	0.506 	&		&		&		\\
3110	&	SR	&	214.4446898	&	110.82 	&	176.23 	&	0.705 	&	0.630 	&		&		&		&	92.22 	&	0.630 	&	0.603 	&		&		&		\\
3157	&		&		&	129.65 	&		&		&		&		&		&		&		&		&		&	180.35 	&	3.233 	&	0.684 	\\
3203	&	SR	&	342	&	145.61 	&	368.30 	&	1.138 	&	0.617 	&		&		&		&	122.12 	&	0.902 	&	0.589 	&		&		&		\\
3267	&	SR	&	130.58	&	86.95 	&	185.12 	&	0.943 	&	0.659 	&		&		&		&	90.09 	&	0.588 	&	0.631 	&		&		&		\\
3324	&	SR	&	173.7469153	&	83.53 	&	180.17 	&	1.559 	&	0.677 	&		&		&		&	90.18 	&	0.680 	&	0.670 	&		&		&		\\
3410	&	RCB	&	37.5	&	226.32 	&	194.33 	&	0.859 	&	0.778 	&		&		&		&	194.33 	&	0.859 	&	0.778 	&	194.92 	&	1.271 	&	0.869 	\\
3590	&	M	&	555.8	&		&		&		&		&		&		&		&		&		&		&	551.19 	&	0.894 	&	0.652 	\\
\hline
\end{tabular}
\begin{flushleft}
$^0$Only 16 columns are shown in this example table. In the data file, there are eight more columns:
GM-W2(P), GM-W2(A), GM-W2($R^2$), GN-W2(P), GN-W2(A), GN-W2($R^2$), Subgroup, and IRAS PSC.
$^1$The CAGB-IRAS identifier (see Tables~\ref{tab:tab1} and \ref{tab:tab10}). See Section~\ref{sec:catalog}.
\end{flushleft}
\end{table*}
\end{longrotatetable}

\begin{longrotatetable}
\begin{table*}
\centering
\scriptsize
\caption{OAGB-WISE objects with obtained variation parameters from WISE light curves (16 columns; 2468 rows)\label{tab:tab15}}
\begin{tabular}{llllllllllllllll}
\hline \hline
OW-N$^1$	&	Subgroup	&	A-Type	&	A-Period	&	G-W1(P)	&	G-W1(A)	&	G-W1($R^2$)	&	GM-W1(P)	&	GM-W1(A)	&	GM-W1($R^2$)	&	G-W2(P)	&	G-W2(A)	&	G-W2($R^2$)	&	GM-W2(P)	&	GM-W2(A)	&	GM-W2($R^2$)	\\
\hline
21	&	OW-ME	&		&		&	597.79 	&	0.605 	&	0.707 	&		&		&		&		&		&		&		&		&		\\
30	&	OW-ME	&		&		&		&		&		&		&		&		&	455.23 	&	0.672 	&	0.683 	&		&		&		\\
43	&	OW-ME	&	M:	&		&		&		&		&		&		&		&	325.59 	&	0.500 	&	0.741 	&		&		&		\\
48	&	OW-ME	&		&		&		&		&		&		&		&		&	541.79 	&	1.082 	&	0.740 	&		&		&		\\
82	&	OW-ME	&		&		&	313.27 	&	0.490 	&	0.610 	&		&		&		&		&		&		&		&		&		\\
86	&	OW-ME	&		&		&	485.32 	&	0.521 	&	0.639 	&		&		&		&	492.84 	&	0.632 	&	0.739 	&		&		&		\\
95	&	OW-ME	&		&		&	467.50 	&	0.562 	&	0.817 	&		&		&		&	467.50 	&	0.591 	&	0.840 	&		&		&		\\
142	&	OW-ME	&		&		&	535.74 	&	0.554 	&	0.708 	&		&		&		&	526.87 	&	0.567 	&	0.623 	&		&		&		\\
149	&	OW-ME	&		&		&	139.94 	&	0.656 	&	0.773 	&		&		&		&	139.83 	&	0.645 	&	0.806 	&		&		&		\\
154	&	OW-ME	&		&		&	444.66 	&	0.614 	&	0.772 	&		&		&		&	443.63 	&	0.628 	&	0.805 	&		&		&		\\
177	&	OW-ST	&		&		&	551.13 	&	0.628 	&	0.734 	&		&		&		&	555.94 	&	0.690 	&	0.777 	&		&		&		\\
190	&	OW-ST	&		&		&	664.94 	&	0.698 	&	0.678 	&		&		&		&	143.78 	&	0.760 	&	0.632 	&		&		&		\\
202	&	OW-ST	&		&		&	601.97 	&	0.739 	&	0.730 	&		&		&		&	603.87 	&	0.688 	&	0.670 	&		&		&		\\
203	&	OW-ST	&		&		&	655.68 	&	0.675 	&	0.644 	&		&		&		&		&		&		&		&		&		\\
210	&	OW-ST	&		&		&	908.00 	&	0.917 	&	0.829 	&		&		&		&	908.00 	&	1.088 	&	0.745 	&		&		&		\\
251	&	OW-ST	&		&		&	506.20 	&	0.588 	&	0.721 	&		&		&		&	504.86 	&	0.550 	&	0.724 	&		&		&		\\
255	&	OW-ST	&		&		&	697.99 	&	0.981 	&	0.732 	&		&		&		&	245.42 	&	1.161 	&	0.792 	&		&		&		\\
257	&	OW-ST	&		&		&	522.81 	&	0.769 	&	0.719 	&		&		&		&	528.60 	&	0.919 	&	0.727 	&		&		&		\\
265	&	OW-ST	&		&		&	242.93 	&	1.085 	&	0.706 	&		&		&		&		&		&		&		&		&		\\
272	&	OW-ST	&		&		&	593.72 	&	0.606 	&	0.873 	&		&		&		&	590.05 	&	0.780 	&	0.869 	&		&		&		\\
282	&	OW-ST	&		&		&	676.37 	&	1.058 	&	0.826 	&		&		&		&	248.58 	&	1.001 	&	0.799 	&		&		&		\\
283	&	OW-ST	&		&		&	706.43 	&	0.851 	&	0.638 	&		&		&		&		&		&		&		&		&		\\
289	&	OW-ST	&		&		&		&		&		&		&		&		&	609.51 	&	0.815 	&	0.606 	&		&		&		\\
294	&	OW-ST	&		&		&	141.24 	&	0.685 	&	0.619 	&		&		&		&		&		&		&		&		&		\\
296	&	OW-ST	&		&		&	644.39 	&	0.736 	&	0.627 	&		&		&		&		&		&		&		&		&		\\
298	&	OW-ST	&		&		&	141.65 	&	0.780 	&	0.706 	&		&		&		&	646.58 	&	0.776 	&	0.685 	&		&		&		\\
299	&	OW-ST	&		&		&	637.94 	&	0.623 	&	0.697 	&		&		&		&	253.85 	&	0.679 	&	0.757 	&		&		&		\\
305	&	OW-ST	&		&		&	468.83 	&	0.800 	&	0.872 	&		&		&		&	468.83 	&	0.880 	&	0.891 	&		&		&		\\
325	&	OW-ST	&		&		&	136.49 	&	0.665 	&	0.717 	&		&		&		&	136.68 	&	0.791 	&	0.778 	&		&		&		\\
339	&	OW-ST	&		&		&	104.57 	&	1.685 	&	0.634 	&		&		&		&		&		&		&		&		&		\\
-	&	-	&	-	&	-	&	-	&	-	&	-	&	-	&	-	&	-	&	-	&	-	&	-	&	-	&	-	&	-	\\
5287	&	OW-OG	&	M	&	336.6	&	341.92 	&	0.513 	&	0.855 	&	341.92 	&	0.513 	&	0.855 	&	342.53 	&	0.362 	&	0.813 	&	342.53 	&	0.362 	&	0.813 	\\
5288	&	OW-OG	&	M	&	283.5	&	278.15 	&	0.498 	&	0.621 	&	278.15 	&	0.498 	&	0.621 	&	278.97 	&	0.394 	&	0.790 	&	278.97 	&	0.394 	&	0.790 	\\
5291	&	OW-OG	&	M	&	245.3	&	236.17 	&	0.192 	&	0.637 	&	236.17 	&	0.192 	&	0.637 	&		&		&		&		&		&		\\
5292	&	OW-OG	&	M	&	305.1	&	305.29 	&	0.342 	&	0.828 	&	305.29 	&	0.342 	&	0.828 	&	304.80 	&	0.255 	&	0.832 	&	304.80 	&	0.255 	&	0.832 	\\
5295	&	OW-OG	&	M	&	339.4	&	339.49 	&	0.464 	&	0.849 	&	339.49 	&	0.464 	&	0.849 	&	339.49 	&	0.350 	&	0.913 	&	339.49 	&	0.350 	&	0.913 	\\
5296	&	OW-OG	&	M	&	229.2	&	870.31 	&	0.406 	&	0.825 	&	228.82 	&	0.385 	&	0.819 	&	866.36 	&	0.284 	&	0.784 	&	228.82 	&	0.274 	&	0.789 	\\
5297	&	OW-OG	&	M	&	197.57	&	200.46 	&	0.237 	&	0.770 	&	200.46 	&	0.237 	&	0.770 	&		&		&		&		&		&		\\
5298	&	OW-OG	&	M	&	263.2	&	271.05 	&	0.506 	&	0.719 	&	271.05 	&	0.506 	&	0.719 	&	271.43 	&	0.468 	&	0.722 	&	271.43 	&	0.468 	&	0.722 	\\
5299	&	OW-OG	&	M	&	298.8	&		&		&		&		&		&		&	479.25 	&	0.373 	&	0.808 	&	292.20 	&	0.363 	&	0.781 	\\
5300	&	OW-OG	&	M	&	199.55	&		&		&		&		&		&		&	205.65 	&	0.277 	&	0.693 	&	205.65 	&	0.277 	&	0.693 	\\
\hline
\end{tabular}
\begin{flushleft}
$^1$The OAGB-WISE identifier (see Tables~\ref{tab:tab2} and \ref{tab:tab11}). See Section~\ref{sec:catalog}.
\end{flushleft}
\end{table*}
\end{longrotatetable}

\begin{longrotatetable}
\begin{table*}
\centering
\scriptsize
\caption{CAGB-WISE objects with obtained variation parameters from WISE light curves (16 columns; 216 rows)\label{tab:tab16}}
\begin{tabular}{llllllllllllllll}
\hline \hline
CW-N$^1$	&	Subgroup	&	A-Type	&	A-Period	&	G-W1(P)	&	G-W1(A)	&	G-W1($R^2$)	&	GM-W1(P)	&	GM-W1(A)	&	GM-W1($R^2$)	&	G-W2(P)	&	G-W2(A)	&	G-W2($R^2$)	&	GM-W2(P)	&	GM-W2(A)	&	GM-W2($R^2$)	\\
\hline
7	&	CW-GC	&	L	&		&	62.76 	&	0.158 	&	0.680 	&		&		&		&		&		&		&		&		&		\\
9	&	CW-GC	&	M:	&	259	&	245.74 	&	0.611 	&	0.933 	&		&		&		&	245.74 	&	0.345 	&	0.941 	&		&		&		\\
65	&	CW-GC	&		&		&		&		&		&		&		&		&	151.26 	&	0.117 	&	0.657 	&		&		&		\\
69	&	CW-GC	&	SR	&	335	&	333.84 	&	0.465 	&	0.828 	&		&		&		&	117.96 	&	0.269 	&	0.832 	&		&		&		\\
94	&	CW-GC	&	M	&	365	&	121.72 	&	0.588 	&	0.878 	&	365.25 	&	0.890 	&	0.873 	&	122.03 	&	0.436 	&	0.932 	&	367.85 	&	0.393 	&	0.924 	\\
176	&	CW-GC	&		&		&	502.38 	&	0.102 	&	0.673 	&		&		&		&		&		&		&		&		&		\\
197	&	CW-GC	&	M	&	250	&		&		&		&		&		&		&	259.83 	&	0.335 	&	0.618 	&	259.83 	&	0.335 	&	0.618 	\\
252	&	CW-GC	&	L:	&	234	&	293.99 	&	0.172 	&	0.785 	&		&		&		&		&		&		&		&		&		\\
261	&	CW-GC	&		&		&	149.61 	&	0.068 	&	0.710 	&		&		&		&		&		&		&		&		&		\\
326	&	CW-GC	&		&		&		&		&		&		&		&		&	176.59 	&	0.354 	&	0.631 	&		&		&		\\
371	&	CW-GC	&	SR	&	356.4753836	&		&		&		&		&		&		&	177.57 	&	0.371 	&	0.705 	&		&		&		\\
392	&	CW-GC	&	SR	&	276	&		&		&		&		&		&		&	283.23 	&	0.175 	&	0.712 	&		&		&		\\
398	&	CW-GC	&	SR:	&		&		&		&		&		&		&		&	427.38 	&	0.119 	&	0.672 	&		&		&		\\
403	&	CW-GC	&		&		&	178.72 	&	0.460 	&	0.615 	&		&		&		&		&		&		&		&		&		\\
459	&	CW-GC	&	CST	&		&	178.72 	&	0.234 	&	0.678 	&		&		&		&		&		&		&		&		&		\\
483	&	CW-GC	&	L:	&		&		&		&		&		&		&		&	1173.07 	&	0.103 	&	0.623 	&		&		&		\\
520	&	CW-GC	&	M	&	374	&	372.89 	&	0.658 	&	0.778 	&	372.89 	&	0.658 	&	0.778 	&	380.26 	&	0.406 	&	0.681 	&	380.26 	&	0.406 	&	0.681 	\\
523	&	CW-GC	&	SR	&	329	&	401.80 	&	0.359 	&	0.696 	&		&		&		&	334.13 	&	0.205 	&	0.856 	&		&		&		\\
552	&	CW-GC	&	VAR	&		&		&		&		&		&		&		&	191.38 	&	0.093 	&	0.613 	&		&		&		\\
586	&	CW-GC	&	I:	&		&	186.36 	&	0.103 	&	0.760 	&		&		&		&	186.91 	&	0.163 	&	0.698 	&		&		&		\\
598	&	CW-GC	&	SR	&	271.08671	&	64.47 	&	0.282 	&	0.613 	&		&		&		&		&		&		&		&		&		\\
604	&	CW-GC	&	SR	&	421	&		&		&		&		&		&		&	71.49 	&	0.164 	&	0.750 	&		&		&		\\
608	&	CW-GC	&	VAR	&		&		&		&		&		&		&		&	1905.48 	&	0.102 	&	0.757 	&		&		&		\\
633	&	CW-GC	&	VAR	&		&	300.98 	&	0.486 	&	0.673 	&		&		&		&	302.40 	&	0.326 	&	0.806 	&		&		&		\\
648	&	CW-GC	&	L	&		&	829.16 	&	0.130 	&	0.630 	&		&		&		&	818.57 	&	0.102 	&	0.642 	&		&		&		\\
680	&	CW-GC	&	SR	&	292	&	95.34 	&	0.275 	&	0.640 	&		&		&		&	112.81 	&	0.246 	&	0.696 	&		&		&		\\
696	&	CW-GC	&		&		&		&		&		&		&		&		&	1029.98 	&	0.095 	&	0.615 	&		&		&		\\
762	&	CW-GC	&		&		&	566.58 	&	0.133 	&	0.672 	&		&		&		&		&		&		&		&		&		\\
786	&	CW-GC	&		&		&		&		&		&		&		&		&	176.11 	&	0.276 	&	0.741 	&		&		&		\\
802	&	CW-GC	&	VAR	&		&		&		&		&		&		&		&	178.07 	&	0.174 	&	0.608 	&		&		&		\\
-	&	-	&	-	&	-	&	-	&	-	&	-	&	-	&	-	&	-	&	-	&	-	&	-	&	-	&	-	&	-	\\
3557	&	CW-OG	&	M	&	242.5	&		&		&		&		&		&		&	147.04 	&	0.324 	&	0.718 	&	240.94 	&	0.268 	&	0.651 	\\
3558	&	CW-OG	&	M	&	308	&	113.41 	&	0.313 	&	0.848 	&	301.91 	&	0.319 	&	0.844 	&		&		&		&		&		&		\\
3560	&	CW-OG	&	M	&	229.3	&	101.42 	&	0.214 	&	0.662 	&	229.92 	&	0.204 	&	0.611 	&	101.42 	&	0.160 	&	0.664 	&	229.64 	&	0.152 	&	0.617 	\\
3564	&	CW-OG	&	M	&	145.71	&	146.70 	&	0.113 	&	0.664 	&	146.70 	&	0.113 	&	0.664 	&		&		&		&		&		&		\\
3568	&	CW-OG	&	M	&	111.86	&	113.21 	&	0.252 	&	0.772 	&	113.21 	&	0.252 	&	0.772 	&	113.35 	&	0.202 	&	0.784 	&	113.35 	&	0.202 	&	0.784 	\\
3569	&	CW-OG	&	RCB:	&		&	2500.00 	&	0.222 	&	0.775 	&		&		&		&		&		&		&		&		&		\\
3570	&	CW-OG	&	M	&	231.9	&	229.64 	&	0.360 	&	0.891 	&	229.64 	&	0.360 	&	0.891 	&	229.09 	&	0.287 	&	0.856 	&	229.09 	&	0.287 	&	0.856 	\\
3571	&	CW-OG	&	M	&	359.7	&	73.28 	&	0.429 	&	0.906 	&	363.40 	&	1.396 	&	0.902 	&	362.71 	&	0.975 	&	0.929 	&	362.71 	&	0.975 	&	0.929 	\\
3573	&	CW-OG	&	M	&	501.2	&	300.46 	&	0.510 	&	0.908 	&	460.72 	&	0.498 	&	0.872 	&	300.94 	&	0.485 	&	0.915 	&	459.61 	&	0.480 	&	0.897 	\\
3576	&	CW-OG	&	M	&	263.1	&	257.16 	&	0.315 	&	0.849 	&	257.16 	&	0.315 	&	0.849 	&	611.18 	&	0.198 	&	0.788 	&	258.21 	&	0.204 	&	0.777 	\\
\hline
\end{tabular}
\begin{flushleft}
$^1$The CAGB-WISE identifier (see Tables~\ref{tab:tab2} and \ref{tab:tab12}). See Section~\ref{sec:catalog}.
\end{flushleft}
\end{table*}
\end{longrotatetable}

\end{document}